# Covalent Welding of Two-Dimensional Molecular Sheets into Ultra-Stiff, Super-Tough Polymeric Networks

Xiaoli Gong[1†], Guohu Luo[2†], Jiaxin Wu[1†], Jiayu Chen[2], Zhengqiao Yin[1], Zhipeng Yan[1], Jingran Chen[1], Bowen Tang[1], Le Tang[3], Yuxiao Wu[1], Guanling Wu[1], Li Peng[1], Qihang Wang[5], Yinglong Hu[6], Yue Zhu[6], Yuan Yu[6], Keyan Liu[1], Xiaozhou Liu[7], Huanhuan Wu[7], Yakai Lin[7], Yifei Xu[1], Zhiliang Pan[6], Hao Ma[6], Man Li[6], Hongjun Zhang[5], Dacheng Wei[1], Youpeng Zhang[4], Yao Fu[3], Yongxiang Hu[2*], Yuwen Zeng[1*]

[1]State Key Laboratory of Molecular Engineering of Polymers, Department of Macromolecular Science, Fudan University, Shanghai, 200438, China

[2]State Key Laboratory of Micro and Nano Engineering Science, School of Mechanical Engineering, Shanghai Jiao Tong University, Shanghai 200240, China

[3]Department of Chemistry, Fudan University, Shanghai, 200438, China

[4]Institute of Modern Physics, Fudan University, Shanghai 200433, China

[5]State Key Laboratory of Particle Detection and Electronics, University of Science and Technology of China, Hefei 230026, China

[6]Institute of Micro-Nano Thermal Energy Transport and Conversion, School of Engineering Science, University of Science and Technology of China, Hefei, Anhui 230027, China

[7]State Key Laboratory of Chemical Engineering and Low-Carbon Technology, Department of Chemical Engineering, Tsinghua University, Beijing 100084, China

[†]These authors contributed equally to this work.

## Abstract

**Since the inception of macromolecular science, polymers have been constructed from zero-dimensional monomers, a paradigm that funnels all stress through discrete one-dimensional chains[1]. Driven by conformational entropy, polymer chains inevitably coil in three-dimensional space[2], precluding effective stress transfer, thereby compromising the ultimate mechanical performance[3]. Two-dimensional (2D) molecular sheets, with intrinsic in-plane rigidity[4], could in principle circumvent this constraint. Yet in existing assemblies, where stress transfers efficiently within each molecular sheet but poorly between them[5], their exceptional properties fail to translate to the macroscale. Here, we demonstrate a 2D highly oriented molecular network (2D HOMN), an ultra-stiff, super-tough polymeric material constructed by the solid-state covalent welding of assembled 2D polyaramid sheets with flexible linkers pre-grafted at their peripheries. Enabled by efficient intersheet load sharing and energy dissipation, the welded films rival the modulus and strength of Kevlar 29 fibers, reaching 69.8 and 3.9 GPa, with impact toughness up to an order of magnitude beyond Kevlar composites. This designed intersheet covalent connectivity further renders the 2D HOMN impermeable to gas molecules and ions, resilient to radiation and efficient in thermal transport. This work establishes 2D molecular motifs as heterogeneous monomeric building blocks and provides a general strategy for constructing multifunctional polymeric networks.**

## Introduction

The mechanical performance of polymeric materials is dictated by the topology of their polymer networks[6,7]. Built from zero-dimensional monomers, polymer chains channel all stress along discrete one-dimensional (1D) pathways[2], and load transfer is thus intrinsically serial. Driven by conformational entropy[8], 1D chains adopt three-dimensional coils, and their strands must straighten before bearing load. Even taut, each strand transmits stress without attenuation, concentrating the full load on the most vulnerable sites, with no outlet but localized bond rupture. Conventional strategies, including crystallization[9], chain alignment[9], and covalent cross-linking[10], merely redistribute stress among amorphous and ordered domains without altering this series topology. Extreme stiffness and toughness are therefore structurally incompatible[11].

Two-dimensional (2D) molecular sheets offer a topological solution to this serial constraint[4,12-14]. Their strong, rigid covalent grids bear load without pre-straightening, while their intramolecular 2D connectivity delocalizes concentrated loads along redundant in-plane pathways, converting series topologies into parallel ones[15,16]. Assemblies of inorganic 2D materials such as graphene have indeed demonstrated outstanding mechanical performance, but load transfer between adjacent sheets relies predominantly on weak van der Waals forces, leaving the exceptional microscopic properties largely unrealized at the macroscale. Hydrogen-bonded assemblies represent an alternative strategy for superior intersheet connection, and the irreversible synthesis of 2D polyaramid (2DPA)[4] has made them scalable (Supplementary Figs. 1 and 2), avoiding the chemical instability[17] that has plagued reversible 2D covalent frameworks[18]. Such hydrogen-bonded assemblies exhibit denser packing[19] and more efficient out-of-plane stress transfer than van der Waals assemblies, and accordingly improve the realization of the microscopic properties at the macroscale. However, the rigid 2D backbones and cooperative interlayer hydrogen bonding leave these assemblies deeply glassy ($T_g > T_d \approx 310$ °C, Supplementary Figs. 23 and 24). The hydrogen bonds are thus kinetically frozen, with their dynamic nature lost and energy dissipation suppressed[20]. Consequently, the nanofilms are intrinsically brittle, and their practical applications remain limited.

In these hydrogen-bonded assemblies, stress transfer requires load to traverse the rigid covalent grid, cross the hydrogen-bonded interface, and enter the adjacent sheet, with interlayer hydrogen bonds serving as the sole and fragile mechanical link (Extended Data Fig. 1a), limiting further modulus enhancement. Beyond this, the architecture also lacks soft domains to buffer deformation[21], and the frozen hydrogen bonds cannot shuffle to dissipate energy[22]. To this end, we introduce covalent welding, in which 2D molecular sheets are first grafted with flexible covalent linkers at their peripheries, then assembled into hydrogen-bonded films, and connected by covalent bonds in the solid state, creating 2D highly oriented molecular networks. These linkers establish direct load-transfer pathways between adjacent sheets, enabling additional stress transfer alongside the fragile hydrogen-bonded interfaces (Extended Data Fig. 1b). In the welded network, the covalent linkers directly share the load, raising the stiffness and strength over the unwelded assembly by 3.3- and 4.3-fold. Furthermore, the flexible linkers act as soft domains, deforming under load to trigger hydrogen bond reorganization prior to covalent scission, with both processes dissipating substantial energy. Under extreme dynamic loading, the welded films sustain frictional sliding at the hydrogen-bonded interfaces, further improving the impact toughness, exceeding 5 $MJ \cdot kg^{-1}$ at 1000 $m \cdot s^{-1}$. The intersheet connectivity underlying these mechanical improvements also enables the welded films to block gas molecules and ions, withstand irradiation and conduct heat efficiently.

## Results and discussion

### Molecular design and material synthesis

Single-step covalent welding of two molecular sheets using a linker is virtually impossible in the solid state, because all components are kinetically frozen, precluding the requisite three-molecule reaction. Meanwhile, free linkers would also disrupt hydrogen-bonded assembly, compromising the formation of oriented nanofilms. We therefore developed a two-step covalent welding approach (Fig. 1a, b) in which pendant chains are first grafted onto individual 2DPA sheets in solution, and with the derivatized sheets assembled into hydrogen-bonded films, intermolecular covalent welds are subsequently established via thermally induced coupling. Here the pendant chain was designed as a three-segment construct comprising a nucleophilic terminus for low-temperature condensation with the electrophilic edges of 2DPA, an electrophilic terminus for high-temperature condensation with residual amine groups at the periphery of adjacent 2DPA sheets, and a flexible covalent spacer bridging these two reactive ends.

We found that reacting 2DPA with amino esters of varying chain lengths, either by direct quenching or solution-phase condensation, efficiently installed the desired pendant chains (Fig. 1a), as confirmed by nuclear magnetic resonance (NMR) and Fourier transform infrared (FTIR) spectroscopy (Supplementary Figs. 5–10 and 21, respectively). High-resolution atomic force microscopy (HR-AFM) revealed that the lateral dimensions and thickness of individual sheets remained unchanged after peripheral modification (Supplementary Fig. 29), and gel permeation chromatography (GPC) showed no detectable shift in the hydrodynamic volume (Supplementary Fig. 15), confirming that the 2D character of the sheets was preserved. Importantly, edge functionalization did not interfere with the hydrogen-bonded assembly of molecular sheets, and films prepared by spin-coating or slot-die coating from dimethylformamide (DMF) solution remained highly oriented and layered. Heating the as-prepared films to 280 °C induced aminolysis of the ester moieties, establishing intermolecular covalent connections both within and across the layers and converting the hydrogen-bonded assembly into a two-dimensional highly oriented molecular network (2D HOMN, Fig. 1b). The welded films retained this morphology. Polarized fluorescence confirmed that the molecular sheets in the welded film lie flat with no preferred in-plane orientation (Fig. 1c), as evidenced by the $C_2$ symmetry of the out-of-plane polar plot. The modulation contrast was essentially preserved upon welding (0.967 versus 0.981 for the unwelded nanofilms, Extended Data Fig. 2). Transmission electron microscopy (TEM) of a torn film edge further revealed a terrace-like layered morphology (Supplementary Fig. 32). These welded nanofilms became insoluble in DMF, whereas the unwelded hydrogen-bonded films dissolved readily, consistent with the formation of a covalently welded architecture (Supplementary Fig. 16).

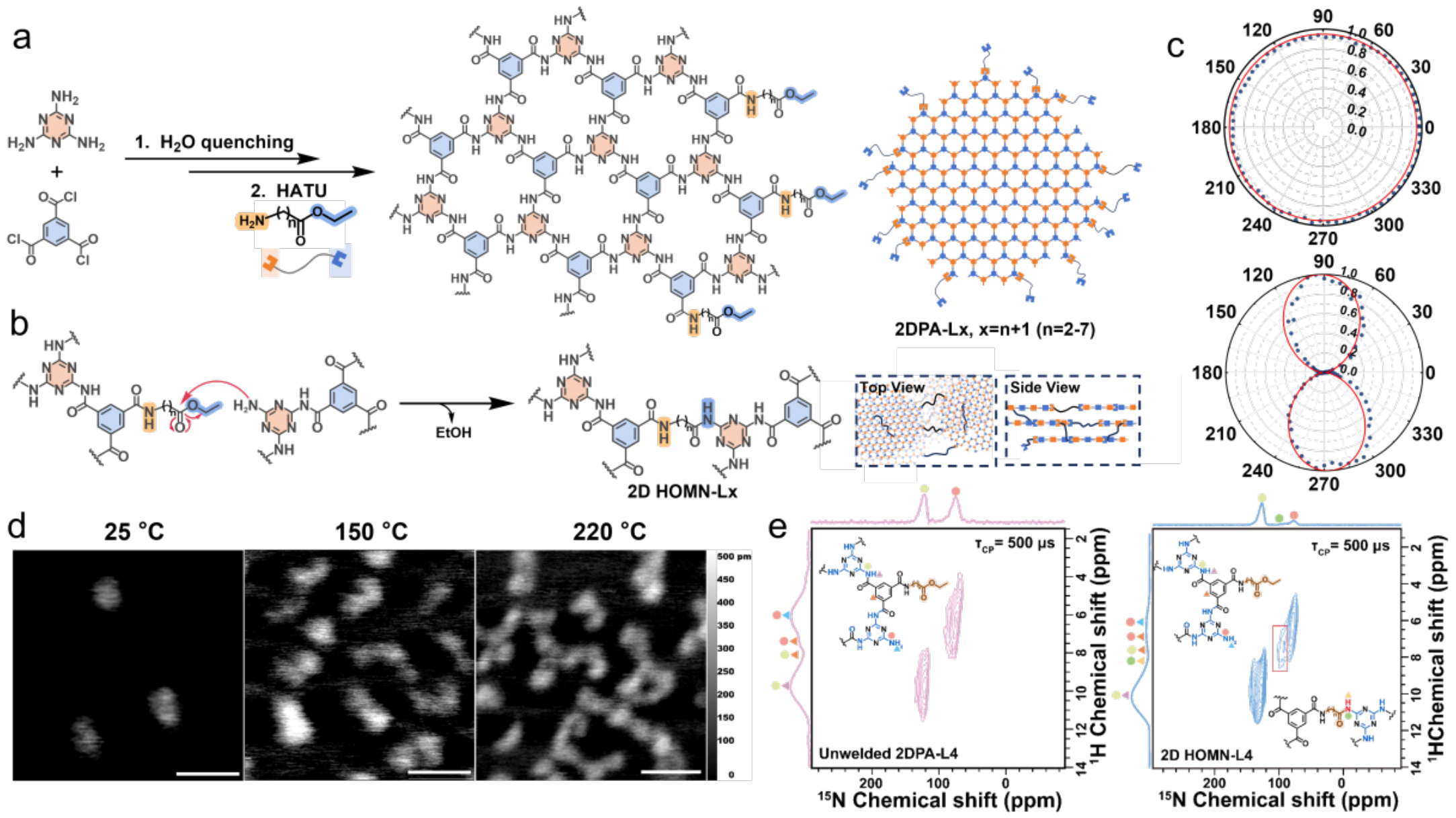

**Fig. 1|Molecular design, synthesis and covalent welding of 2D HOMN nanofilms. a,** Two-step synthesis of amino-ester-functionalized welding precursors (2DPA-Lx). **b,** Thermally induced intersheet covalent welding into 2D HOMN-Lx networks (top and side views). **c,** Polar plots of polarized fluorescence intensity of a welded 2D HOMN film, with laser incidence normal to (top) and parallel to (bottom) the film plane. Data for the unwelded films are provided in Extended Data Fig. 2. **d,** AFM images of discrete sheets (25 °C), sparsely welded aggregates (150 °C) and fused clusters (220 °C). **e,** $^{1}H$–$^{15}N$ solid-state NMR correlation spectra of unwelded 2DPA-L4 films (left) and 2D HOMN-L4 films (right). Colored circles mark distinct $^{15}N$ atoms, whereas colored triangles mark distinct $^{1}H$ atoms.

## Mechanistic validation of covalent welding

Differential scanning calorimetry (DSC) of the unwelded films showed an exothermic peak from 180 to 250 °C peaking at 220 °C (Supplementary Fig. 24), and thermogravimetric-mass spectrometric (TGA-MS) analysis detected the ethanol byproduct of aminolysis upon heating (Supplementary Fig. 26), together confirming that intermolecular covalent welding proceeds in the solid state. More decisively, heating and redispersing the welding precursor for HR-AFM (Fig. 1d) revealed a transition from discrete sheets at room temperature, to sparsely welded aggregates at 150 °C (well below full welding), to fused clusters at 220 °C (the DSC exothermic peak). To elucidate the reaction details of covalent welding, we prepared films using melamine selectively $^{15}N$-enriched (99%) at the amino ($NH_2$) groups and investigated them by solid-state NMR spectroscopy. After welding, the $^{15}N$ resonance of the free $NH_2$ groups at the 2DPA peripheries diminished, alongside the emergence of a new amide resonance (Extended Data Fig. 3). These assignments were further supported by 2D $^{1}H$–$^{15}N$ correlation spectra (Fig. 1e) and density functional theory (DFT) calculations (Supplementary Fig. 20). Consistent with these spectral changes, X-ray photoelectron spectroscopy (XPS) and FTIR of the films before and after welding indicated a marked decrease in ester content and a concomitant increase in amide species (Supplementary Figs. 21, 22).

## Mechanical performance and impact toughness

To systematically evaluate how linker length affects mechanical performance, we screened welded films across a series of amino esters (L0 to L8) using strain-induced elastic buckling instability for mechanical measurements (SIEBIMM)[23], a technique that enables rapid modulus screening of nanoscale films, with each welded film co-measured against the unwelded 2DPA-IPA reference[4] on the same elastomer substrate (Fig. 2a, Extended Data Fig. 4, and Supplementary Figs. 34-41). All welded samples exhibited higher moduli than the unwelded 2DPA-IPA assembly. Even 2D HOMN-L0, obtained by thermally welding 2DPA-IPA directly without any flexible segment, afforded a 4.63-fold increase in modulus, reflecting the dominant contribution of rigid direct covalent connections to the film modulus. With no flexible chain to bridge adjacent sheets, however, most reactive groups in this analogue remain geometrically inaccessible, which would limit the attainable welding density. Moreover, with no pliable segment to buffer deformation, the film remains brittle, failing to self-support at centimeter scale (Supplementary Fig. 47). For the flexible linkers, the effective modulus is therefore governed by the interplay between welding density and chain compliance. Relative to L0, the short L3 linker improves the accessibility of reactive groups, but the resulting connections are compliant, and this loss in stiffness outweighs the gain in welding density, leaving a smaller net increase (3.00-fold). In contrast, longer linkers (L5 to L8) introduce increasingly compliant spacers that deform preferentially under load, progressively diminishing the modulus enhancement (from 4.17-fold for L5 to 2.57-fold for L8). The medium-length linker β-alanine ethyl ester (β-AEE, L4) represents an optimal balance between these two factors, affording the largest modulus enhancement of the series (6.93-fold).

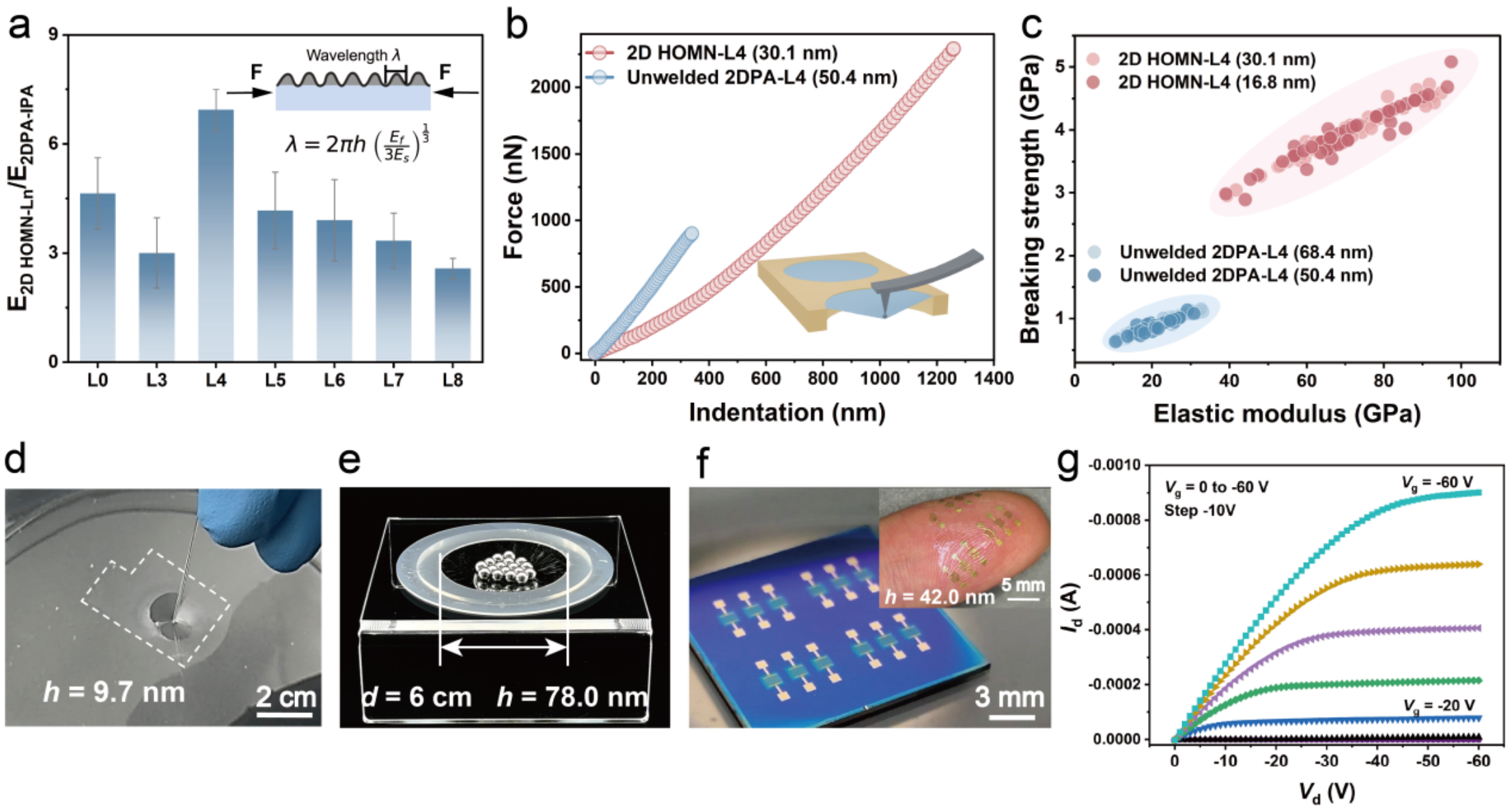

**Fig. 2 | Static mechanical properties and macroscopic robustness of 2D HOMN nanofilms. a,** SIEBIMM modulus screening of films with varying spacer lengths (L0 and L3–L8). Inset, wrinkle-wavelength analysis. **b,** Representative nanoindentation force-displacement curves for 2D HOMN-L4 and unwelded 2DPA-L4 films. Inset, schematic of the nanoindentation measurement. **c,** Breaking strength versus elastic modulus for unwelded 2DPA-L4 and 2D HOMN-L4 films. **d,** A 9.7-nm-thick free-standing 2D HOMN-L8 film on water resisting puncture by a glass capillary. Scale bar, 2 cm. **e,** A 78-nm-thick 2D HOMN-L8 film suspended across a 6-cm-diameter frame, supporting 17 metal spheres (0.3 g each) without rupture. **f,** Optical photograph of organic transistor arrays fabricated on a 42-nm-thick 2D

HOMN-L8 film on a silicon wafer. Inset, the film conformably transferred onto a finger. **g,** Output characteristics ($I_d$–$V_d$) of a representative transistor on the film.

We therefore selected 2D HOMN-L4 for detailed characterization. Force–displacement curves from nanoindentation of suspended films (n = 184, thickness 16.8–68.4 nm) exhibited substantially higher fracture loads and indentation depths upon welding (Fig. 2b). The elastic modulus increased from 21.4 ± 5.4 GPa to 69.8 ± 13.0 GPa, and the breaking strength increased from 0.9 ± 0.1 GPa to 3.9 ± 0.4 GPa (Fig. 2c), representing 3.26- and 4.33-fold improvements over the unwelded 2DPA-L4 assembly, respectively. These improvements reflect the establishment of covalent stress-transfer channels as the primary load-bearing framework between adjacent sheets, alongside the fragile hydrogen-bonded interfaces. These channels simultaneously elevate the effective modulus and delay fracture, as covalent linkers are stiffer than hydrogen bonds and require substantially higher stress to break. The welded films rival the modulus and strength of high-performance fibers such as E-glass (~76 GPa modulus; ~3.45 GPa strength)[24] and Kevlar 29 (~70.5 GPa; ~3.6 GPa)[25], yet as two-dimensional architectures they combine high stiffness and strength with solution processability, in-plane isotropy, and area scalability[4].

Under quasi-static conditions, the covalently welded films exhibit mechanical robustness at the macroscale. Even 9.7-nm-thick films of 2D HOMN-L8 (the most compliant variant of the series) are self-supporting on water and withstand puncture by a glass capillary (0.3 mm in diameter; Fig. 2d), whereas the unwelded films cannot be self-supported under the same conditions. Such macroscopic load-bearing capacity is rarely achieved with conventional 2D material nanofilms, including van der Waals[26] and hydrogen-bonded[27] assemblies. As a further demonstration, a 78-nm-thick, 6-cm-diameter 2D HOMN-L8 film (aspect ratio of 7.7 × $10^5$) holds 17 metal spheres (0.3 g each) without failure (Fig. 2e and Supplementary Video 1), a capacity also demonstrated for 2D HOMN-L4 films (Supplementary Video 2). Crucially, this load-bearing capacity persists even after the spheres locally pierce the film, with the damage remaining localized (Supplementary Video 3), underscoring the tear resistance of the welded network. This robustness also enables the fabrication of organic field-effect transistor arrays on a 42-nm-thick 2D HOMN-L8 film (Fig. 2f), which exhibit well-behaved gate-modulated output characteristics (Fig. 2g). Once released intact from the wafer, the film can be either suspended across a frame (Extended Data Fig. 5) or conformably mounted on human skin (Fig. 2f, inset).

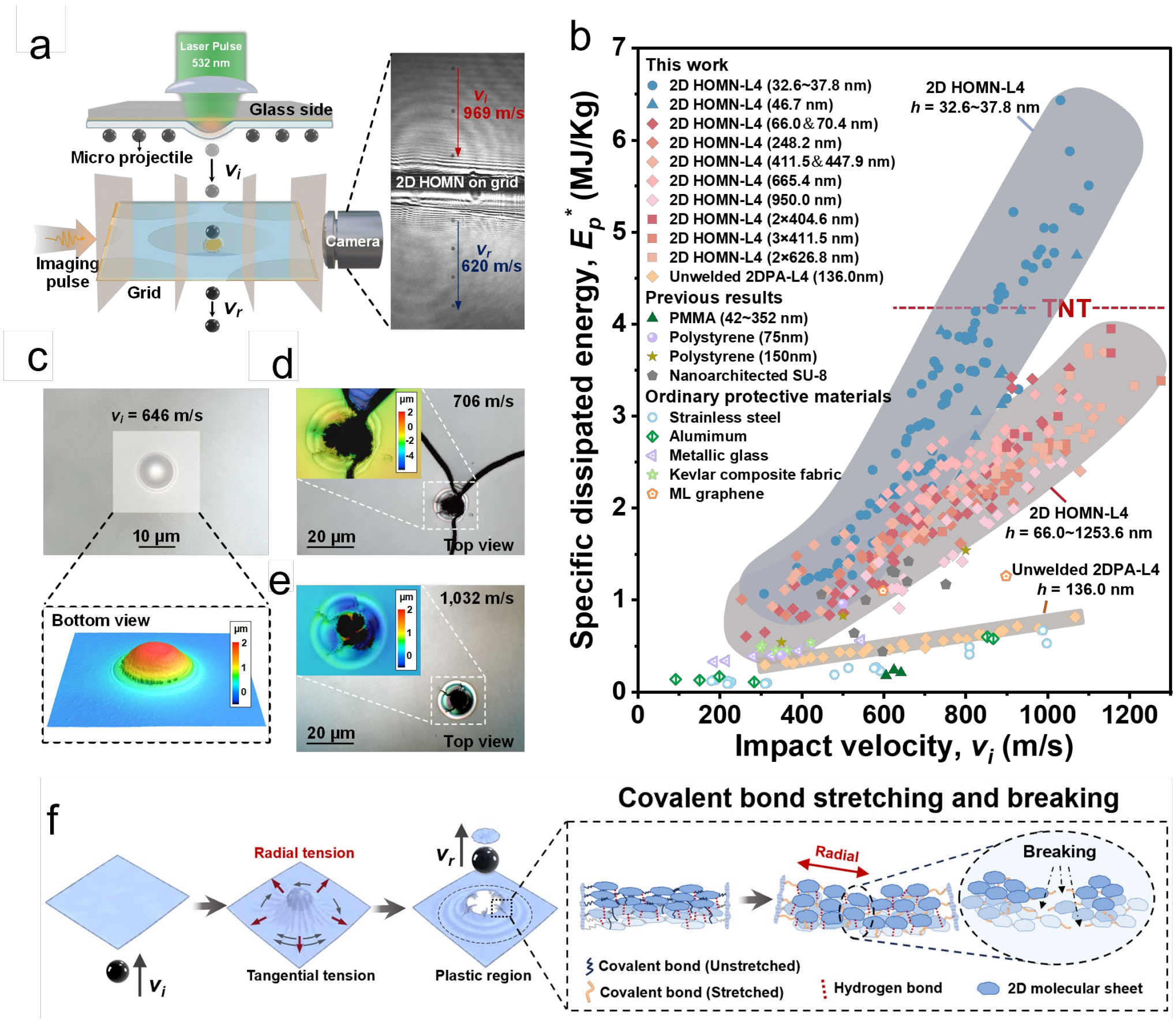

**Fig. 3|Dynamic impact response of 2D HOMN-L4 nanofilms. a,** Schematic of the laser-induced projectile impact testing (LIPIT) setup. **b,** Specific dissipated energy of 2D HOMN-L4 films (32.6–1,253.6 nm) and unwelded 2DPA-L4 films (136.0 nm) compared with conventional protective materials and engineering polymers. See Supplementary Information for data and references. **c,** Optical micrograph and bottom-view 3D height profile of a 665.4-nm-thick film after impact at 646 $m \cdot s^{-1}$, showing the completely captured projectile. **d-e,** Confocal micrographs of impact sites in a 665.4-nm-thick film at 706 and 1,032 $m \cdot s^{-1}$, showing a velocity-dependent transition in the failure mode. **f,** Schematic of the impact-deformation and energy-dissipation mechanism of 2D HOMN-L4.

To evaluate the energy dissipation capacity under extreme dynamic loading (strain rates of ~$10^7$–$10^8$ $s^{-1}$), we performed laser-induced projectile impact testing (LIPIT)[28] on 2D HOMN-L4 nanofilms (332 data points from 14 films). Laser-launched $SiO_2$ microprojectiles struck suspended films in vacuum at incident velocities of 300–1,200 $m \cdot s^{-1}$, and time-resolved trajectories provided the incident and residual velocities used to calculate the mass-normalized specific dissipated energy, $E_p$* (Fig. 3a). For both unwelded and welded films, $E_p$* increased with impact velocity, reminiscent of strain-rate-dependent energy dissipation in 1D polymers[29]. While the unwelded 2DPA-L4 film (136.0 nm) dissipated 0.53 $MJ \cdot kg^{-1}$ at 600–800 $m \cdot s^{-1}$, the welded films (66.0-665.4 nm) showed a roughly four-fold enhancement over this velocity range and reached ~3 $MJ \cdot kg^{-1}$ near 1,000 $m \cdot s^{-1}$ (Fig. 3b). Stacking two 626.8-nm films, two 404.6-nm films, or three 411.5-nm films yielded the same specific dissipated energy and velocity dependence as individual layers, confirming that the performance is preserved upon lamination and amenable to thickness scaling. Across all tested velocities, 2D HOMN-L4 exhibited specific dissipated energy up to an order of magnitude greater than that of conventional protective materials such as Kevlar composite fabric, stainless steel, and engineering polymers, demonstrating an architecture-driven route to extreme impact resistance[30-32]. Notably, the thinnest films (32.6–37.8 nm) displayed the most pronounced velocity-dependent response, with the specific dissipated energy reaching 5.2 $MJ\cdot kg^{-1}$ at an impact velocity of ~1000 $m\cdot s^{-1}$. This magnitude surpasses the specific detonation energy of 2,4,6-trinitrotoluene (TNT, 4.2 $MJ\cdot kg^{-1}$), though by absorption rather than release, which points to a distinct energy dissipation behavior in the ultrathin film regime.

Post-impact imaging revealed a velocity-dependent transition in the deformation and failure modes of the 665.4-nm-thick film. At 646 $m\cdot s^{-1}$, the film withstood the impact and completely captured a projectile whose diameter (8.11 μm) is more than ten times the film thickness, localizing the permanent deformation to an axisymmetric bulge and a narrow annular region around the impact site, without perforation (Fig. 3c). At 706 $m\cdot s^{-1}$, deformation was distributed around the impact site and accompanied by long radial cracks (Fig. 3d), whereas at 1,032 $m\cdot s^{-1}$, deformation became localized and produced a near-circular perforation (Fig. 3e). In both regimes, the fractured edges exhibit a rough, serrated morphology (insets in Fig. 3d, e). The transition from long radial cracks to near-circular perforation suggests that at higher impact velocities, the shorter projectile–film interaction time and inertial confinement suppress lateral stress redistribution and long-range radial-crack growth, localizing failure to the projectile footprint[33].

At the molecular level, the welded films dissipate impact energy through a mechanism (Fig. 3f) fundamentally distinct from the pull-out and plastic drawing of entangled chains that toughen conventional 1D polymer networks[32-34]. Upon impact, membrane tension shears adjacent sheets against each other, and this interlayer sliding dissipates energy through friction at the hydrogen-bonded interfaces[35]. Held together by the covalent linkers, the welded films sustain this sliding, whereas their unwelded counterparts delaminate catastrophically at the onset of sliding. Once the finite length of the linkers constrains sliding in one region, more regions are drawn into sliding sequentially. At failure, the linkers undergo forced covalent scission, providing additional irreversible dissipation. The staggered hard–soft architecture of rigid sheets bridged by flexible linkers deflects propagating cracks into tortuous paths, so that each advancing crack ruptures more linkers and produces irregular sheet pull-out consistent with the serrated fracture edges. The further enhancement in the specific dissipated energy of the thinnest films (32.6–37.8 nm) is attributed to a net out-of-plane ordering of the amide dipoles within the sheets, imprinted by the polar $SiO_2$ substrate during film formation. This dipole ordering is corroborated by Kelvin probe force microscopy (KPFM). The contact-potential difference between the film's two faces decreases with increasing thickness and vanishes at approximately 60 nm (Extended Data Fig. 6), in line with the thickness dependence of the specific dissipated energy. Such ordering is expected to increase the hydrogen-bond density and interfacial shear resistance, thereby enhancing impact energy dissipation.

**Beyond mechanical robustness**

In hydrogen-bonded 2DPA nanofilms, the tightly packed, staggered molecular sheets block continuous diffusion pathways, conferring intrinsic gas barrier properties[4,19]. However, centimeter-scale 2DPA-IPA films supported on polyimide membranes show oxygen permeability indistinguishable from that of the bare polyimide substrate (Supplementary Fig. 81), presumably because the brittle hydrogen-bonded assembly develops microcracks during fabrication and transfer. Furthermore, its efficacy against small permeants remains limited even at the microscopic scale, particularly for hydrogen[19]. Covalent welding eliminates this fragility, enabling macroscopic free-standing films with hydrogen permeability ($1.73\times10^{-5}$ Barrer) three orders of magnitude lower than that of ethylene-vinyl alcohol (EVOH), the benchmark polymer barrier[36,37] (Fig. 4a, Supplementary Table 5). The same welded architecture extends barrier performance from gases to ions. Across monovalent and divalent cations ($K^+$, $Na^+$, $Li^+$, $Ca^{2+}$, $Mg^{2+}$), the ion permeabilities of the welded films are on the order of $10^{-11}$ $cm^2\cdot s^{-1}$ (Fig. 4b, Supplementary Figs. 83-87), the lowest among state-of-the-art reduced graphene oxide[38], MXene[39] and COF membranes[40] and three to four orders of magnitude below those of the commercial membrane Nafion 212[41]. Even without welding, ultrathin 2DPA-IPA layers confine ion migration in perovskite solar cells, enabling thermally durable modules (>90% efficiency retention after 1,000 h at 100 °C) that operate in orbit aboard the Tianyan-24 satellite with a voltage T80 lifetime exceeding 8,000 h[42].

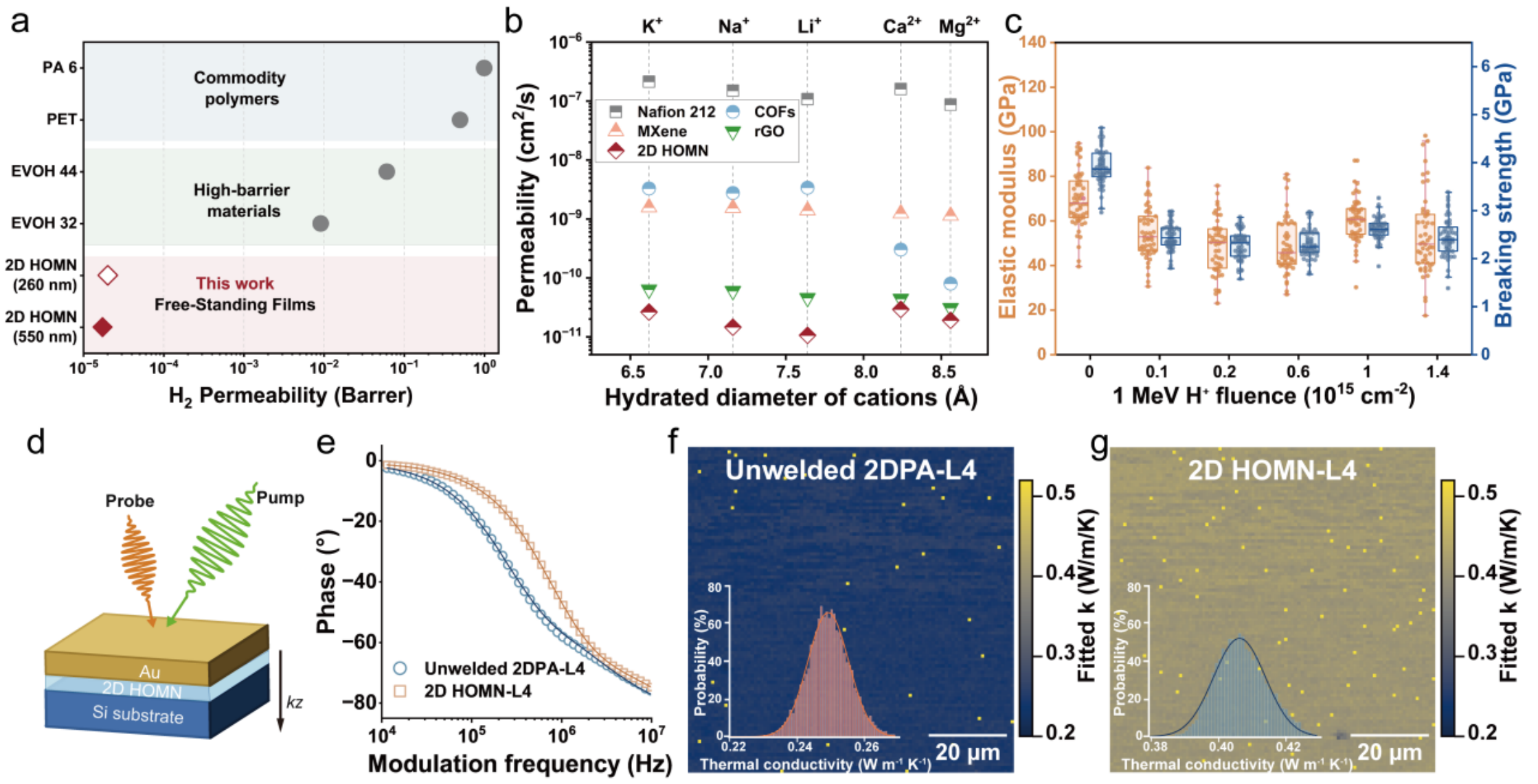

**Fig. 4|Properties of 2D HOMN-L4 films beyond mechanical robustness. a,** Hydrogen permeability of free-standing 2D HOMN-L4 films compared with commodity polymers and representative high-barrier materials. **b,** Ion permeabilities as a function of hydrated-ion diameter, compared with representative membrane materials. **c,** Elastic modulus and breaking strength of 2D HOMN films as a function of 1 MeV proton fluence. **d,** Schematic of the frequency-domain thermoreflectance (FDTR) measurement. **e,** Frequency-dependent FDTR phase responses (dots) and the corresponding fits (lines) for unwelded 2DPA-L4 and 2D HOMN-L4 films. **f,g,** Out-of-plane thermal conductivity maps of unwelded 2DPA-L4 (**f**) and 2D HOMN-L4 (**g**) films.

Covalent welding establishes intersheet connectivity in the 2D HOMNs, which further enhances their radiation resilience and thermal transport. Under irradiation, the welded network tolerates random bond scission through its redundant bonding pathways. Indeed, ~33-nm-thick 2D HOMN-L4 films retain 77% of their elastic modulus and 62% of their breaking strength after exposure to 1 MeV protons at fluences up to $1.4\times10^{15}$ protons$\cdot cm^{-2}$ (Fig. 4c, Extended Data Fig.

7a-b). This fluence exceeds the annual background of >1 MeV protons in low Earth orbit by over five orders of magnitude[43]. This resilience extends to 300 keV electron irradiation, with a 68-nm-thick 2D HOMN-L4 film showing no detectable structural change at doses exceeding 6,000 $e^{-}\cdot Å^{-2}$ (Extended Data Fig. 7c-e). Interlayer welding also enhances phonon transport across the interfaces originally held by hydrogen bonds alone. Accordingly, the out-of-plane thermal conductivity measured by frequency-domain thermoreflectance (FDTR) rises by 64%, from 0.25 $W\cdot m^{-1}\cdot K^{-1}$ in the unwelded assembly to 0.41 $W\cdot m^{-1}\cdot K^{-1}$ in the welded films (Fig. 4d–g), consistent with time-domain thermoreflectance (TDTR) measurements (from 0.21 to 0.42 $W\cdot m^{-1}\cdot K^{-1}$, Supplementary Fig. 96) and approaching the upper bound of conventional bulk polymers such as high-density polyethylene (0.38–0.51 $W\cdot m^{-1}\cdot K^{-1}$)[44].

## Conclusion

In conclusion, we have demonstrated a covalent welding strategy that interconnects 2D molecular sheets into highly oriented molecular networks. By forming covalent links between rigid molecular sheets, the architecture provides additional covalent load-transfer pathways beyond the fragile hydrogen-bonded interfaces, establishing a hierarchical energy landscape for extreme dissipation. The resulting films overcome the intrinsic brittleness of the hydrogen-bonded precursor, achieving ultra-high stiffness, strength and impact toughness alongside exceptional macroscopic load-bearing capability. Beyond mechanical robustness, the covalently welded architecture integrates molecular and ionic impermeability, radiation resilience and efficient thermal transport within a single material platform, revealing a versatility rooted in the welded 2D network. This strategy establishes a general framework for constructing polymer networks from 2D molecular sheets with chemically tunable peripheries, opening a broad design space for high-performance polymeric materials.

# Methods

**Synthesis of 2D molecular building blocks**

To a 40 ml vial equipped with a stir bar were added trimesoyl chloride (265 mg, 1 mmol, 1 equiv.) and melamine (126 mg, 1 mmol, 1 equiv.), followed by N-methyl-2-pyrrolidone (NMP, 9 ml) and pyridine (1 ml). The mixture was stirred at room temperature for 48 h, during which the reaction mixture gelled. β-alanine ethyl ester hydrochloride (553 mg, 3.6 equiv.) was then added, and the mixture was stirred for an additional 24 h. The resulting mixture was poured dropwise into diethyl ether to precipitate the product. The precipitate was washed successively with acetone, ethanol and deionized water, and collected by centrifugation. The solid was lyophilized at −90 °C for 48 h and subsequently dried in a vacuum oven at 80 °C for 12 h, affording 2DPA-β-alanine as a pale-yellow powder (245 mg, 87%).

A two-step approach is also viable for the synthesis of 2DPA-β-alanine ethyl ester. In the first step, trimesoyl chloride (265 mg, 1 mmol, 1 equiv.) and melamine (126 mg, 1 mmol, 1 equiv.) were polymerized in NMP (9 ml) with pyridine (1 ml) at room temperature for 48 h, during which the reaction mixture gelled. The reaction was then quenched with water (0.2 ml, 11.1 equiv.), to hydrolyse the peripheral acyl chloride groups to carboxylic acids. The precipitate was collected by centrifugation, washed with ethanol and deionized water, and dried under vacuum to afford 2DPA-COOH as a pale-yellow solid (253 mg, 90% yield). In the second step, 2DPA-COOH (500 mg, 1 equiv.) was dispersed in DMF (10 ml) and cooled in an ice bath. After the addition of DIPEA (692 µl, 3.0 equiv.), the mixture was stirred until homogeneous. O-(7-Azabenzotriazol-1-yl)-N,N,N',N'-tetramethyluronium hexafluorophosphate (HATU, 514 mg, 1.2 equiv.) was subsequently added, and the reaction was maintained in an ice bath for 30 min. β-Alanine ethyl ester hydrochloride (461 mg, 3 equiv.) was then added; the ice bath was removed, and the mixture was stirred at room temperature for 10–12 h. The product was isolated by precipitation into diethyl ether, washed with ethanol and deionized water, and lyophilized to afford 2DPA-β-alanine as a pale-yellow powder (350 mg, 70%).

**Film fabrication and covalent welding**

The edge-modified 2DPA precursor was dissolved in DMF at concentrations ranging from 3 to 400 $mg \cdot ml^{-1}$. The solution was spin-coated onto a silicon wafer with a 300 nm oxide layer at 1,000–3,000 rpm, yielding hydrogen-bonded nanofilms with thicknesses ranging from 4-1000 nm. The as-prepared films were annealed at 90 °C for 30 min to remove residual solvent, and then heated at 280 °C for 15 min to effect covalent welding.

For film transfer, a 10% (v/v) HF solution was applied to the wafer edge to etch the underlying $SiO_2$ layer. Once the film detached from the substrate, the wafer was immersed in a glass container filled with ultrapure water, and the water was exchanged three times to ensure complete removal of residual HF, affording free-standing 2D HOMN films.

**High-resolution atomic force microscopy**

High-resolution AFM imaging was performed on a Cypher VRS 1250 instrument (Asylum Research, Oxford Instruments) in tapping mode using Multi75DLC probes (Budget Sensors). Grade A highly oriented pyrolytic graphite (HOPG) was used as the substrate to provide an atomically flat surface. For partially welded samples, 2DPA-L4 powder was heated at 150 °C for 5 min, dissolved in DMF (0.1 mg·ml$^{-1}$) and sonicated for 10 min. The dilute solution (20 µl) was drop-cast onto freshly cleaved HOPG, followed by rinsing with DMF. For individual-molecule imaging, 2DPA-L4 was dissolved in DMF (0.1 mg·ml$^{-1}$), and the same deposition and rinsing procedure was used. Samples were imaged after complete evaporation of DMF. Data were processed using Asylum Research AR software (version 16.33.234) and Gwyddion.

**Strain-induced elastic buckling instability for mechanical measurements (SIEBIMM)**

Polydimethylsiloxane (PDMS, Sylgard 184, Dow Corning) was prepared by mixing the base and curing agent at a mass ratio of 10:1. The mixture was stirred with a glass rod for 10 min, allowed to stand at room temperature for 4 h to remove bubbles, and then cured at 80 °C for 4 h. The cured PDMS was cut into 10 mm × 10 mm × 5 mm pieces. To minimize substrate-to-substrate variation, unwelded 2DPA-IPA and 2D HOMN films were transferred onto the same PDMS specimen. After transfer from the water surface, the films on PDMS were left at room temperature for 3 h until the residual water had evaporated, and then heated in a vacuum oven at 60 °C for 12 h to ensure intimate contact between the films and the substrate. Wrinkling was induced by applying 10% uniaxial strain to the PDMS substrate. The wrinkle patterns were observed by optical microscopy, and the wrinkle wavelength was measured from the micrographs.

**Nanoindentation**

Nanoindentation tests were carried out using the same Cypher VRS 1250 instrument. Holey copper grids were used to support the suspended films. Diamond-like-carbon spherical probes (NT_B300_v0010, Biosphere), Tap300GB-G (purchased from Budget Sensors) were used for indentation. For thickness determination, a fragment of the same film was transferred onto a silicon wafer, and the thickness was determined by averaging measurements from three distinct locations. After force calibration, the probe tip was positioned at the suspended region above a grid aperture, and measurements were conducted at varying trigger forces. Force–displacement curves were extracted and analyzed to obtain the Young's modulus and fracture strength of the films.

**Laser-induced projectile impact testing (LIPIT)**

High-velocity micro-projectile impact tests were performed using an advanced laser-induced projectile impact testing (LIPIT) apparatus in a vacuum chamber (pressure < 260 mTorr) to eliminate air drag effects. The launch pad comprised a glass substrate (8 mm thick, 50 mm diameter) bearing a sputtered sacrificial gold layer (50 nm) and a polyimide film (15 µm thick), with the polyimide adhered to the gold via a thin PDMS interlayer (Sylgard 184, 10:1). $SiO_2$ microspheres (diameter 8.11 ± 0.47 µm, density 2.2 g·cm$^{-3}$) were suspended in ethanol and spin-coated onto the polyimide surface. A single microsphere was selected as the projectile. A pulsed Nd:YAG laser (532 nm, 6 ns pulse width) was then focused onto the gold layer directly beneath the selected particle, causing local ablation of the gold film, expansion of the polyimide layer and subsequent acceleration of the microsphere. Impact velocity was tuned by adjusting the laser pulse energy over the range 300–1,200 m·s$^{-1}$. The particle trajectory was illuminated by a supercontinuum laser (6 ps, SC-PRO-M) and recorded via a global-exposure CMOS camera (MV-GED500M, 10 µs exposure) triggered by a digital delay generator, with the supercontinuum laser repetition rate setting the inter-frame interval (0.2-2 µs). Typically, <20 pulses were used per sequence to maintain image contrast. Impact and rebound velocities were determined from the recorded multi-exposure images by dividing the pixel inter-particle spacing by the pulse interval.

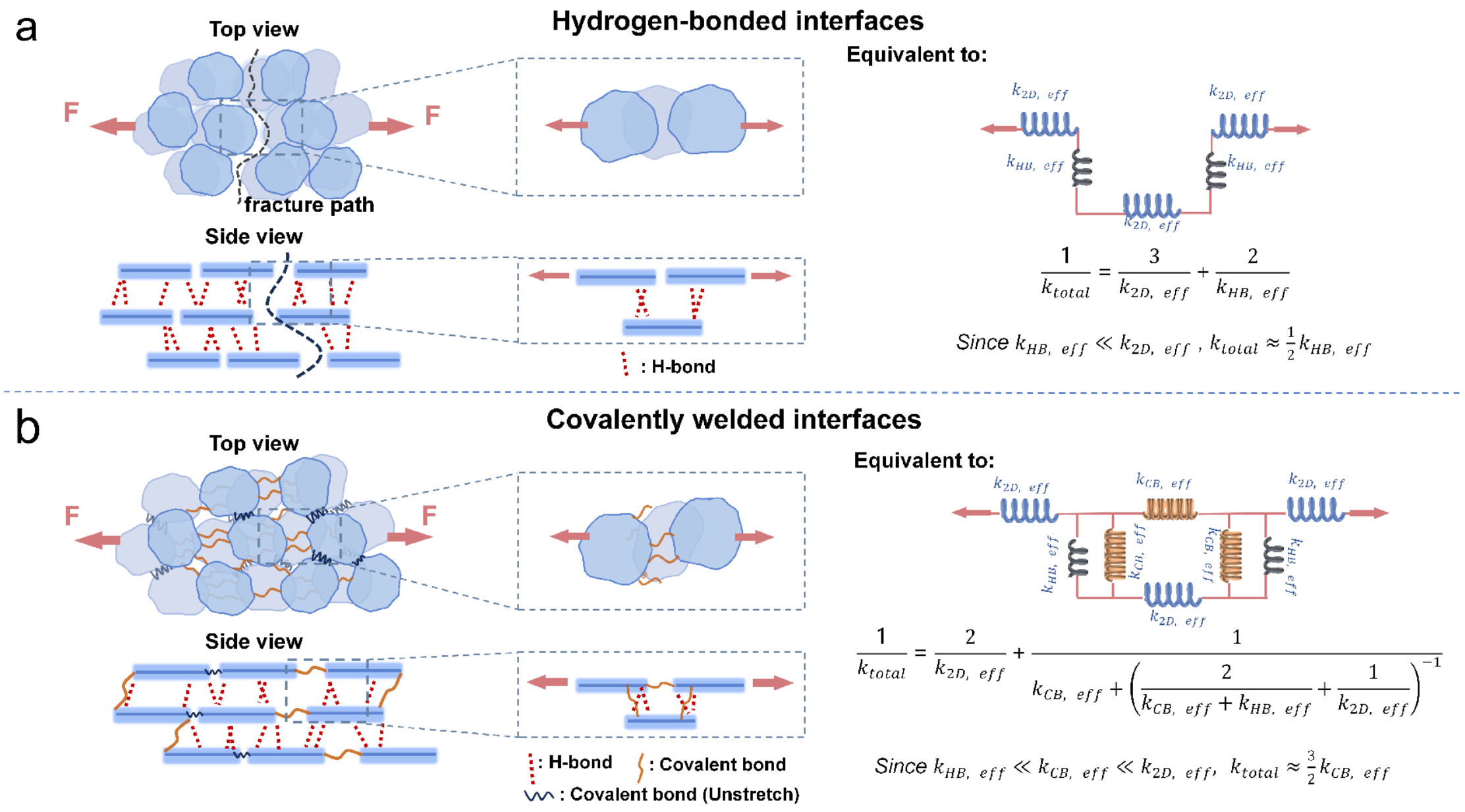

$$\frac{1}{k_{total}} = \frac{3}{k_{2D,\ eff}} + \frac{2}{k_{HB,\ eff}}$$

Since $k_{HB,\ eff} \ll k_{2D,\ eff}$, $k_{total} \approx \frac{1}{2} k_{HB,\ eff}$

$$\frac{1}{k_{total}} = \frac{2}{k_{2D,\ eff}} + \frac{1}{k_{CB,\ eff} + \left(\frac{2}{k_{CB,\ eff} + k_{HB,\ eff}} + \frac{1}{k_{2D,\ eff}}\right)^{-1}}$$

Since $k_{HB,\ eff} \ll k_{CB,\ eff} \ll k_{2D,\ eff}$, $k_{total} \approx \frac{3}{2} k_{CB,\ eff}$

**Extended Data Fig. 1|Spring model for load transfer in hydrogen-bonded and covalently welded 2DPA assemblies. a,** Load transfer in the unwelded hydrogen-bonded assembly. Top and side views (upper left) show the molecular packing, together with the equivalent spring circuit in which the 2D sheets ($k_{2D,\ eff}$) and the hydrogen bonds ($k_{HB,\ eff}$) are connected in series. Under an applied force F, rupture occurs preferentially at the hydrogen-bonded interfaces, and the crack propagates along these interfaces (fracture path, bottom views), leading to brittle failure of the film. **b,** Load transfer in the covalently welded molecular network. The covalent bridges ($k_{CB,eff}$) connect adjacent sheets in parallel with the surviving hydrogen bonds, so that the overall stiffness is instead set by the covalent bridges ($k_{total} \approx 3/2\ k_{CB,eff}$). Under the same applied force, the covalent bridges bear the load and buffer deformation, so the welded film resists fracture.

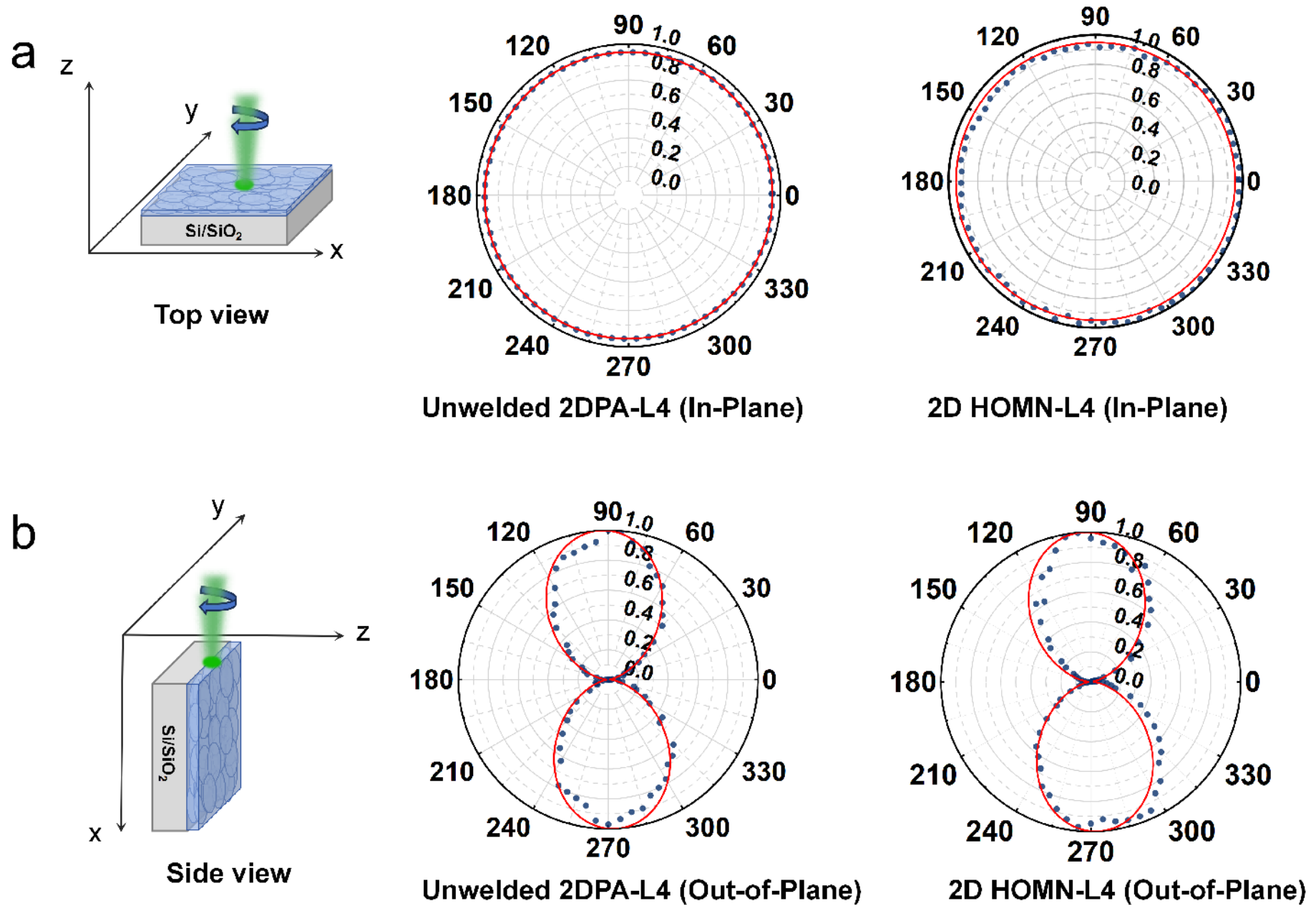

**Extended Data Fig. 2|Polarized photoluminescence of 2D HOMN-L4 and unwelded 2DPA-L4 films. a,** PL measurement with laser incidence normal to the film plane. **b,** PL measurement with laser incidence parallel to the film plane. Left, schematic illustrations; middle and right, polar plots of normalized integrated PL intensity as a function of the excitation-polarization angle for the unwelded (middle) and welded (right) films. Blue points represent experimental data and red curves are twofold-symmetry fits. The response is nearly isotropic under normal incidence, whereas pronounced two-lobed anisotropy appears under parallel incidence; the anisotropy is comparable between the unwelded and welded films (modulation contrast of 0.981 and 0.967, respectively).

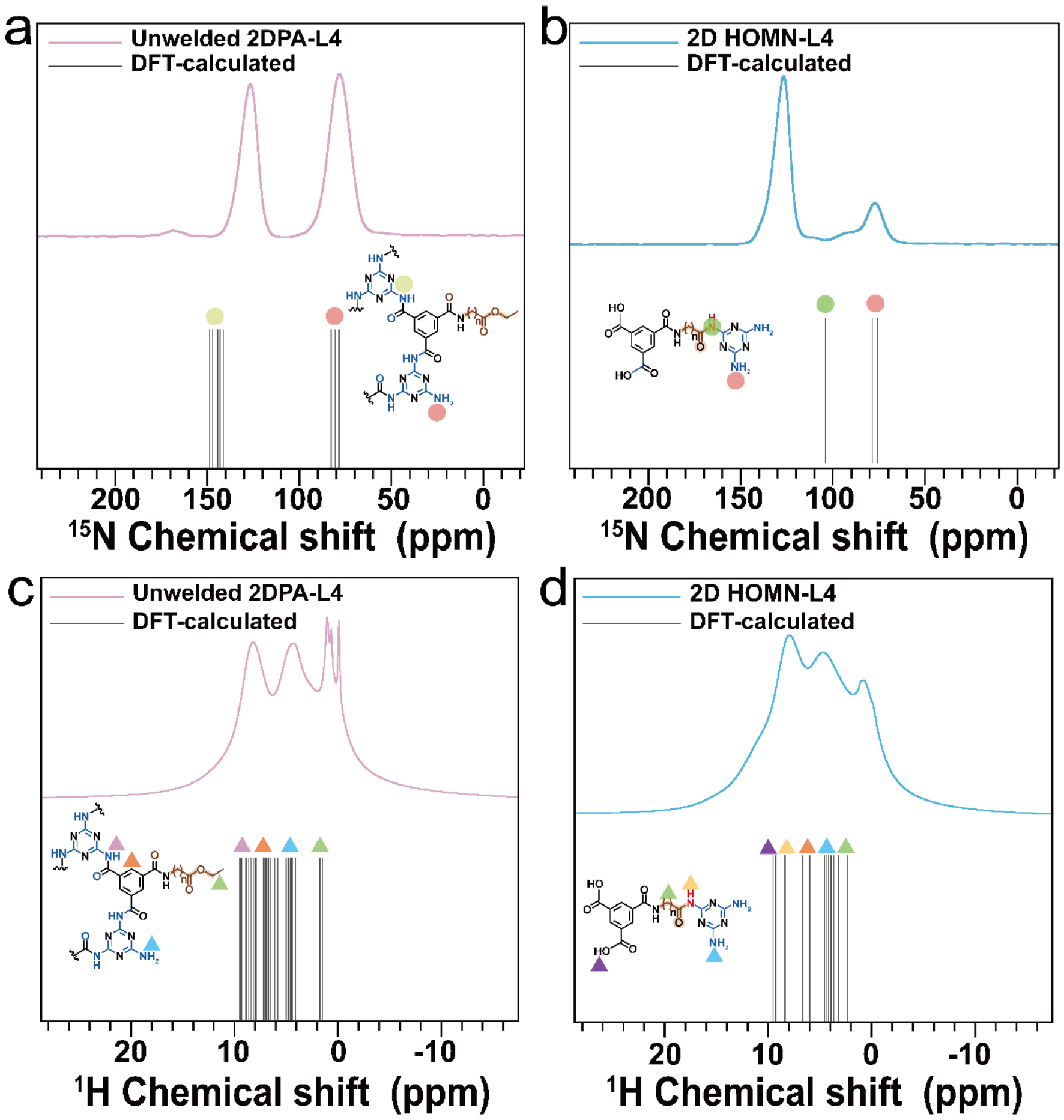

**Extended Data Fig. 3| Solid-state NMR characterization of unwelded 2DPA-L4 and 2D HOMN-L4 films and comparison with DFT-calculated chemical shifts.** Experimental $^{15}$N CPMAS NMR spectra of **(a)** unwelded 2DPA-L4 and **(b)** 2D HOMN-L4 films, and $^{1}$H MAS NMR spectra of **(c)** unwelded 2DPA-L4 and **(d)** 2D HOMN-L4 films, together with their corresponding DFT-calculated chemical shifts. Vertical sticks indicate the calculated chemical shifts of individual nitrogen and proton sites, as highlighted in the molecular structures. Colored circles mark distinct $^{15}$N environments, whereas colored triangles mark distinct $^{1}$H environments.

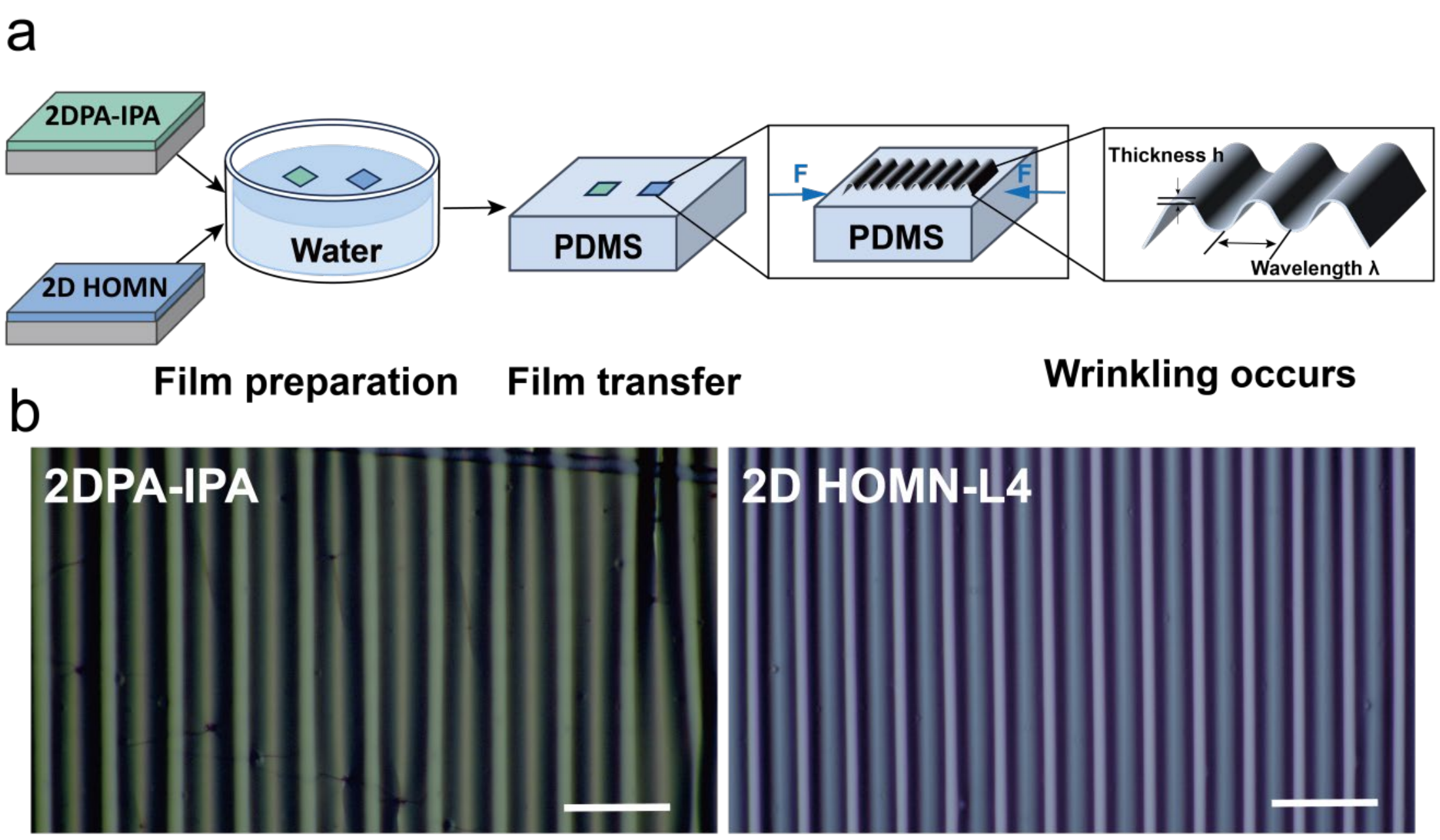

**Extended Data Fig. 4| Rapid modulus screening by SIEBIMM. a,** Schematic of the strain-induced elastic buckling instability for mechanical measurements (SIEBIMM) procedure, in which the unwelded reference and welded films are floated on water, transferred together onto the same PDMS substrate and compressed uniaxially; the film modulus is extracted from the wrinkle wavelength $\lambda$ and the film thickness $h$. **b,** Optical microscopy images of the buckling wrinkles of unwelded 2DPA-IPA (159.2 nm) and 2D HOMN-L4 (64.0 nm) films on a PDMS substrate. Scale bars, 40 µm.

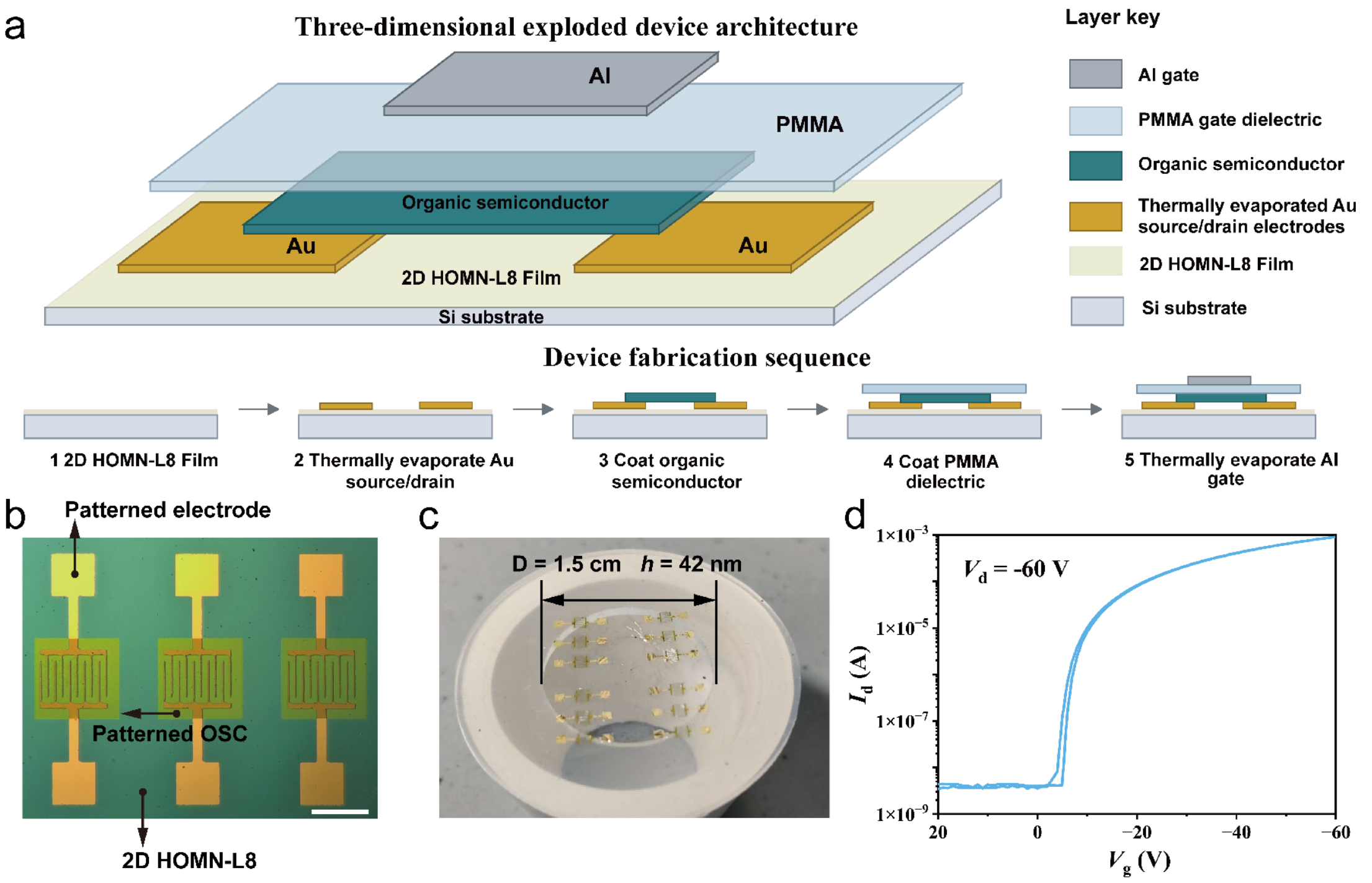

**Extended Data Fig. 5| Organic field-effect transistors on a 42-nm-thick 2D HOMN-L8 film and their electrical performance. a,** Exploded device architecture and fabrication sequence. The 2D HOMN-L8 film serves as the mechanical substrate and PMMA as the gate dielectric. **b,** Optical micrograph of the patterned Au electrodes and the organic semiconductor (OSC) on the film. Scale bar, 200 µm. **c,** Photograph of the free-standing 2D HOMN-L8 film (diameter, 1.5 cm) released from the Si wafer, bearing multiple devices. **d,** Transfer characteristics (forward and reverse sweeps) measured at $V_d$ = −60 V over a gate-voltage cycle from 20 V to −60 V and back.

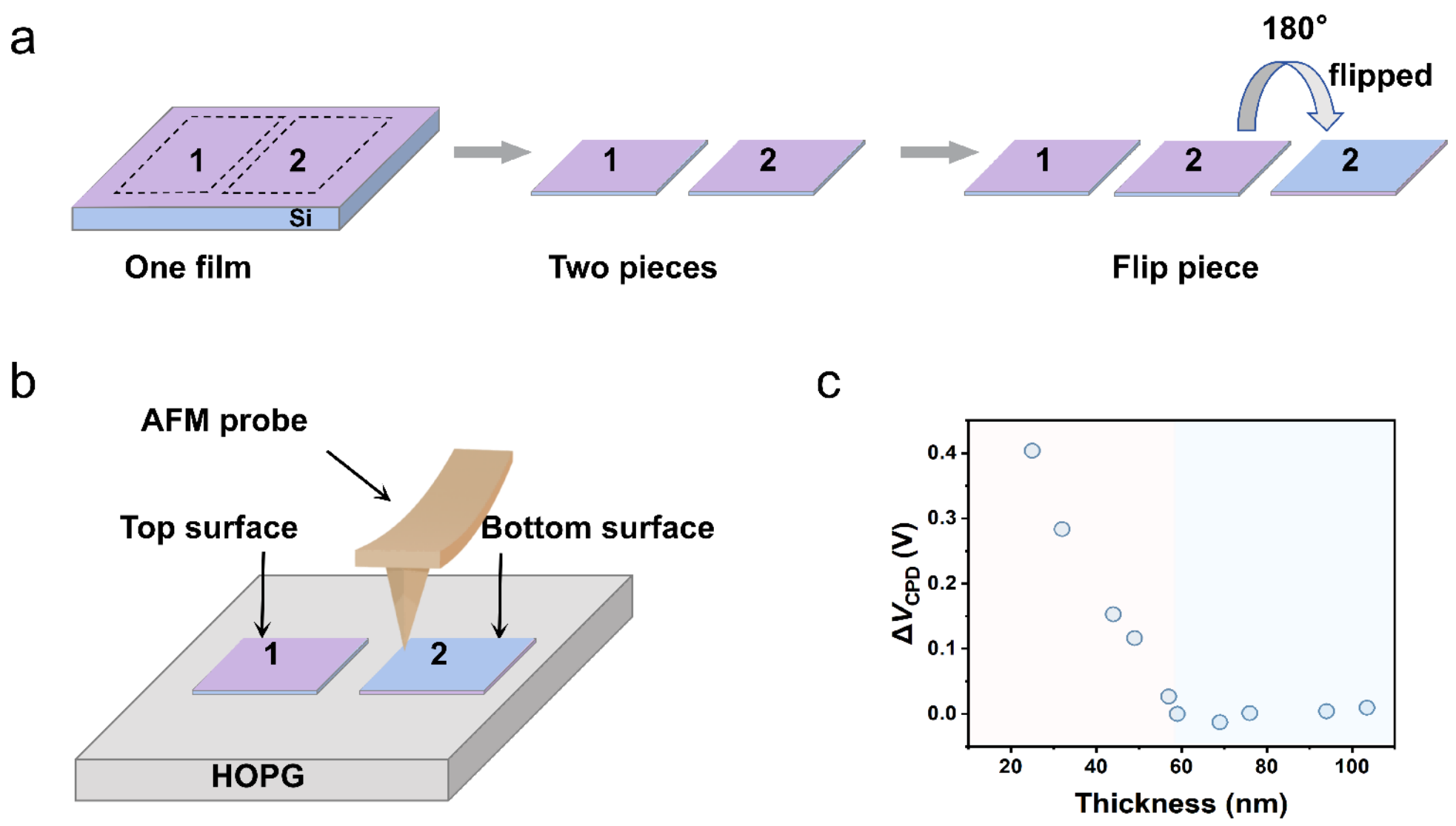

**Extended Data Fig. 6| Surface contact potential difference of 2D HOMN-L4 films. a,** Schematic of sample preparation for top and bottom surface measurements. **b,** Schematic of the Kelvin probe force microscopy (KPFM) measurement of the contact potential difference ($V_{CPD}$) on the top and bottom surfaces of the same film supported on HOPG. **c,** Difference in contact potential ($\Delta V_{CPD}$) between the top and bottom surfaces as a function of film thickness.

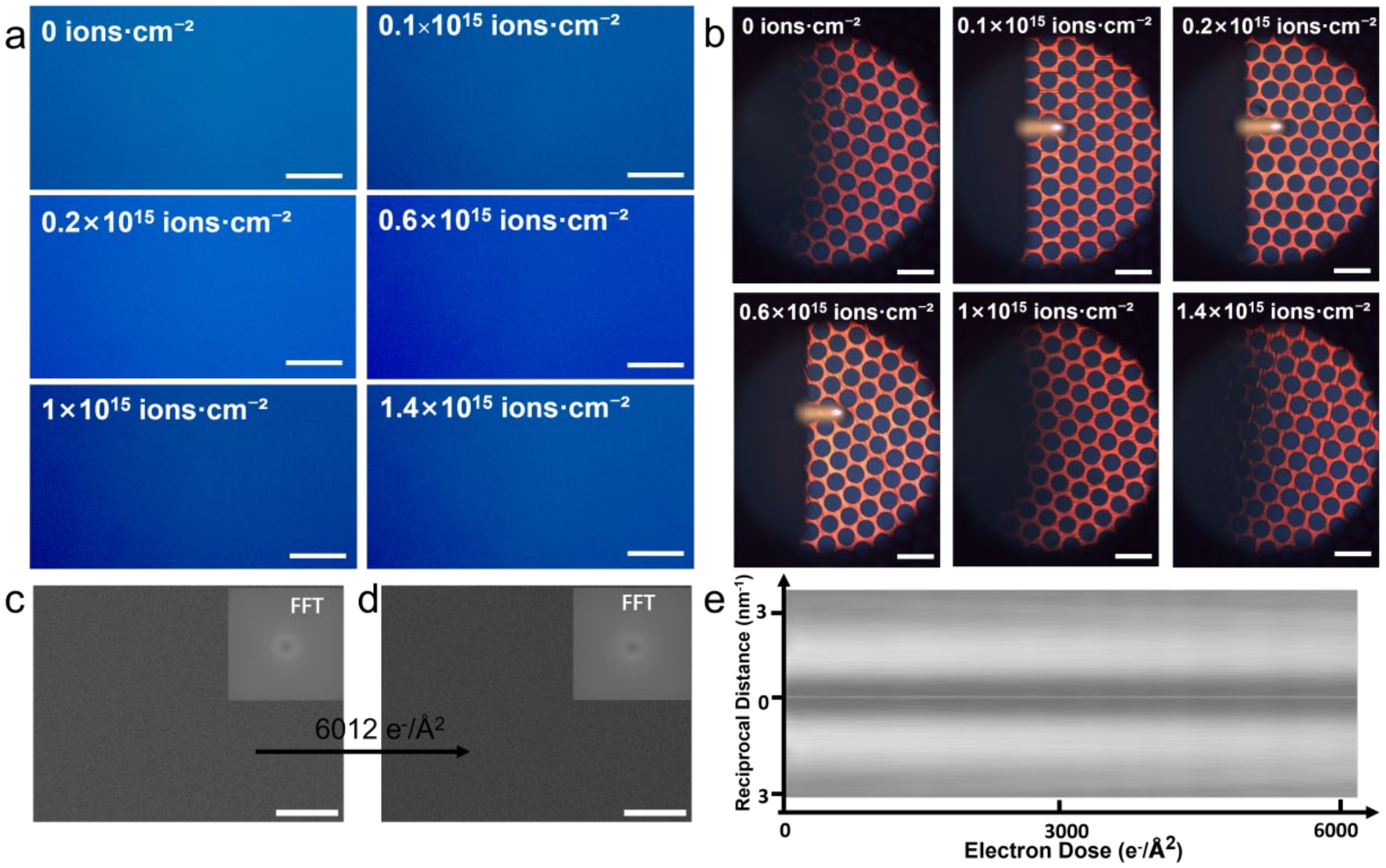

**Extended Data Fig. 7| Structural integrity of 2D HOMN-L4 films under electron and proton irradiation. a,** Optical microscopy images of the films on $Si/SiO_2$ substrates at increasing 1-MeV proton fluences. **b,** Optical microscopy images of the corresponding films transferred onto copper TEM grids. **c, d,** TEM images of a 68-nm-thick film before **(c)** and after **(d)** irradiation with 300-keV electrons at a dose of 6,012 $e^{-}\cdot Å^{-2}$. Insets, corresponding fast Fourier transform (FFT) patterns. Scale bars, 50 nm. **e,** FFT pattern evolution as a function of electron dose.

**Data availability**

The data that support the findings of this study are available from the corresponding author upon reasonable request. All data are available in the main text or the Supplementary Information.

**Acknowledgments**

This work was supported by the Start-up Fund from Fudan University and the Fundamental Research Funds for the Central Universities (No. WK2090250112). H.M. acknowledges the funding support from the National Natural Science Foundation of China (No. 52576100) and the USTC Startup Program (No. KY2090000117). M. L. gratefully acknowledges the support from the National Natural Science Foundation of China (No. 12504227). X. L. acknowledges the support from Beijing Natural Science Foundation (2252036). We acknowledge the State Key Laboratory of Molecular Engineering of Polymers at Fudan University and the State Key Laboratory of Mechanical System and Vibration at Shanghai Jiao Tong University for access to experimental facilities. We thank Oxford Instruments Technology (Shanghai), especially Hongyang Su and Zhonghui Zhang, for assistance with polarized photoluminescence measurements. We also thank Dr. Yuchuan Tian and Dr. Junyu Li at the Sinopec Shanghai Research Institute of Petrochemical Technology for discussions and technical assistance.

**Author Contributions**

Y. Zeng conceived and supervised the project, wrote the initial manuscript with input from X. Gong, J. Wu and Y. Hu. X. Gong synthesized materials (with Z. Yin and G. Wu), performed HR-AFM, nanoindentation, KPFM (with B. Tang), and SIEBIMM (with J. Wu). G. Luo and J. Chen designed and performed LIPIT experiments, and analyzed impact data under the supervision of Y. Hu. J. Wu performed covalent welding for film preparation (with Y. Wu and X. Gong) and characterization (polarized PL). L. Tang and Y. Fu performed solid-state NMR measurements and analyzed the spectra. Z. Yan measured gas-barrier properties. J. Chen, X. Liu, H. Wu and Y. Lin performed ion-barrier measurements. Q. Wang and H. Zhang performed positron-annihilation measurements and analyzed the data. K. Liu and Y. Xu performed TEM characterization and FFT analysis. Y. Hu, Y. Zhu, Y. Yu, Z. Pan, M. Li and H. Ma carried out thermal conductivity measurements and analyzed data. Y. Zhang performed proton-irradiation experiments. L. Peng and B. Tang fabricated the organic transistor devices under the supervision of D. Wei. All authors discussed the results and approved the final manuscript.

**Corresponding authors**

Correspondence to Yongxiang Hu, Yuwen Zeng.

**Competing interests**

The authors declare no competing interests.

**Additional information**

Correspondence and requests for materials should be addressed to Y.Z.

# Supplementary Information

## Covalent Welding of Two-Dimensional Molecular Sheets into Ultra-Stiff, Super-Tough Polymeric Networks

Xiaoli Gong[1†], Guohu Luo[2†], Jiaxin Wu[1†], Jiayu Chen[2], Zhengqiao Yin[1], Zhipeng Yan[1], Jingran Chen[1], Bowen Tang[1], Le Tang[3], Yuxiao Wu[1], Guanling Wu[1], Li Peng[1], Qihang Wang[5], Yinglong Hu[6], Yue Zhu[6], Yuan Yu[6], Keyan Liu[1], Xiaozhou Liu[7], Huanhuan Wu[7], Yakai Lin[7], Yifei Xu[1], Zhiliang Pan[6], Hao Ma[6], Man Li[6], Hongjun Zhang[5], Dacheng Wei[1], Youpeng Zhang[4], Yao Fu[3], Yongxiang Hu[2*], Yuwen Zeng[1*]

[1]State Key Laboratory of Molecular Engineering of Polymers, Department of Macromolecular Science, Fudan University, Shanghai, 200438, China

[2]State Key Laboratory of Micro and Nano Engineering Science, School of Mechanical Engineering, Shanghai Jiao Tong University, Shanghai 200240, China

[3]Department of Chemistry, Fudan University, Shanghai, 200438, China

[4]Institute of Modern Physics, Fudan University, Shanghai 200433, China

[5]State Key Laboratory of Particle Detection and Electronics, University of Science and Technology of China, Hefei 230026, China

[6]Institute of Micro-Nano Thermal Energy Transport and Conversion, School of Engineering Science, University of Science and Technology of China, Hefei, Anhui 230027, China

[7]State Key Laboratory of Chemical Engineering and Low-Carbon Technology, Department of Chemical Engineering, Tsinghua University, Beijing 100084, China

[†]These authors contributed equally to this work.

# Contents

## Supplementary Methods

The Supplementary Methods provide the reproducible experimental and analytical details supporting the chemical, structural, mechanical, impact and multifunctional claims in the main Article.

### Materials

Unless otherwise stated, all chemicals were obtained commercially and used without further purification. Anhydrous N-methyl-2-pyrrolidone (NMP) was purchased from Bailingwei Technology. Melamine and anhydrous N, N-dimethylformamide (DMF) were purchased from Sinopharm Chemical Reagent. Anhydrous ethanol was purchased from Shanghai Dahe Chemical. Diethyl ether was purchased from Greagent (Shanghai Titan Technology). Benzene-1,3,5-tricarbonyl trichloride (TMC), pyridine, acetone and the amino-ester hydrochlorides were purchased from Adamas (Shanghai Titan Technology). HATU and DIPEA (≥99%) were obtained from Sigma-Aldrich. Si wafers bearing a 300-nm thermal $SiO_2$ layer were purchased from Lijin Electronics.

### Characterization techniques

Proton nuclear magnetic resonance (Solution-state $^1$H NMR) spectra were recorded on a Bruker AVANCE NEO spectrometer operating at 600 MHz. $^{15}$N and $^1$H solid-state NMR experiments were performed on a Bruker Avance III HD 400 MHz NMR spectrometer. Transmission electron microscopy (TEM) images were acquired using a field-emission transmission electron microscope (JEM-F200, JEOL). Fourier-transform infrared (FTIR) measurements were performed using a Nicolet 6700 spectrometer. Raman spectra were collected on a Horiba XploRA spectrometer equipped with a 532 nm Coherent Sapphire laser. Gel permeation chromatography (GPC) was performed on an Agilent system to determine the molecular weight distribution of 2DPA. Solid-state NMR experiments were conducted on a Bruker AVANCE NEO 600WB spectrometer to investigate the molecular structure of unwelded 2DPA films and 2D HOMN films. Atomic force microscopy (AFM) images were acquired using an Oxford Instruments Cypher VRS1250 microscope in tapping mode and analyzed with Gwyddion or Asylum Research (AR) software (v16.33.234). Film thickness and lateral dimensions were measured using a KLA-Tencor AlphaStep D-600 profilometer. Differential scanning calorimetry (DSC) was carried out on a TA Instruments DSC 250 under a nitrogen atmosphere, with scans from 40 to 290 °C at a rate of 10 °C/min. Thermogravimetric analysis (TGA) was performed on a Mettler Toledo instrument under a constant $N_2$ flow from 50 to 800 °C at a heating rate of 10 °C/min. Simultaneous Thermogravimetric analysis coupled with differential scanning calorimetry and mass spectrometry (TGA-DSC-MS) measurements were performed on a NETZSCH STA 449 F3 Jupiter thermal analyser coupled to a NETZSCH QMS 403 D quadrupole mass spectrometer. The powder sample was heated from room temperature to 600 °C at a heating rate of 5 K $min^{-1}$ under continuous $N_2$ purge, while the mass change, heat-flow signal and selected ion currents were recorded simultaneously. Scanning electron microscopy (SEM) images were collected on a Zeiss Ultra 55 microscope. X-ray photoelectron spectroscopy (XPS) spectra were acquired using a Thermo Scientific K-Alpha spectrometer equipped with a monochromatic Al Kα source (1486.6 eV, 150 W). Optical microscopy images were captured using a Motic BA310 halogen microscope and analyzed with ImageJ software. Freeze-drying was conducted using a BIOCOOL FD-1A-50 freeze dryer.

**Sample nomenclature**

**Supplementary Table 1 |** L-series linker nomenclature and spacer definitions.

| Code | Pendant linker | Spacer definition |
|---|---|---|
| L0 (n=0) | / | 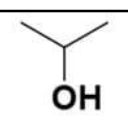 |
| L3 (n=2) | glycine ethyl ester or corresponding linker | 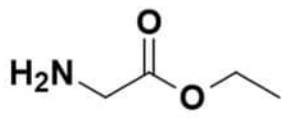 |
| L4 (n=3) | β-alanine ethyl ester | 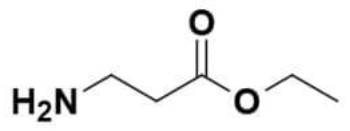 |
| L5 (n=4) | ethyl 4-aminobutyrate | 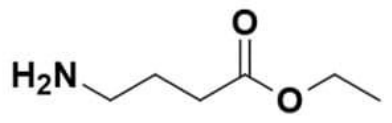 |
| L6 (n=5) | ethyl 5-aminovalerate | 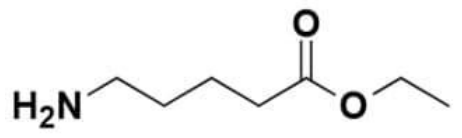 |
| L7 (n=6) | ethyl 6-aminohexanoate | 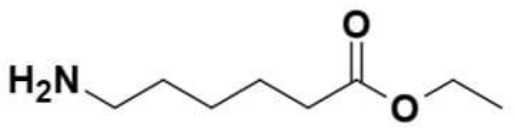 |
| L8 (n=7) | ethyl 7-aminoheptanoate | 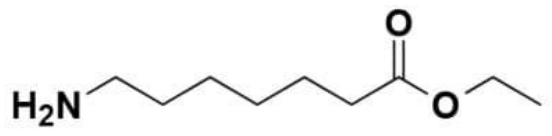 |

**Synthesis of two-dimensional molecular building blocks**

**Synthesis of edge-modified 2DPA -Ln**

Two synthetic schemes for edge-modified 2DPA precursors. One-pot in-situ amino-ester grafting route; Two-step sequential synthesis via 2DPA-COOH intermediate precursor. All core reactive functional groups (acyl chloride, carboxylic acid, primary amine, ester, amide) are labeled on molecular structures.

**One-pot grafting synthesis of L-series 2DPA precursors**

TMC (265 mg, 1.00 mmol) and melamine (126 mg, 1.00 mmol) were combined in a reaction flask. NMP (9 mL) and pyridine (1 mL) were added, and the mixture was stirred at room temperature in air for 48 h until gelation. Amino-ester hydrochloride (3.6 equiv) was added, and stirring was continued for 24 h. The mixture was added dropwise to excess diethyl ether. The precipitate was isolated by centrifugation and washed sequentially with acetone, ethanol and deionized water. The product was freeze-dried at −90 °C for 48 h and then dried under vacuum at 80 °C for 12 h to yield a pale-yellow powder.

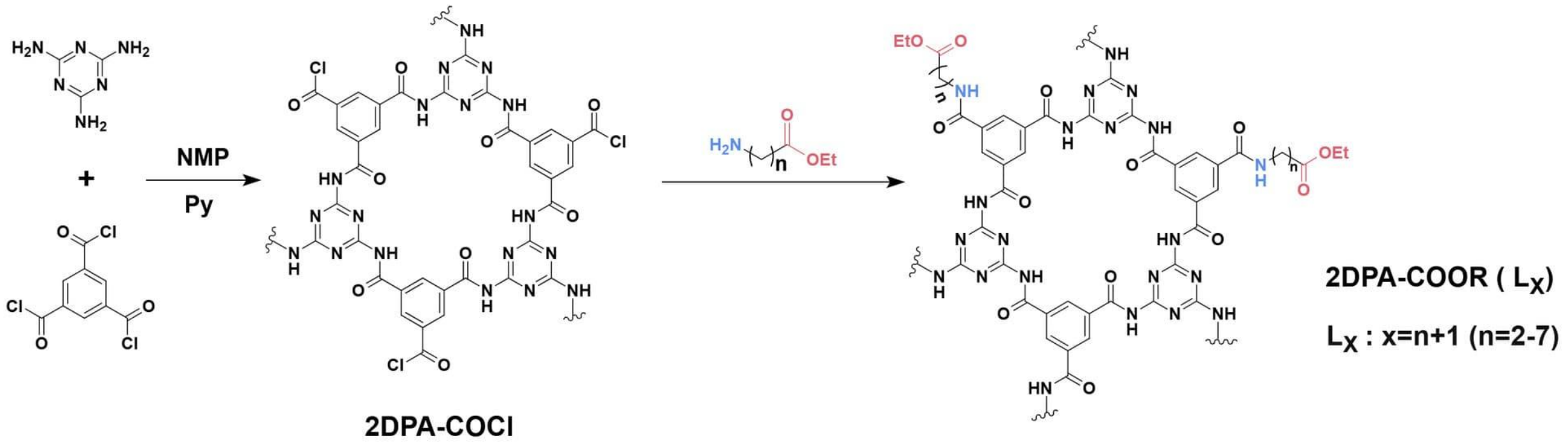

**Supplementary Fig. 1 |** One-pot grafting synthesis of L-series 2DPA precursors.

**Two-step synthesis through the 2DPA-COOH substrate**

Two-step sequential synthesis of 2DPA-Ln

Step 1: Synthesis of carboxyl-terminated 2DPA-COOH. TMC and melamine underwent 48 h polycondensation in NMP/pyridine solvent as described above. 0.2 mL deionized water was added to hydrolyze peripheral acyl chloride groups. Crude precipitates were collected by centrifugation, washed and vacuum-dried to yield 2DPA-COOH powder.

Step 2: Amide coupling grafting. 500 mg 2DPA-COOH powder was dispersed in 10 mL anhydrous DMF and cooled in an ice bath. DIPEA (3.0 eq) and HATU (1.2 eq) were added sequentially with 30 min stirring, followed by Amino-ester hydrochloride (3.0 eq). The ice bath was removed, and the suspension reacted at room temperature for 10–12 h. Product isolation adopted identical precipitation, washing and drying protocols as the one-pot route.

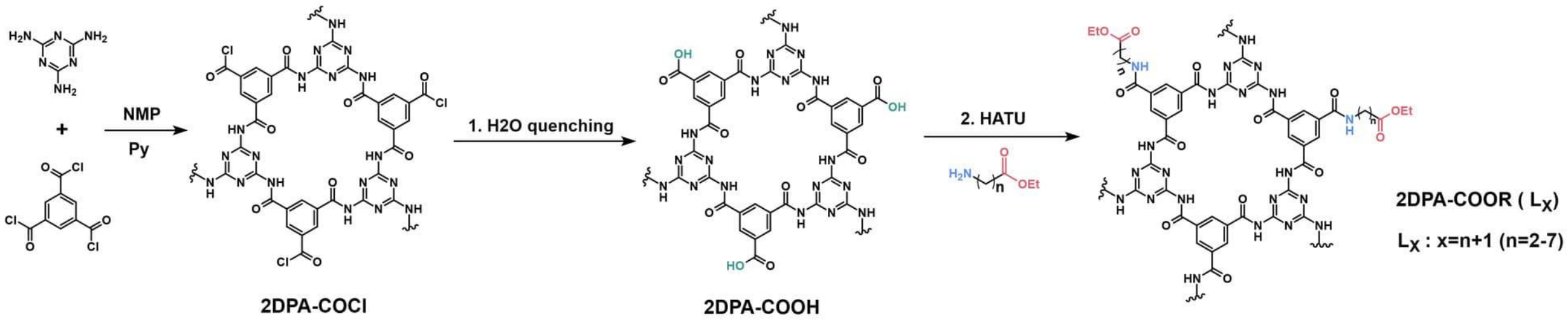

**Supplementary Fig. 2 |** Two-step synthesis of L-series 2DPA precursors.

## Batch synthesis of 2DPA precursors

The batch synthesis was performed in a nitrogen-purged reactor. 1,3,5-Benzenetricarbonyl trichloride (TMC; 212 g, 0.8 mol, 1.00 equiv) was added to anhydrous N-methyl-2-pyrrolidone (NMP; 2.5L) and dissolved by mechanical stirring at 150 r.p.m. The reagent bottle and addition funnel were rinsed twice with an additional 500 ml of NMP, and the rinses were transferred to the reactor. After stirring for 10 min, melamine (MA; 100.8 g, 0.8 mol, 1.00 equiv) was added slowly in small portions, with each portion allowed to disperse and react before the next addition. The container and funnel were rinsed with NMP (600 ml), and the rinses were added to the reaction mixture. After confirming the absence of visible melamine aggregates, the stirring rate was increased to 180 r.p.m. After 10 min, anhydrous pyridine (800 ml) was added slowly through a pressure-equalizing dropping funnel. The nitrogen inlet was then closed, all reactor ports were sealed, and the reaction mixture was stirred at room temperature and ambient pressure for 72 h. The viscosity of the mixture gradually increased during polymerization.

The reaction was terminated by adding ultrapure water (160 ml) under nitrogen, followed by stirring at 500 r.p.m. overnight at room temperature to complete the end-capping reaction. The mixture was subsequently allowed to stand for approximately 10 h to sediment the 2DPA product. The supernatant was removed using a peristaltic pump, and the sediment was collected by centrifugation at 4,000 r.p.m. for 10 min. The resulting yellow crude product was purified by successive centrifugation and washing with deionized water and ethanol, followed by a final wash with deionized water. The purified product was prefrozen in liquid nitrogen and freeze-dried using a lyophilizer equipped with a −80 °C cold trap for 3–5 days, yielding approximately 200 g of dry 2DPA precursors. The product was ground into a fine powder and stored under nitrogen in a sealed container.

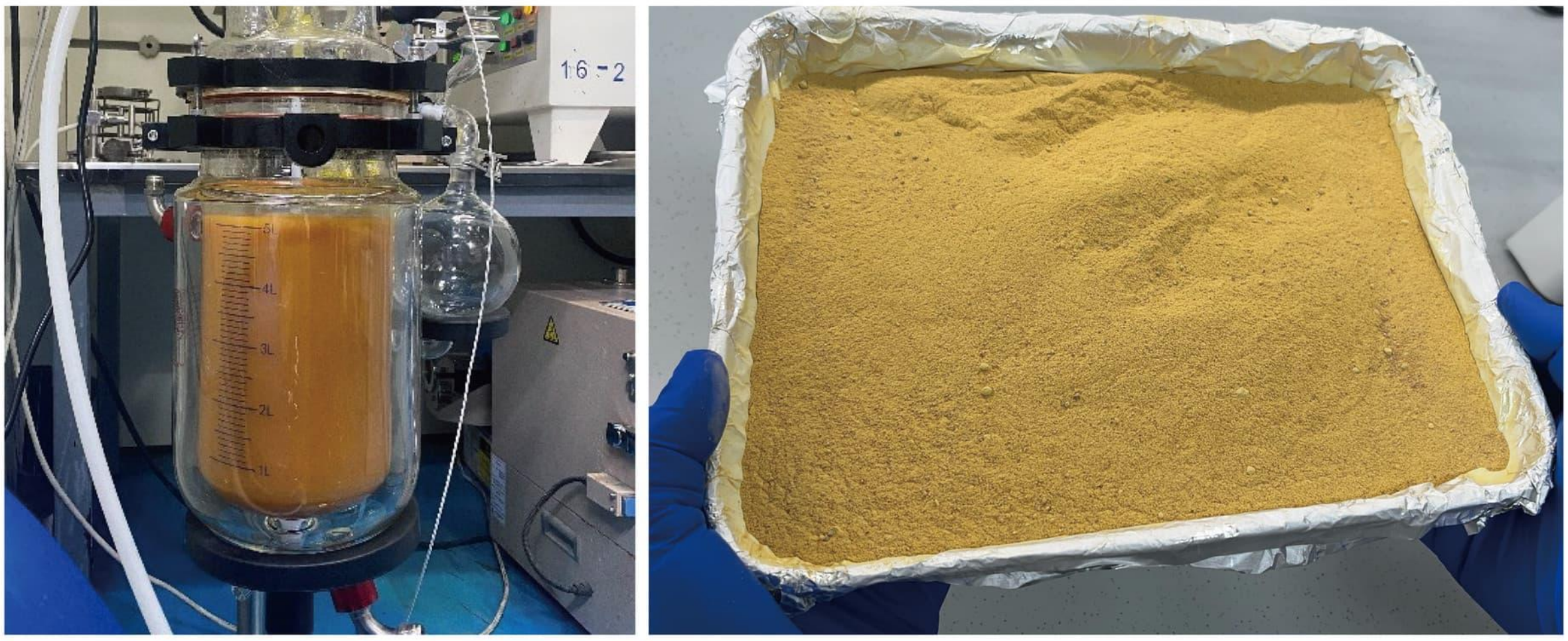

**Supplementary Fig. 3 |** Photographs of the reactor used for the synthesis of 2DPA precursors (left) and dried 2DPA precursors powder (right).

### Film fabrication and solid-state covalent welding

### Spin-Coating & Large-Area Film Fabrication Protocols

Film thickness was calibrated by varying the precursor concentration in DMF (5–350 mg $ml^{-1}$) and spin speed (1,000–3,000 r.p.m.). Centimetre-scale continuous films were prepared using the optimized coating conditions. Free-standing films were released by etching the 300-nm $SiO_2$ sacrificial layer with aqueous HF (10 wt%). The released films were rinsed thoroughly with ultrapure water (18.2 MΩ cm).

### Preparation and Transfer of Free-Standing 2DPA Films

Following welding, the precursor-coated $SiO_2$/Si wafers were immersed in hydrofluoric acid (10% (v/v)) to selectively dissolve the sacrificial oxide layer. Upon complete $SiO_2$ etching, the liberated 2DPA film spontaneously detached and floated onto the surface of an aqueous surfactant bath (sodium dodecyl sulfate), which was employed to minimize interfacial surface tension and prevent film curling or mechanical rupture during release. The floating membrane was carefully lifted from the liquid interface using a support frame (e.g., PTFE, soft silicone rubber or iron sheet) and allowed to air-dry under ambient conditions, yielding a uniformly adhered, relaxed suspended film. All HF-handling and transfer procedures were strictly conducted under certified fume-hood conditions with appropriate personal protective equipment.

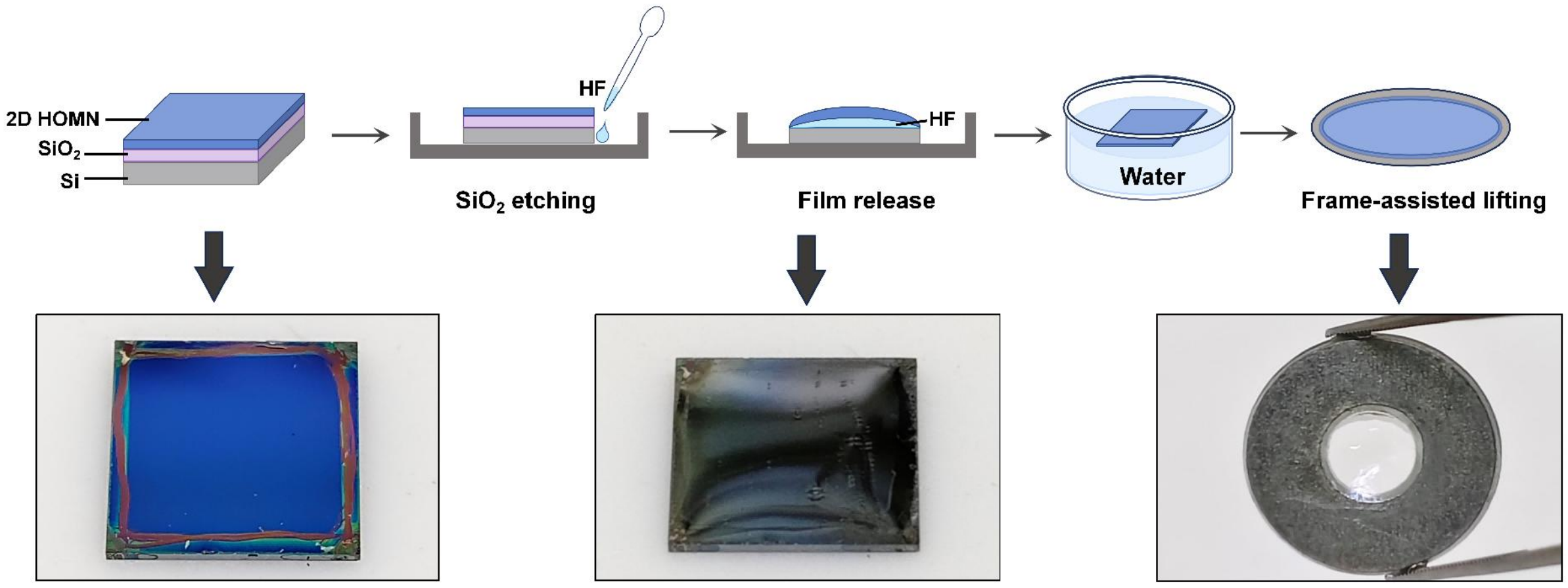

**Supplementary Fig. 4 |** Schematic flow diagram of the etching and transfer process for free-standing 2D HOMN-Ln films.

## Chemical characterization of edge-modified 2DPA

### Solution-state $^1$H NMR

Solution-state $^1$H NMR spectra were recorded using a Bruker AVANCE NEO 600-MHz spectrometer. Powder precursors were dissolved in DMSO-$d_6$. The grafting density of pendant amino-ester linkers was estimated from the integrated areas of the linker methylene resonances relative to the selected invariant framework resonances.

Full-range $^1$H NMR spectra (0-12 ppm) of L3-L8 precursors dissolved in DMSO-$d_6$. Insets show magnified alkyl proton regions for grafting density calculation.

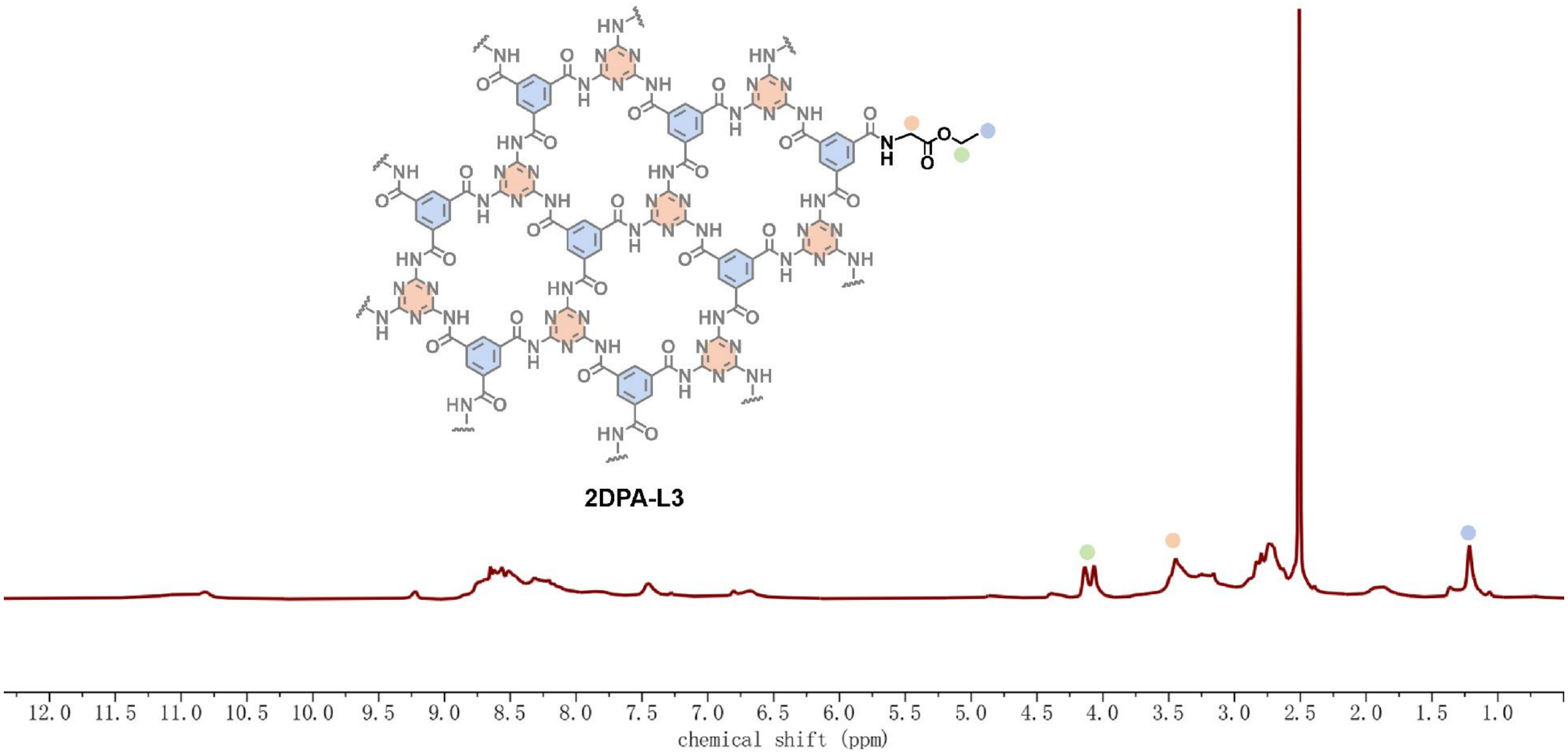

**Supplementary Fig. 5 |** $^1$H NMR spectrum of 2DPA-L3.

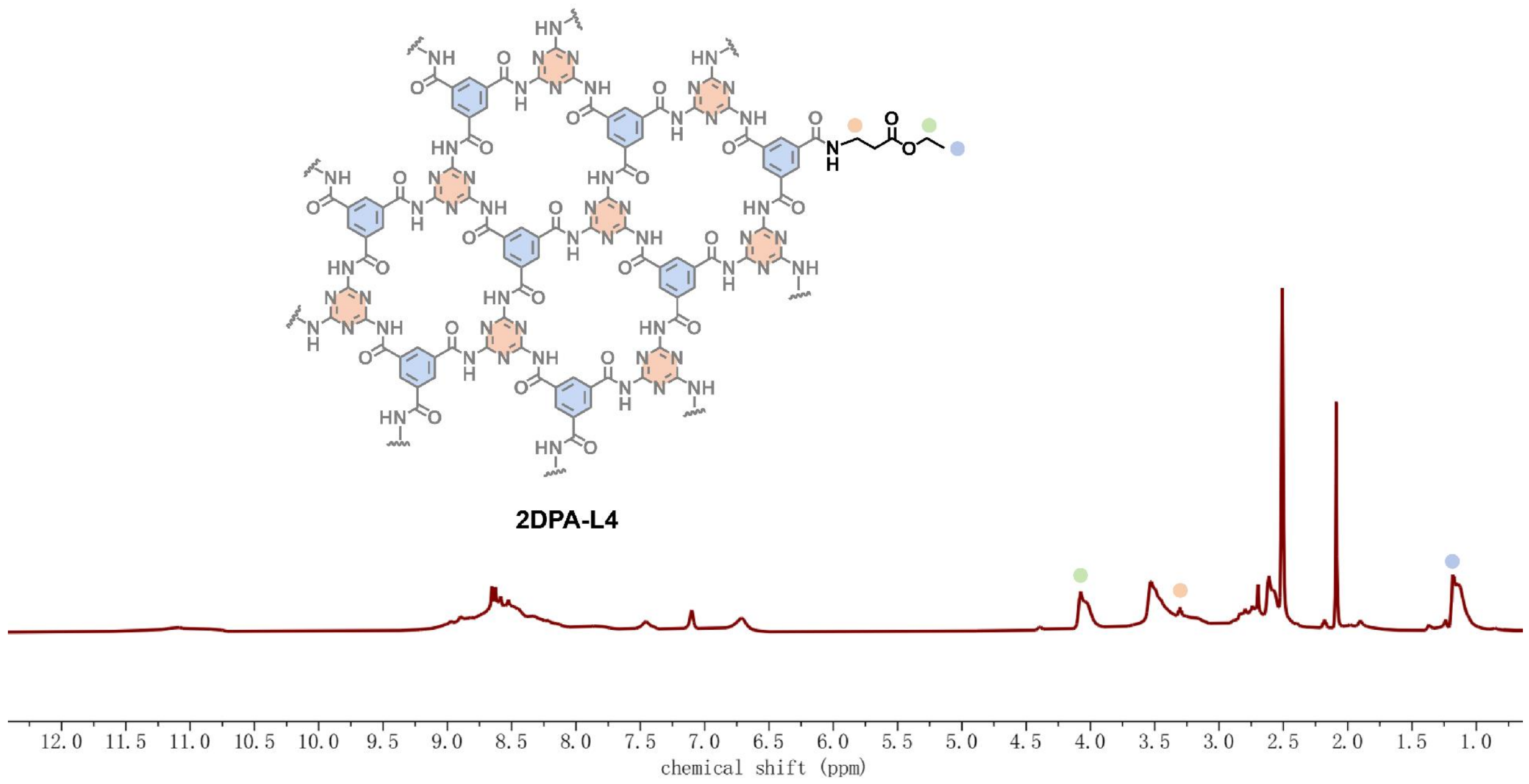

**Supplementary Fig. 6 |** $^{1}$H NMR spectrum of 2DPA-L4.

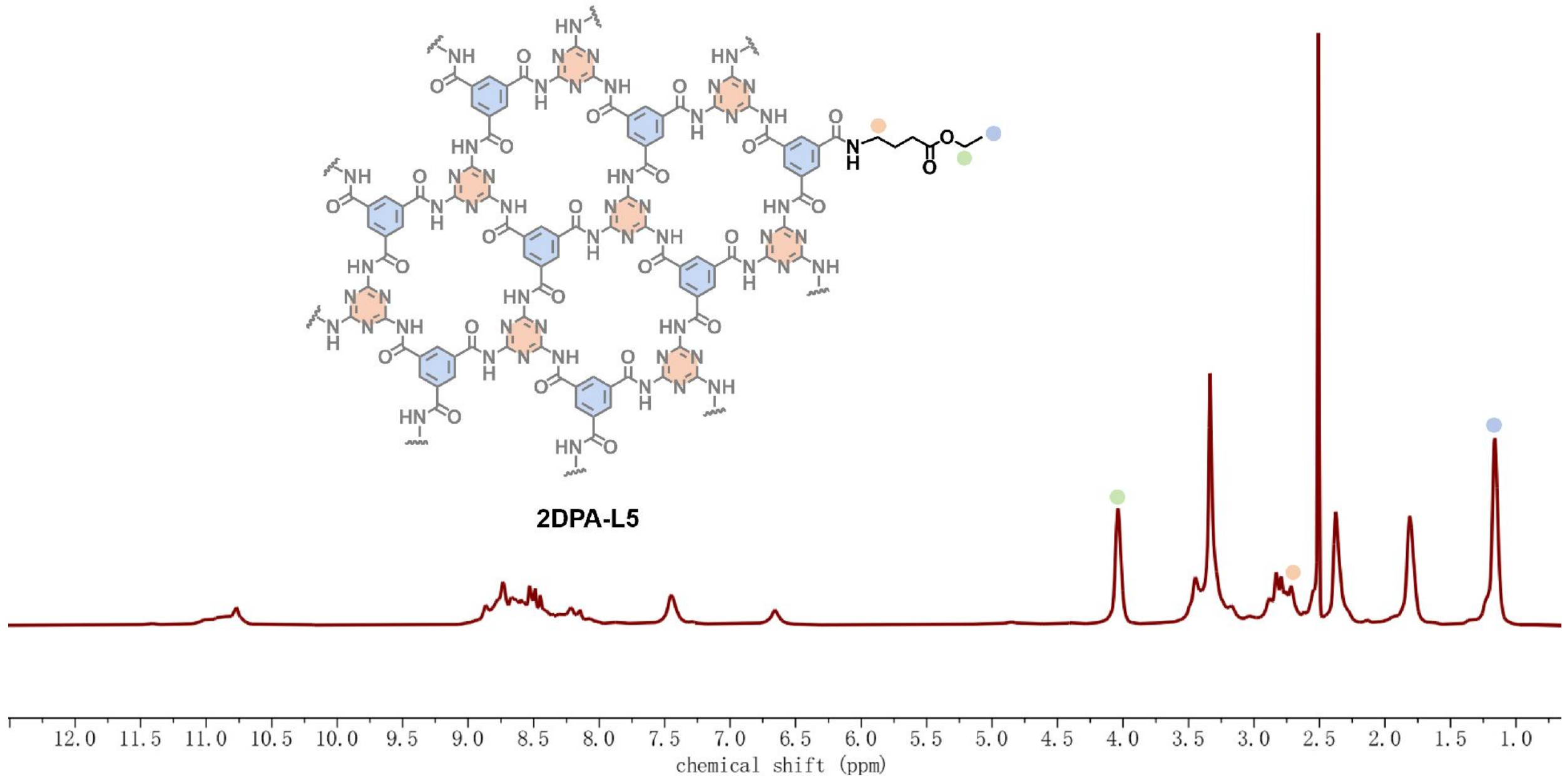

**Supplementary Fig. 7 |** $^1$H NMR spectrum of 2DPA-L5.

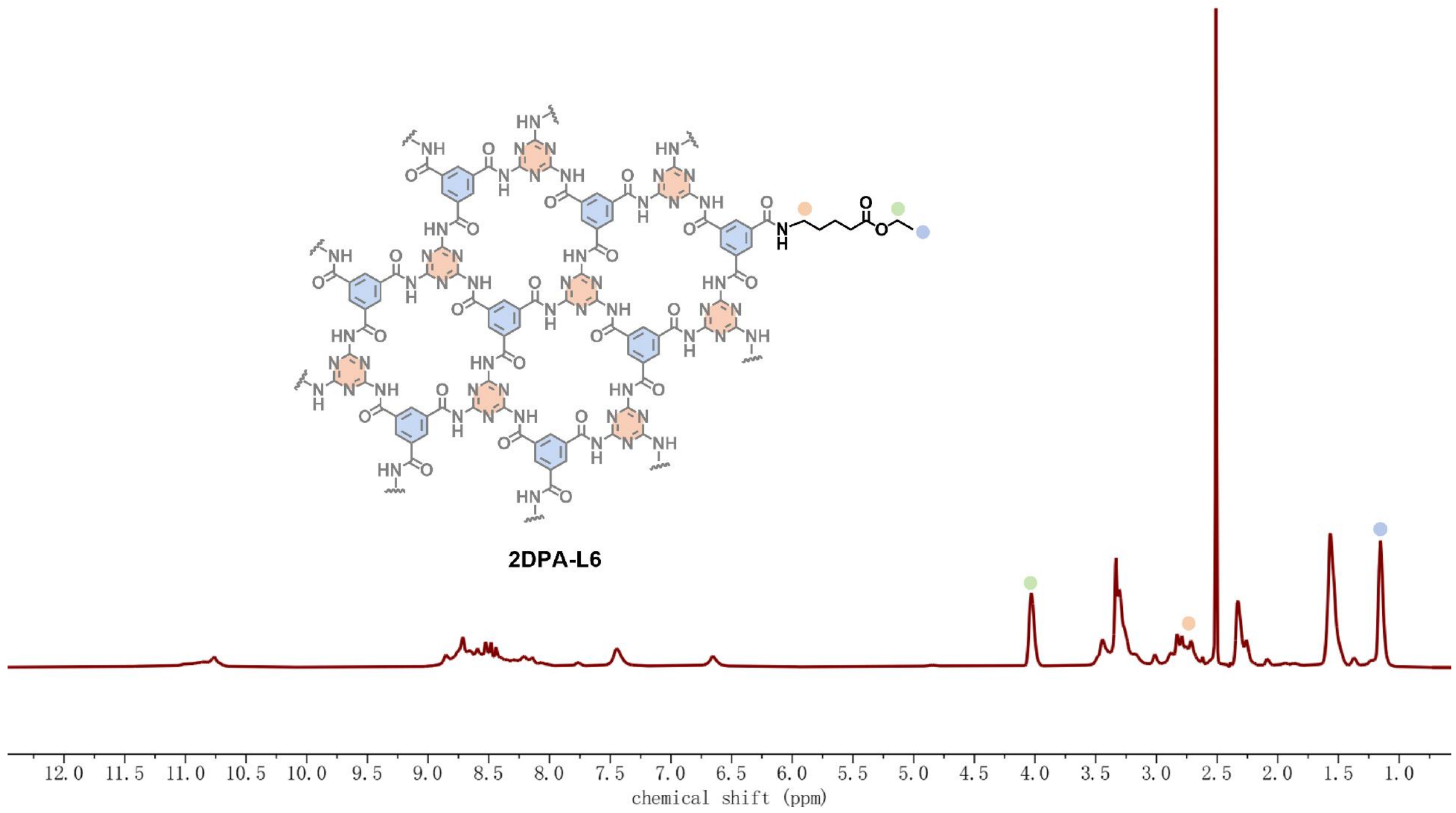

**Supplementary Fig. 8 |** $^1$H NMR spectrum of 2DPA-L6.

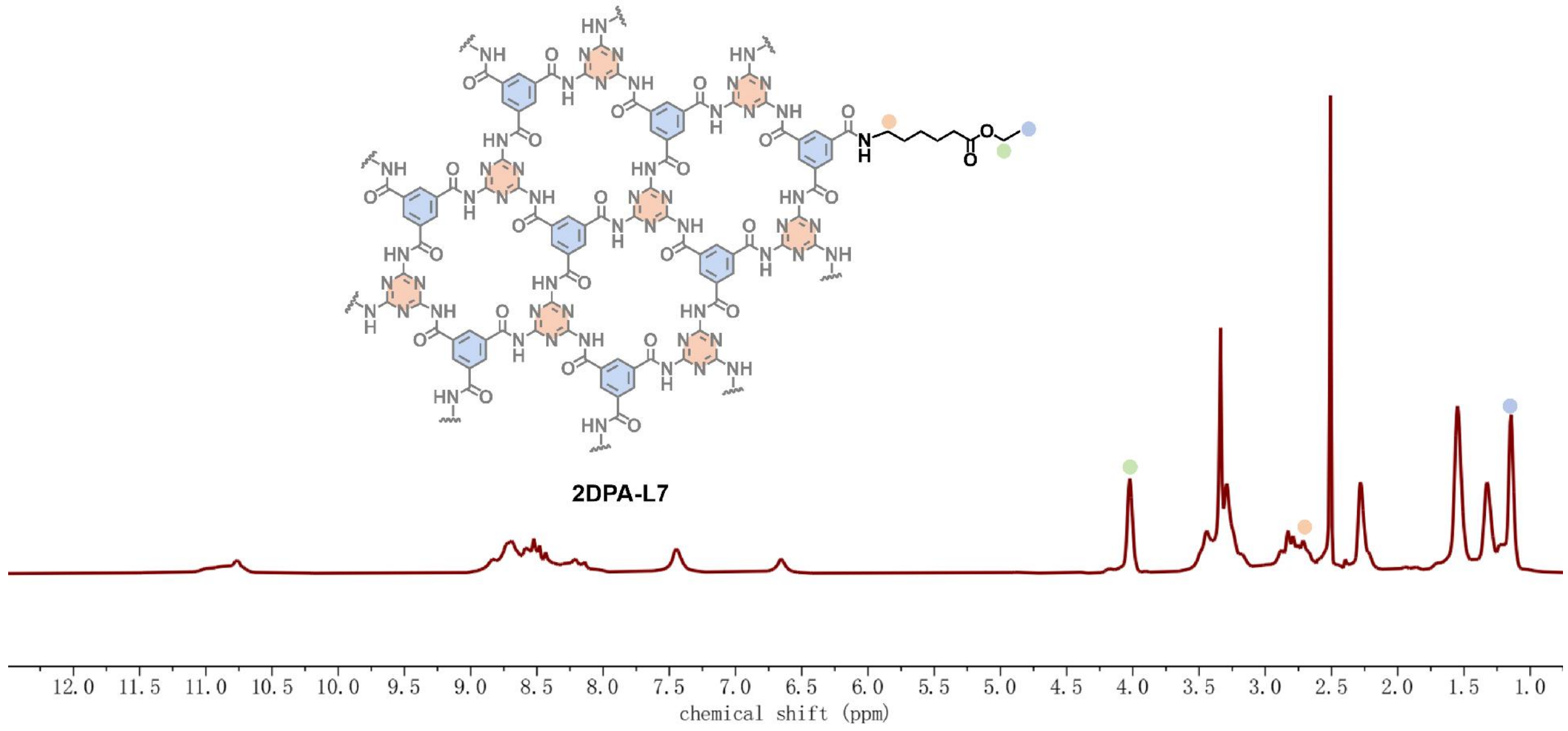

**Supplementary Fig. 9 |** $^1$H NMR spectrum of 2DPA-L7.

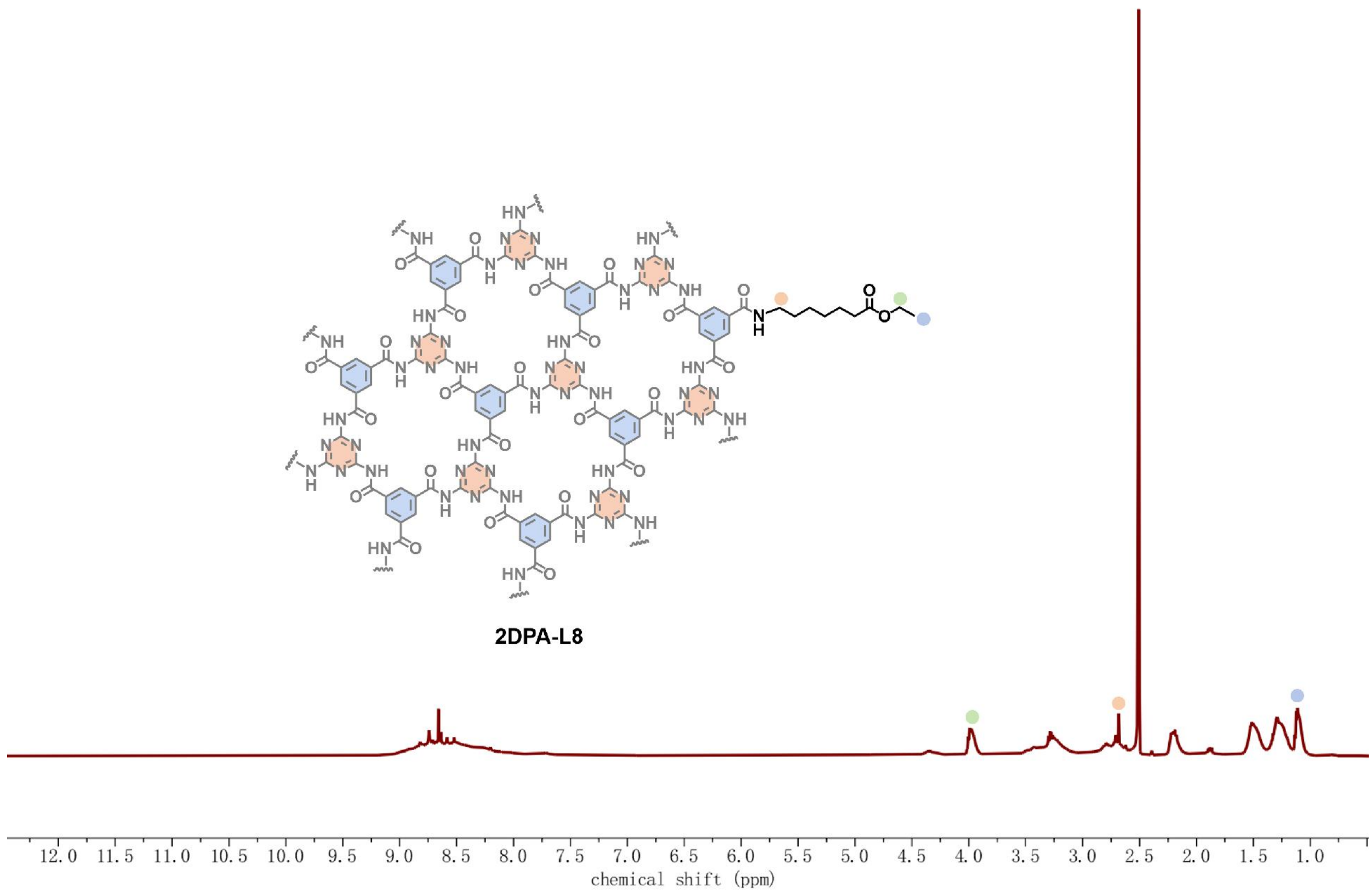

**Supplementary Fig. 10 |** $^{1}$H NMR spectrum of 2DPA-L8.

**Grafting density quantitative calculation**

The grafting density of pendant amino-ester linkers on 2DPA sheet edges was calculated via integrated proton peak area ratio:

$$\text{Grafting Density} = \frac{\text{Integra (NH} - \text{CH}_2\text{) / 2}}{\text{Integral Ar / 9}} \tag{1}$$

The benzene ring of Benzene-1,3,5-tricarbonyl trichloride proton signals at 6.0–8.0 ppm served as internal invariant reference. All integral calculations were averaged over three independent NMR scans of the same sample powder.

$$\text{Degree of Modification} = \frac{\text{Integra (NH} - \text{CH}_2\text{) / 2}}{\text{Integral (CH}_2 - \text{CH}_3\text{) / 2 + Integra (COOH)}} \tag{2}$$

**Supplementary Table 2 |** Grafting density and degree of modification of pendant amino-ester linkers on 2DPA sheet edges quantified by $^1$H NMR peak-area integration

| Sample | Grafting Density | Degree of Modification |
| --- | --- | --- |
| 2DPA-L3 | 0.21 | 0.73 |
| 2DPA-L4 | 0.11 | 0.72 |
| 2DPA-L5 | 0.17 | 0.68 |
| 2DPA-L6 | 0.12 | 0.48 |
| 2DPA-L7 | 0.14 | 0.67 |
| 2DPA-L8 | 0.08 | 0.84 |

**Polarized Photoluminescence (PL) Characterization**

**Optical configuration and acquisition**

Photoluminescence (PL) spectra were recorded with an Oxford Instruments alpha300 R confocal Raman microscope equipped with a Zeiss 100× objective (numerical aperture, 0.9), a UHTS600 spectrometer and a 300 grooves $mm^{-1}$ grating. Excitation was provided by a 532 nm laser operated at 0.3 mW at the sample, and each spectrum was integrated for 1 s. All measurements were performed in air at room temperature.

The specimen remained fixed while the half-wave plate was mechanically rotated through 180°, corresponding to a 360° rotation of the incident linear polarization. Here, theta denotes the excitation-polarization angle rather than the mechanical waveplate angle. No analyzer was used in the emission path, so the experiment measured excitation-polarization-dependent PL intensity rather than emission polarization.

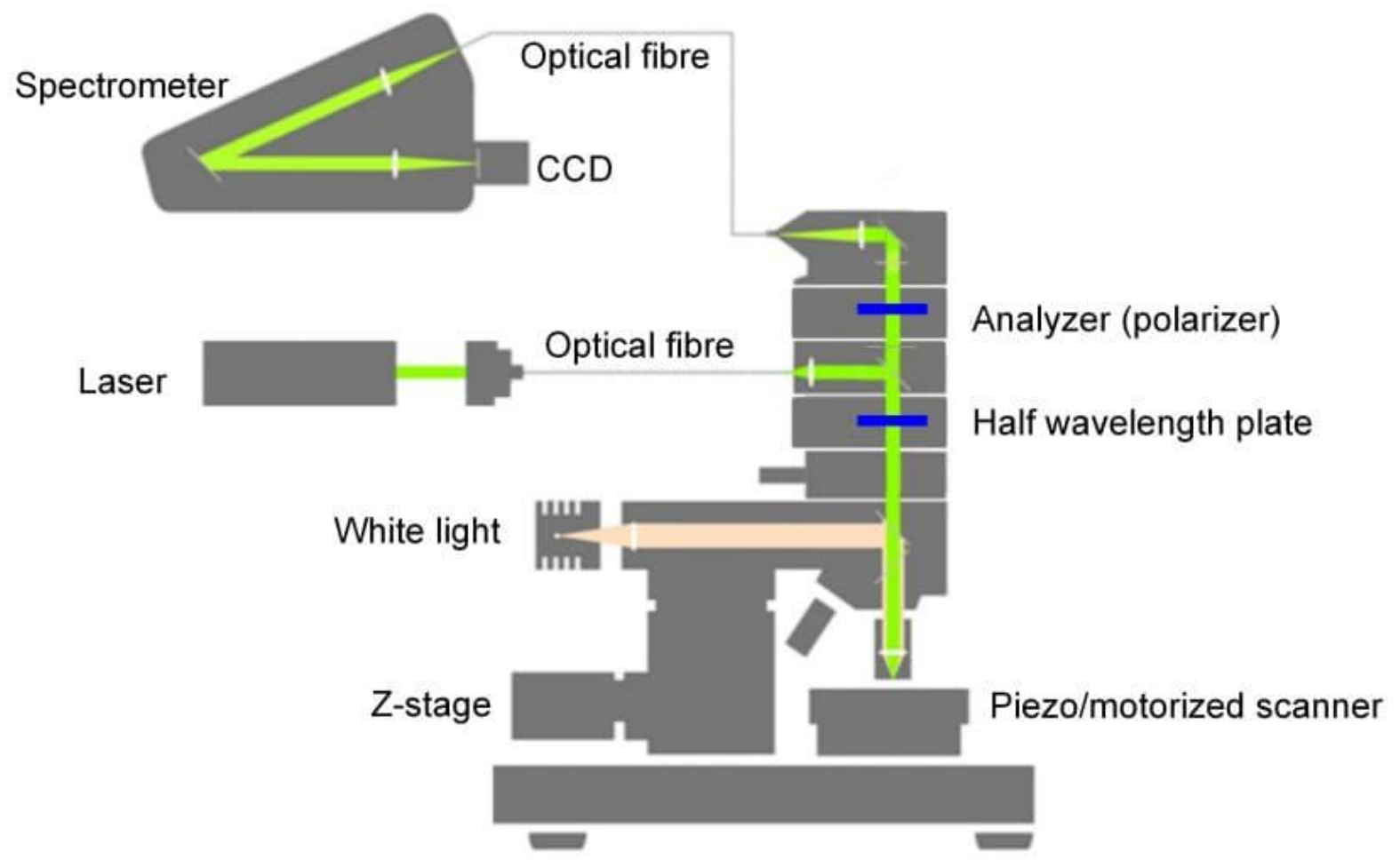

**Supplementary Fig. 11 |** Optical configuration for polarization-dependent PL acquisition. The half-wave plate rotated the incident polarization. No emission analyzer was used.

## Sample preparation and measurement geometries

Unwelded 2DPA-L4 and welded 2D HOMN-L4 films were spin-coated onto $SiO_2$/Si substrates. For face-on measurements, the specimen was mounted flat on a glass slide and excited through the film surface. For edge-view measurements, the film was sectioned and a freshly cut edge was mounted perpendicular to the glass slide. These two configurations probe the angular response from complementary sample orientations.

## Normalization and angular fitting

The PL intensity recorded at each excitation angle was normalized to the maximum value within the corresponding angular scan. The normalized response was fitted with a twofold model derived from the $\cos^2$ dependence of linear excitation coupling:

$$I_{norm}(\theta) = y_0 + (1 - y_0)cos^2[\pi(\theta - \theta_0)/180] \quad (3)$$

Here, $I_{norm}$ is the normalized integrated PL intensity, $y_0$ is the fitted minimum-to-maximum intensity ratio and $\theta_0$ is the excitation angle at maximum intensity. The adjusted $R^2$ value was used to describe the agreement between the measured angular trace and the twofold model.

## Geometry dependent photoluminescence response

The $\cos^2$ dependence describes photoselection when an anisotropic ensemble couples preferentially to one direction of the incident electric field. A two-lobed pattern with 180° periodicity therefore indicates a preferred projection of the absorption transition dipole moments onto the polarization-rotation plane. A nearly angle-independent trace indicates weak anisotropy in the probed projection and can arise from orientational averaging or a nearly isotropic in-plane distribution.

Unwelded 2DPA-L4

$$I_{\mathrm{norm}}(\theta)=0.0097+0.99cos^2[\pi(\theta\text{-}92.77°)/180°] \quad (4)$$

2D HOMN-L4

$$I_{\mathrm{norm}}(\theta)=0.0169+0.98cos^2[\pi(\theta\text{-}92.81°)/180°] \quad (5)$$

The unwelded 2DPA-L4 film displayed a pronounced twofold response in the edge-view geometry (Supplementary Fig. 12). The fit gave a maximum-intensity angle of 92.77°, a normalized minimum intensity of 0.0097 and an adjusted $R^2$ of 0.977, indicating a pronounced twofold polarization response. By contrast, the face-on trace remained close to unity throughout the angular scan.

The welded 2D HOMN-L4 film retained the same geometry-dependent pattern (Supplementary Fig. 13). Its edge-view response gave a maximum-intensity angle of 92.81°, a normalized minimum intensity of 0.0169 and an adjusted $R^2$ of 0.955. The face-on response again showed only weak angular modulation.

## Apparent modulation contrast

The apparent modulation contrast was calculated from the fitted extrema, $I_{min} = y_0$ and $I_{max} = 1$:

$$M_{app} = \frac{I_{max} - I_{min}}{I_{max} + I_{min}} = \frac{1 - y_0}{1 + y_0} \quad (6)$$

$M_{app}$ was 0.981 for the unwelded 2DPA-L4 film and 0.967 for 2D HOMN-L4 film.

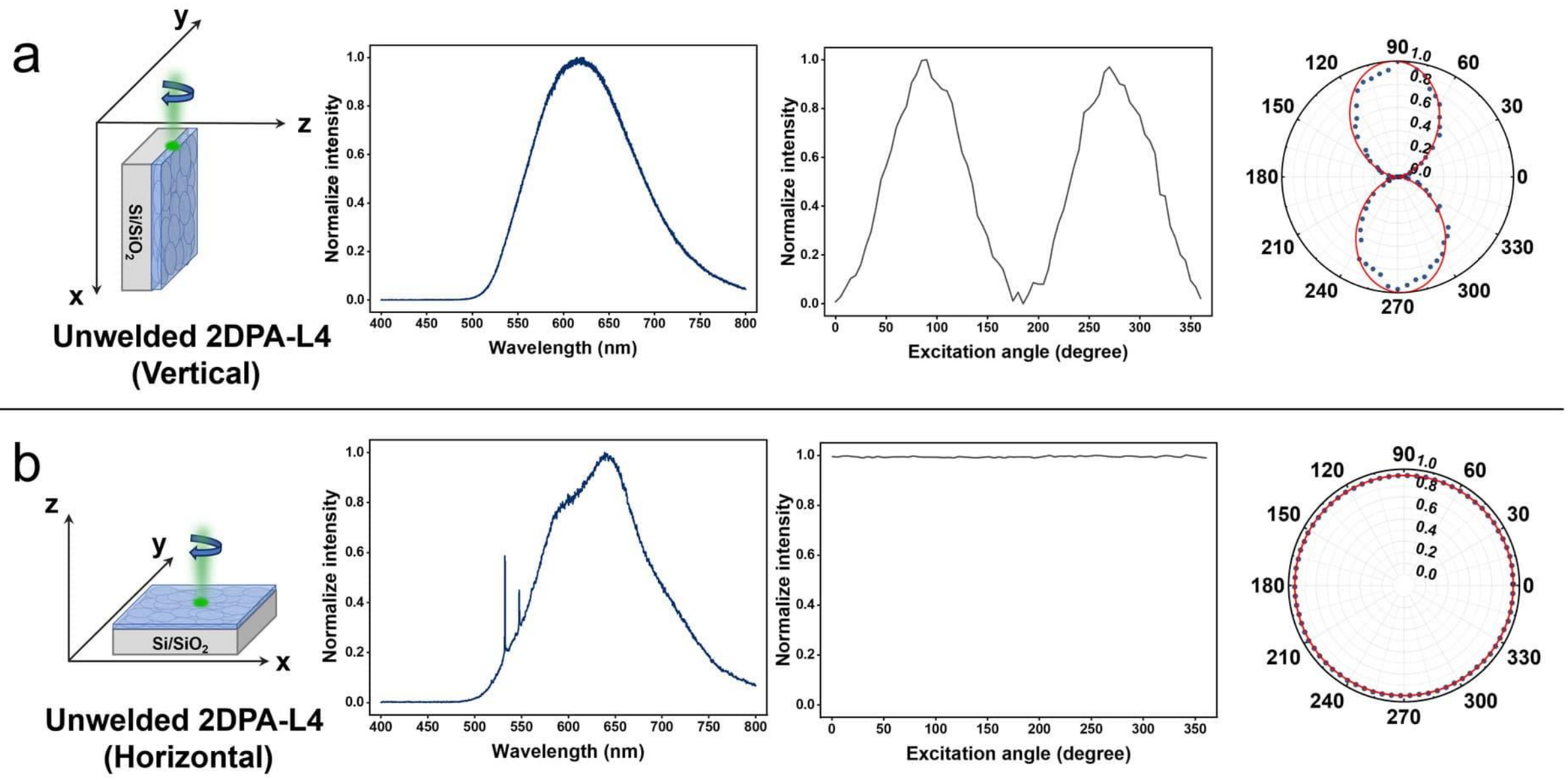

**Supplementary Fig. 12 |** Polarized photoluminescence of an unwelded 2DPA-L4 film measured in vertical and horizontal geometries. a, Edge-view (vertical) measurement. b, Face-on (horizontal) measurement. Each row shows the measurement geometry, PL spectrum, normalized angular trace and polar representation; blue points are measured intensities and red curves are $\cos^2$ fits.

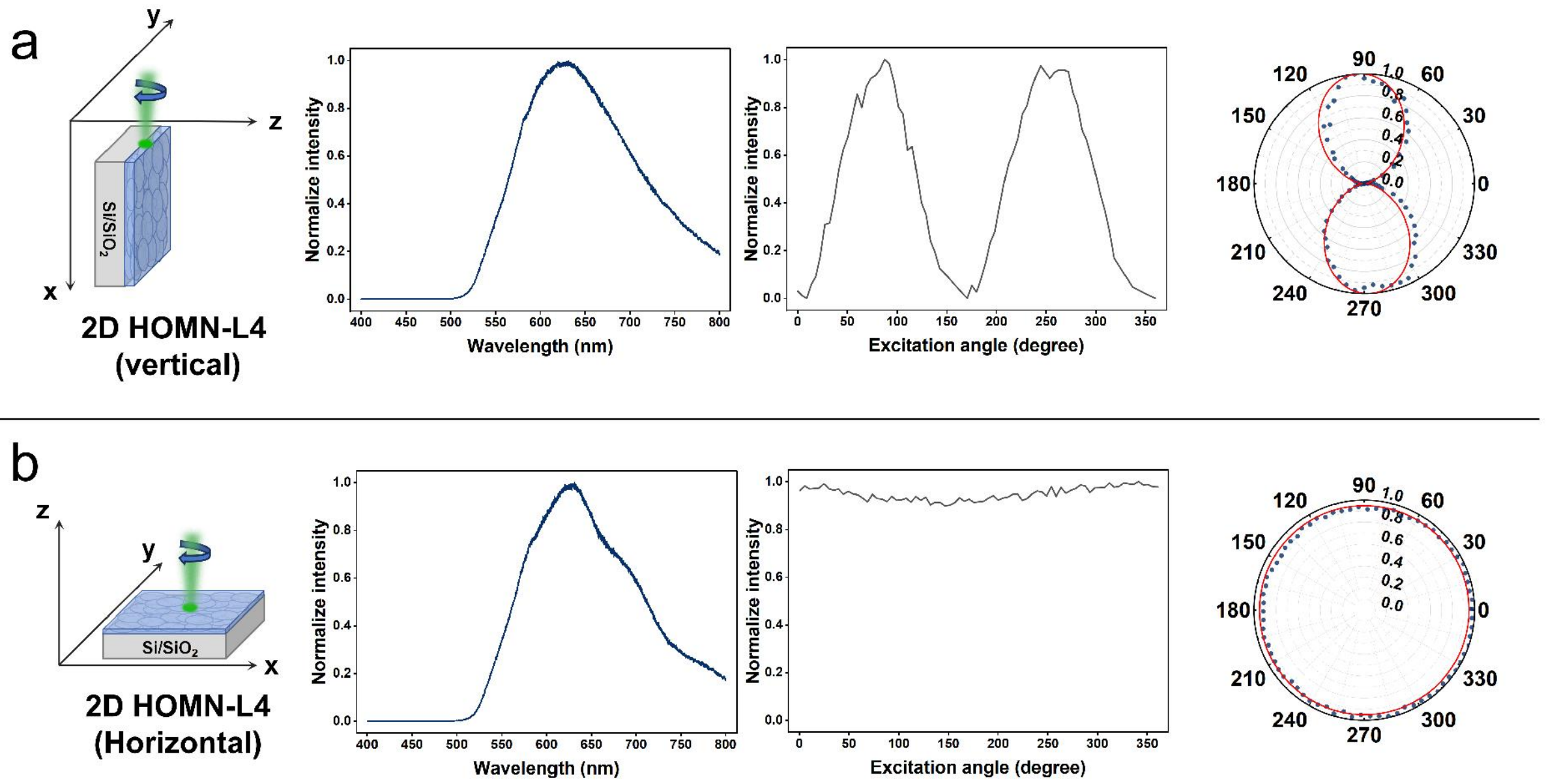

**Supplementary Fig. 13 |** Polarized photoluminescence of a 2D HOMN-L4 film measured in vertical and horizontal geometries. a, Edge-view (vertical) measurement. b, Face-on (horizontal) measurement. Each row shows the measurement geometry, PL spectrum, normalized angular trace and polar representation; blue points are measured intensities and red curves are $cos^2$ fits.

**Raman spectrum**

Compared with the unwelded 2DPA-L4 film, the welded 2D HOMN-L4 film exhibited an enhanced Raman response in both the amide III region at approximately 1200–1300 cm$^{-1}$ and the higher-wavenumber amide I–II region at approximately 1500–1700 cm$^{-1}$. The concurrent enhancement of the amide III and amide I–II bands provides mutually consistent spectroscopic evidence for an increased contribution from amide linkages after welding[19].

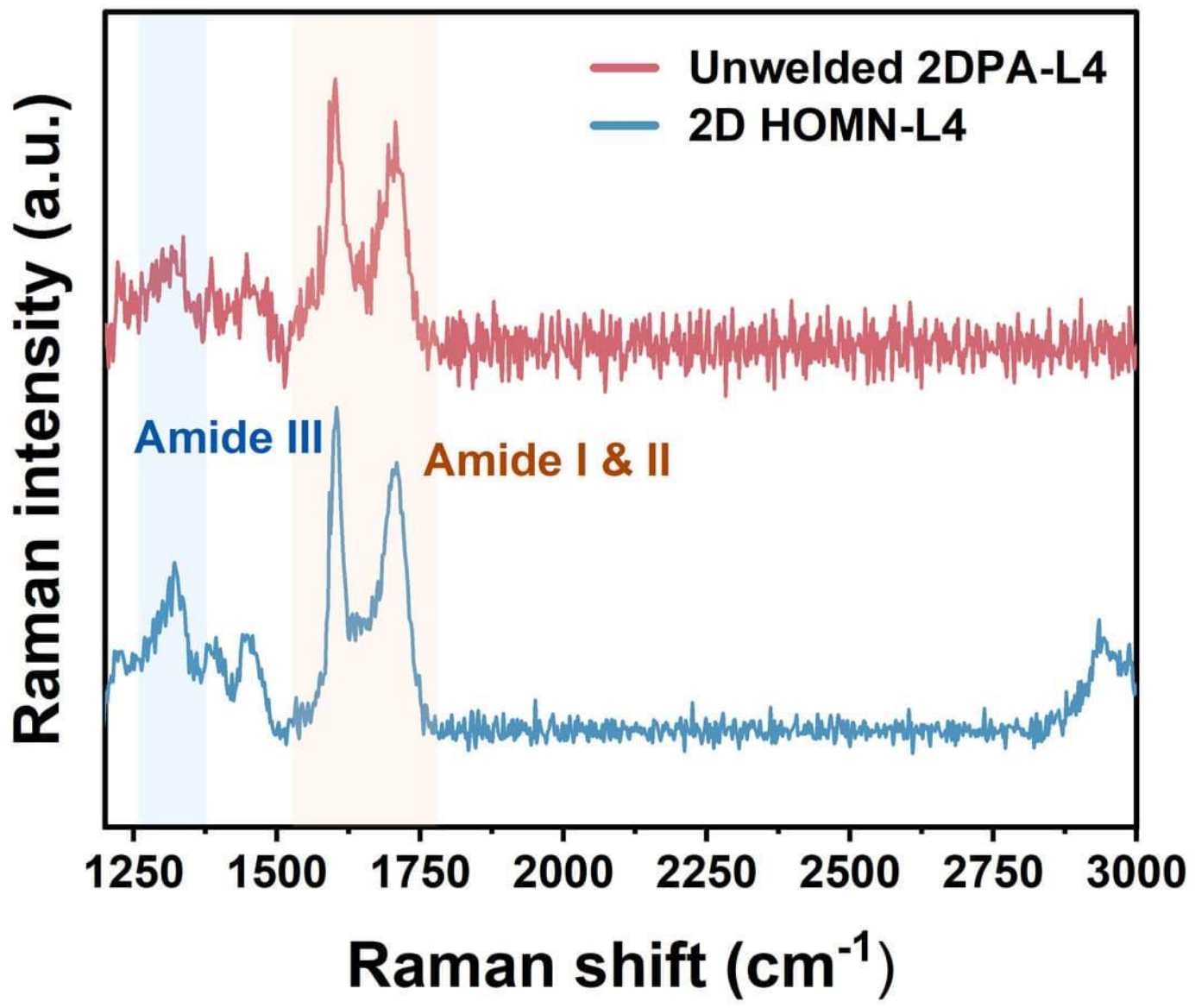

**Supplementary Fig. 14 |** Raman spectra of the unwelded 2DPA-L4 film and 2D HOMN-L4 film.

**Gel permeation chromatography**

Gel permeation chromatography (GPC) was used to compare the elution behaviour of 2DPA-IPA and 2DPA-L4 (Supplementary Fig. 15). The chromatograms showed closely overlapping principal elution peaks, with peak retention times of 9.97 min (598.2 s) for 2DPA-IPA and 9.96 min (597.6 s) for 2DPA-L4. The difference was only 0.01 min (0.6 s), indicating no appreciable shift in the principal elution peak.

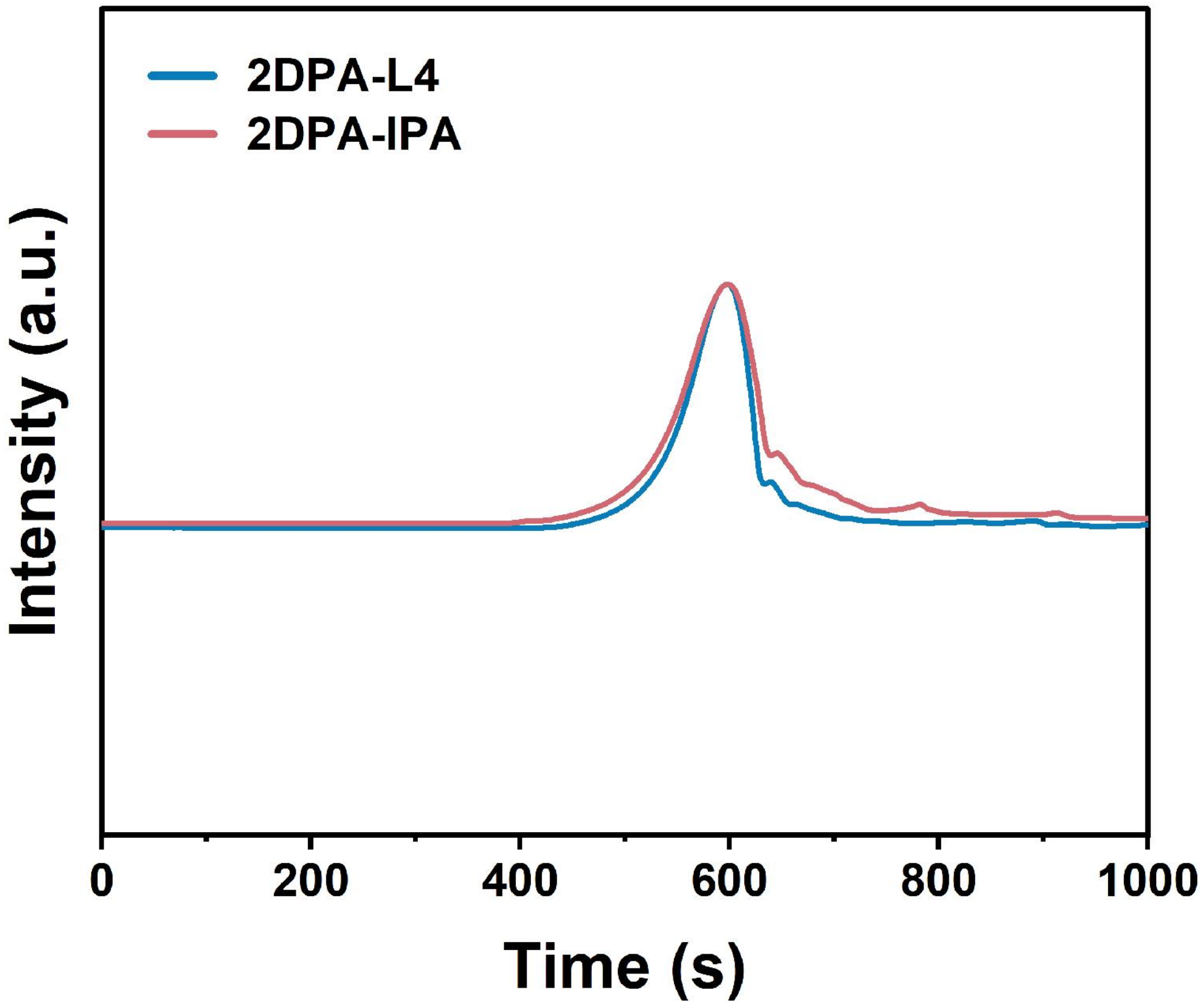

**Supplementary Fig. 15 |** Gel-permeation chromatography (GPC) of 2DPA-L0 films and 2DPA-L4 films.

**Solubility Test**

To evaluate the solvent resistance of the covalently welded network, 2D HOMN-L4 films fragments were immersed in a range of polar aprotic, chlorinated, ketone, ester, ether and fluorinated alcohol solvents. The films remained macroscopically visible as solid fragments in all tested solvents, with no complete dissolution observed. This qualitative resistance to chemically diverse solvents is consistent with the formation of a covalently interconnected network.

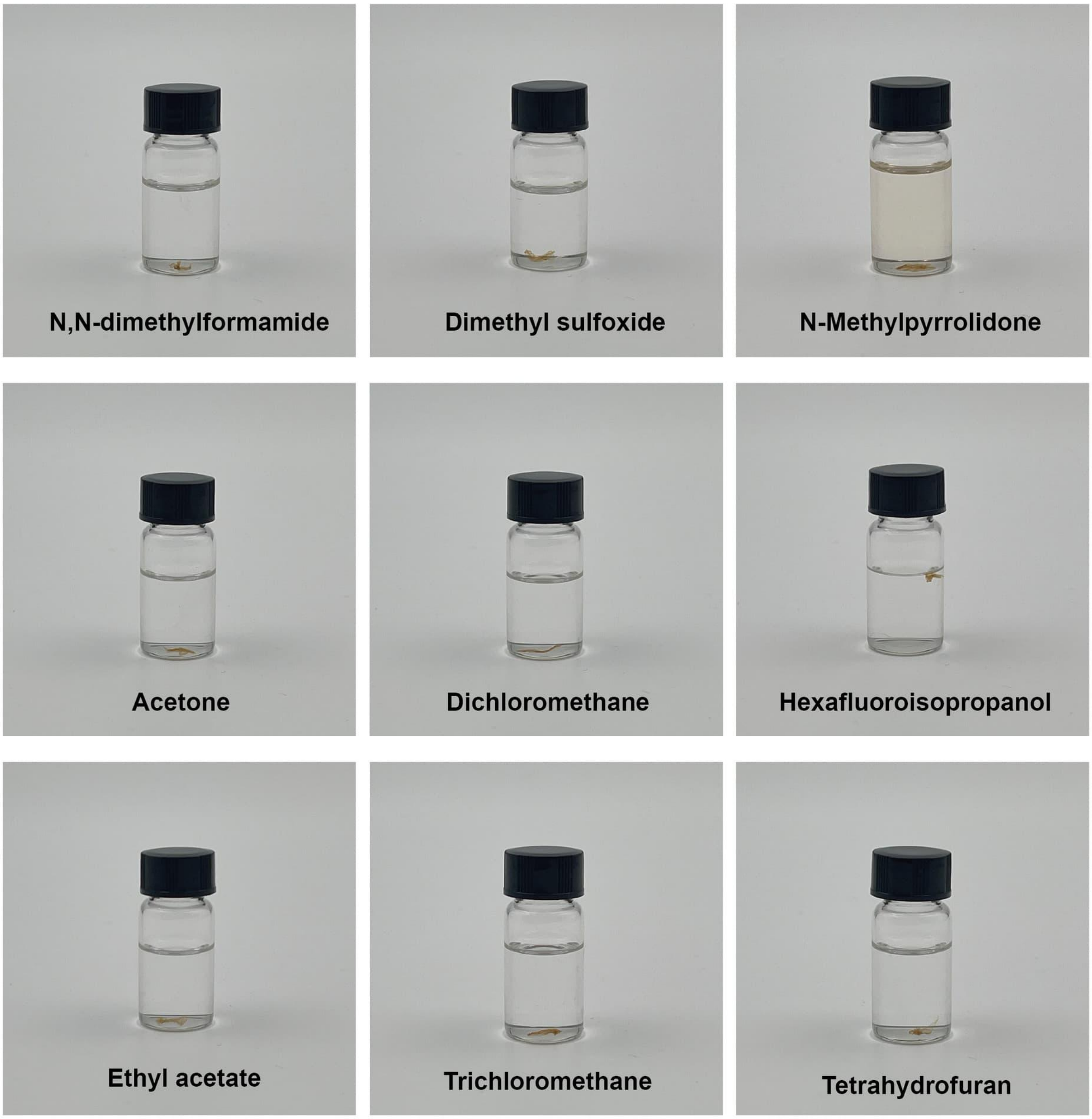

**Supplementary Fig. 16 |** Solubility tests of 2D HOMN-L4 films in various solvents.

### Step Profilometry

Film thickness was measured by stylus profilometry using a KLA-Tencor AlphaStep D-600 profilometer. This method was applied to relatively thick films that exceeded the accessible range of atomic force microscopy. The film was gently scratched with a scalpel on the supporting substrate (silicon wafer or quartz glass) to locally expose the underlying substrate and create a well-defined step edge. The stylus was then scanned across the scratch, and the film thickness was determined from the height difference between the film surface and the exposed substrate region.

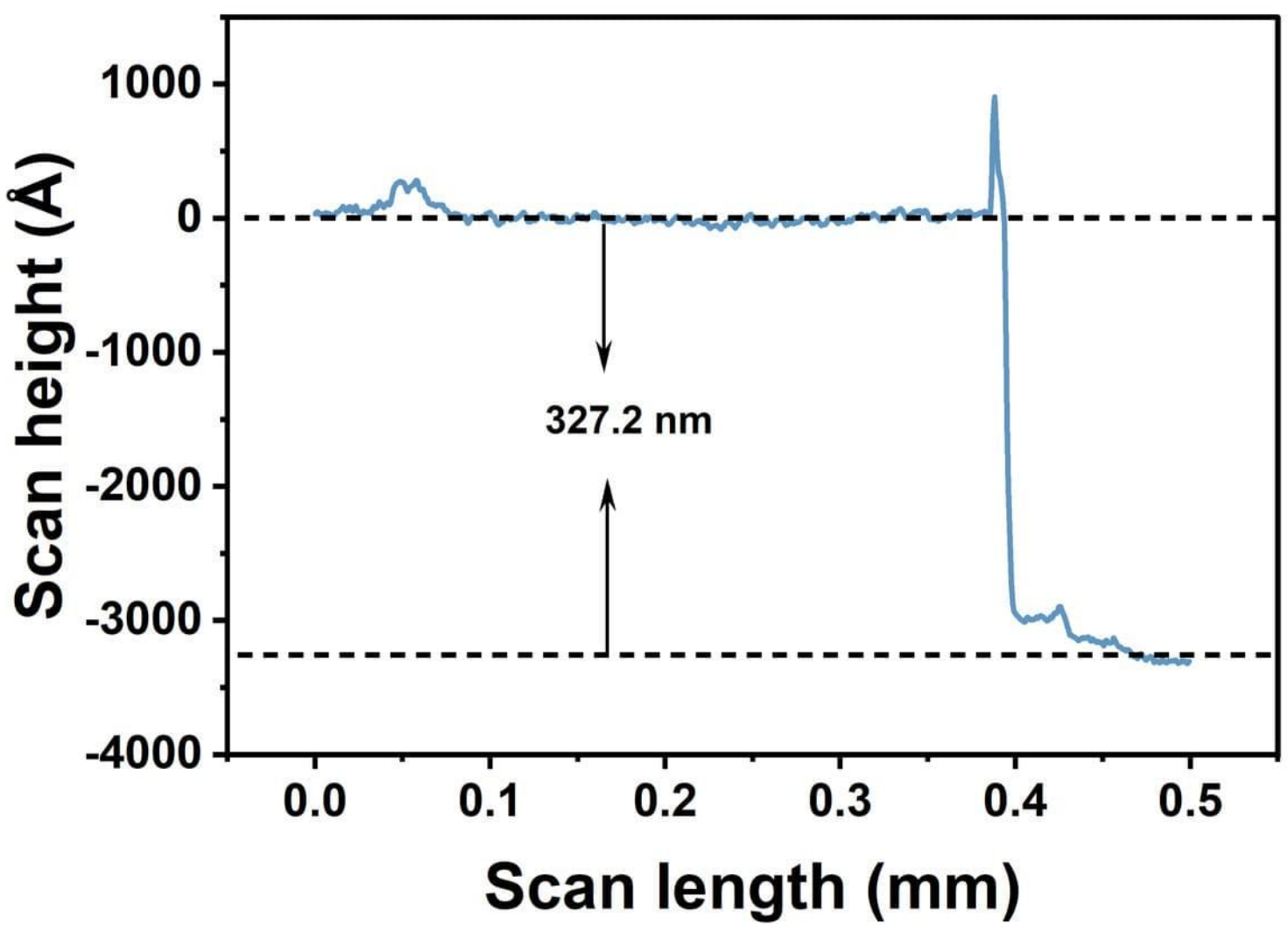

**Supplementary Fig. 17 |** Step-profiler measurement of 2D HOMN-L4 films thickness.

### Atomic Force Microscopy (AFM)

AFM imaging was conducted on an Oxford Cypher VRS1250 instrument in tapping mode with Multi75DLC probes. The lateral dimension, monolayer thickness and surface morphology of individual 2DPA molecular sheets and layered nanofilms were extracted from height maps. Raw AFM data were processed with native AR 16.33.234 software or open-source Gwyddion software. The morphology of AFM tips was inspected via scanning electron microscopy prior to imaging tests.

Film thickness was predominantly determined by atomic force microscopy (AFM). The film was first spin-coated onto a silicon wafer. A scratch was then introduced through the film using a sharp blade to expose the underlying silicon substrate. AFM height mapping was performed across the step edge between the exposed substrate and the adjacent intact film, and the thickness was extracted from the resulting height profile.

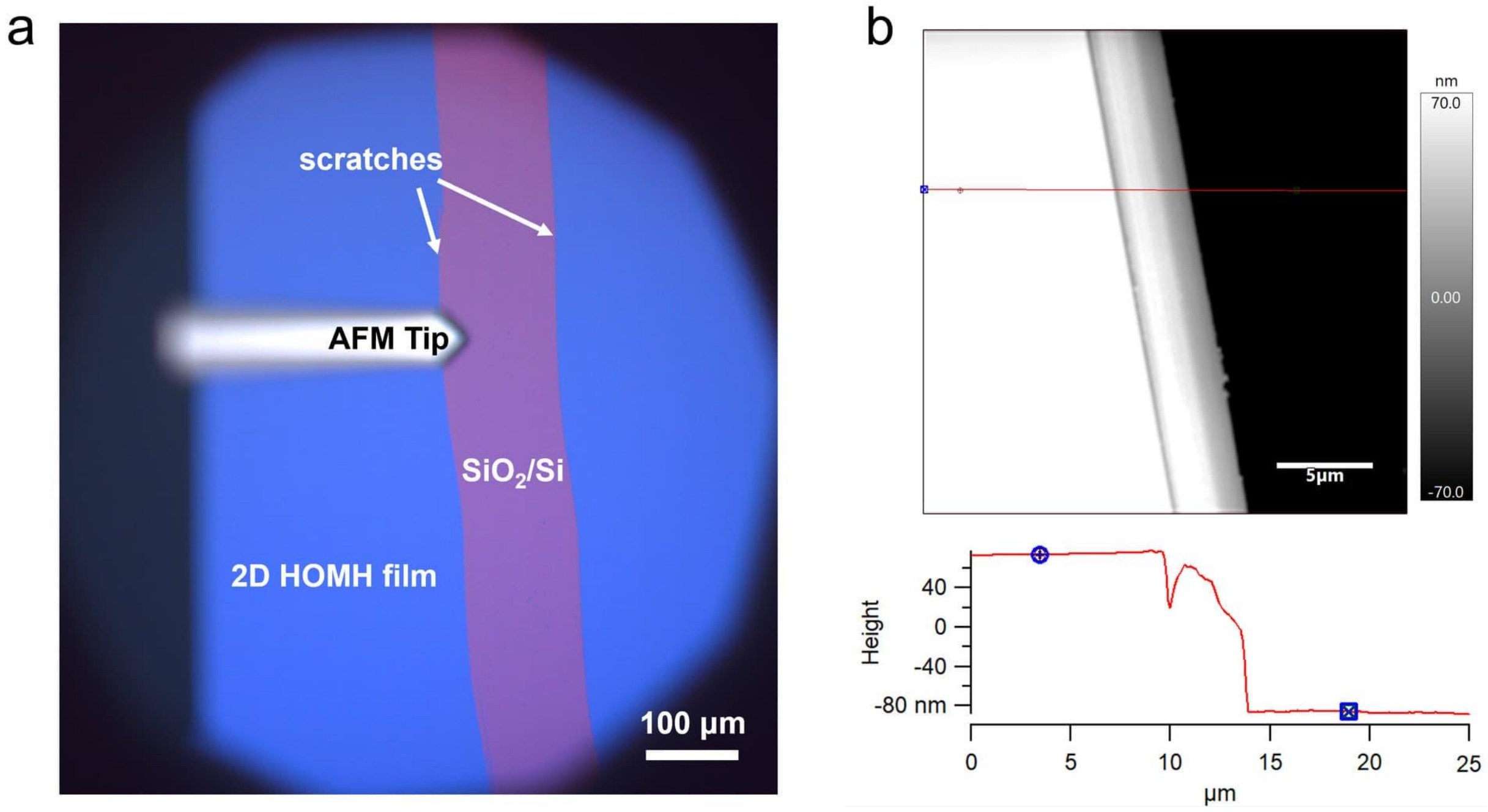

**Supplementary Fig. 18 | a,** Optical micrograph of a scratched film on a silicon wafer. **b,** AFM topography and corresponding height profile along the red line.

## Surface roughness measurement

AFM height images revealed continuous surfaces with sub-nanometre average roughness before and after covalent welding. Over the scan area, the root-mean-square roughness Rq decreased from 0.836 nm for unwelded 2DPA-L4 films to 0.636 nm for 2D HOMN-L4 films, while the arithmetic mean roughness Ra decreased from 0.624 to 0.517 nm. The reduced prominence of surface protrusions indicates modest surface smoothing following welding, consistent with structural densification and local rearrangement of the stacked molecular sheets.

**Supplementary Table 3 |** AFM roughness of the unwelded 2DPA-L4 films and 2D HOMN-L4 films.

| Sample | $R_a$ (nm) | $R_q$ (nm) |
|---|---|---|
| Unwelded 2DPA-L4 films | 0.624 | 0.836 |
| 2D HOMN-L4 films | 0.517 | 0.636 |

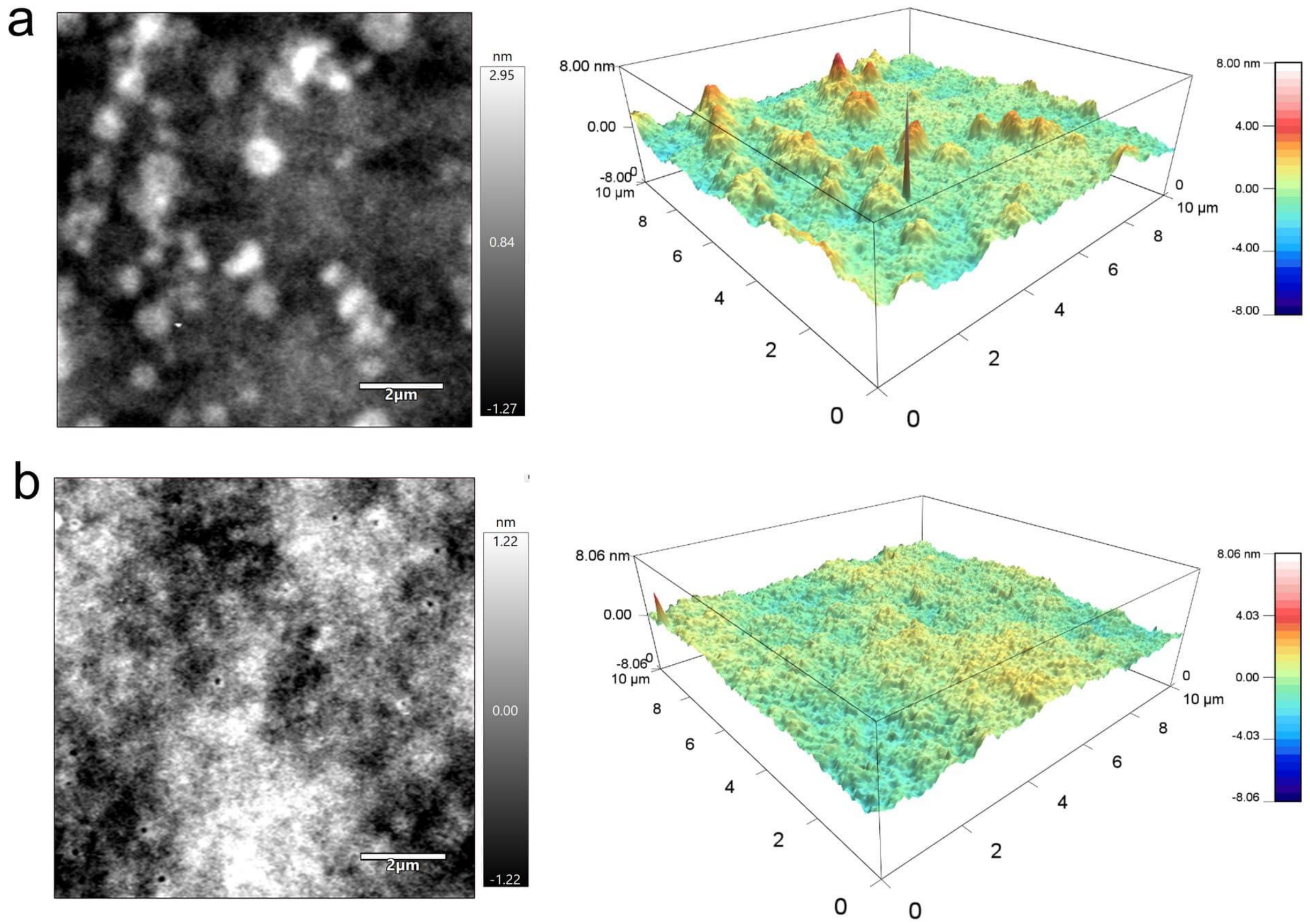

**Supplementary Fig. 19 |** Surface morphology and roughness of unwelded 2DPA-L4 films **(a)** and 2D HOMN-L4 films **(b)**.

## Validation of covalent welding

### Solid-State Nuclear Magnetic Resonance (Solid-State NMR)

We prepared films using melamine selectively enriched with $^{15}N$ (99%) at the amino($NH_2$) sites and investigated them by solid-state NMR spectroscopy. $^{15}N$ and $^{1}H$ solid-state NMR experiments of unwelded L4 films and fully welded 2D HOMN-L4 films were performed on a Bruker Avance III HD 400 MHz NMR spectrometer ($^{1}H$, 400.13 MHz; $^{15}N$, 40.55 MHz) using a 3.2 mm magic angle spinning (MAS) probe[46]. $^{1}H$ spectra were acquired under the spinning rate of 15 kHz, using a hahnecho sequence, with a recycle delay of 3-4 s, which is slightly 5 times longer than their spin-lattice relaxation times. $^{15}N$ spectra were collected using cross-polarization (CP) under MAS of 15 kHz. The recycle delay for CP was set to 0.9-1 s. Two-dimensional $^{15}N$-$^{1}H$ heteronuclear correlation (HETCOR) solid-state NMR measurements were acquired with a cross-polarization-based pulse sequence (CP contact time = 500 μs). FSLG homonuclear decoupling was applied during the $^{1}H$ evolution period under magic-angle spinning at 15 kHz. The $^{1}H$ was referenced to those of adamantane at 1.8 ppm. The $^{15}N$ signals were referenced to the α-amino nitrogen ($N_{\alpha}$, $^{15}NH_3^{+}$) resonance of uniformly $^{15}N$-labeled L-arginine at 33.0 ppm.

**DFT calculations of NMR chemical shifts.**

All quantum-chemical calculations were performed using Gaussian 16, Revision C.01, ES64L-G16RevC.01[47]. Molecular structures were fully optimized at the TPSSh/def2-SVP level without imposing symmetry constraints. Tight geometry and SCF convergence criteria were used together with an ultrafine numerical integration grid. The Gaussian route section used for geometry optimization and frequency calculations was TPSSh/def2-SVP Opt = (CalcFC, Tight, MaxCycles = 300) Freq NoSymm Int = UltraFine SCF = (XQC, Tight, MaxCycle = 512). Harmonic vibrational-frequency calculations were performed at the same level of theory to characterize the optimized structures. The optimized geometries were subsequently used for NMR calculations without further structural relaxation.

The $^{15}N$ and $^{1}H$ magnetic shielding tensors were calculated using the GIAO method at the TPSSh/def2-SVP level with no symmetry constraints, an ultrafine integration grid, and tight SCF convergence criteria. The isotropic shielding constant was calculated according to $\sigma_{iso} = \frac{(\sigma 11+\sigma 22+\sigma 33)}{3}$. Ammonia ($NH_3$) was used as the reference compound for $^{15}N$, with $\delta(^{15}N) = 0$ ppm, and tetramethylsilane (TMS) was used as the reference compound for $^{1}H$, with $\delta\ (^{1}H) = 0$ ppm. The isotropic shielding constants of NH3 and TMS were calculated using the GIAO-B3LYP/6-311+G (2d, p) method. The calculated chemical shifts were obtained from $\sigma_{cal.} = \sigma_{ref.} - \sigma_{iso}$. The calculated shifts were used primarily to compare relative chemical-environment trends and to assist in the assignment of the experimental solid-state NMR resonances.

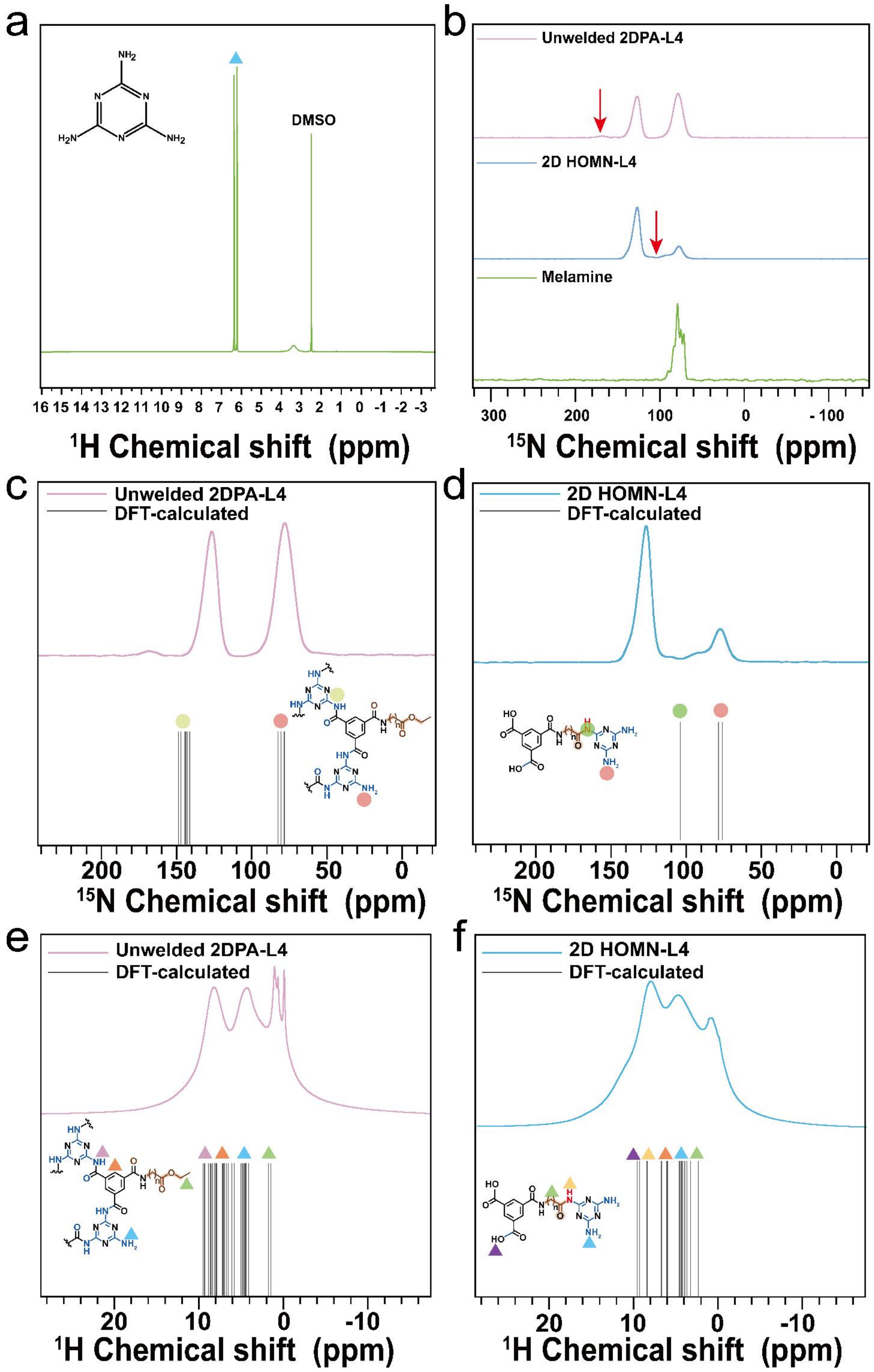

**Supplementary Fig. 20 | Solution-state and solid-state NMR characterization of unwelded 2DPA films and 2DHOMN-L4 films and comparison with DFT-calculated chemical shifts**. $^{1}H$ solution-state NMR spectra of melamine (a). CPMAS $^{15}N$ NMR spectra of $^{15}N$- melamine (b), Experimental $^{15}N$ CPMAS NMR spectra of (c) unwelded 2DPA-L4 and (d) 2D HOMN-L4 films, and $^{1}H$ MAS NMR spectra of (e) unwelded 2DPA-L4 and (f) 2D HOMN-L4 films, together with their corresponding DFT-calculated chemical shifts. Colored circles mark distinct $^{15}N$ environments, whereas colored triangles mark distinct $^{1}H$ environments. Vertical sticks indicate the calculated chemical shifts of individual nitrogen and proton sites, as highlighted in the molecular structures.

To establish the spectral assignment of the nitrogen-containing environments, $^{1}H$ and $^{15}N$ NMR measurements were performed using melamine as a molecular reference. The $^{1}H$ NMR spectrum of melamine dissolved in DMSO-d6 shows a characteristic resonance at 6.22 ppm, which is assigned to the protons of the exocyclic amino groups ($–NH_2$). The $^{15}N$ solid-state NMR spectrum of $^{15}N$-labeled melamine was subsequently compared with those of the unwelded 2DPA films and 2D HOMN-L4 films, to confirm the signal associated with the unreacted $–NH_2$ groups.

After welding, the intensity or position of this signal changes, accompanied by the appearance of an additional $^{15}N$ resonance. This spectral evolution indicates that the nitrogen atoms do not remain exclusively in the original amino environment after welding, but experience a different local chemical environment.

To further determine the origin of the newly observed resonance, the experimentally measured $^{15}N$ chemical shifts were compared with density functional theory (DFT)-calculated isotropic chemical shifts for the proposed molecular structures. The calculated shifts agree with the experimentally observed resonances of both the unreacted amino groups and the newly formed amide nitrogen environments, supporting their respective assignments. Together, these results provide molecular-level evidence for thermally induced amidation through the reaction of peripheral amino groups with pendant ester functionalities. This transformation converts amino nitrogen sites into amide linkages, consistent with the proposed covalent welding mechanism in the 2DPA films.

**Fourier-Transform Infrared Spectroscopy (FTIR)**

ATR-FTIR spectra were acquired using a Nicolet 6700 spectrometer equipped with a diamond ATR accessory. Each spectrum was collected from 4,000 to 500 $cm^{-1}$ at a resolution of 4 $cm^{-1}$ and averaged over 64 scans. Because of the extensive overlap between the vibrational bands of the 2DPA framework, the fitted components were used only to evaluate relative spectral changes and were not regarded as uniquely resolved chemical species[19].

After welding, the contribution near 1655–1660 $cm^{-1}$, to which peripheral primary-amino-group deformation may contribute, decreased substantially. Meanwhile, the ester-carbonyl component at approximately 1727 $cm^{-1}$ was markedly attenuated, together with a redistribution of the amide-I/amide-II envelope[48–50].

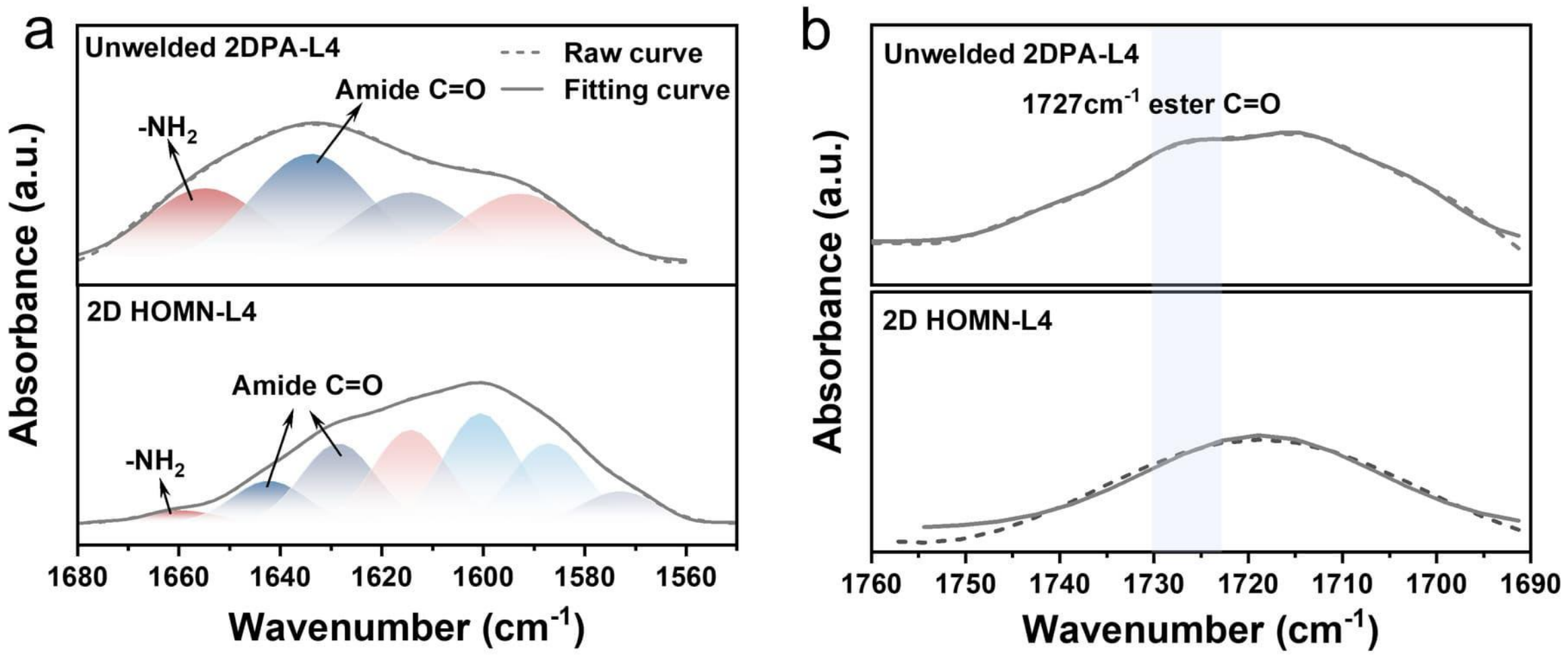

**Supplementary Fig. 21 | FTIR spectra of unwelded 2DPA films and 2DHOMN-L4 films. a**, FTIR spectra in the 1550-1680 $cm^{-1}$ region for the unwelded 2DPA-L4 film and 2D HOMN-L4 film. **b,** FTIR spectra in the 1690-1760 $cm^{-1}$ region for the unwelded 2DPA-L4 films and 2D HOMN-L4 films.

## X-ray Photoelectron Spectroscopy (XPS)

XPS spectra were acquired on a ULVAC-PHI PHI Genesis 500 spectrometer equipped with Thermo Kalpha monochromatic Al Kα X-ray source (1486.6 eV, $\lambda$ = 0.8 nm, power = 150 W). Core-level C 1s and N 1s spectra were deconvoluted to quantify the attenuation of ester functional groups and generation of amide species before and after solid-state covalent welding.

After welding, the fitted C 1s contribution near 286 eV, assigned to C–O/C–N species, decreased, whereas the carbonyl-related contribution near 288 eV increased. In the N 1s region, the component assigned to terminal $–NH_2$ near 399.6 eV was no longer independently resolved[51,52]. These changes are consistent with the consumption of terminal amino and ester-related groups and the formation of additional amide linkages.

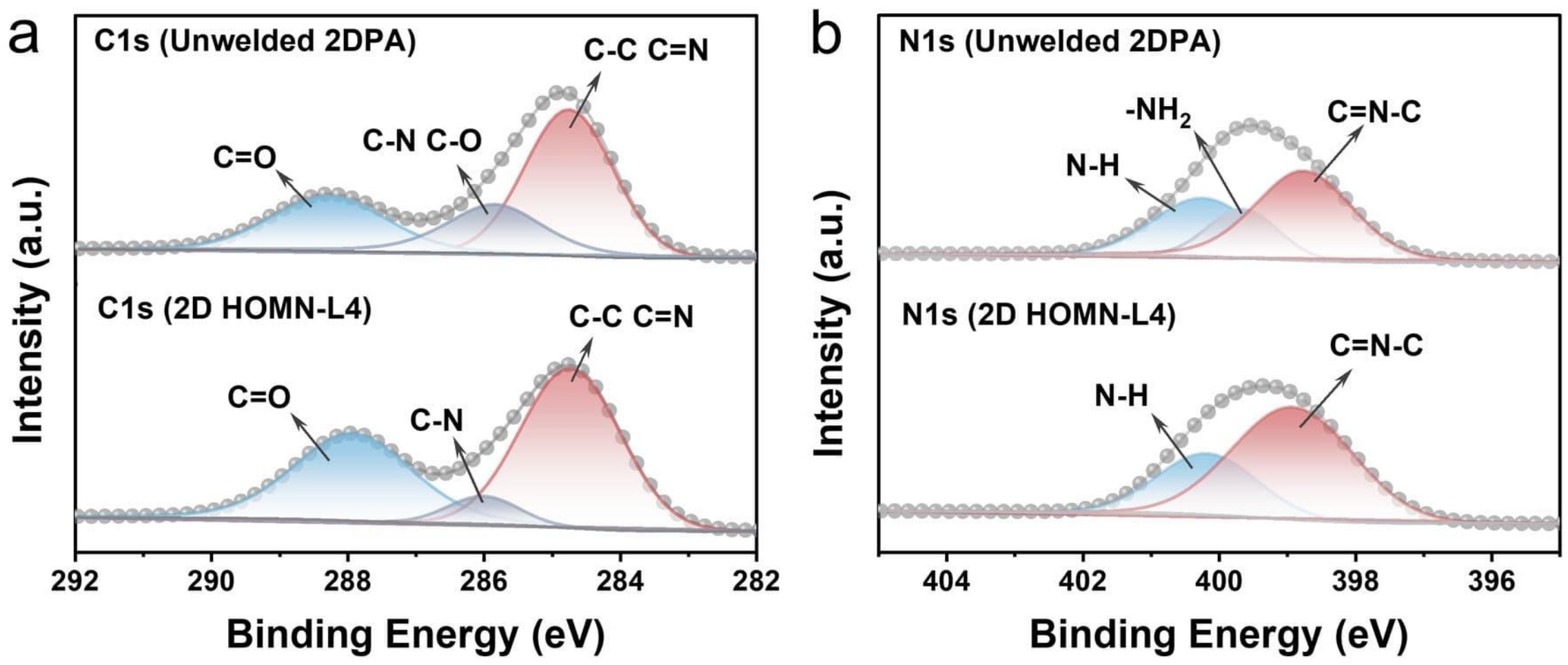

**Supplementary Fig. 22 | The XPS C1s and N1s spectra of unwelded 2DPA-L4 films and 2D HOMN-L4 films.**

**Thermogravimetric Analysis (TGA)**

TGA tests were conducted on a Mettler Toledo thermal analyzer under continuous $N_2$ purge. As-synthesized 2DPA-L4 powder and 2D HOMN-L4 films were heated from 50 °C to 800 °C at a heating rate of 10 °C $min^{-1}$ to evaluate thermal decomposition temperature and residual char yield.

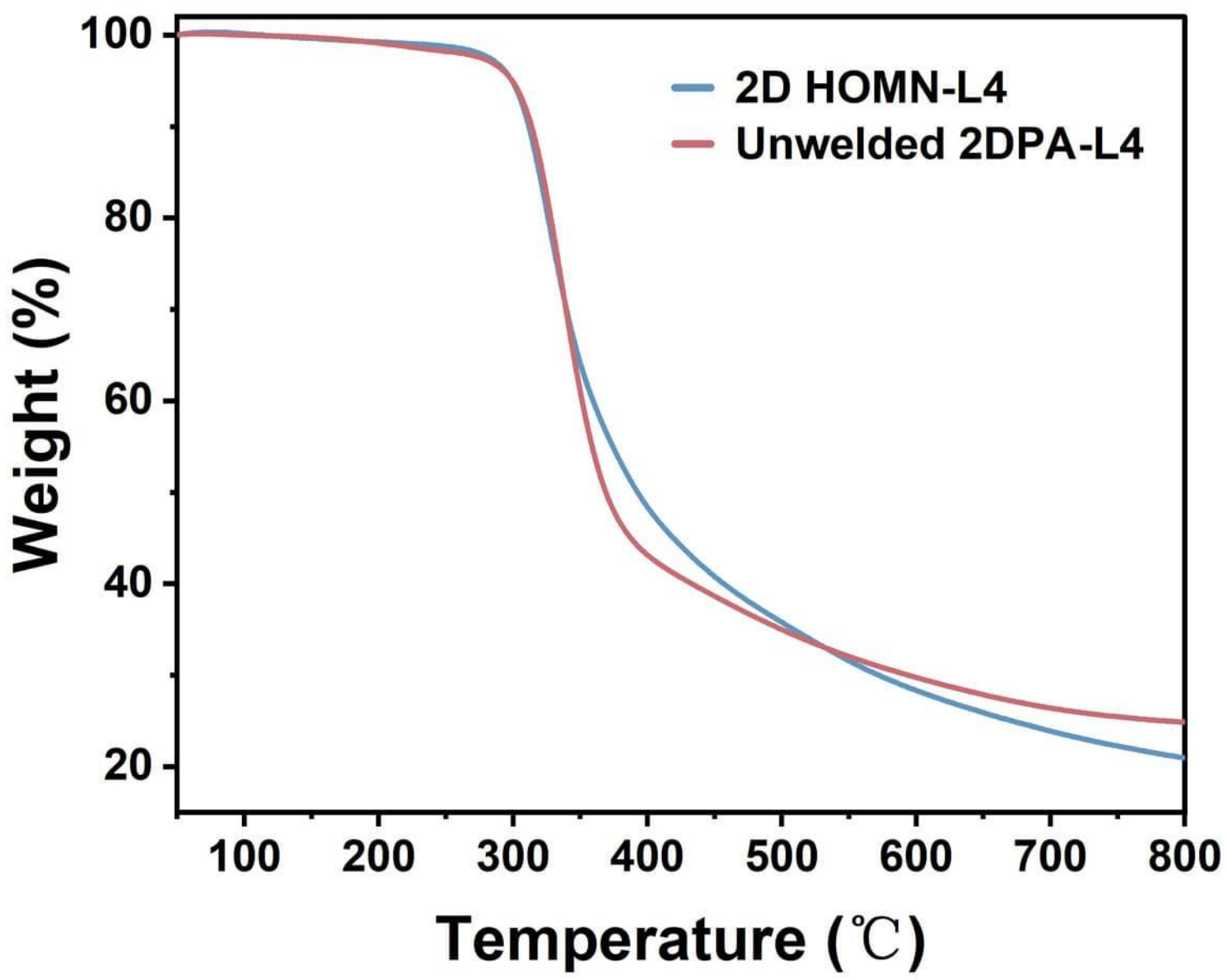

**Supplementary Fig. 23 |** Thermogravimetric analysis (TGA) curves of 2D HOMN-L4 films and unwelded 2DPA-L4 powder.

## Differential Scanning Calorimetry (DSC)

DSC measurements were performed on a TA DSC 250 instrument under nitrogen atmosphere. Samples were scanned from 40 °C to 290 °C at a constant heating rate of 10 °C min$^{-1}$ to capture welding exothermic signals of amino-ester modified precursors.

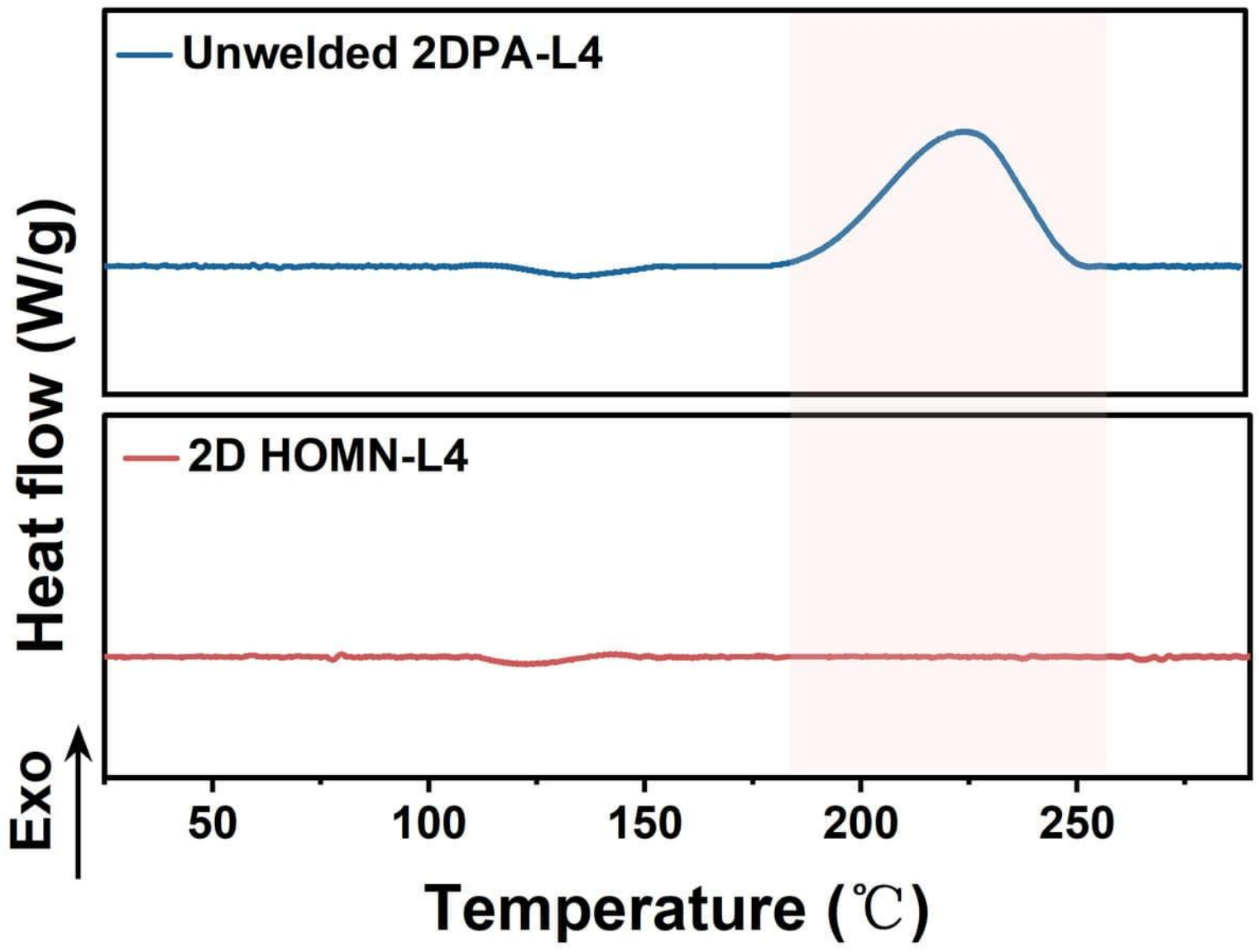

**Supplementary Fig. 24 |** DSC thermograms of unwelded 2DPA-L4 films and 2D HOMN-L4 films.

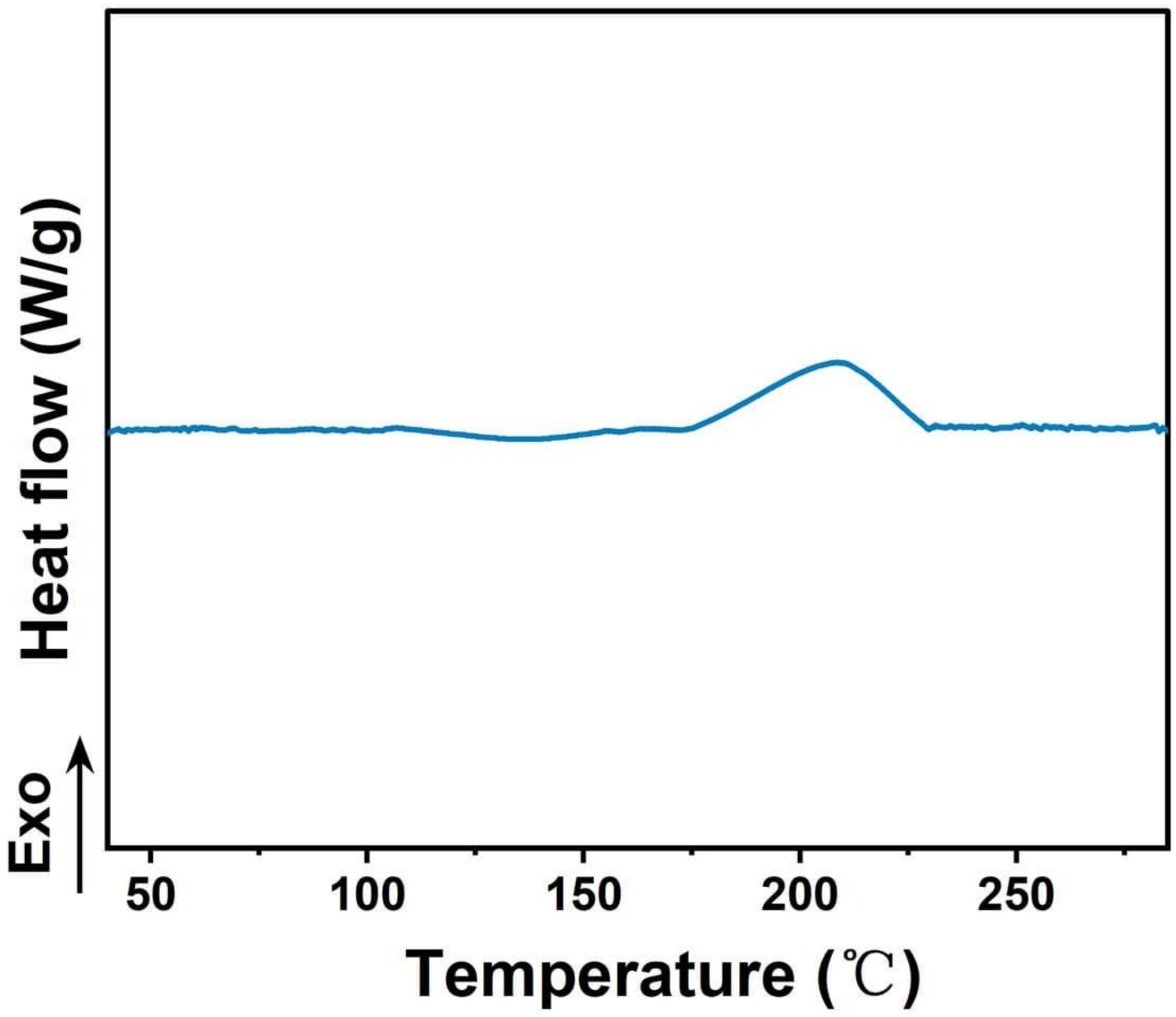

**Supplementary Fig. 25 |** DSC thermograms of 2DPA-L0 (2DPA-IPA) films.

**Thermogravimetric Analysis-Differential Scanning Calorimetry-Mass Spectrometry (TGA-DSC-MS)**

The powder sample was heated from room temperature to 600 °C at a heating rate of 5 K $min^{-1}$ under continuous $N_2$ purge, while the mass change, heat-flow signal and selected ion currents were recorded simultaneously. To examine possible ethanol-related volatile species, the ion currents at m/z 31 ($CH_2OH^+$), 45 ($C_2H_5O^+$) and 46 ($[C_2H_6O]^{+\bullet}$) were analysed. The three signals showed concurrent evolution and reached their maxima near 307 °C. At 307.2 °C, the relative intensities of m/z 31, 45 and 46, normalized to m/z 31, were approximately 100:50:18. The coincident evolution and relative fragment distribution are consistent with an ethanol-related fragmentation pattern, supporting a tentative assignment of the evolved species to ethanol.

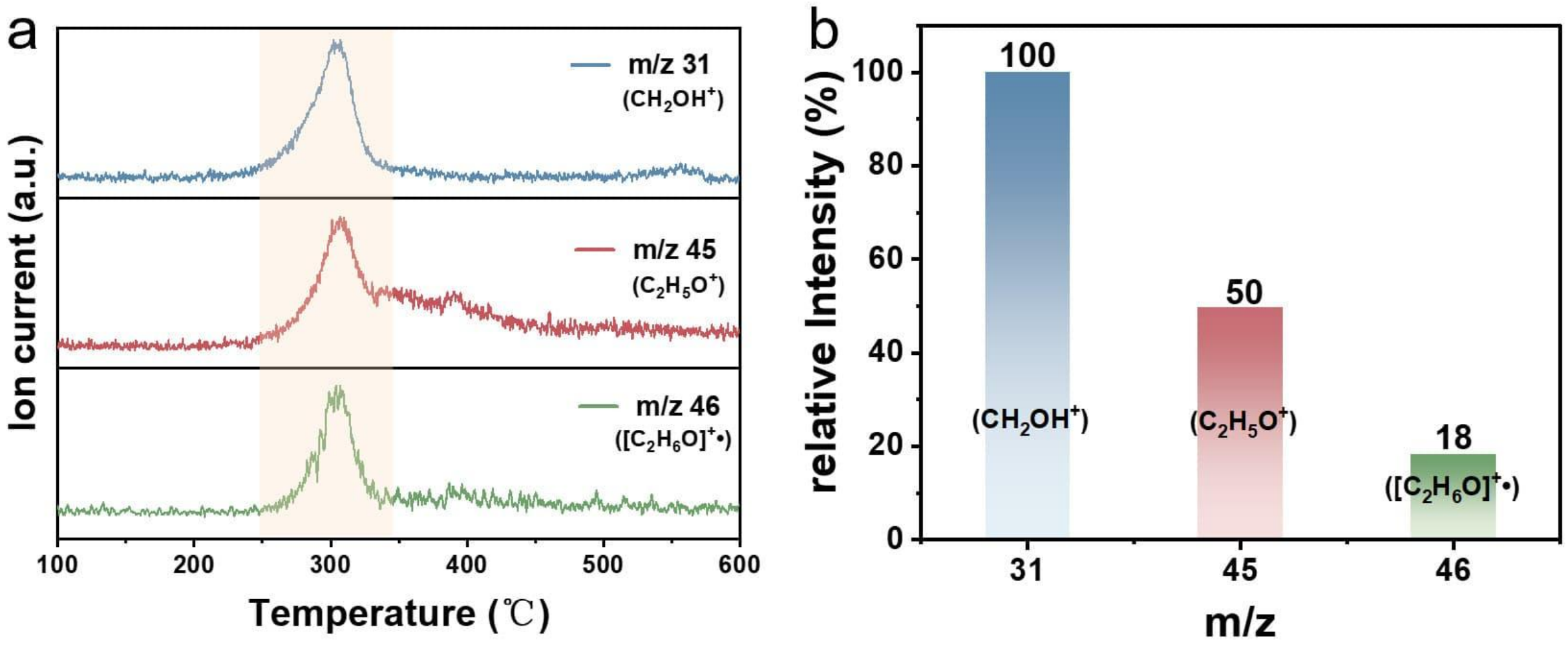

**Supplementary Fig. 26 |** a, Extracted ion current profiles of the characteristic ethanol-related fragments at m/z 31, 45 and 46 as a function of temperature. The shaded region highlights the main release window. b, Relative intensities of m/z 31, 45 and 46 at 307.2 °C, normalized to m/z 31.

**Density Measurement of welded 2D HOMN-L4 Films**

2DPA-L4 powder was dissolved in DMF to prepare a solution with a concentration of 200 mg/mL. The 2DPA solution was spin-coated onto a 5×5 cm quartz glass substrate at a rotational speed of 2000 rpm to form a film. The film was dried on a hotplate at 120 °C for 5 hours to remove excess solvent. Subsequently, it was heated on a hotplate at 280 °C for 15 minutes to obtain a covalently welded film. The thickness of the above-mentioned film was measured and recorded using a step profilometer. The film was then scribed into a regular shape with a knife, and its area was calculated. Afterwards, a 10% (v/v) HF solution was dropped onto the edge of the quartz glass for etching. When the entire film was detached from the quartz glass substrate, the substrate was immersed in a glass container filled with a large amount of ultrapure water, and the ultrapure water was replaced three times to ensure that the HF solution on the film surface was completely washed off. The washed film was gently lifted with a clean silicon wafer and collected in a glass bottle. It was placed in a vacuum oven at 80 °C for 12 hours to remove residual water, and the dried film was weighed to obtain its mass. The density of the film was calculated using the equation provided below:

$$\rho = m/v \tag{7}$$

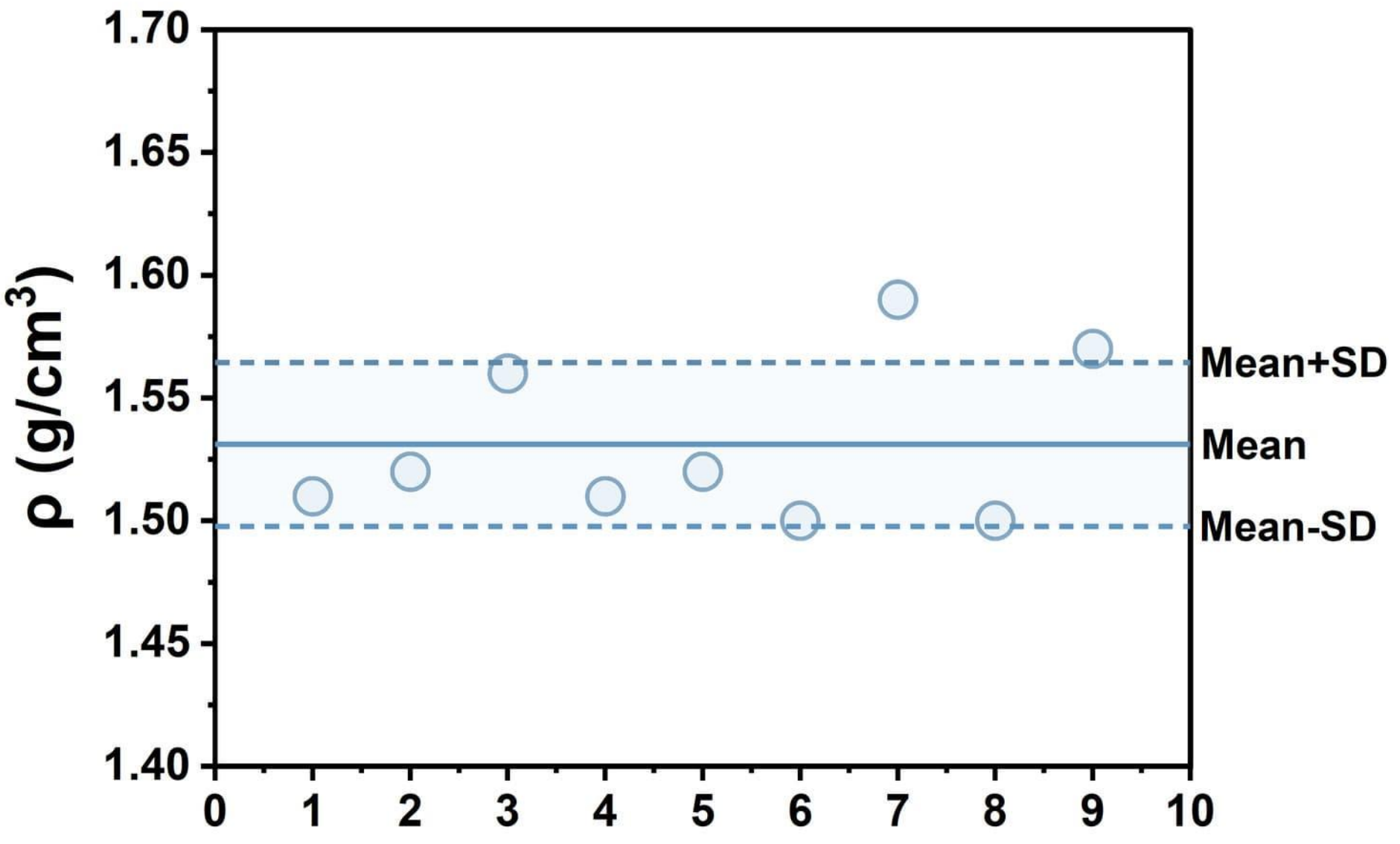

**Supplementary Fig. 27 |** The density of the welded films.

The apparent bulk density of the 2D HOMN-L4 films was estimated using a two-liquid flotation test. The film sank in chloroform ($\rho$ = 1.48 g/cm$^3$) but floated in carbon tetrachloride ($\rho$ = 1.595 g/cm$^3$). These observations bracket the film density between 1.48 and 1.595 g cm$^{-3}$, consistent with the formation of a densely packed covalently welded network.

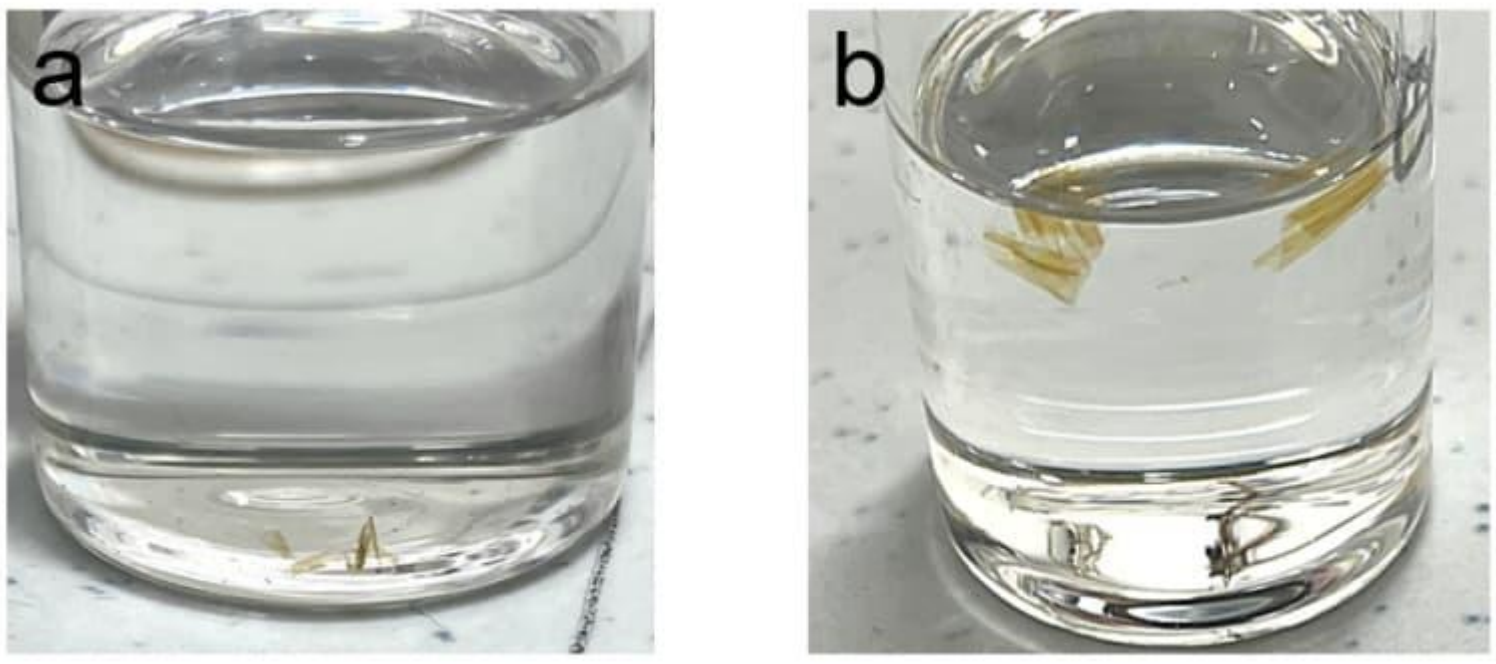

**Supplementary Fig. 28 |** The state of the welded films in (a) trichloromethane ($\rho$ = 1.48 g/cm$^3$) and (b) carbon tetrachloride ($\rho$ = 1.595 g/cm$^3$).

## Morphology, stacking and molecular orientation

### High-resolution atomic force microscopy

**Sample preparation:**

To achieve sub-nanometer resolution, the sample had to be extremely flat. Therefore, A-grade highly oriented pyrolytic graphite (HOPG) was used as the base material for high-resolution atomic force microscopy (AFM) measurements. The 2DPA-L4 powder was heated at 220 °C for 5 minutes and then dissolved in DMF to prepare a solution with a concentration of 0.1 mg/mL. The above solution was ultrasonicated for 10 minutes to ensure uniform dispersion. Subsequently, 20 μL of the uniformly dispersed solution was dropped onto freshly exposed HOPG, and the sample was immediately rinsed with DMF solution. After the DMF on the sample surface had completely evaporated, the sample was ready for AFM testing. For the high-resolution testing of single-layer two-dimensional molecules, 2DPA-L4 powder was dissolved in a DMF solution to prepare a mixed solution with a concentration of 1 μg/mL. The remaining operations were the same as those performed in the step of partial covalently welded molecule testing.

**Method:**

High-resolution AFM images were acquired in AC mode on a Cypher VRS 1250 instrument (Asylum Research/Oxford Instruments) using Multi75DLC probes (BudgetSensors). Data were analysed using Asylum Research software (version 16.33.234) and Gwyddion. A zero-order polynomial background correction was applied to the selected height profiles.

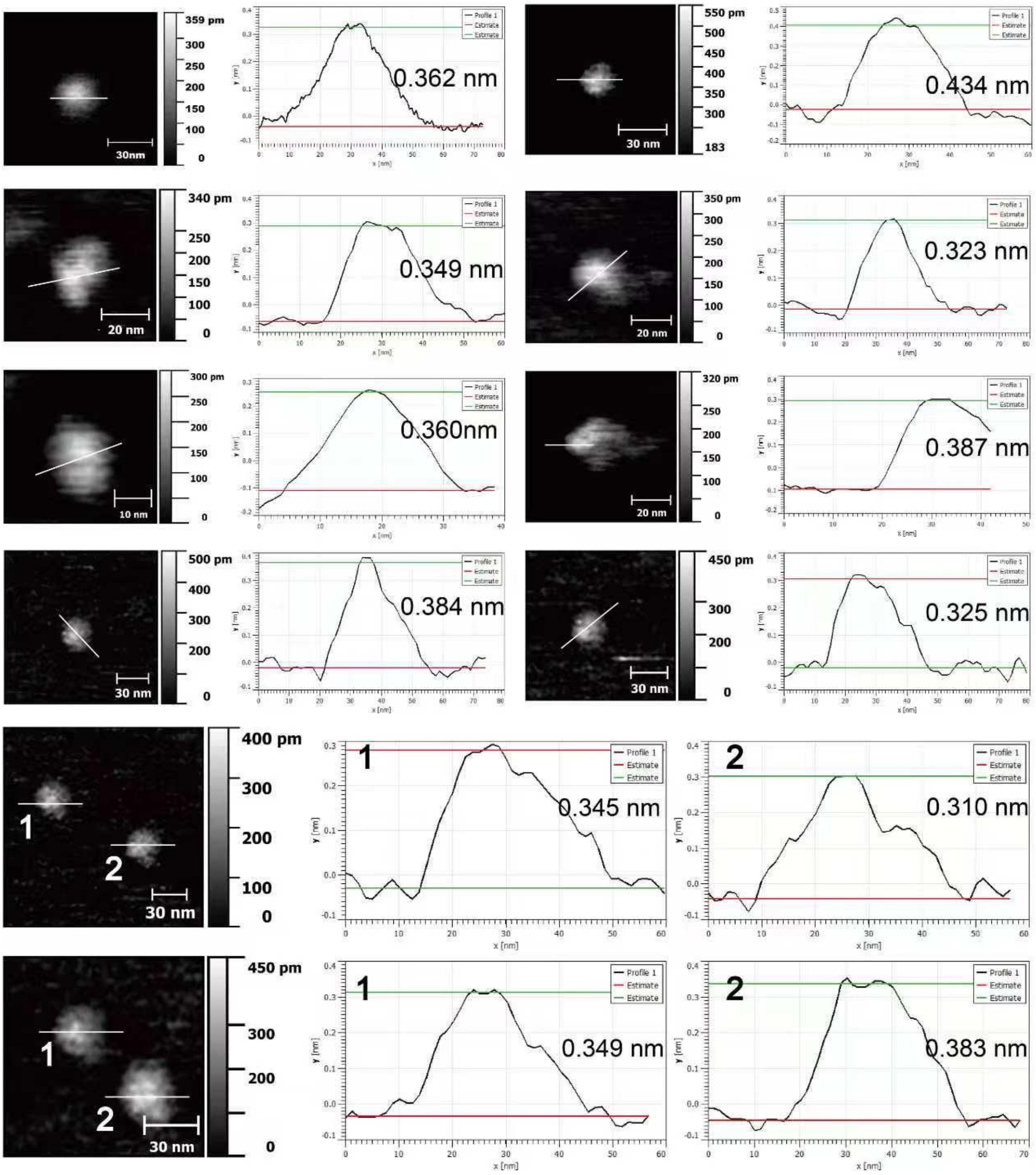

**Supplementary Fig. 29 |** Monolayers and their thicknesses.

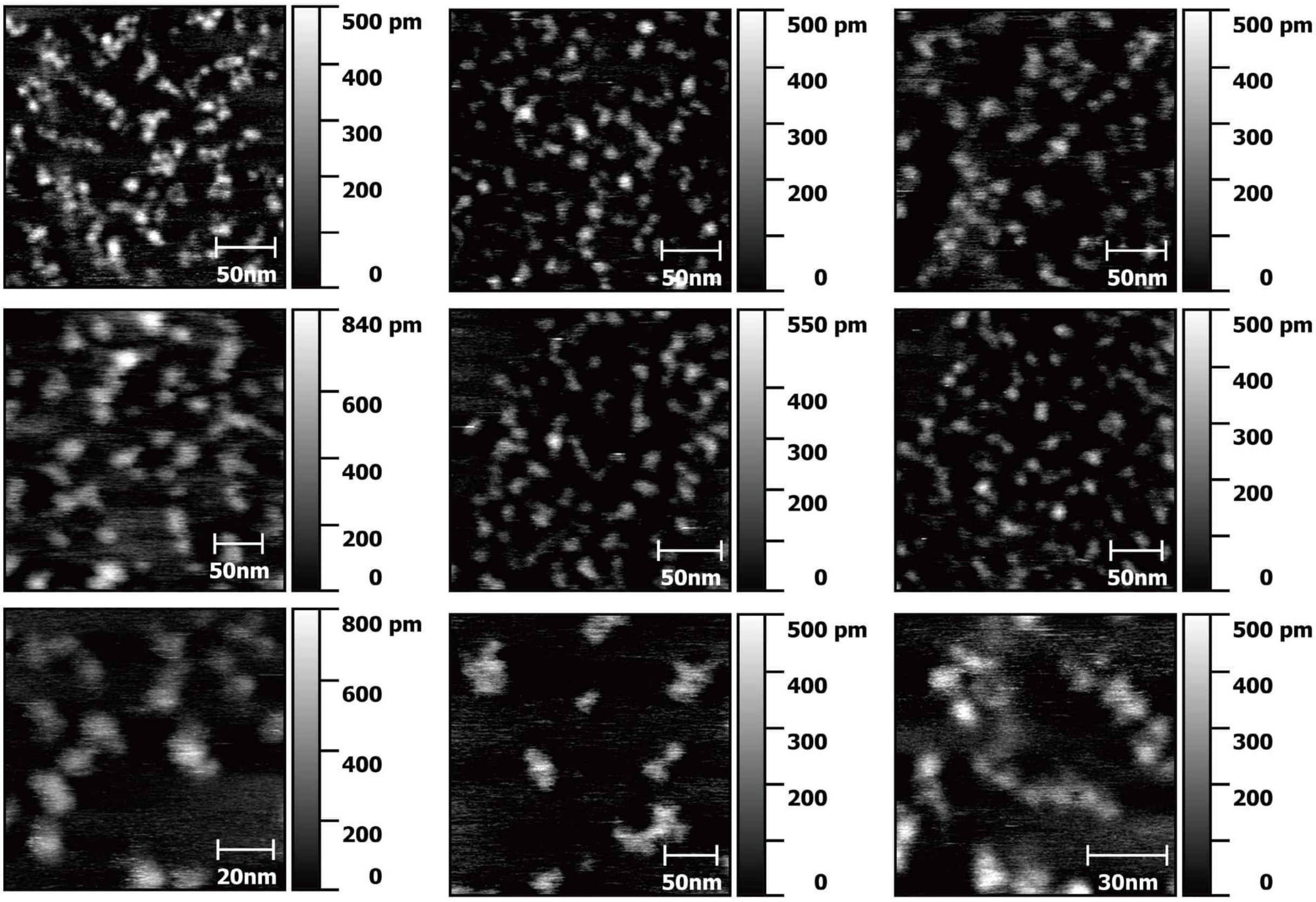

**Supplementary Fig. 30 |** High-resolution morphological images of partially welded molecules (150°C).

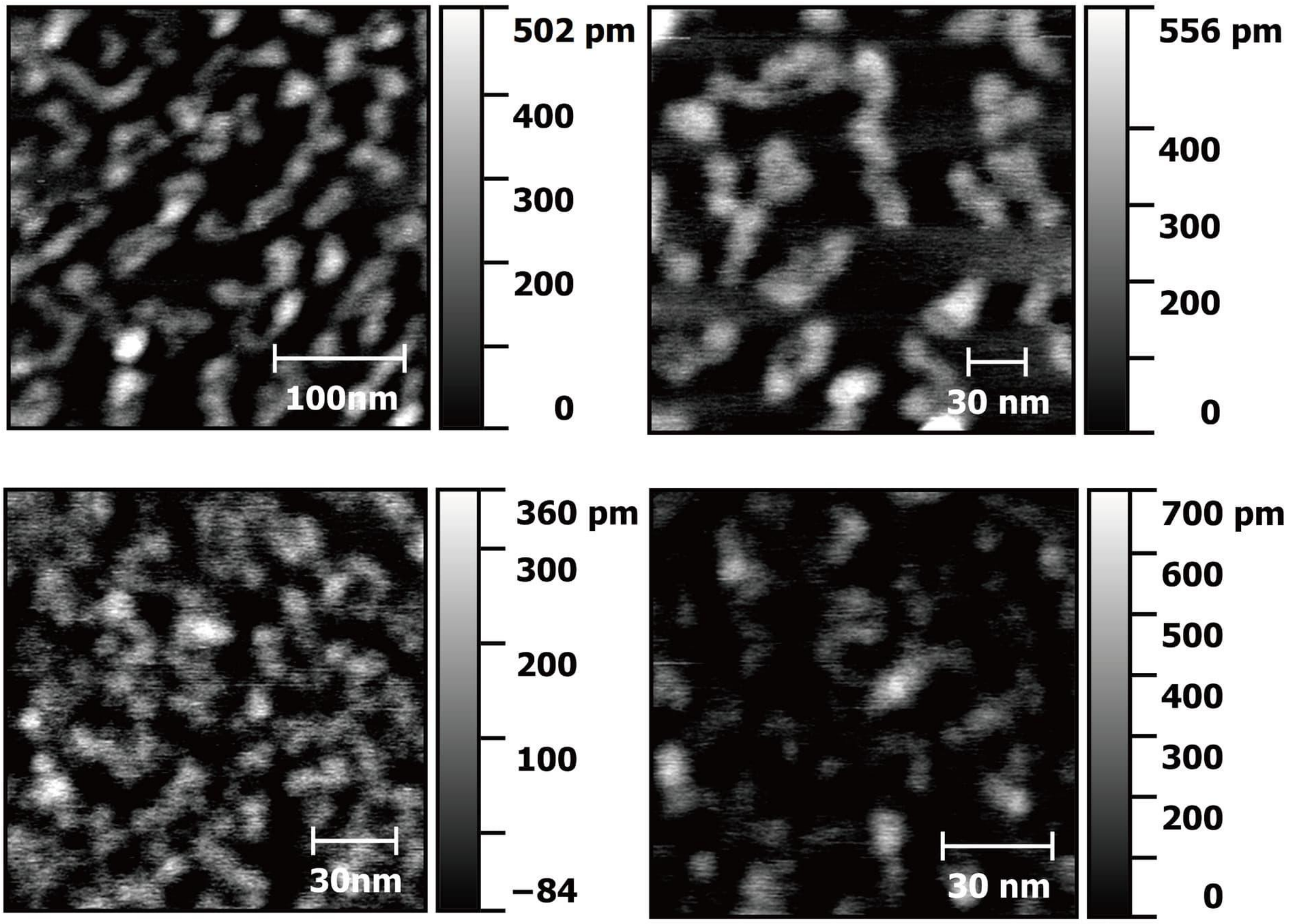

**Supplementary Fig. 31 |** High-resolution morphological images of partially welded molecules (220℃).

**Transmission electron microscopy**

**Transmission Electron Microscopy Characterization**

Bright-field TEM imaging was performed using a field-emission transmission electron microscope operated at 200 kV. Free-standing 2D HOMN-L4 films with a thickness of approximately 32 nm were transferred onto holey-carbon-coated copper grids using a water-assisted floating-transfer method and dried before imaging.

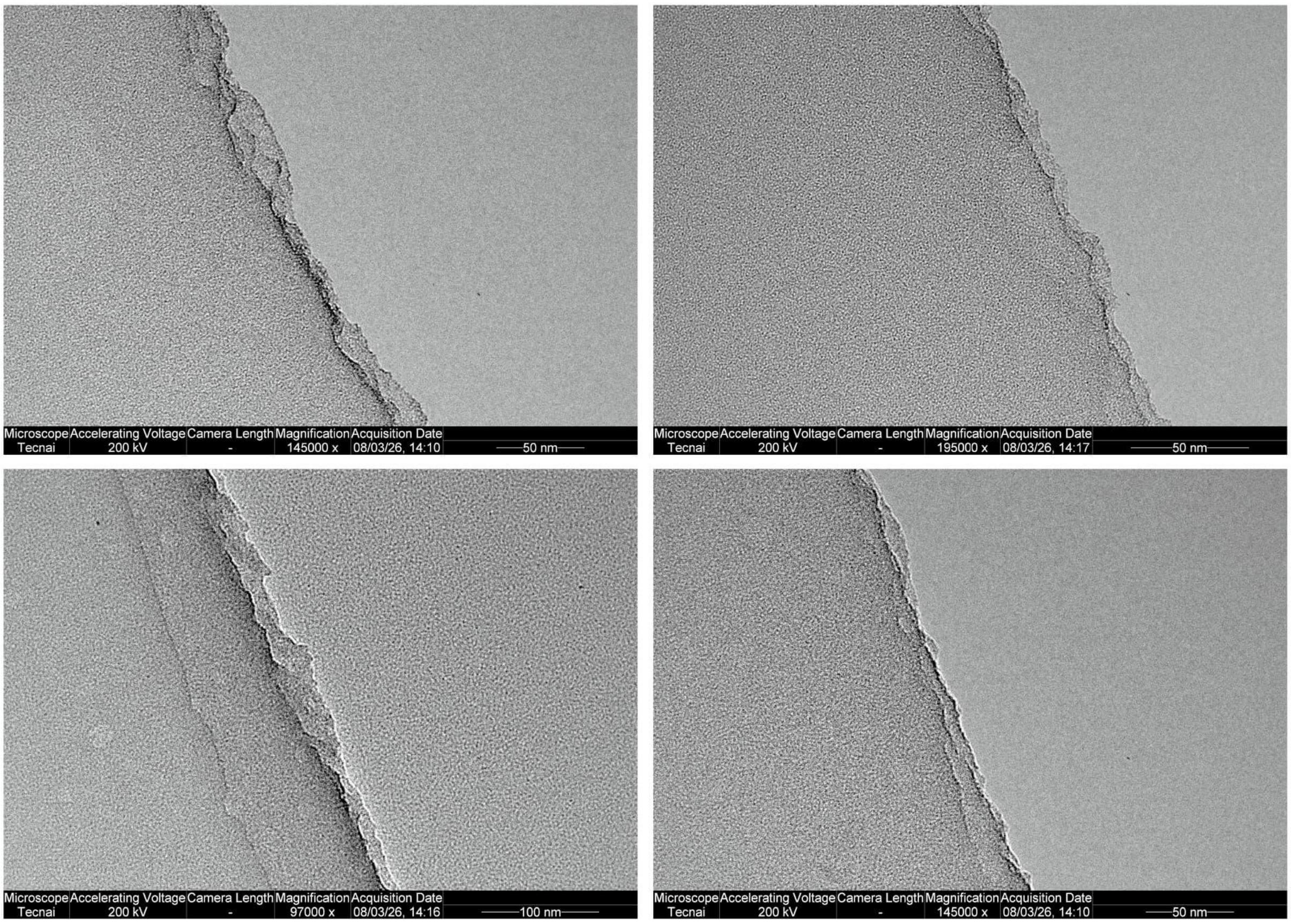

**Supplementary Fig. 32 |** Transmission electron micrograph of a 2D HOMN-L4 films.These images show edge regions of the films.

### Scanning electron microscopy

SEM characterization was carried out on a Zeiss Ultra 55 instrument. Cross-sectional scanning electron microscopy (SEM) was used to examine the continuity of the 2D HOMN-L4 films and its interface with the substrate. The film formed a continuous layer on the $SiO_2$/Si substrate, with a compact cross-sectional morphology and a well-defined film–substrate interface. No obvious cracks, delamination or large voids were observed within the imaged regions.

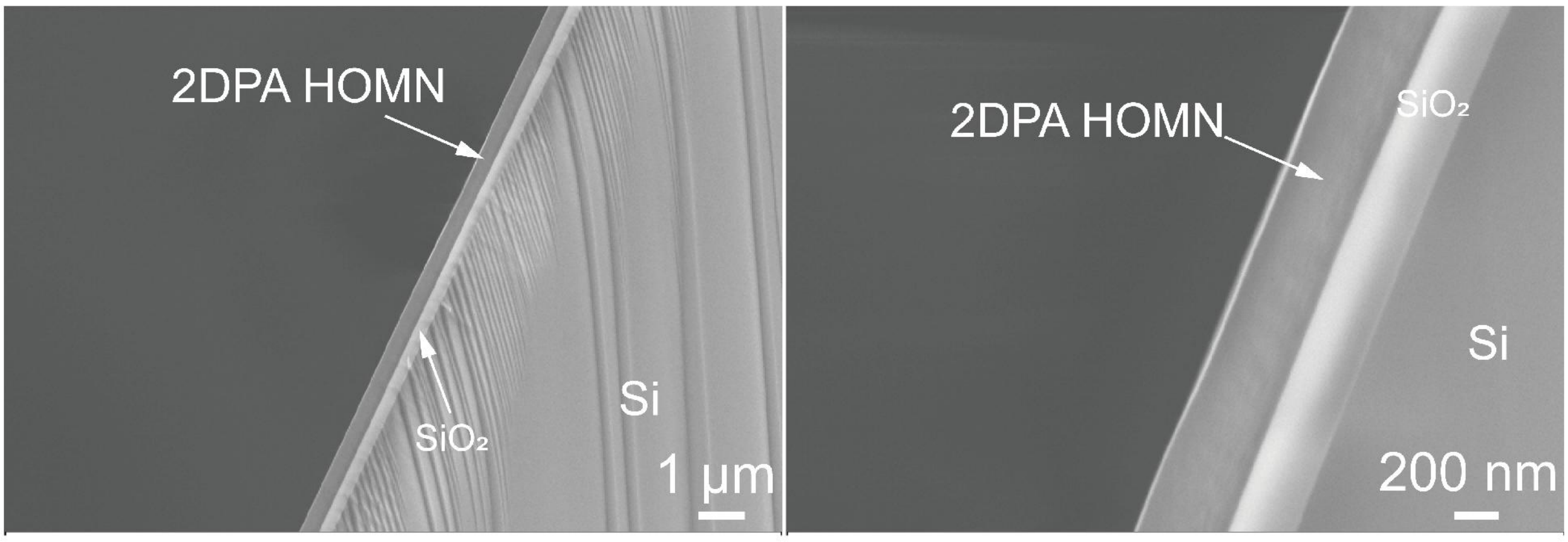

**Supplementary Fig. 33 |** Cross-sectional morphology of 2D HOMN-L4 films.

**Linker-length-dependent structure-property relationships**

The L-series was designed to examine how the length of the amino-ester spacer regulates intersheet covalent welding. Increasing the spacer length simultaneously increases the conformational freedom and spatial reach of the terminal amino group, thereby increasing its probability of accessing an ester group on an adjacent, imperfectly registered molecular sheet. However, linker elongation also introduces a more compliant aliphatic segment and reduces the density of load-bearing covalent connections per unit volume.

Within this series, L4 was selected as the principal sample for quantitative structural and mechanical characterization because it provided a suitable balance between efficient intersheet welding, film uniformity and mechanical stiffness. L8 was used for selected large-area handling and load-bearing demonstrations because its longer spacer is expected to tolerate greater local structural mismatch and deformation. These demonstrations do not imply that L8 possesses a higher intrinsic modulus than L4; rather, they highlight the capacity of a more compliant welded network to preserve macroscopic continuity under large deformation.

**Quasi-static mechanical characterization**

Strain-induced Elastic Buckling Instability for Mechanical Measurements (SIEBIMM)

The wrinkle wavelength, λ, was measured by optical microscopy in at least five independent film regions per sample. The in-plane Young's modulus was calculated using the thin-film buckling relation described by Stafford et al.[23]:

$$\frac{E_f}{\left(1 - v_f^2\right)} = \frac{3E_S}{(1 - v_s^2)}\left(\frac{\lambda}{2\pi h}\right)^3 \tag{8}$$

where $\lambda$ represents the wrinkle wavelength, $h$ is the thickness of the upper 2D HOMN-Ln films, $v$ is Poisson’s ratio, $E_f$ is the Young’s modulus of film and $E_s$ is the Young’s modulus of substrate.

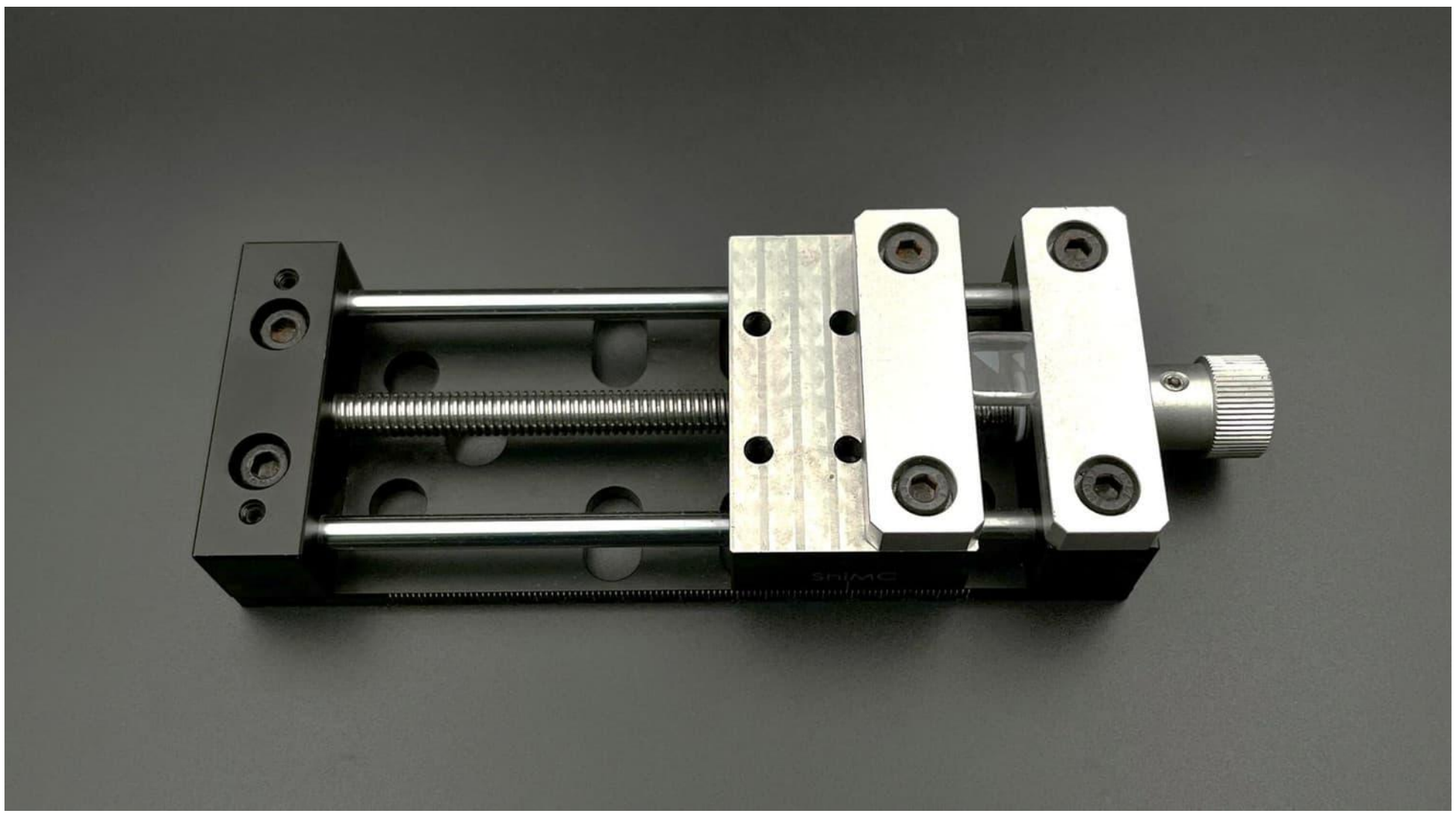

**Supplementary Fig. 34 |** Custom-built stretching device for SIEBIMM measurements.

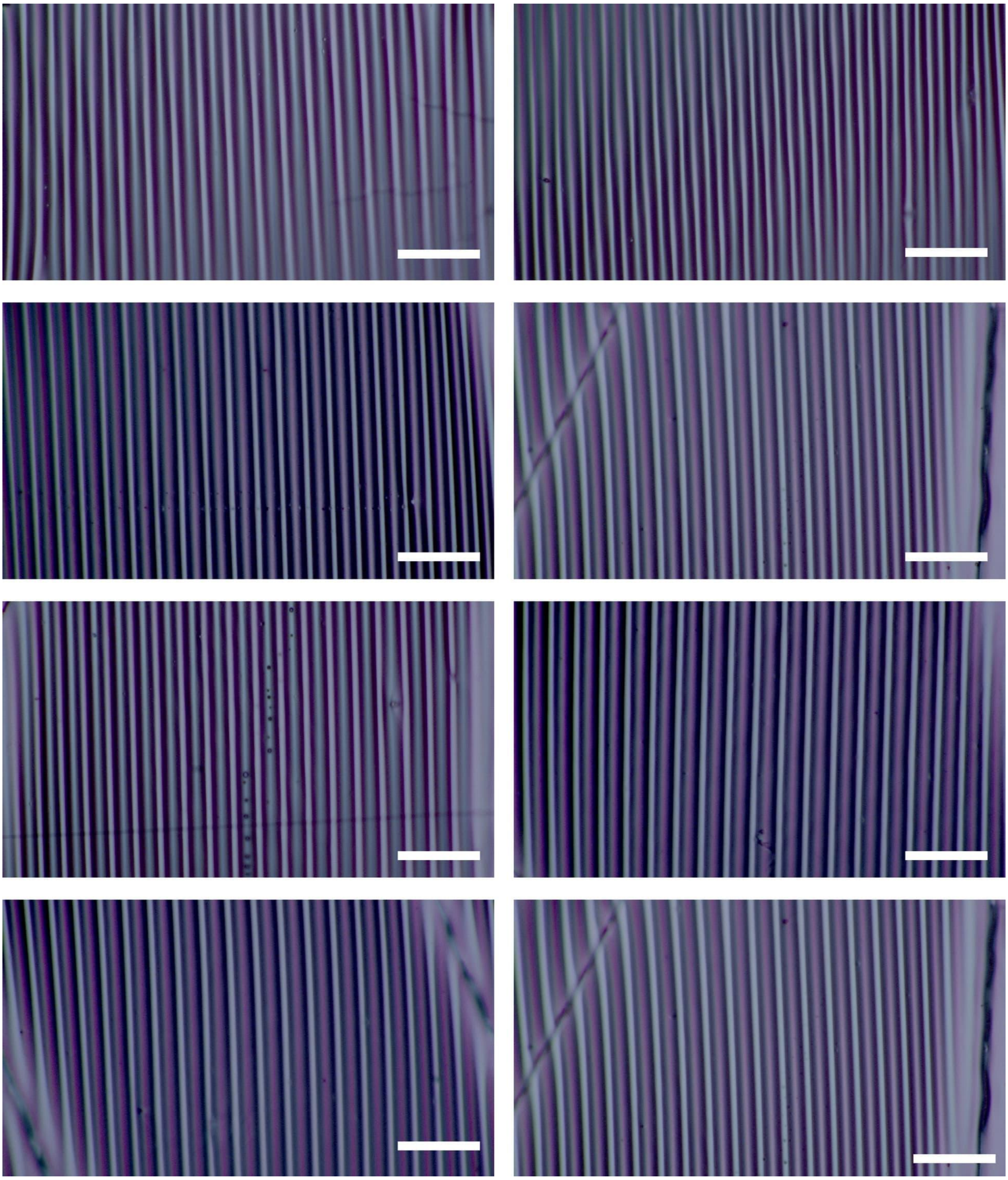

**Supplementary Fig. 35 |** Optical micrographs of the large scaled wrinkle pattern of 2D HOMN-L0 films at 10% compressive strain, Scale bar, 40 µm.

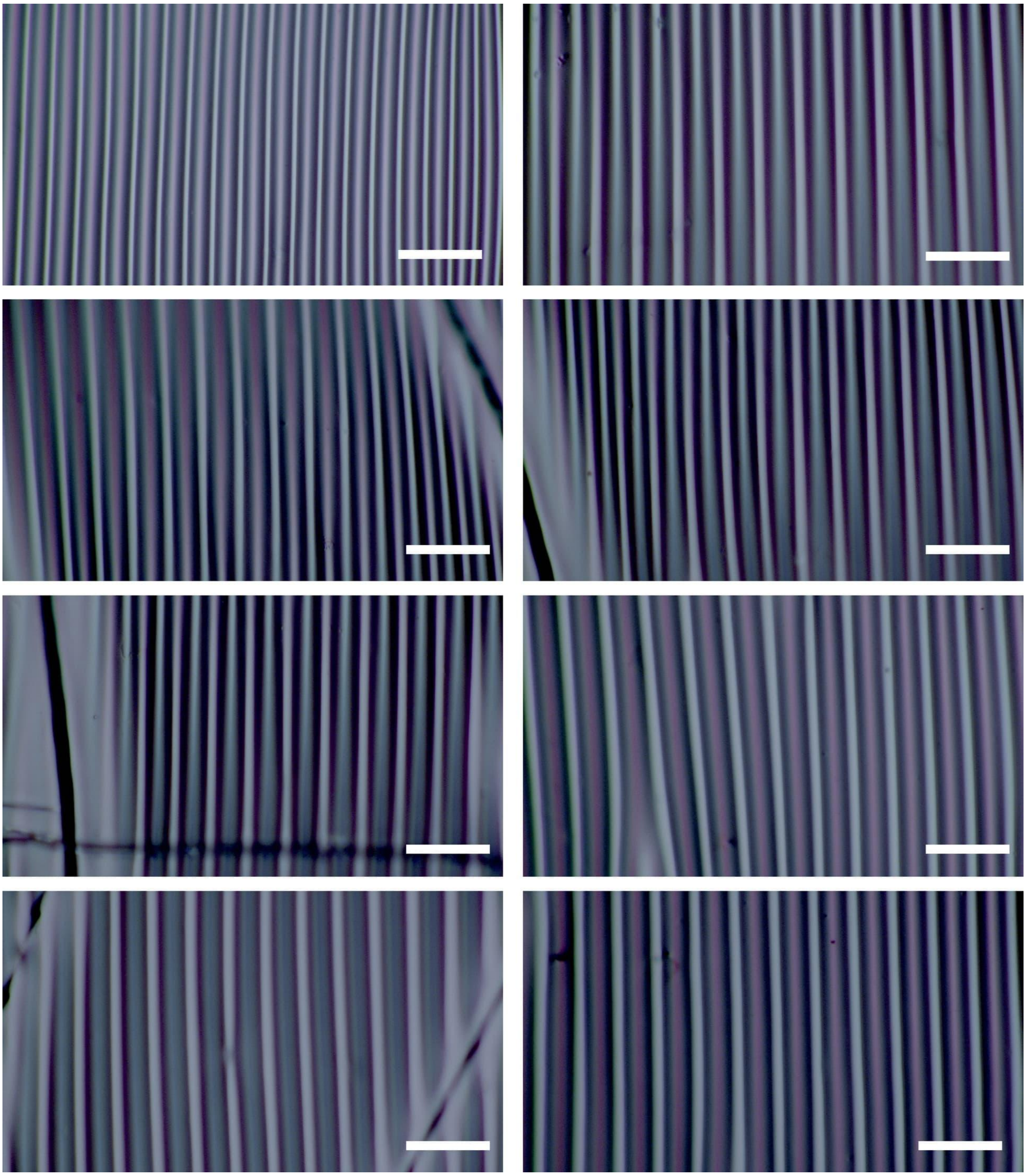

**Supplementary Fig. 36 |** Optical micrographs of the large scaled wrinkle pattern of 2D HOMN-L3 films at 10% compressive strain, Scale bar, 40 μm.

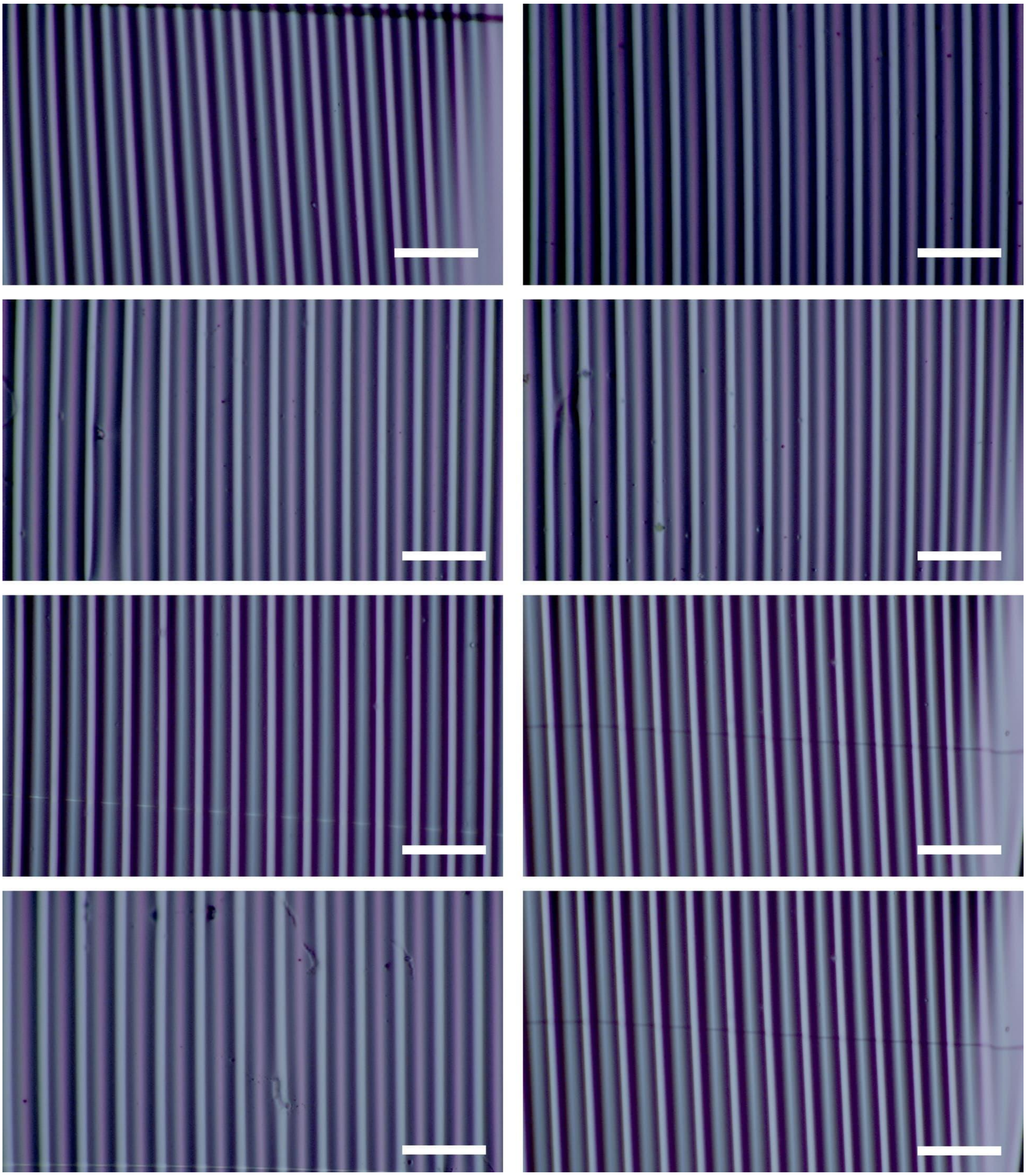

**Supplementary Fig. 37 |** Optical micrographs of the large scaled wrinkle pattern of 2D HOMN-L4 films at 10% compressive strain, Scale bar, 40 μm.

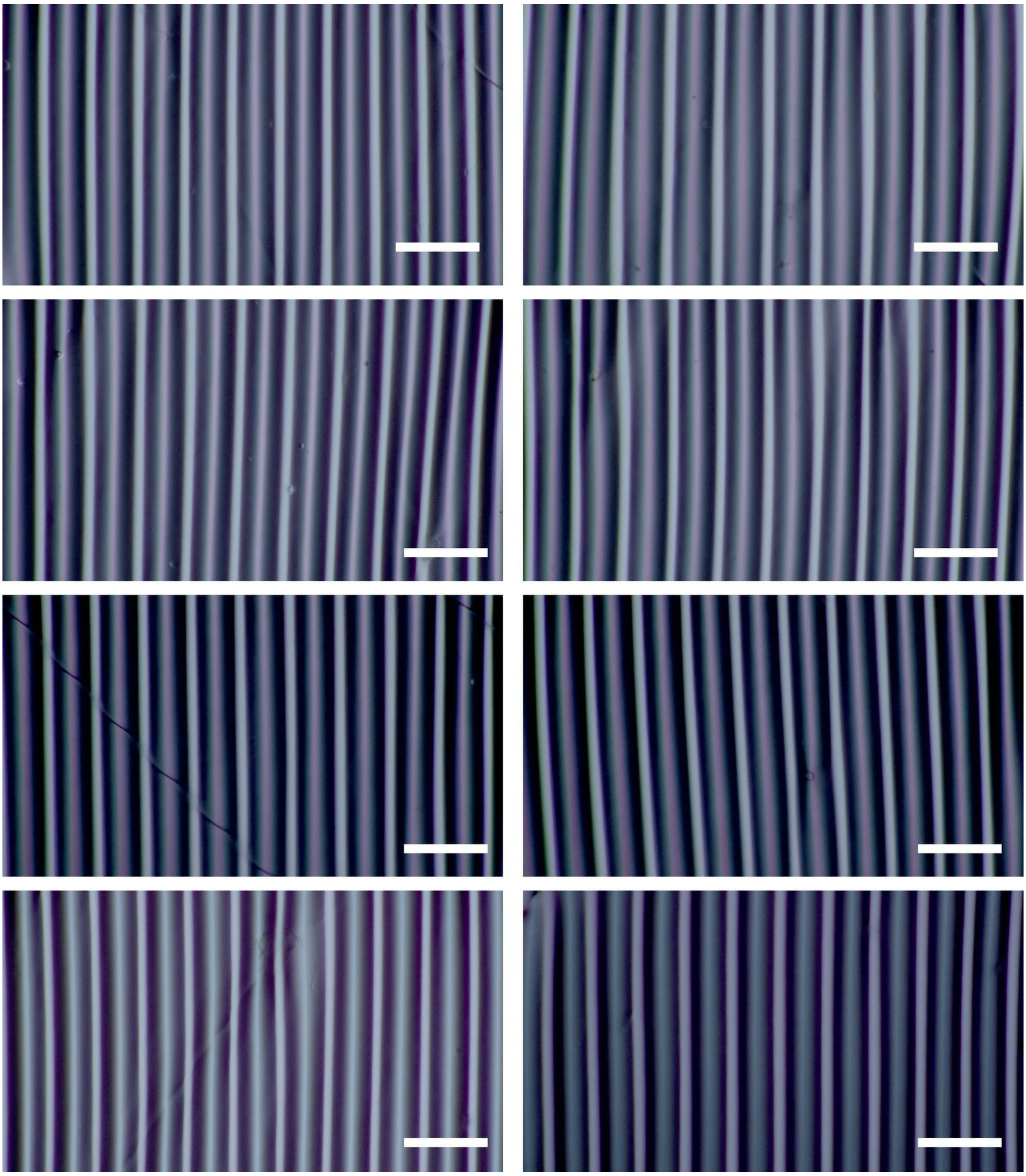

**Supplementary Fig. 38 |** Optical micrographs of the large scaled wrinkle pattern of 2D HOMN-L5 films at 10% compressive strain, Scale bar, 40 μm.

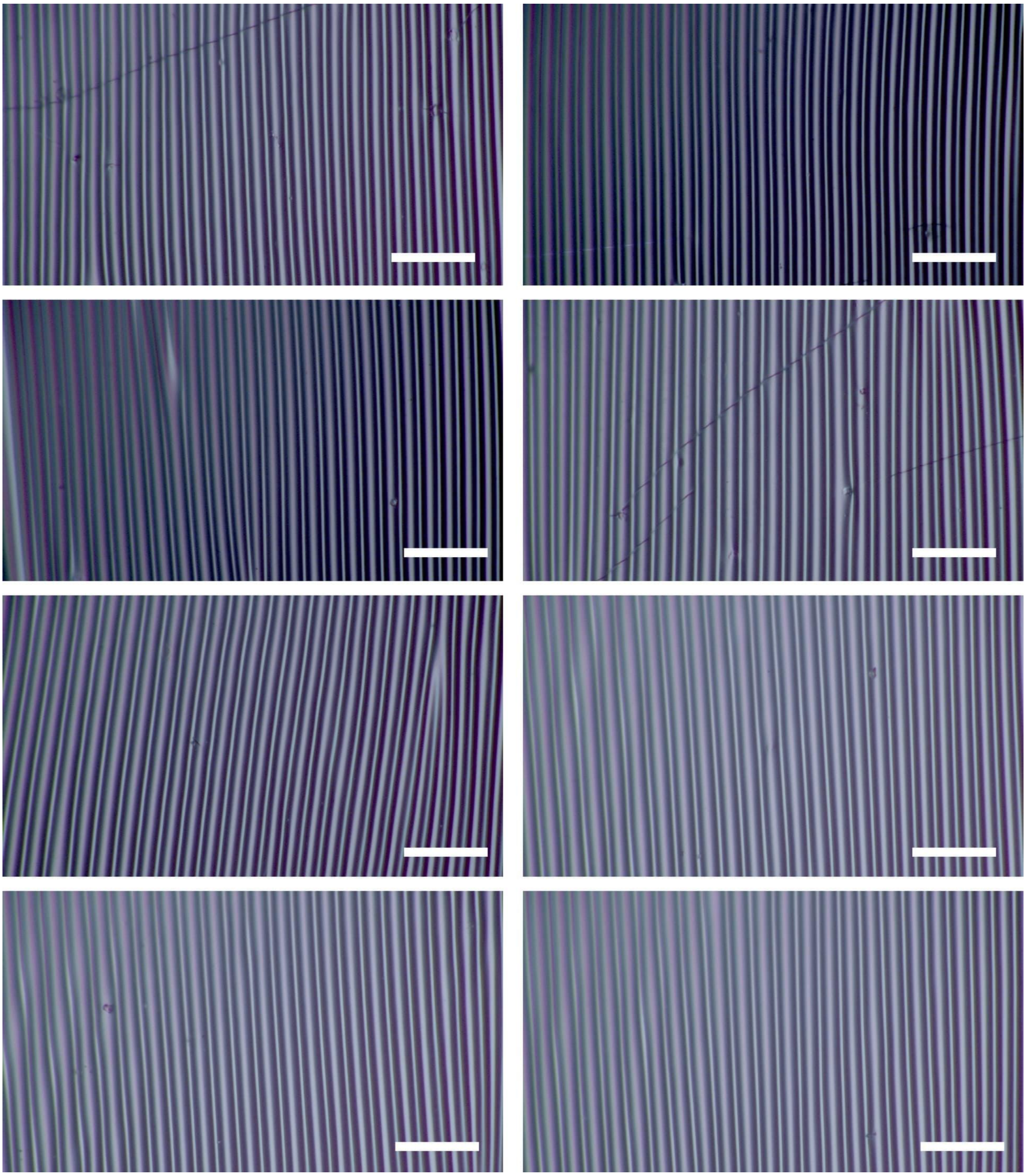

**Supplementary Fig. 39 |** Optical micrographs of the large scaled wrinkle pattern of 2D HOMN-L6 films at 10% compressive strain, Scale bar, 40 μm.

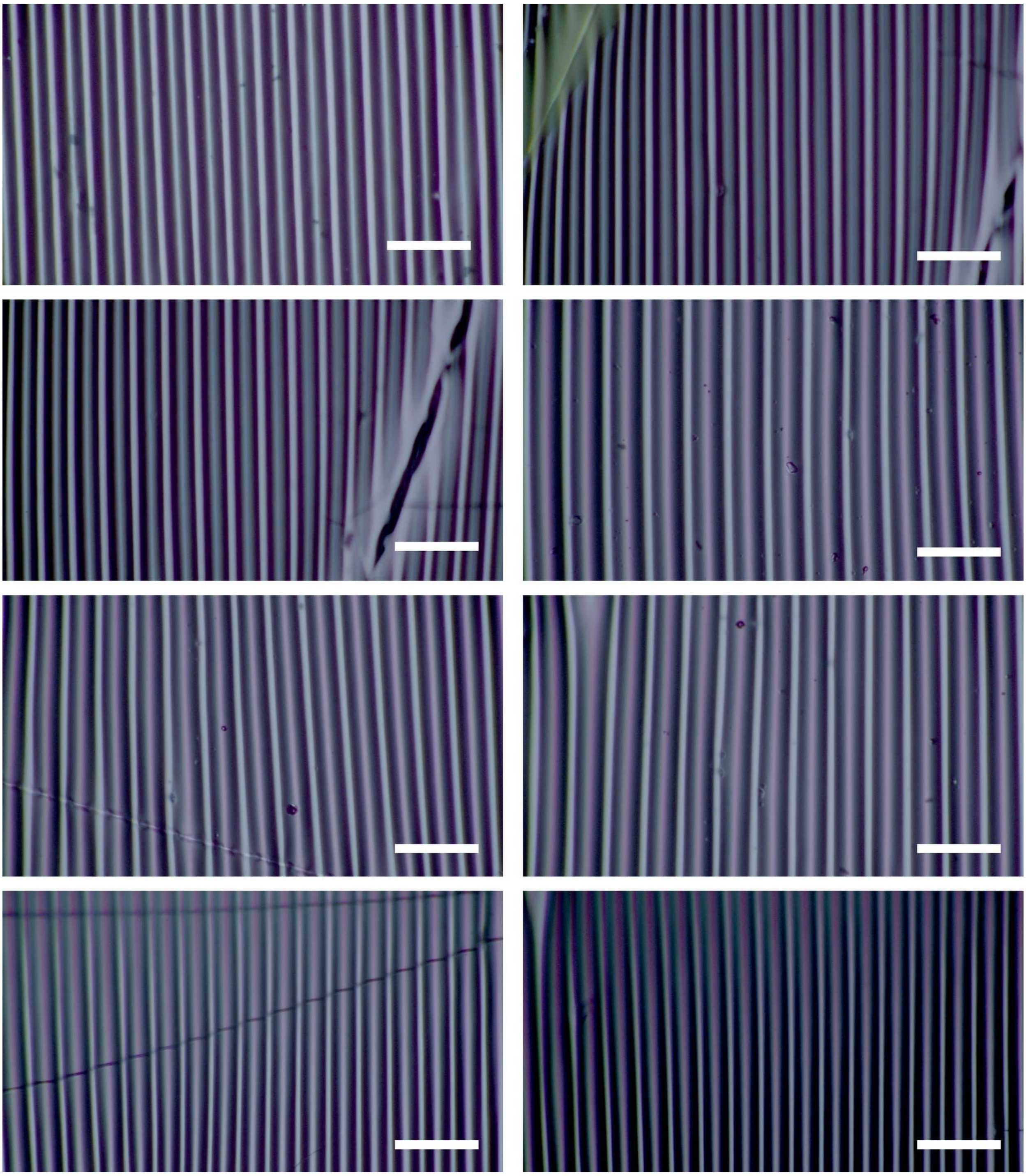

**Supplementary Fig. 40 |** Optical micrographs of the large scaled wrinkle pattern of 2D HOMN-L7 films at 10% compressive strain, Scale bar, 40 μm.

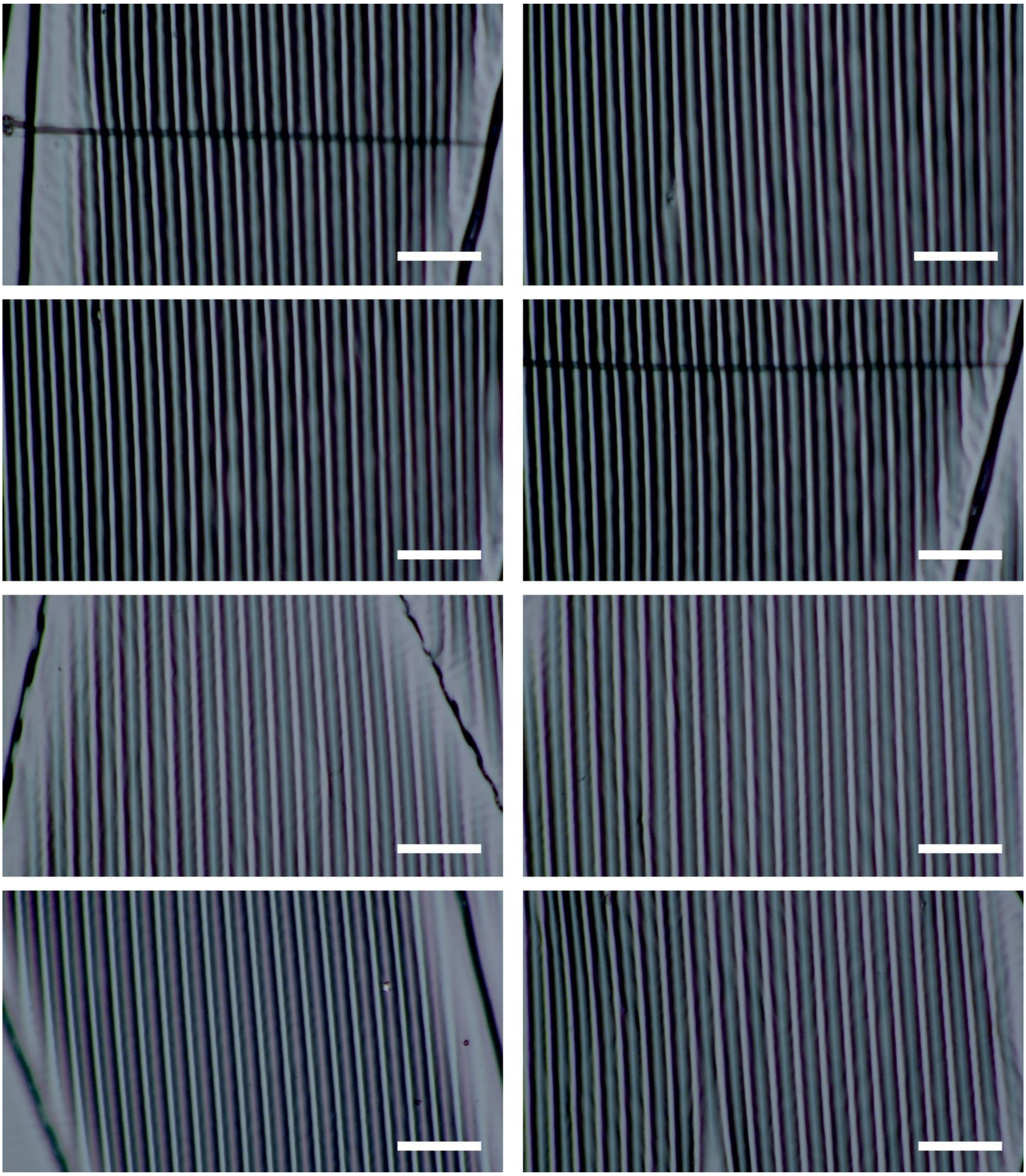

**Supplementary Fig. 41 |** Optical micrographs of the large scaled wrinkle pattern of 2D HOMN-L8 films at 10% compressive strain, Scale bar, 40 μm.

**AFM nanoindentation**

**Materials**

Nanoindentation tests were conducted using a Cypher atomic force microscope. The substrate used in the tests was a 300-mesh copper mesh. Two types of diamond-like spherical probes were purchased from the respective suppliers: Biosphere NT_B300_v0010 (purchased from Nanotool AFM Probes.), which was used to press within the elastic range of the film to obtain modulus data, and Tap300-GB-G (purchased from Budget Sensors), which was used to crush the film to obtain breaking strength data.

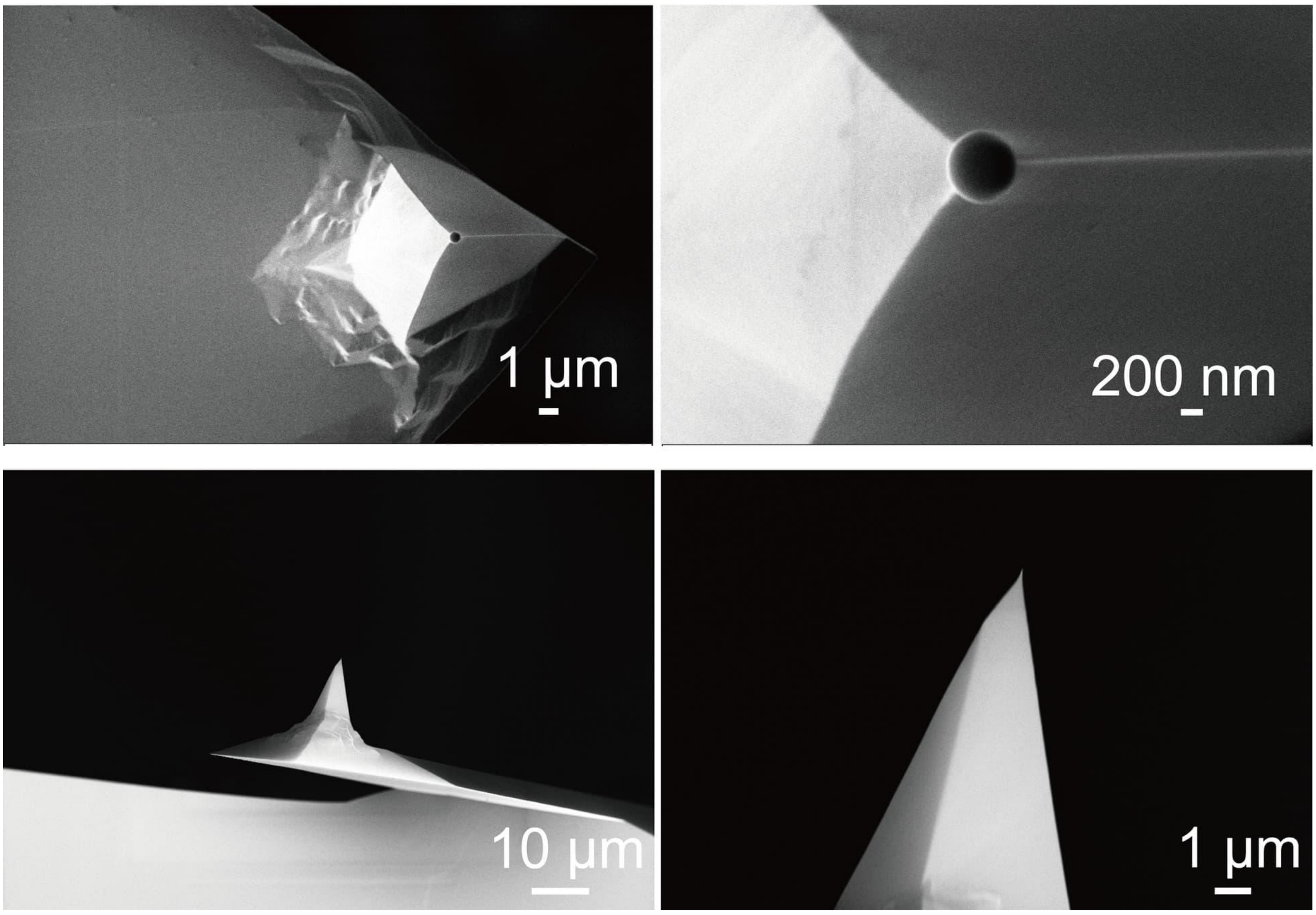

**Supplementary Fig. 42 |** SEM image of AFM tip used for nanoindentation. **(a)** Biosphere NT_B300_v0010 and **(b)** Tap300GB-G.

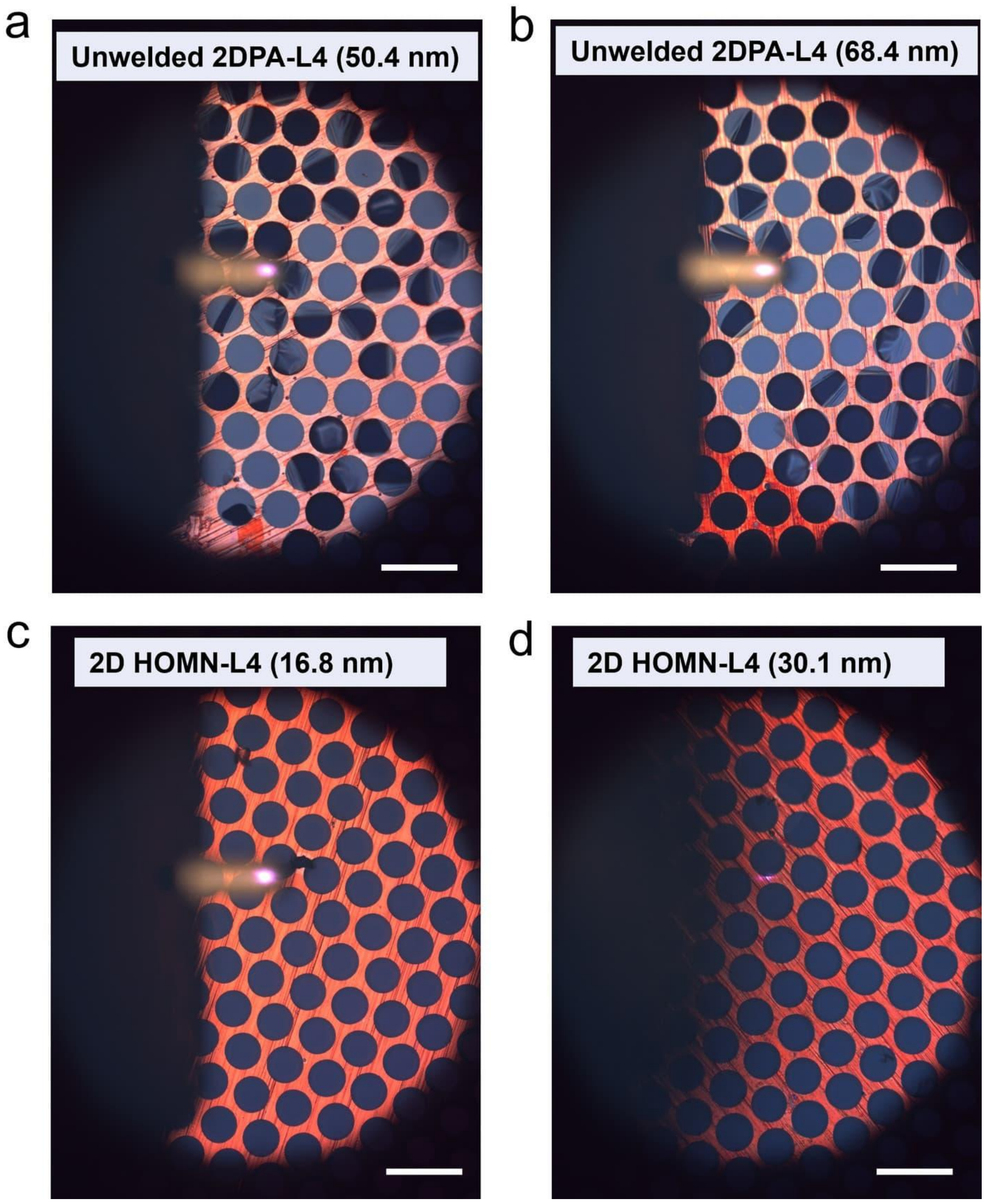

**Supplementary Fig. 43 | a, b,** Unwelded 2DPA film suspended over a 300-mesh copper grid. **c, d,** 2D HOMN-L4 films suspended over a 300-mesh copper grid. Scale bar: 100 µm.

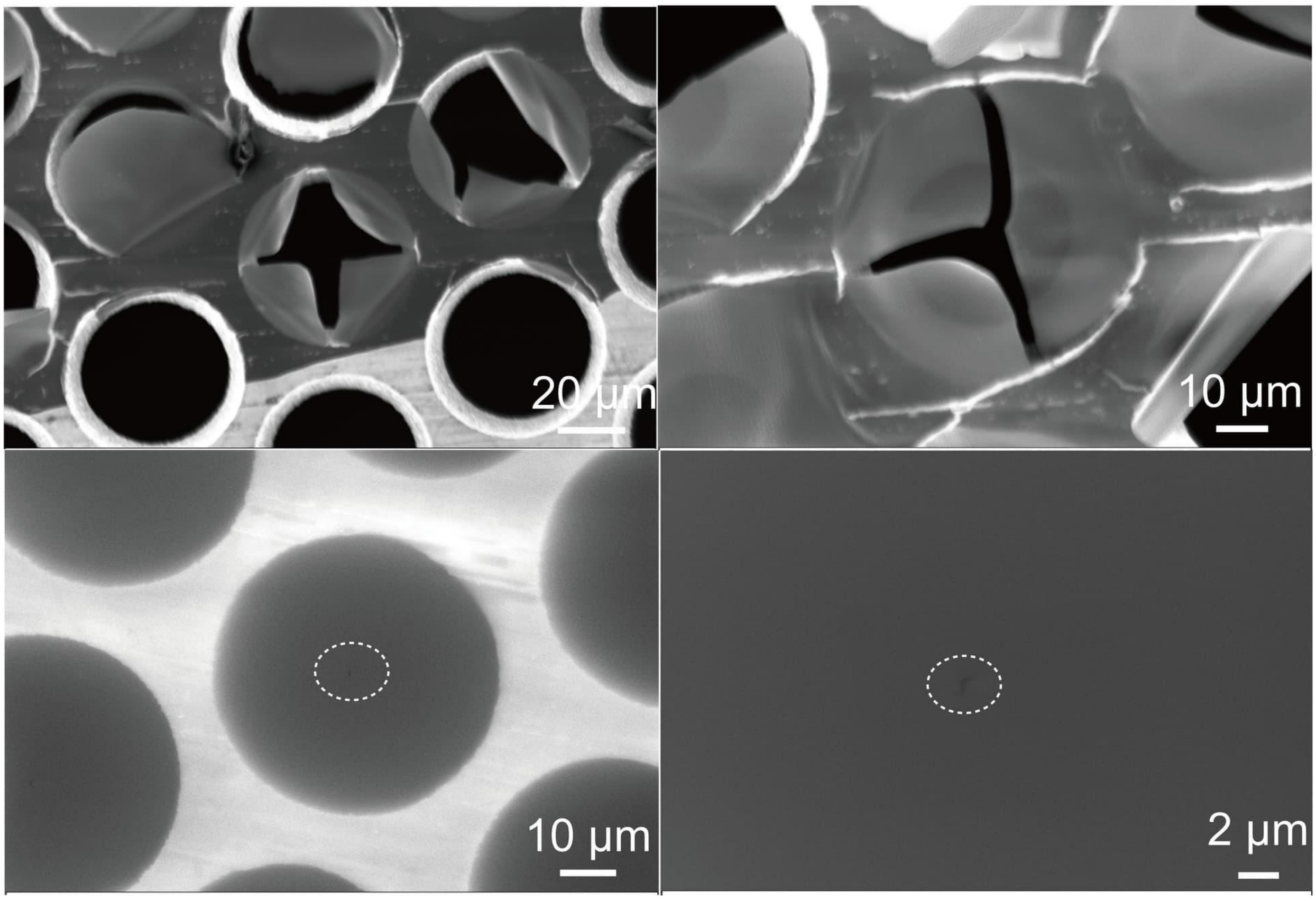

**Supplementary Fig. 44 |** Post-nanoindentation SEM images of unwelded 2DPA films (up) and 2D HOMN-L4 films (down).

**Demonstrations of free-standing films**

To demonstrate their macroscopic continuity and self-supporting capability, approximately 80-nm-thick 2D HOMN-L4 films were mounted across circular apertures with diameters of 4 and 6 cm. Both films remained suspended over the unsupported openings without visible macroscopic rupture, demonstrating that the covalently welded architecture can be fabricated and handled as centimeter-scale free-standing nanofilms.

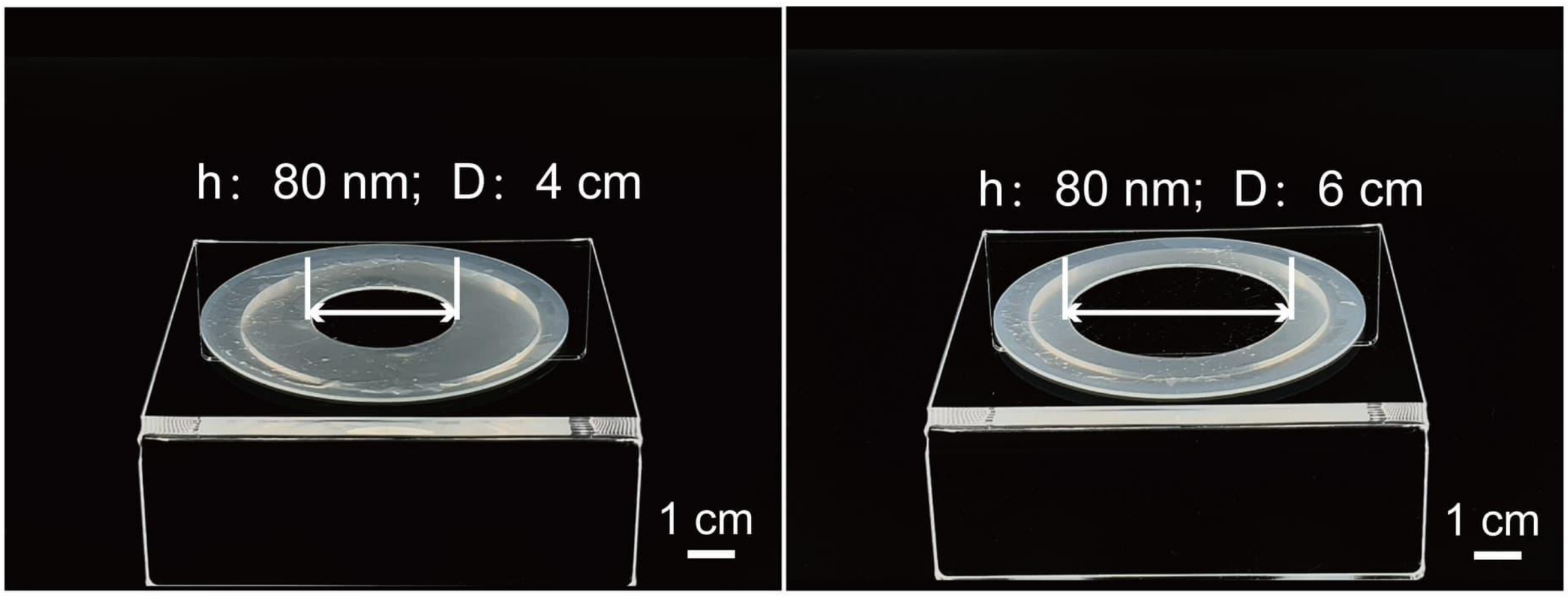

**Supplementary Fig. 45 |** Macroscopic free-standing 2D HOMN-L4 films. Approximately 80-nm-thick films suspended across circular apertures with diameters of 4 cm and 6 cm. Scale bar, 1 cm.

A 9.7-nm-thick free-standing 2D HOMN-L8 films was transferred onto ultrapure water for a qualitative puncture-resistance test. A glass capillary with a diameter of 0.3 mm was pressed vertically against the floating film.

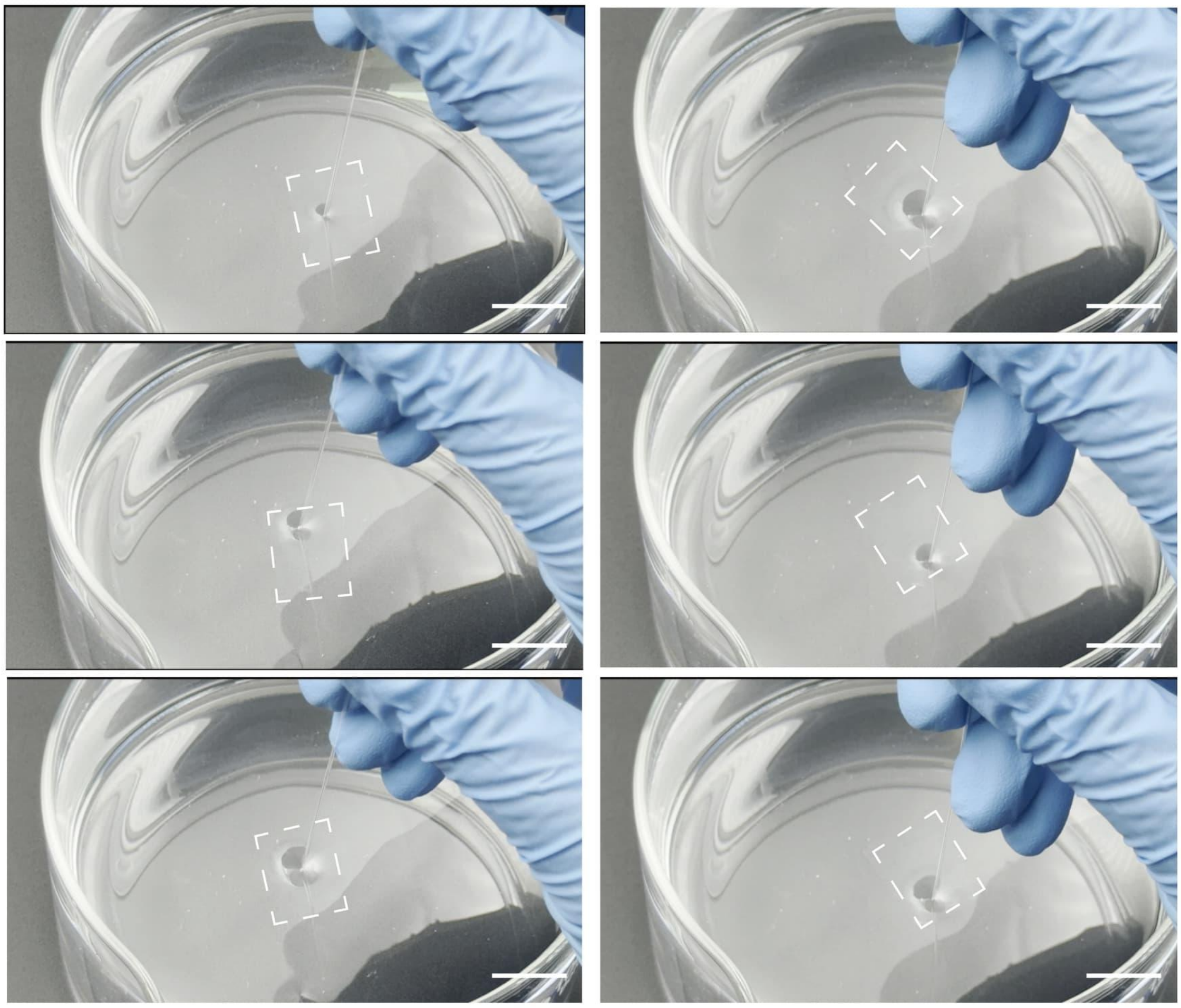

**Supplementary Fig. 46 |** Photographs of a 9.7-nm-thick free-standing 2D HOMN-L8 films resisting puncture by a 0.3-mm-diameter glass capillary on water. Scale bar, 2 cm.

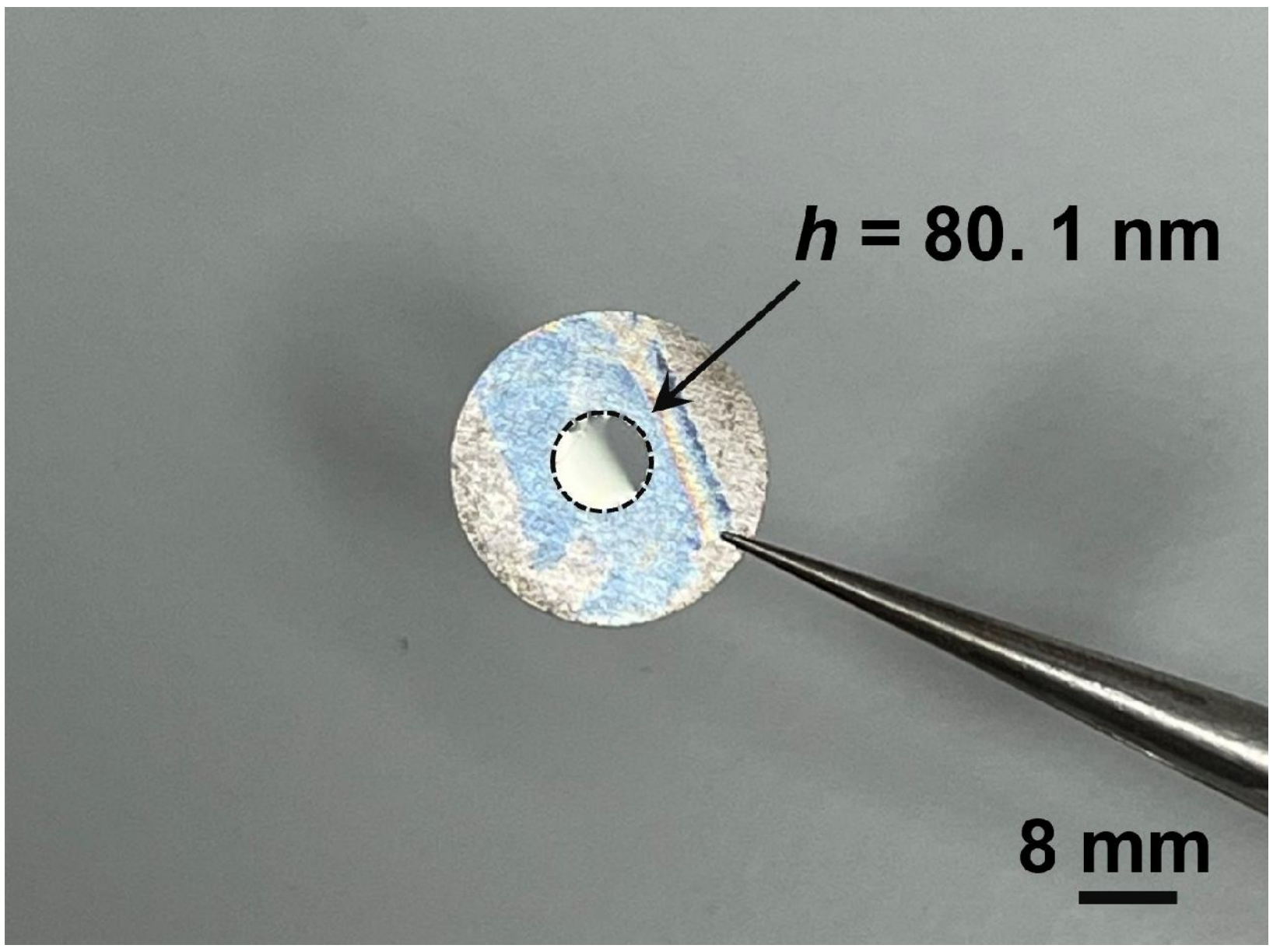

**Supplementary Fig. 47 |** Photographs of a millimeter-scale free-standing 2DHOMN-L0 films.

### Organic field-effect transistor fabrication on ultrathin 2D HOMN-L8 substrates

A 42-nm-thick 2D HOMN-L8 film served as the mechanical substrate for a bottom-contact, top-gate organic field-effect transistor. The device comprised patterned Au source and drain electrodes, a Poly[2,5-(2-octyldodecyl)-3,6-diketopyrrolopyrrole-alt-5,5-(2,5-di(thien-2-yl)thieno[3,2-b]thiophene)] (PDPP-DTT) organic semiconductor, a poly (methyl methacrylate) (PMMA) gate dielectric and an Al gate electrode. The 2D HOMN-L8 film functioned only as the supporting substrate, not as the semiconductor or gate dielectric.

The 42-nm-thick 2D HOMN-L8 film supported the complete Au/organic semiconductor/ PMMA/ Al device stack (Supplementary Fig. 48) and remained macroscopically intact during device fabrication and subsequent handling. The representative device exhibited p-type field-effect behaviour (Supplementary Fig. 49). The forward and reverse transfer sweeps followed similar trajectories, with a limited separation near the turn-on region. The output curves increased systematically with negative gate bias and approached saturation at high V, consistent with transistor operation.

Optical photographs further showed that the completed device structures could be handled on a rigid carrier, suspended across an opening and conformed to a fingertip. These configurations demonstrate the macroscopic continuity, low bending stiffness and transferability of the 42-nm L8 substrate after device integration.

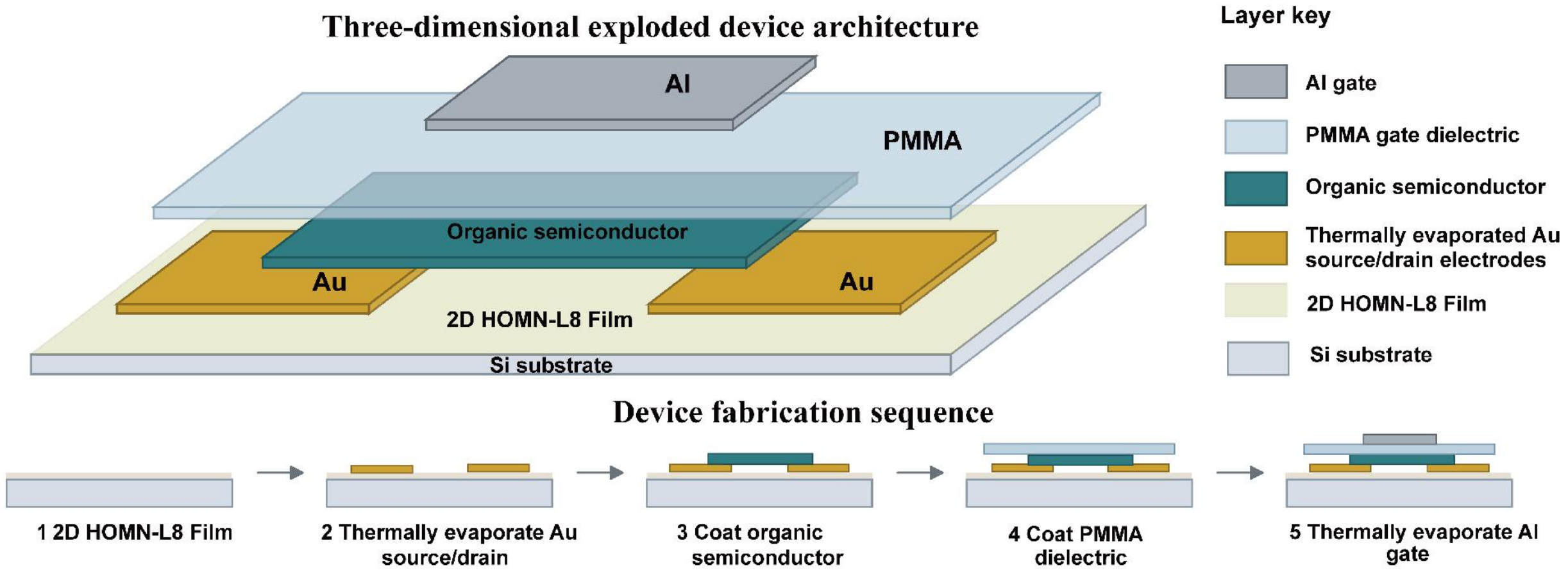

**Supplementary Fig. 48 |** Design and fabrication of an organic transistor on a 42-nm 2D HOMN-L8 substrate. a, Schematic cross-section of the bottom-contact, top-gate device comprising a 42-nm 2D HOMN-L8 supporting substrate, Au source and drain electrodes, an organic semiconductor, a PMMA gate dielectric and an Al gate electrode. b, Optical micrograph of a representative device on its fabrication carrier. The 2D HOMN-L8 film functions as the mechanical substrate and is not part of the semiconducting or dielectric layers.

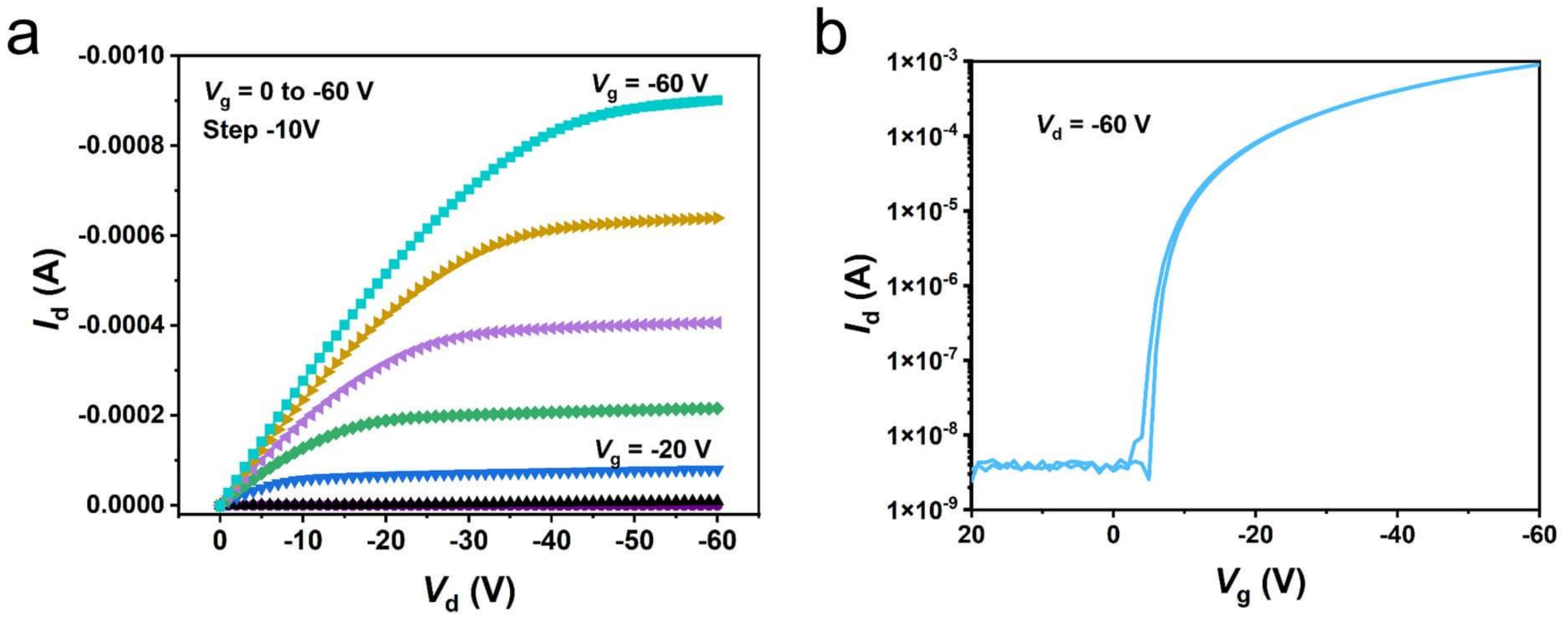

**Supplementary Fig. 49 |** Electrical characteristics of an organic transistor supported by a 42-nm 2D HOMN-L8 film. a, Output characteristics measured as the drain voltage was swept from 0 to −60 V at gate voltages from 0 to −60 V in −10 V steps. b, Forward and reverse transfer characteristics measured at a drain voltage of −60 V. The traces show hysteresis; its microscopic origin was not established by this measurement.

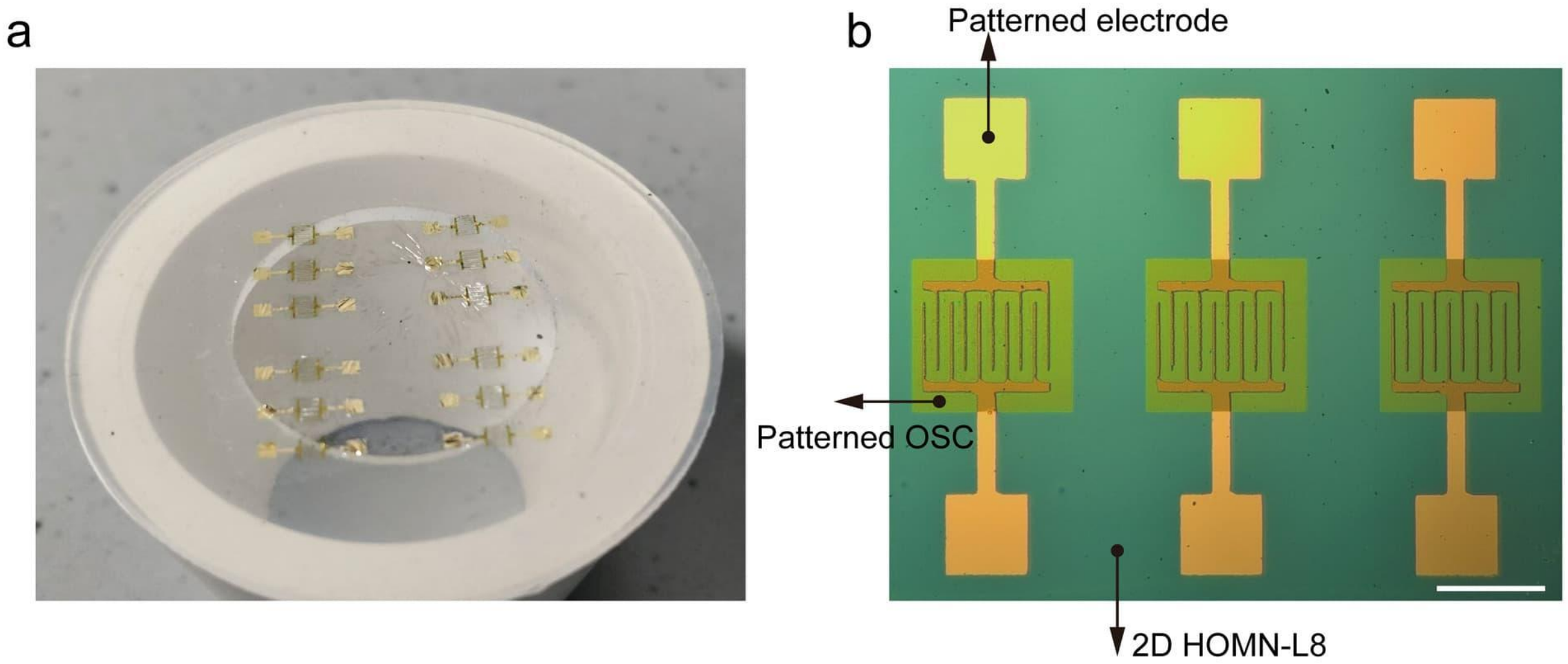

**Supplementary Fig. 50 |** Handling and transfer of transistor structures supported by ultrathin L8 films. a, Photograph of the transferred 42-nm-thick free-standing 2D HOMN-L8 film supporting multiple patterned devices. b, Representative optical micrograph of devices fabricated on the 2D HOMN-L8 film, showing the patterned organic semiconductor (OSC) channels and patterned metal electrodes, with device patterns defined via metal evaporation and standard photolithography.

**Kelvin probe force microscopy of top and bottom film surfaces**

To examine thickness-dependent top–bottom surface-potential asymmetry, 2D HOMN-L4 films of 25, 32, 44, 49, 57, 59, 69, 76, 94 and 103.5 nm were prepared on Si substrates. After selective HF etching, each membrane was divided into two pieces and transferred to highly oriented pyrolytic graphite (HOPG): one with the original air-facing surface exposed to the KPFM tip and the other inverted to expose the original substrate-facing surface. This pairing permits comparison of the two surfaces, subject to the effects of transfer and ambient conditions.

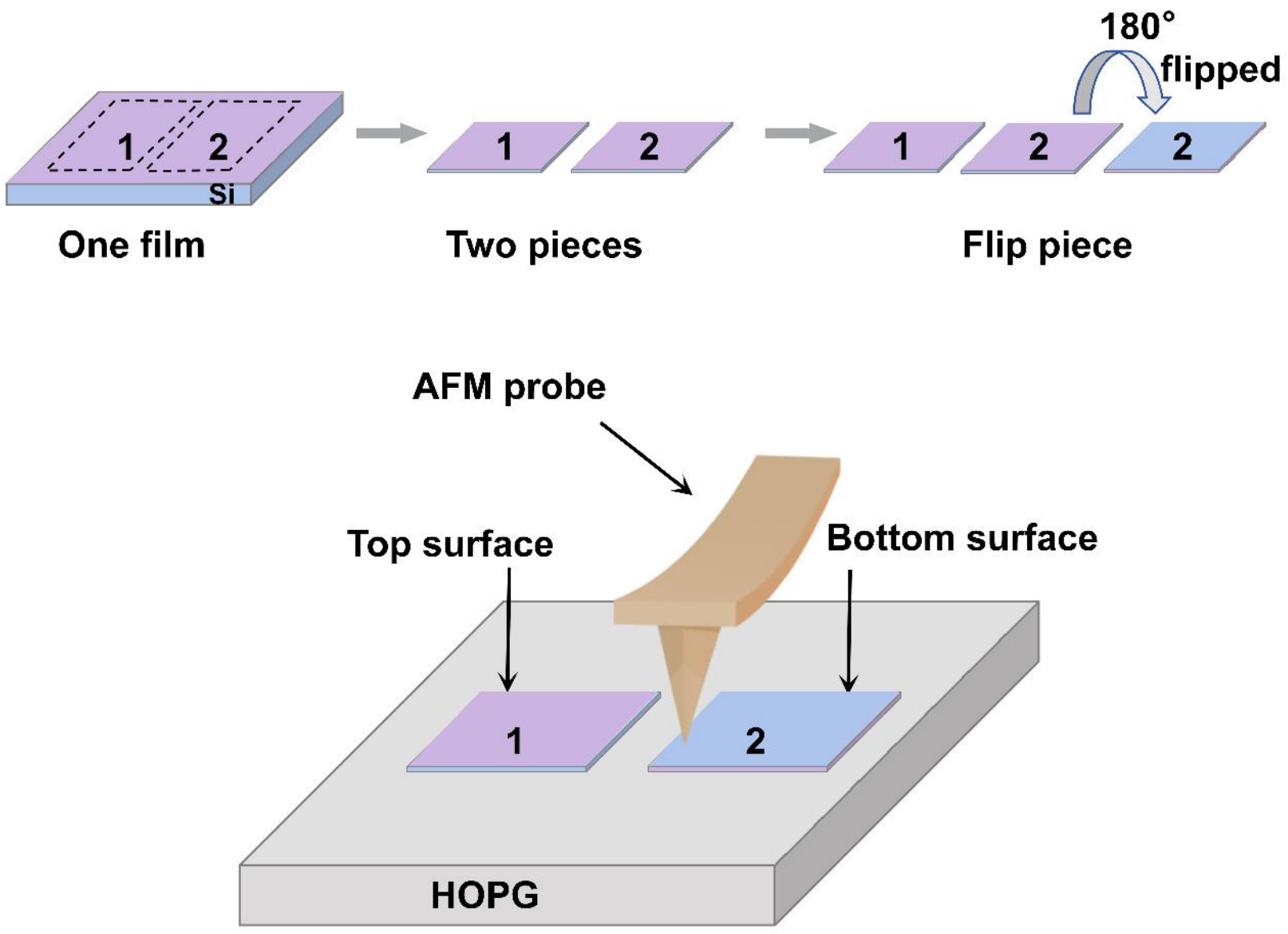

**Supplementary Fig. 51 |** Surface contact potential difference of 2D HOMN-L4 films. a, Schematic of sample preparation for top and bottom surface measurements. b, Schematic of the Kelvin probe force microscopy (KPFM) measurement of the contact potential difference (VCPD) on the top and bottom surfaces of the same film supported on HOPG.

Kelvin probe force microscopy (KPFM) under ambient conditions was employed to map the contact potential difference ($V_{CPD}$) across the top and bottom surfaces of all thickness-variant 2D HOMN specimens. For each acquired VCPD spatial map, the statistical distribution of pixel-wise $V_{CPD}$ values was quantified via relative frequency histograms, and the median $V_{CPD}$ value of each surface was extracted as the representative characteristic potential for subsequent quantitative comparison. (Supplementary Figs. 52–61)

The same Au-coated conductive tip and HOPG support were used throughout. Nevertheless, absolute contact potentials may vary with local HOPG condition, humidity, tip drift and transfer-related surface changes. For each film thickness, the top–bottom offset was calculated as $\Delta V_{CPD} = V_{CPD,\ bottom} - V_{CPD,\ top}$. This within-thickness comparison reduces some between-measurement offsets, but it does not uniquely isolate a substrate-induced amide orientation; complementary orientation-sensitive measurements are required.

The thickness-dependent top–bottom contact-potential offset is consistent with a vertical electrical asymmetry. KPFM measures surface potential rather than the direction of individual amide bonds; the data alone therefore do not establish a unique molecular-orientation mechanism. Substrate-induced orientation, trapped charge, surface chemistry and transfer-related effects remain possible contributors. Independent orientational measurements would be required to distinguish these contributions.

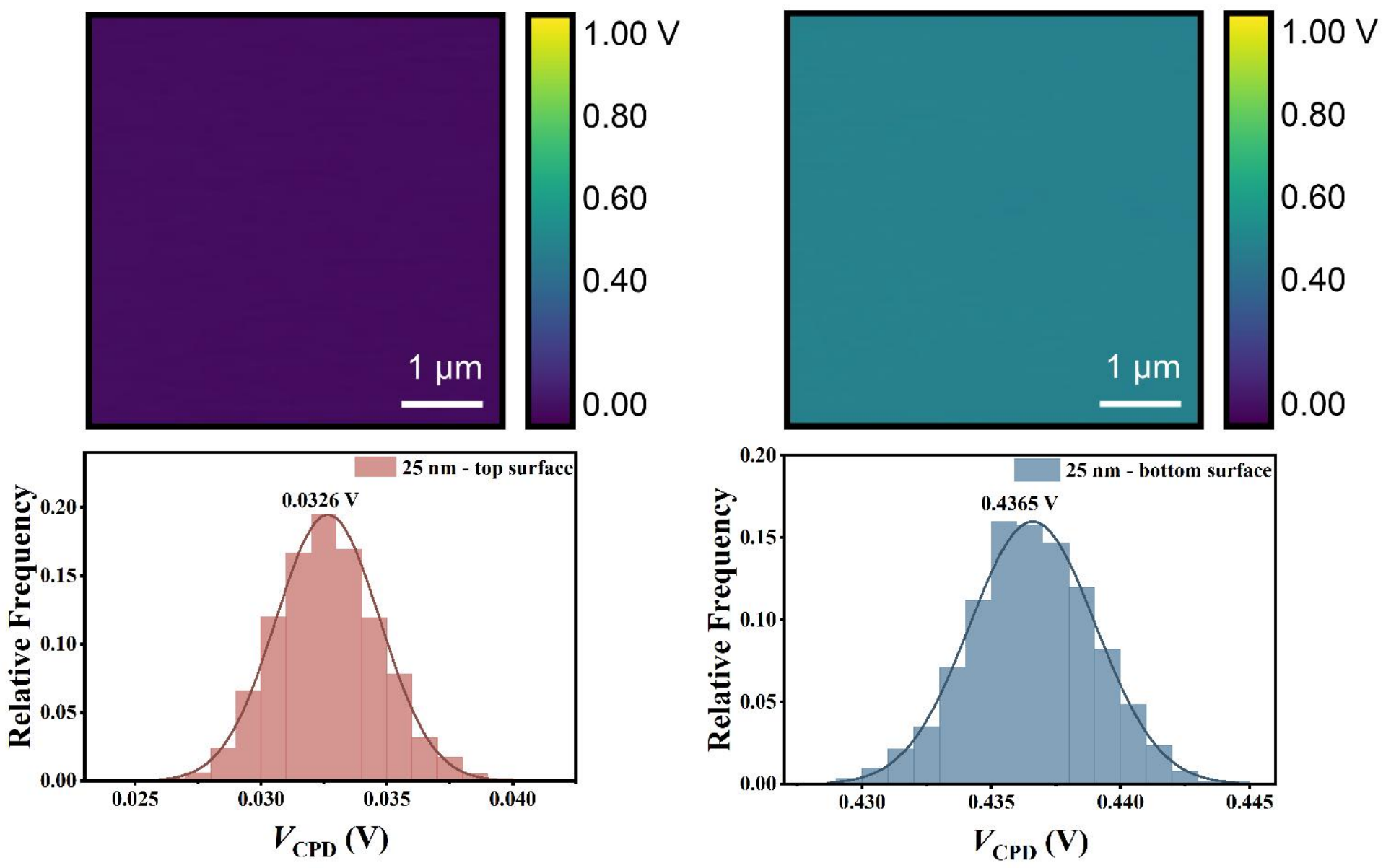

**Supplementary Fig. 52 |** KPFM characterization for the top and bottom surfaces of the 25 nm-thick 2D HOMN-L4 film. Upper panels: Spatial contact potential difference ($V_{CPD}$) maps collected on the top (left column) and bottom (right column) surfaces; the scale bar is 1 μm for all images. Lower panels: Corresponding relative frequency histograms of measured $V_{CPD}$ values, overlaid with Gaussian fitting curves, where the median $V_{CPD}$ value of each surface is labelled.

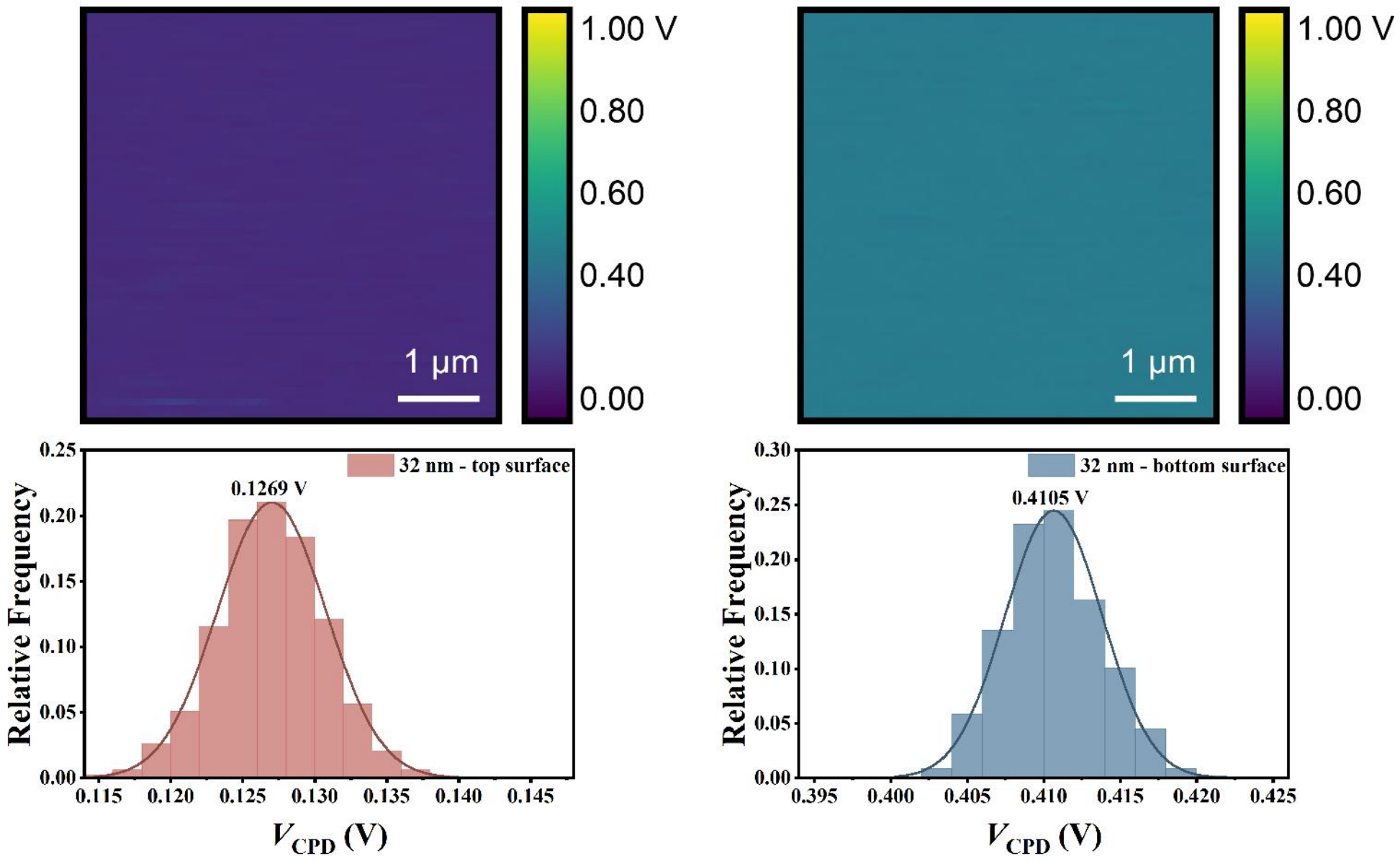

**Supplementary Fig. 53 |** KPFM characterization for the top and bottom surfaces of the 32 nm-thick 2D HOMN-L4 film. Upper panels: Spatial contact potential difference ($V_{CPD}$) maps collected on the top (left column) and bottom (right column) surfaces; the scale bar is 1 μm for all images. Lower panels: Corresponding relative frequency histograms of measured $V_{CPD}$ values, overlaid with Gaussian fitting curves, where the median $V_{CPD}$ value of each surface is labelled.

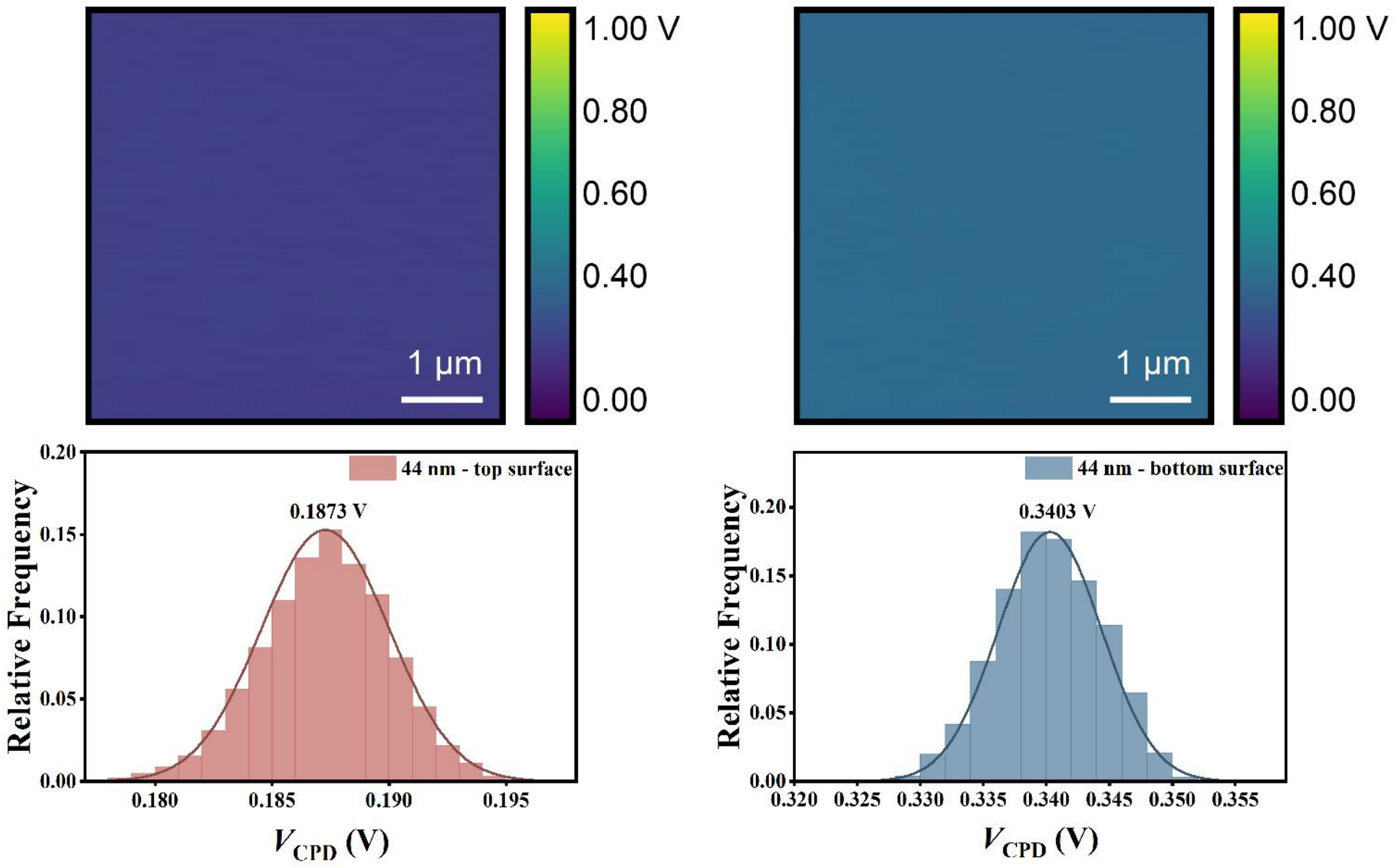

**Supplementary Fig. 54 |** KPFM characterization for the top and bottom surfaces of the 44 nm-thick 2D HOMN-L4 film. Upper panels: Spatial contact potential difference ($V_{CPD}$) maps collected on the top (left column) and bottom (right column) surfaces; the scale bar is 1 μm for all images. Lower panels: Corresponding relative frequency histograms of measured $V_{CPD}$ values, overlaid with Gaussian fitting curves, where the median $V_{CPD}$ value of each surface is labelled.

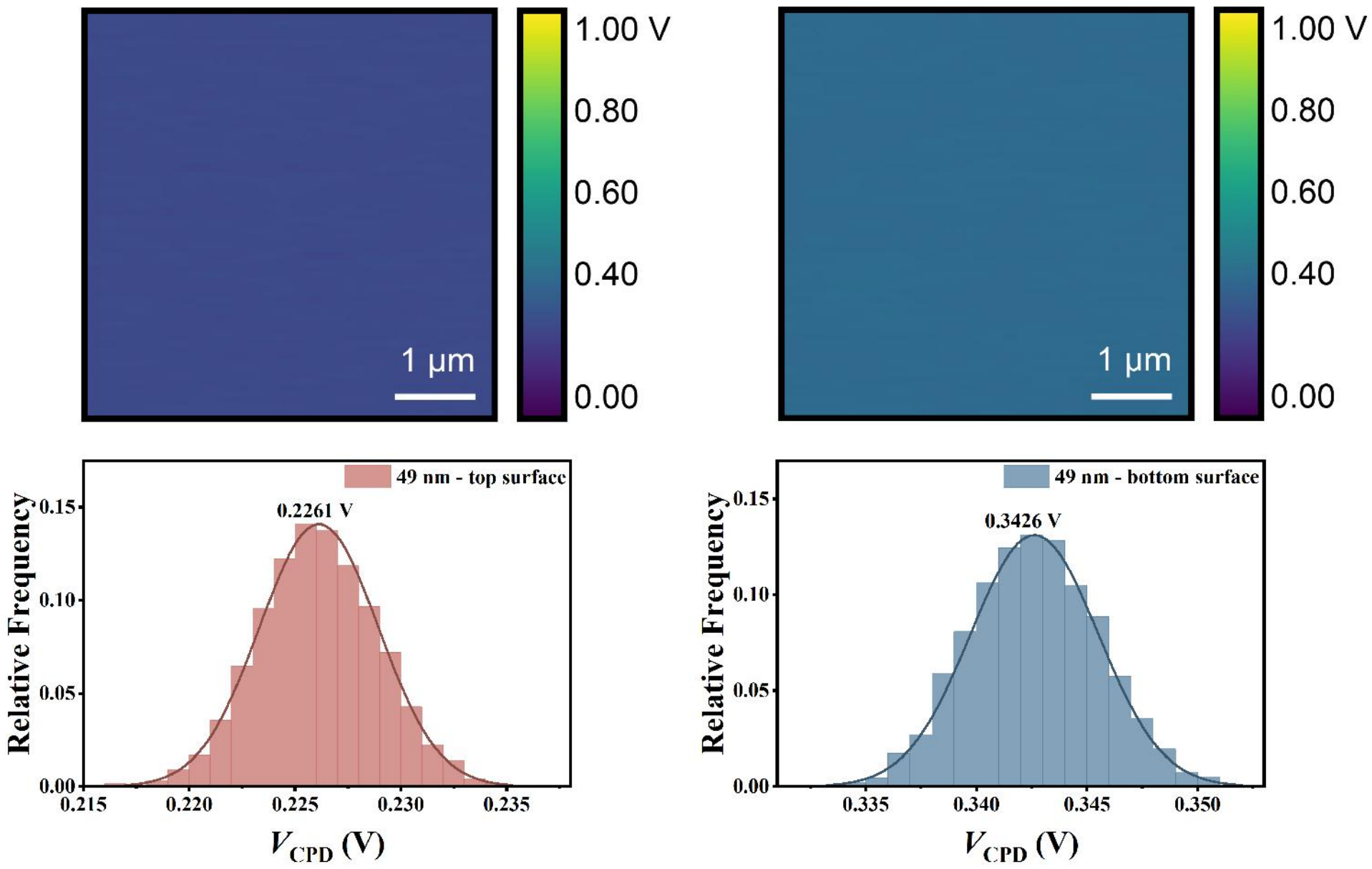

**Supplementary Fig. 55 |** KPFM characterization for the top and bottom surfaces of the 49 nm-thick 2D HOMN-L4 film. Upper panels: Spatial contact potential difference ($V_{CPD}$) maps collected on the top (left column) and bottom (right column) surfaces; the scale bar is 1 μm for all images. Lower panels: Corresponding relative frequency histograms of measured $V_{CPD}$ values, overlaid with Gaussian fitting curves, where the median $V_{CPD}$ value of each surface is labelled.

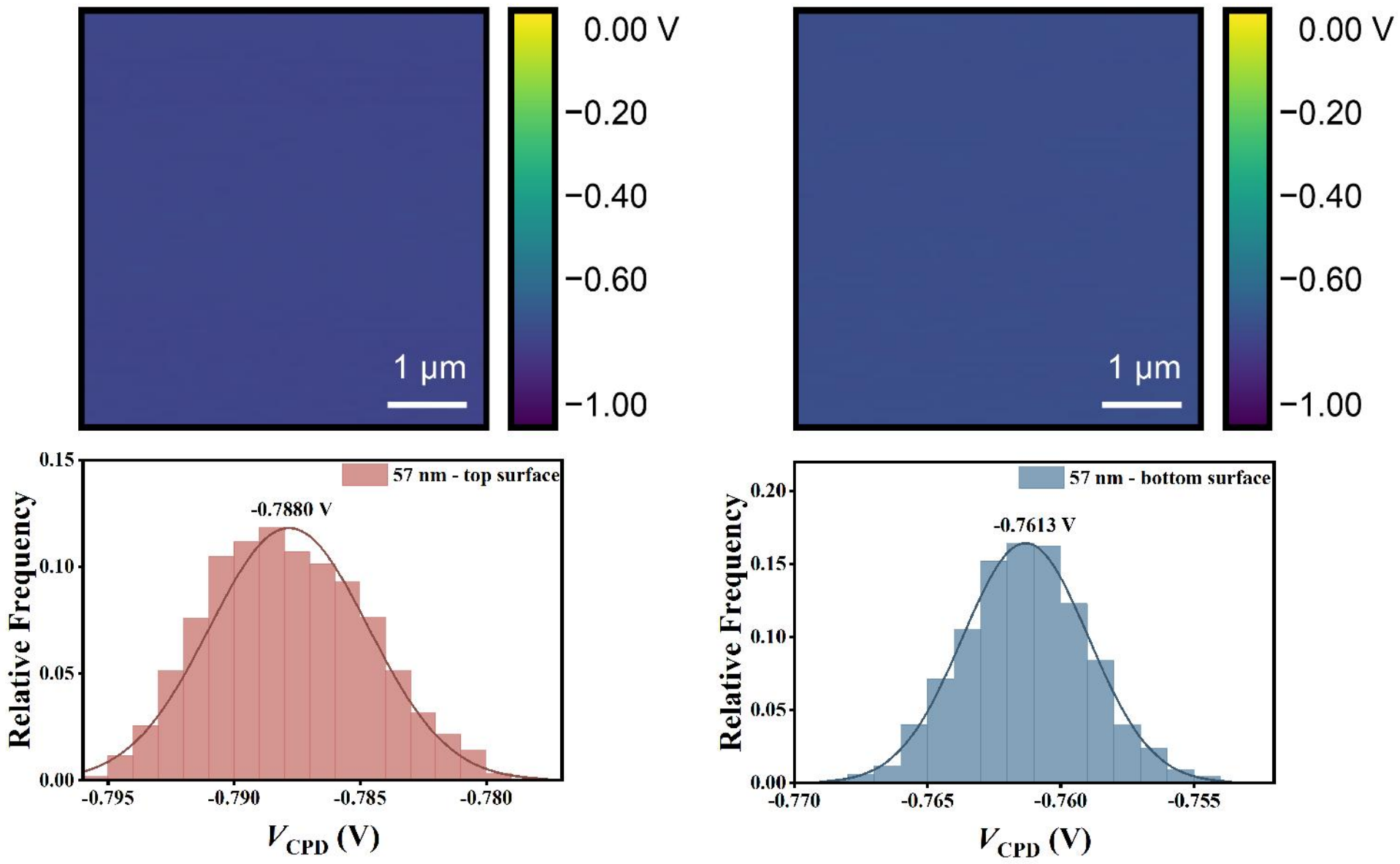

**Supplementary Fig. 56 |** KPFM characterization for the top and bottom surfaces of the 57 nm-thick 2D HOMN-L4 film. Upper panels: Spatial contact potential difference ($V_{CPD}$) maps collected on the top (left column) and bottom (right column) surfaces; the scale bar is 1 μm for all images. Lower panels: Corresponding relative frequency histograms of measured $V_{CPD}$ values, overlaid with Gaussian fitting curves, where the median $V_{CPD}$ value of each surface is labelled.

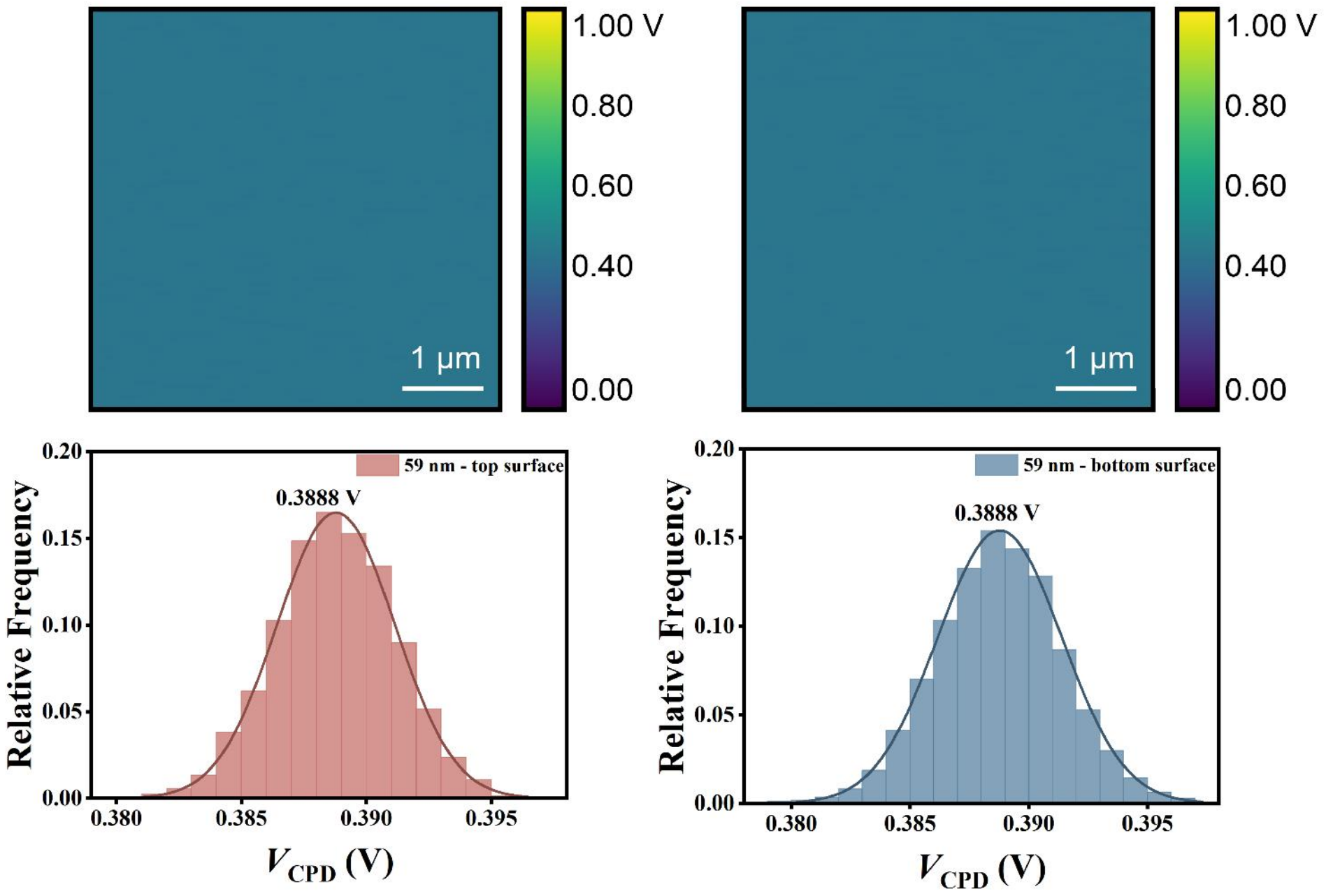

**Supplementary Fig. 57 |** KPFM characterization for the top and bottom surfaces of the 59 nm-thick 2D HOMN-L4 film. Upper panels: Spatial contact potential difference ($V_{CPD}$) maps collected on the top (left column) and bottom (right column) surfaces; the scale bar is 1 μm for all images. Lower panels: Corresponding relative frequency histograms of measured $V_{CPD}$ values, overlaid with Gaussian fitting curves, where the median $V_{CPD}$ value of each surface is labelled.

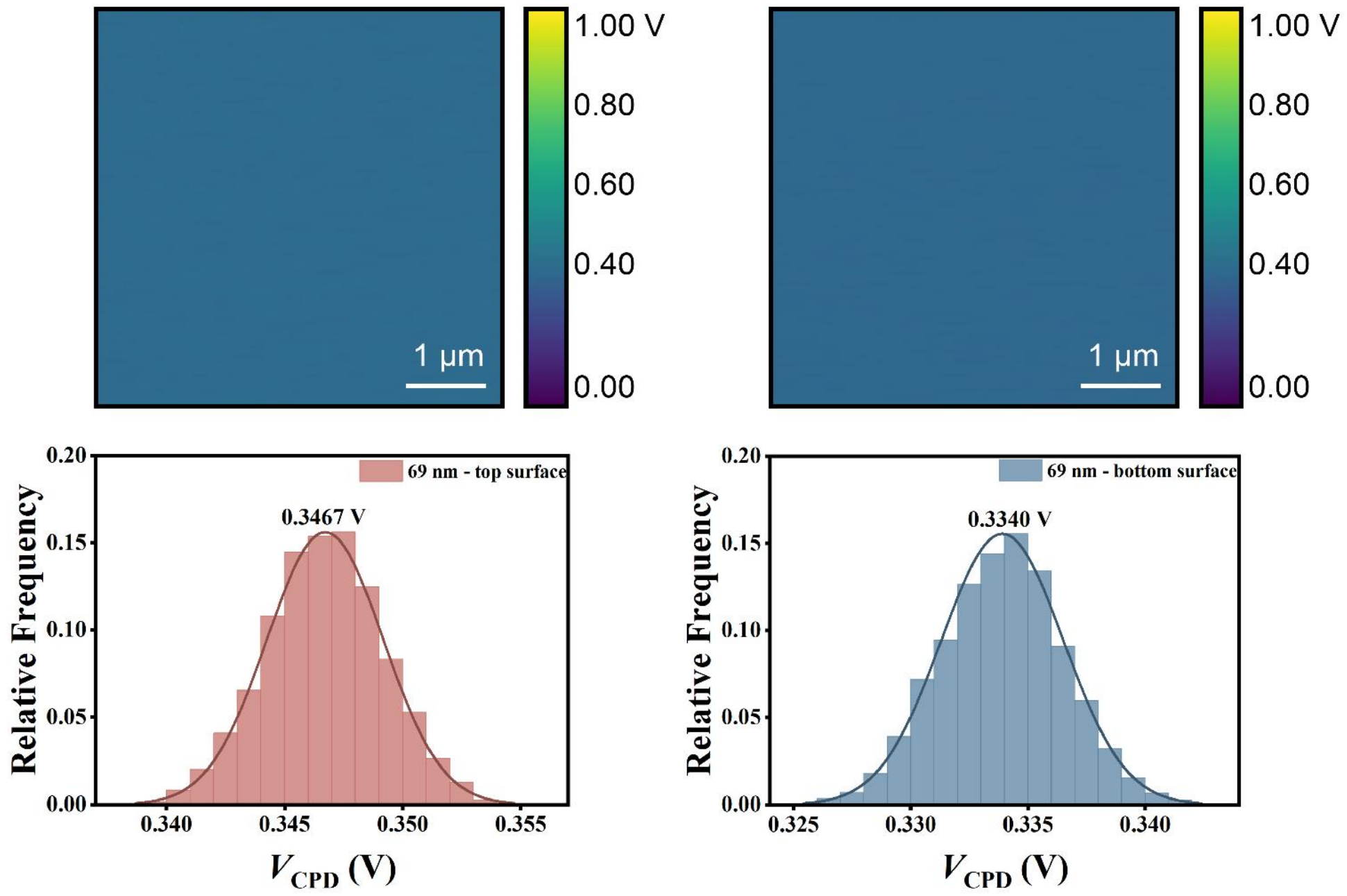

**Supplementary Fig. 58 |** KPFM characterization for the top and bottom surfaces of the 69 nm-thick 2D HOMN-L4 film. Upper panels: Spatial contact potential difference ($V_{CPD}$) maps collected on the top (left column) and bottom (right column) surfaces; the scale bar is 1 μm for all images. Lower panels: Corresponding relative frequency histograms of measured $V_{CPD}$ values, overlaid with Gaussian fitting curves, where the median $V_{CPD}$ value of each surface is labelled.

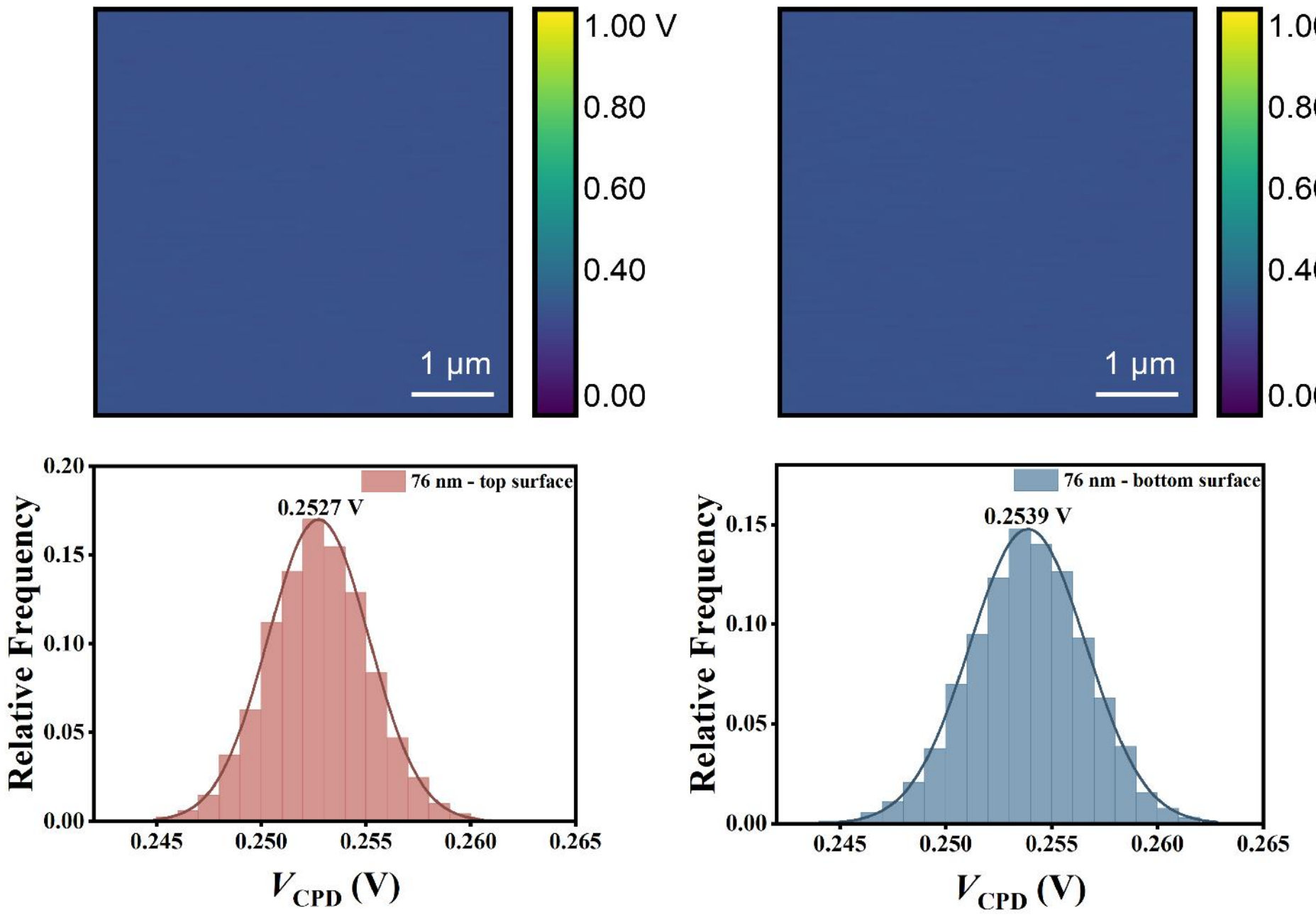

**Supplementary Fig. 59 |** KPFM characterization for the top and bottom surfaces of the 76 nm-thick 2D HOMN-L4 film. Upper panels: Spatial contact potential difference ($V_{CPD}$) maps collected on the top (left column) and bottom (right column) surfaces; the scale bar is 1 μm for all images. Lower panels: Corresponding relative frequency histograms of measured $V_{CPD}$ values, overlaid with Gaussian fitting curves, where the median $V_{CPD}$ value of each surface is labelled.

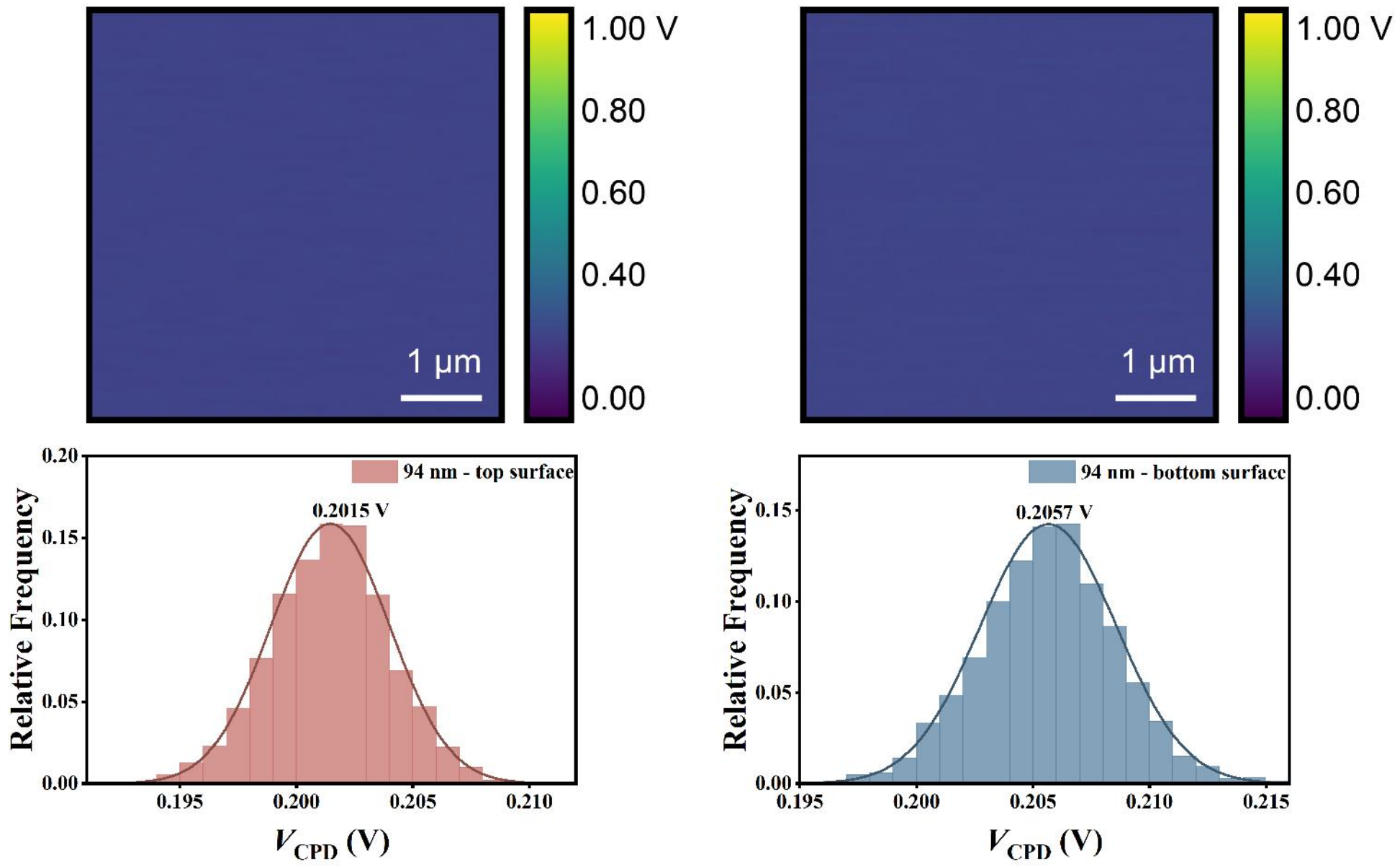

**Supplementary Fig. 60 |** KPFM characterization for the top and bottom surfaces of the 94 nm-thick 2D HOMN-L4 film. Upper panels: Spatial contact potential difference ($V_{CPD}$) maps collected on the top (left column) and bottom (right column) surfaces; the scale bar is 1 μm for all images. Lower panels: Corresponding relative frequency histograms of measured $V_{CPD}$ values, overlaid with Gaussian fitting curves, where the median $V_{CPD}$ value of each surface is labelled.

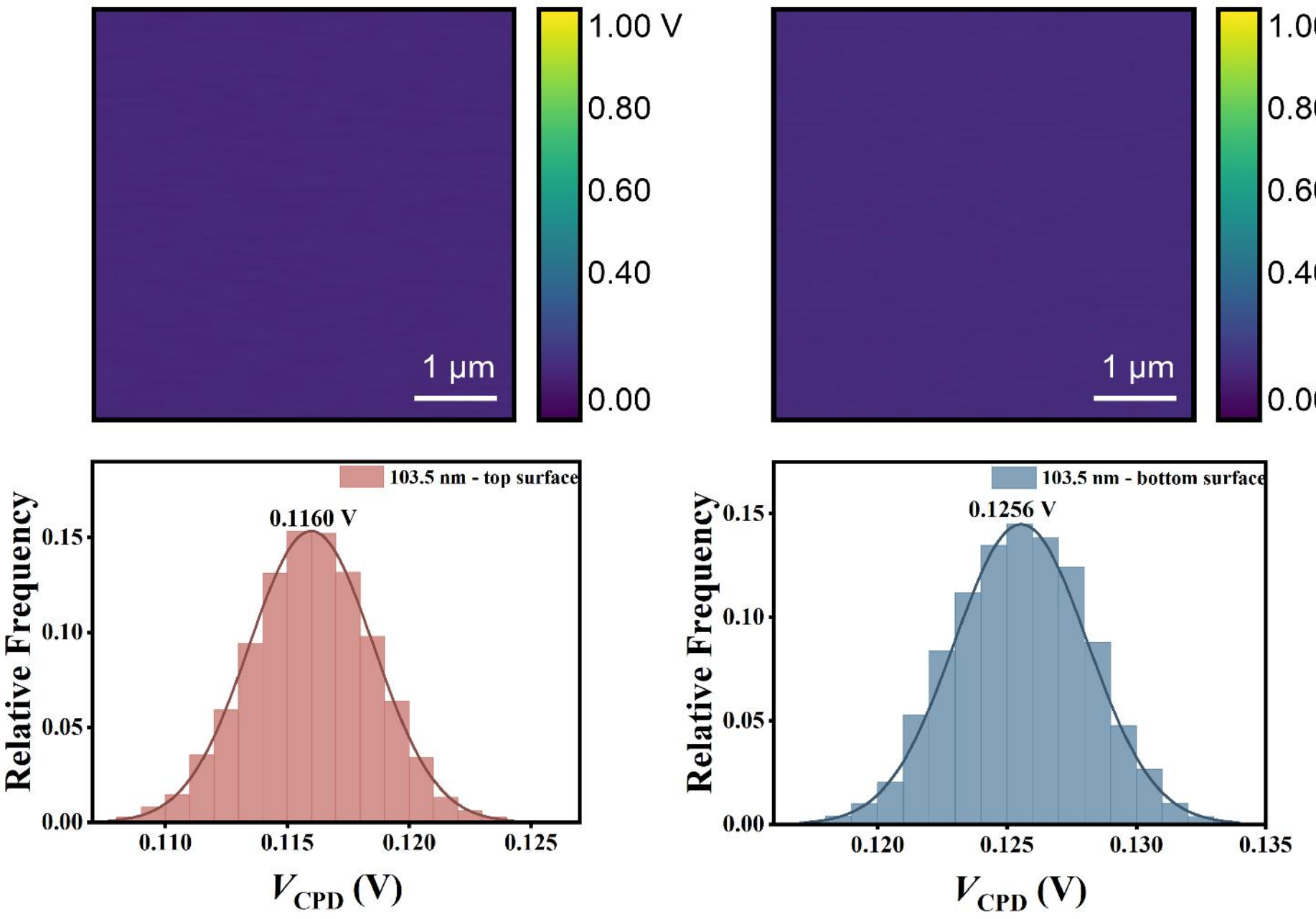

**Supplementary Fig. 61 |** KPFM characterization for the top and bottom surfaces of the 103.5 nm-thick 2D HOMN-L4 film. Upper panels: Spatial contact potential difference ($V_{CPD}$) maps collected on the top (left column) and bottom (right column) surfaces; the scale bar is 1 μm for all images. Lower panels: Corresponding relative frequency histograms of measured $V_{CPD}$ values, overlaid with Gaussian fitting curves, where the median $V_{CPD}$ value of each surface is labelled.

**High-velocity impact characterization**

**High-Velocity Supersonic Impact Toughness**

**Laser-induced microprojectile impact testing (LIPIT)**

**Sample preparation**

First, different amounts of 2DPA-β-Alanine ethyl ester hydrochloride were dissolved in the dimethylformamide (DMF) solvent (99 %, Sigma). then heat the solution at 60 °C until a visually homogeneous solution is obtained. The polymer solution with different 2DPA-β-Alanine ethyl ester hydrochloride concentrations (30-150 mg/mL) were prepared. A silicon wafer (1×1 cm) was used as the substrate to fabricate polymer films in a spin-coater. Polymer solution was dropped on the rotating silicon wafer, then spun for 60 s at 1500 rpm. Polymer films with different thicknesses were fabricated using different concentrations of polymer solution. Then, the films were annealed at 80 °C for 30 minutes and welded at 280 °C for 15 minutes. To separate the film from the silicon wafer, the silicon wafer coated with the polymer film was slowly immersed into 10 % HF for chemical etching. Then using another silicon wafer transferred it to ultra-pure water, and collected it on a copper grid for subsequent LIPIT.

The film thickness was obtained by measuring the height between the substrate and the film top surface across a cut step on the film using an Atomic Force Microscope (Cypher VRS 1250).

**Optical setup of the LIPIT**

Supplementary Fig. 62 shows the key optical elements of the advanced laser induced projectile impact test[28,53–55] (LIPIT) experimental setup. The laser-induced LlPIT ablation laser pulse (6 ns pulse width, 532 nm) was created with a neodymium-doped yttrium aluminum garnet (Nd: YAG) laser (Dawa-300, Beijing Lablaser) to accelerate a single microparticle placed near the focal point of the ablation target. The composition of the target includes: a glass substrate (8 mm thick, 50 mm in diameter, Baojia Quartz from China), a sacrificial gold layer (50 nm) sputtered onto the glass, and a polyimide layer (15 μm, provided by Dongguan Kaipu Adhesive Tape Co., Ltd). The polyimide layer is adhered to the sacrificial gold layer with a few micrometers of PDMS (10:1, Sylgard 184, Dow Chemical). $SiO_2$ particles (mass density = 2.2 g/cm$^3$, diameter = 8.11±0.47 μm, Thermo Fisher Scientific) were suspended in ethanol and the mixed solution was spin-coated onto the surface of the PI film.

In the test, an individual micro-projectile was selected as the target, and a digital delay generator was used to trigger the laser to generate a laser pulse to focus on the gold layer directly above the micro-projectile, resulting in local ablation of the gold film, expansion of the polyimide, and acceleration of the silica microsphere. In addition, the emission velocity of the silica microspheres can be changed by adjusting the energy of the excitation laser pulse (30~1100 m/s). The trajectory of the micro-projectile was captured by global exposure mode cmos camera (MV-GED500M, Shenzhen Medway Vision), triggered for a 10 μs exposure by a digital delay generator (DG645, Stanford Research Systems). The supercontinuum laser (6 ps, 0.5~5 MHz adjustable, SC-PRO-M, Wuhan Anyang laser) was triggered as an illumination source. Each laser pulse of the laser records the current microsphere position on the camera sensor. The fast motion of the microsphere was recorded on a photo by multiple exposures, where the laser repetition rate determines the time interval between frames (0.2~2 μs), and the number of pulses determines how many exposures are captured in a single image. Typically, we use less than 20 pulses because the increase in the number of pulses can leads to a decrease in image contrast. In the test, the background image before the impact and the image during the impact were taken respectively, and the image contrast was improved by gray value difference and image filtering (Matlab). The impact velocity and rebound velocity of the microsphere were calculated by dividing the actual distance between the pixels of the adjacent projectiles in the image by the pulse interval. In addition, the system was equipped with a vacuum chamber that enables impact testing at pressures below 260 m Torr to avoid any influence of air resistance.

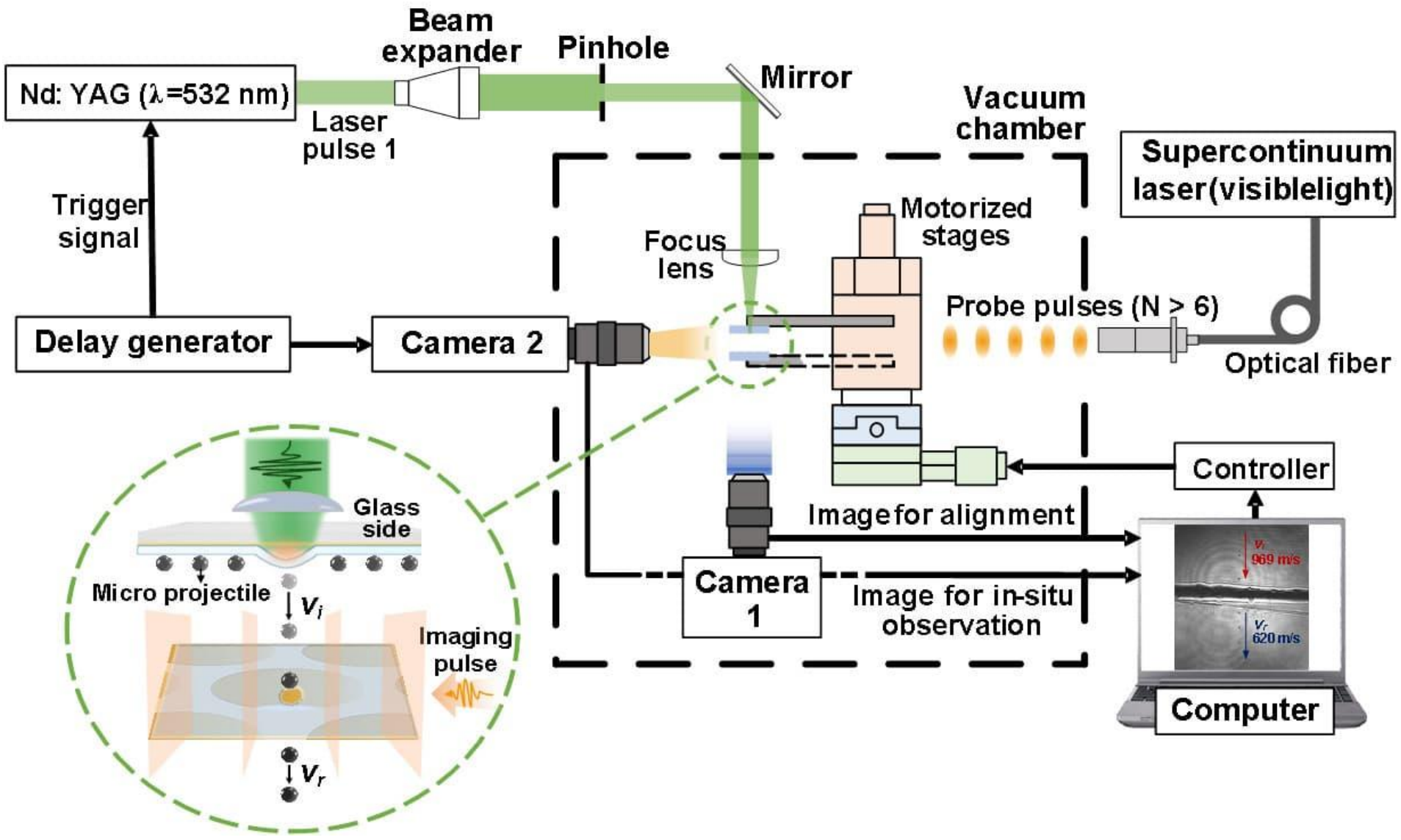

**Supplementary Fig. 62 |** Experimental setup for the laser-induced projectile impact test (LIPIT).

**Strain-rate and energy calculations**

During the film perforation process, the strain rate (Δε/Δt) is a function of the radial expansion rate in the deformation region of the polymer film:

$$\frac{\Delta\varepsilon}{\Delta t} \cong \frac{\varepsilon_{\max}}{t_p} \cong \frac{t_p}{2}\left(\frac{v_i}{R_c}\right)^2 \tag{9}$$

Where $\varepsilon_{\max} \cong \left(v_i t_p / R_c\right)^2 / 2$, based on the 1-D approximation of the maximum tensile strain for isotropic membrane with biaxial deformation, The film is approximated as a conical expansion film, $R_c$ is the deformation cone radius, $t_p \cong R_c / v_c$, $t_p$ is the perforation time of the projectile, which is approximately equal to the ratio of the cone's radius to its velocity. The cone velocity is calculated by $v_c = 1.23c_2\left(v_i / \left(\sqrt{2}c_2\right)\right)^{2/3}$, where $c_2$ is the in-plane sound speed in the materials, which is directly related to the plane-strain elastic modulus and mass density of the material. For the 8.11-μm $SiO_2$ microprojectiles used in this study, impact velocities of 200–1,200 m/s correspond to characteristic membrane strain rates of approximately $10^7$–$10^8$ $s^{-1}$.

**Specific dissipation energy**

The kinetic energy change (($E_k$) is related to the initial and final velocities of the projectile ($v_i$, $v_r$). Ignoring the influence of air resistance, the penetration dissipation energy ($E_p$) of projectile target is equal to the change of projectile kinetic energy (($E_k$).

$$E_p = \Delta E_k = \frac{1}{2}m_p(v_i^2 - v_r^2) \tag{10}$$

Where $m_p$ is the particle mass; $v_i$ is the velocity before the projectile penetrates the target; $v_r$ is the velocity of the projectile after penetrating the target. $E_p$ is a function of the various dissipation mechanisms including elastic stretching of the film, fracture, and adiabatic heating, which is composed of two forms of kinetic energy dissipation ($E_{plug}$) and other dissipation ($E_d$),

$$E_p = E_{plug} + E_d = \rho h A_p v_i^2 / 2 + E_d \tag{11}$$

Where $E_{plug}$ is the kinetic energy transmitted by the projectile to the target deformation area $A_p$; $\rho$ is the target density; $h$ is the thickness of the target. To compare the penetration resistance of different materials under ballistic impact, the penetration dissipation energy is normalized with respect to mass:

$$E_p^* = E_p / \rho h A_p = v_i^2 / 2 + E_d^* \tag{12}$$

Where $E_p^*$ is specific dissipation energy or specific energy absorption, and $E_d^*$ represents inelastic dissipation mechanisms such as impact delocalization or material compression, which can serve as an indicator parameter for comparing the penetration resistance of different thin-film or bulk materials.

**Effect of air resistance**

Residual-gas drag can reduce projectile velocity and affect the energy dissipation inferred from impact and residual velocities. For the vacuum conditions considered here, the estimated speed change over the 1.5-2 mm launch-to-target distance is small. The following calculation uses a free-molecular drag model and explicitly states its assumptions. The one-dimensional equation of motion in a stationary gas is[56,57]:

$$M_p \frac{dv}{dt} = -\frac{1}{2} C_D \rho_g A_p v^2 \quad (13)$$

Here $M_P$ is the projectile mass, v is its speed relative to the gas, $C_D$ is the effective drag coefficient, $\rho_g$ is the residual-gas density, and $A_P$ is the projected cross-sectional area. For a spherical projectile of diameter $d_P = 2R_P$ and density $\rho_P$, $A_P = \pi d_P^2/4$ and $M_P = \pi\rho_P d_P^3/6$, giving

$$a_p = \frac{dv}{dt} = -\frac{3C_D\rho_g}{4\rho_p d_p} v^2 \quad (14)$$

The $SiO_2$ microspheres have a diameter of 8.11 ± 0.47 μm and a density of 2200 kg $m^{-3}$. The calculation below uses the nominal diameter of 8.11 μm and the velocity range of 200-1100 m $s^{-1}$ considered in this section. At the reported chamber pressure of less than 260 mTorr (34.7 Pa), the ideal-gas relation $\rho_g = p/(R_gT)$, with $R_g$ = 287.05 J $kg^{-1}$ $K^{-1}$ and an assumed temperature T = 293.15 K, gives $\rho_g < 4.12 \times 10^{-4}$ kg $m^{-3}$. A larger value of $6.0 \times 10^{-4}$ kg $m^{-3}$ is retained as a conservative estimate with respect to gas density, rather than as the actual chamber density.
Using this estimate and an air dynamic viscosity $\mu = 1.81 \times 10^{-5}$ Pa s gives $Re = \rho_g v d_p/\mu \approx$ 0.054–0.296. The mean free path $\lambda \approx (\mu/\rho_g)[\pi/(2R_gT)]1/2$ is approximately 130 μm, so $Kn = \lambda/d_p \approx 16$. These conditions lie in the free-molecular regime. The atmospheric drag coefficient $C_D$ = 0.7 ± 0.2 reported for hollow glass microspheres at Re ≈ 200 therefore cannot be directly applied to the present low-pressure conditions, and $C_D \leq 1$ is not a justified upper bound[56,57].
For a stationary gas, fully diffuse molecular reflection, and equal gas and particle temperatures, the free-molecular sphere-drag coefficient is[57]

$$S = \frac{v}{\sqrt{2R_gT}}, \quad (15)$$

$$C_{D,fm} = \frac{(1 + 2S^2)e^{-S^2}}{\sqrt{\pi}S^3} + \frac{(4S^4 + 4S^2 - 1)\,\mathrm{erf}(S)}{2S^4} + \frac{2\sqrt{\pi}}{3S} \quad (16)$$

Here S is the molecular speed ratio and erf is the error function. Under these assumptions, $C_{D,fm}$ is approximately 8.88 at 200 m $s^{-1}$ and 2.71 at 1100 m $s^{-1}$. The corresponding initial accelerations are approximately $-9.0 \times 10^3$ and $-8.3 \times 10^4$ m $s^{-2}$, respectively. These are model estimates, rather than experimentally established upper bounds.

To calculate the speed loss over a fixed distance, the velocity-dependent coefficient is retained and the following equation is integrated numerically:

$$\frac{dv}{dx} = -\frac{3\rho_g}{4\rho_p d_p} C_{D,fm}(v)\, v \quad (17)$$

For a flight distance of 2 mm, the calculated speed reduction is 0.090 m $s^{-1}$ at an initial speed of 200 m $s^{-1}$ and 0.150 m $s^{-1}$ at 1100 m $s^{-1}$, corresponding to 0.045% and 0.014% of the initial speeds. The associated losses are approximately 0.090% and 0.027% of the initial projectile kinetic energies, respectively. The short flight time (approximately 1.82 μs at 1100 m $s^{-1}$) explains the small speed change despite the large instantaneous deceleration.

These estimates indicate a small absolute influence of residual-gas drag on projectile speed. Its contribution to the uncertainty in film energy dissipation depends on the difference between the squared impact and residual velocities and should be evaluated using the distances between the velocity measurement positions and the target. If both velocities are measured locally near the target, upstream speed loss is already included in the measured impact velocity. The contribution of the drag correction alone to the inferred film dissipation is so small as to be negligible.

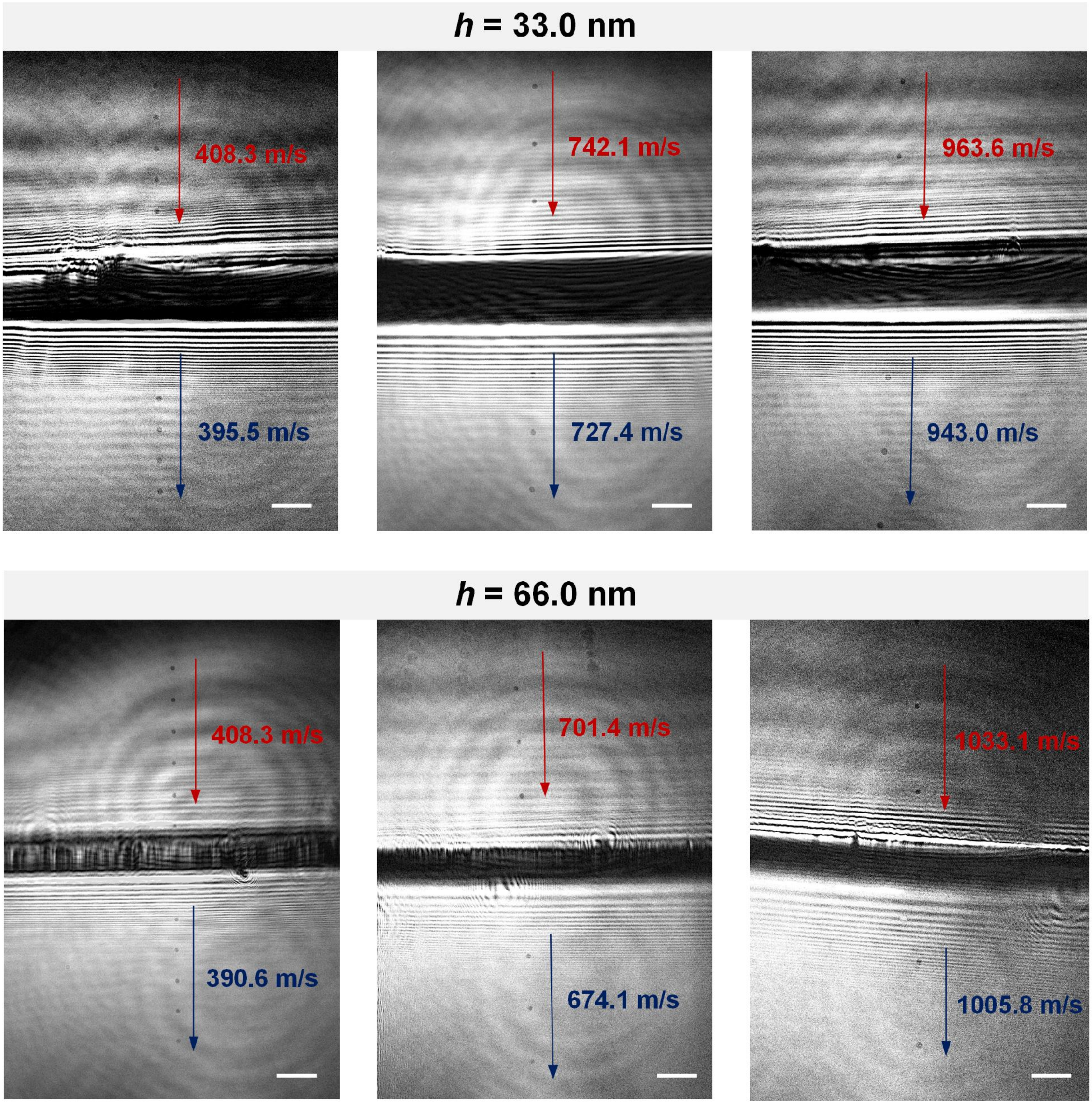

**Supplementary Fig. 63 |** The picture of Projectile penetrating films of 33.0 nm and 66.0 nm thickness. Scale bar, 100 μm

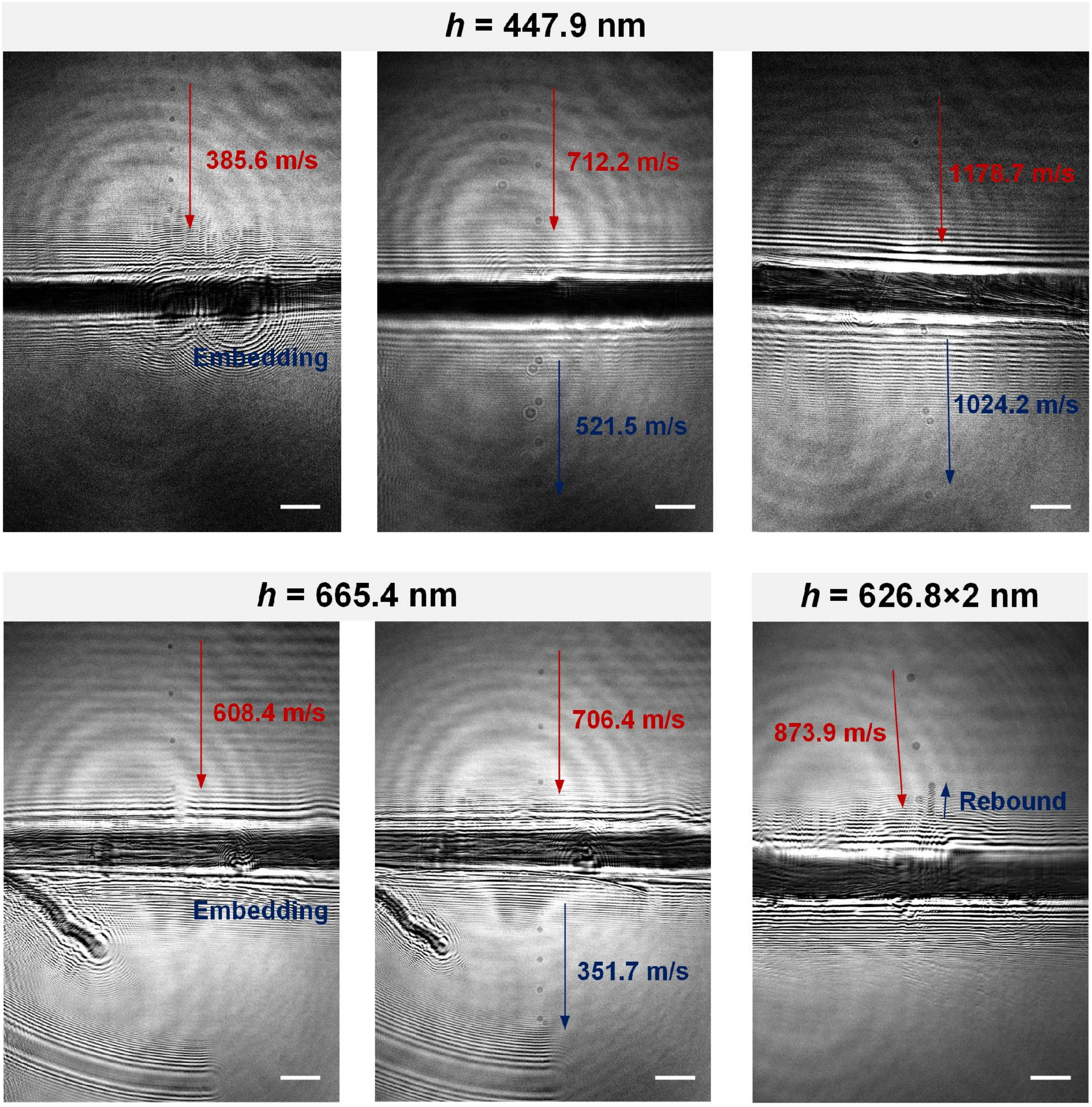

**Supplementary Fig. 64 |** The picture of Projectile penetrating films of 447.9 nm, 665.4 nm and double-layer 626.8 nm thickness. Scale bar, 100 µm.

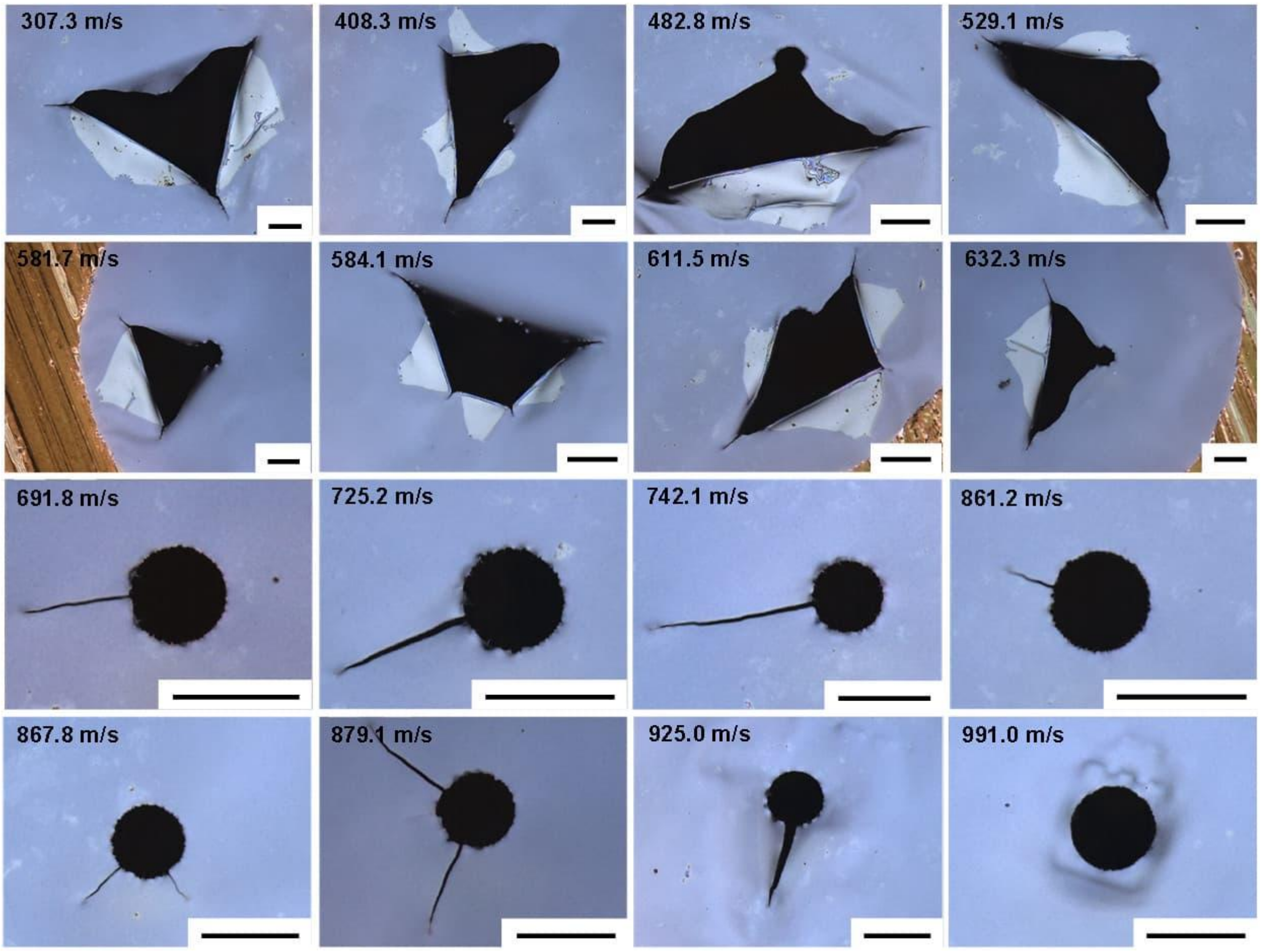

**Supplementary Fig. 65 |** Confocal micrographs of a 33.0-nm-thick 2D HOMN-L4 film after LIPIT. Scale bar, 10 μm

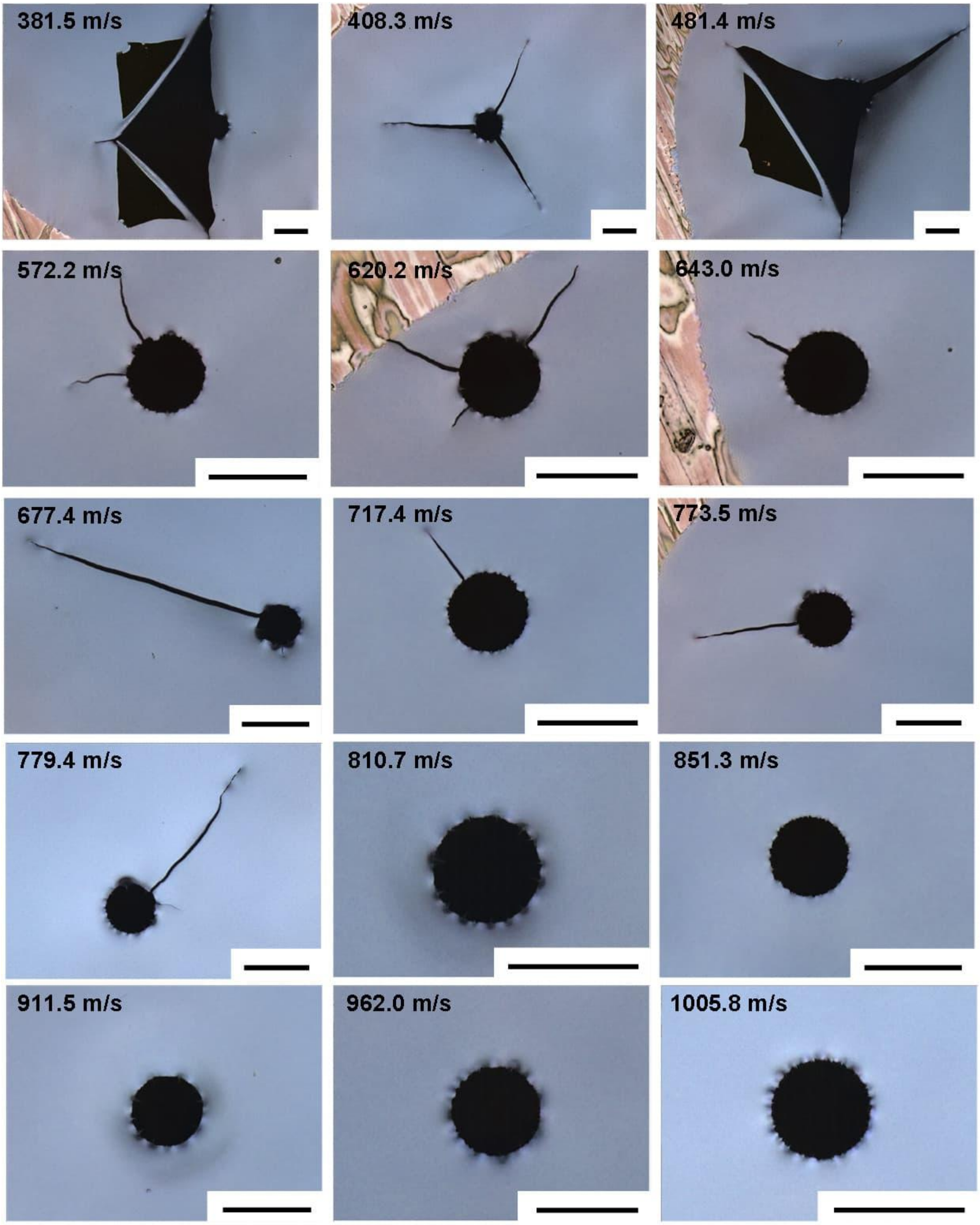

**Supplementary Fig. 66 |** Confocal micrographs of a 66.0-nm-thick 2D HOMN-L4 film after LIPIT. Scale bar, 10 μm

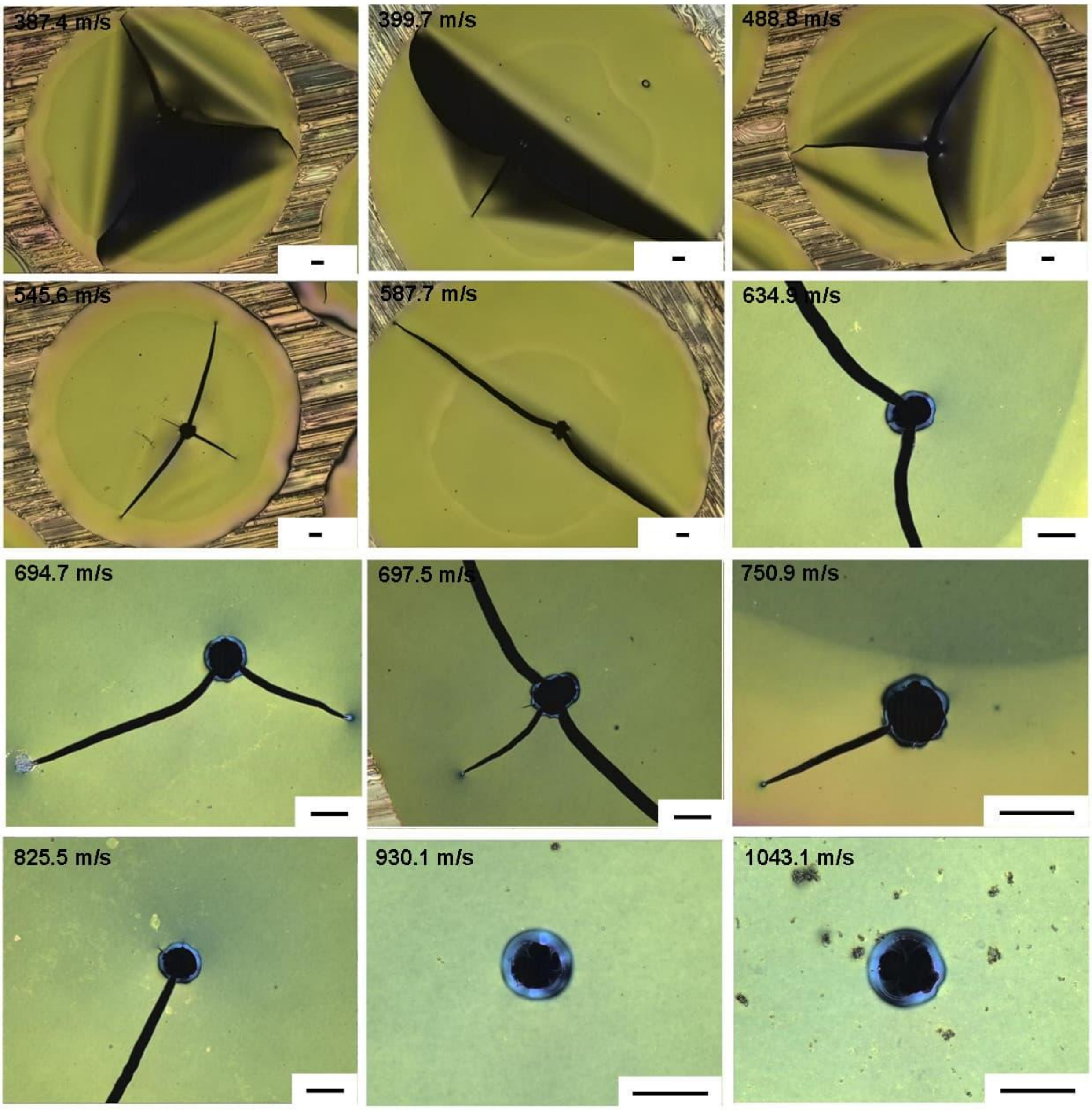

**Supplementary Fig. 67 |** Confocal micrographs of a 248.2-nm-thick 2D HOMN-L4 film after LIPIT. Scale bar, 10 μm.

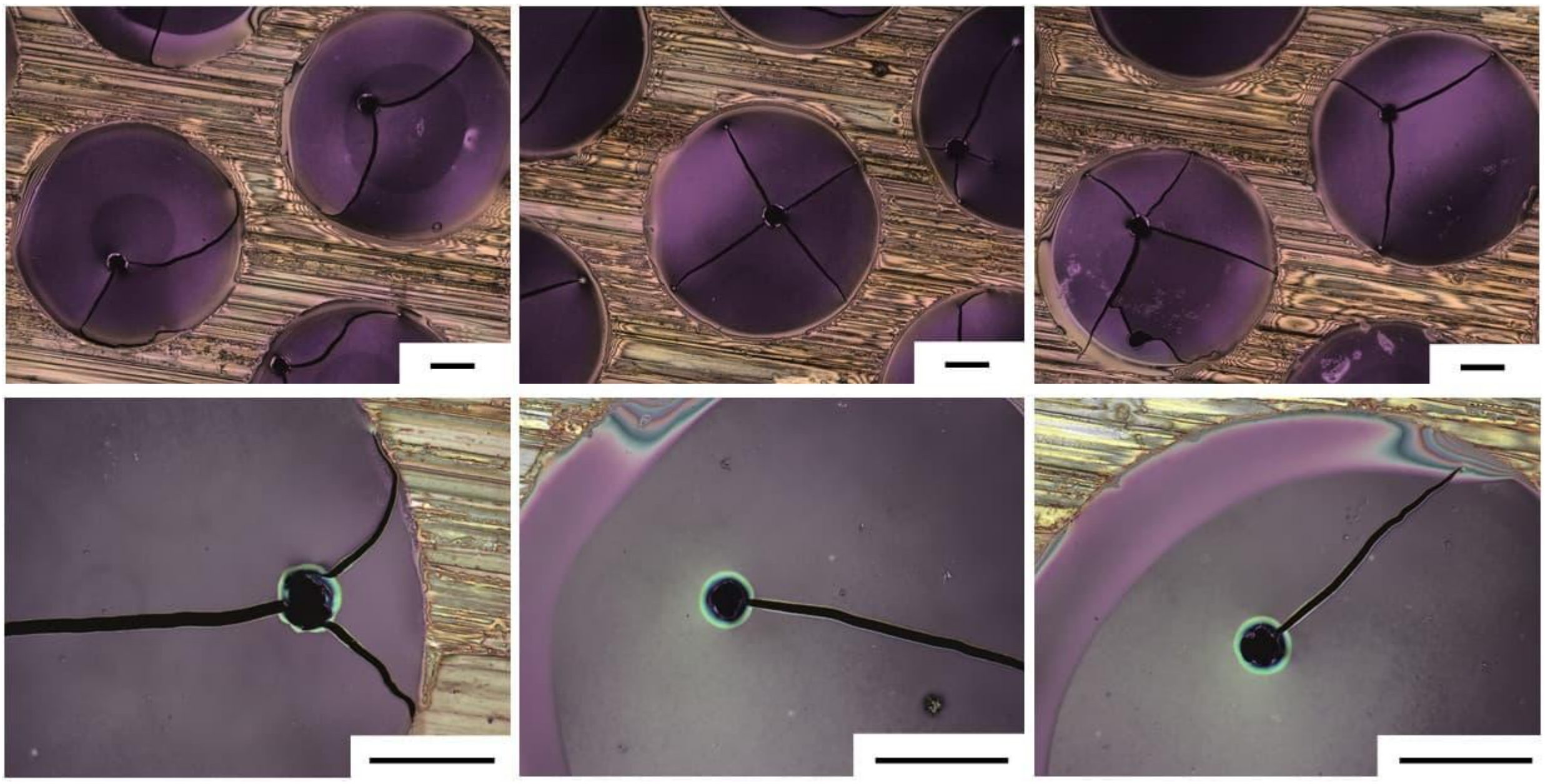

**Supplementary Fig. 68 |** Confocal micrographs of a 447.9-nm-thick 2D HOMN-L4 film after LIPIT. Scale bar, 25 μm.

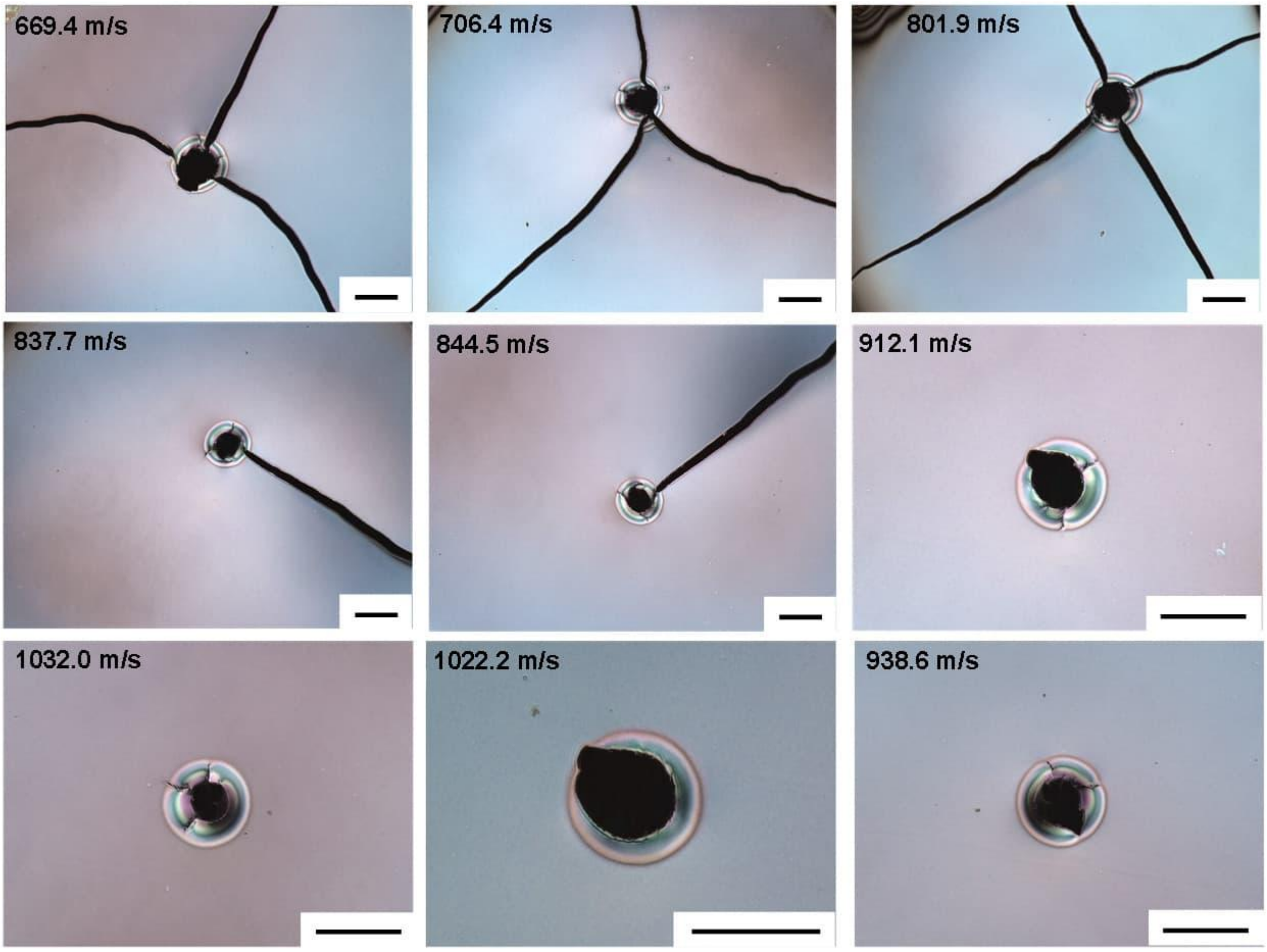

**Supplementary Fig. 69 |** Confocal micrographs of a 665.4-nm-thick 2D HOMN-L4 film after LIPIT. Scale bar, 10 μm.

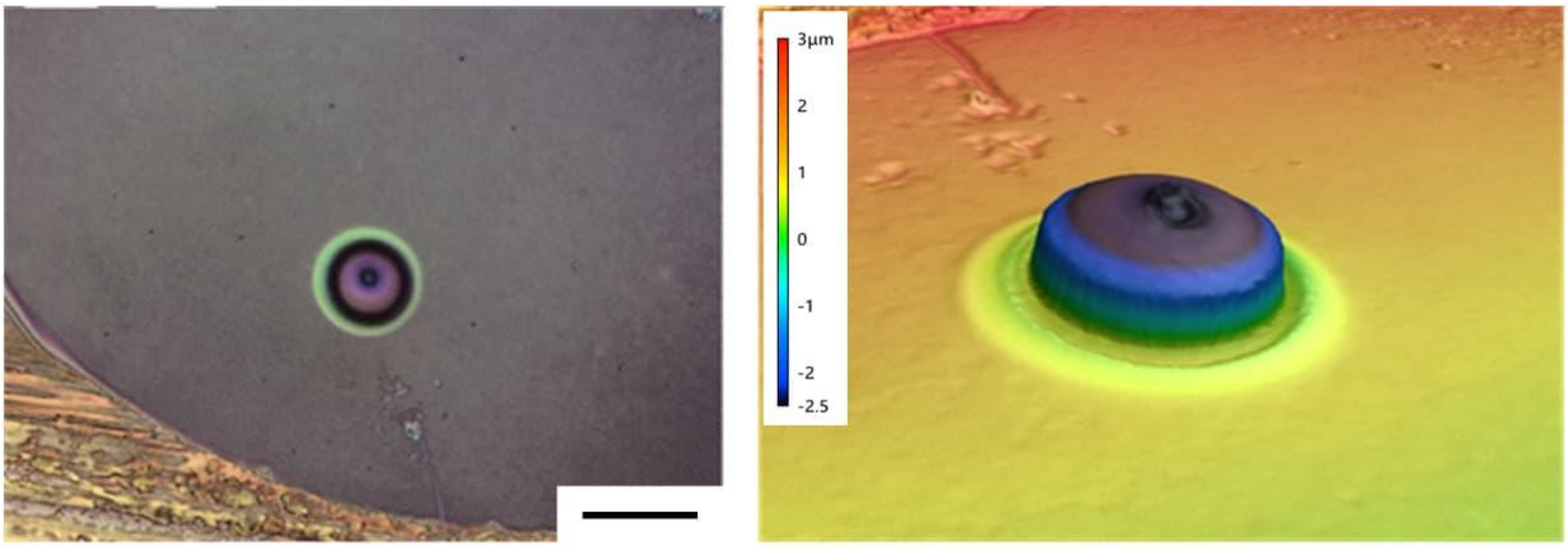

**Supplementary Fig. 70 |** The confocal laser image of the micro-projectile blocked by the 2D HOMN-L4 film (447.9 nm) during the impact process. Scale bar, 10 μm.

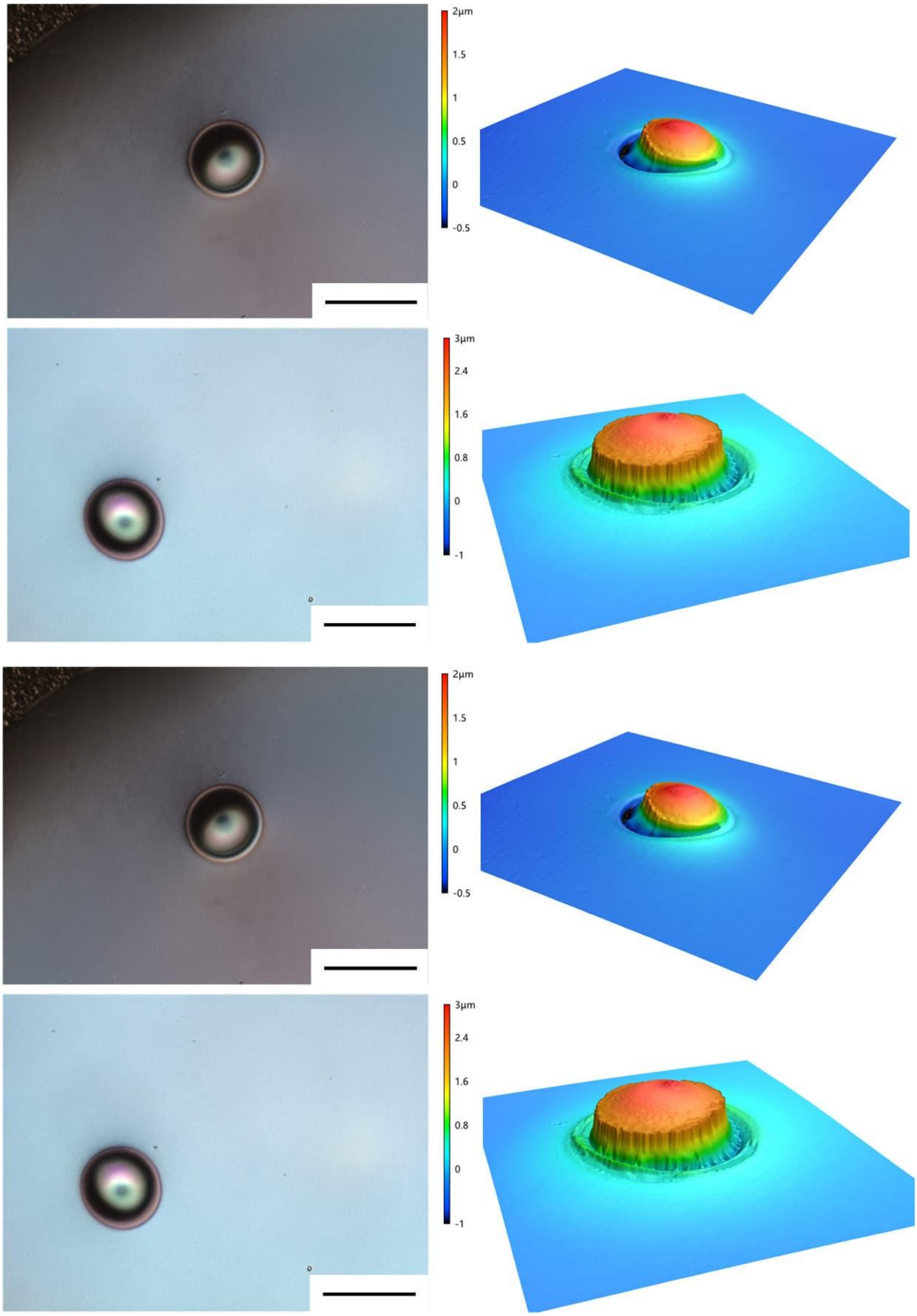

**Supplementary Fig. 71 |** The confocal laser image of the micro-projectile blocked by the 2D HOMN-L4 film (665.4 nm) during the impact process. Scale bar, 10 μm.

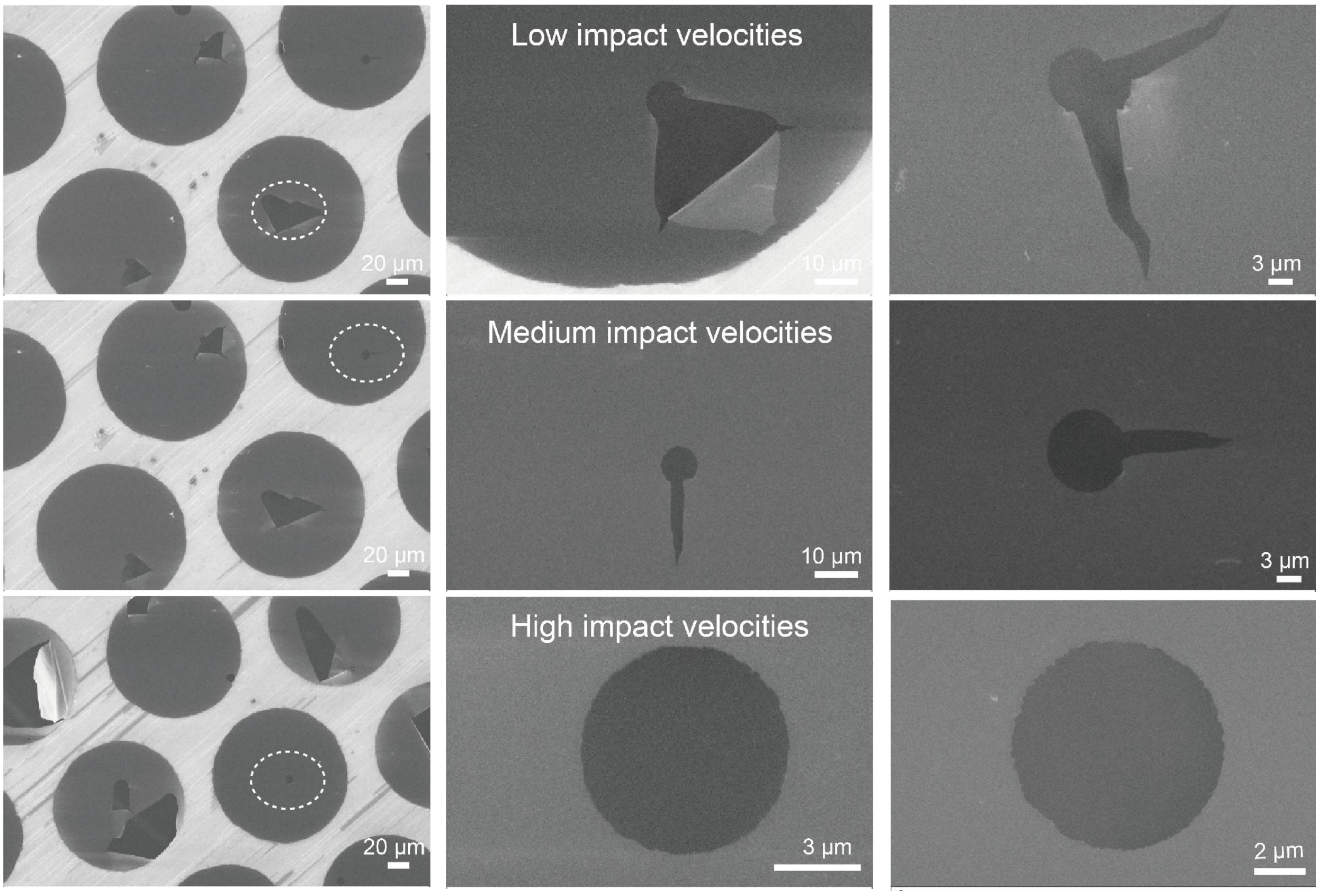

**Supplementary Fig. 72 |** Velocity-dependent structural response of the 33.0 nm 2D HOMN-L4 welded film to LIPIT.

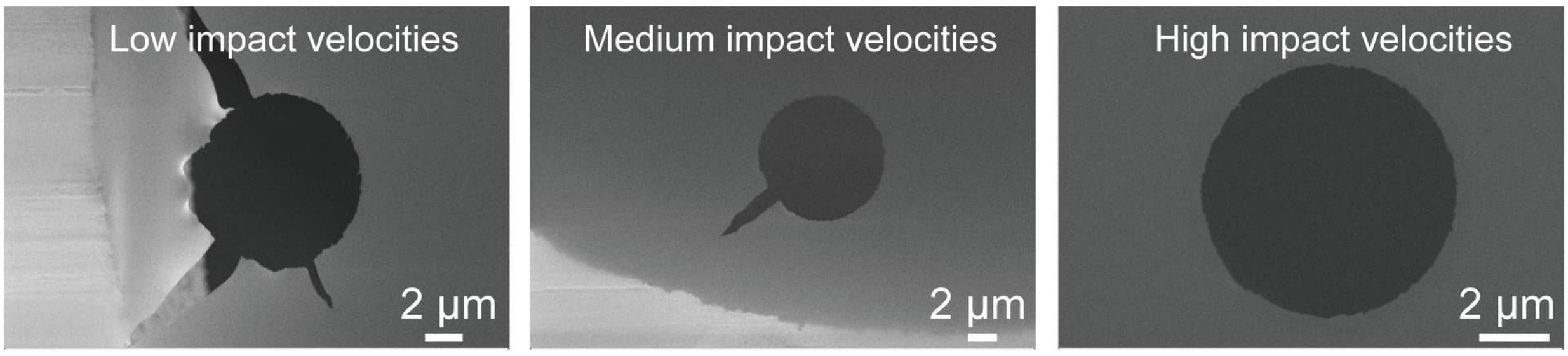

**Supplementary Fig. 73 |** Velocity-dependent structural response of the 66.0 nm 2D HOMN-L4 welded film to LIPIT.

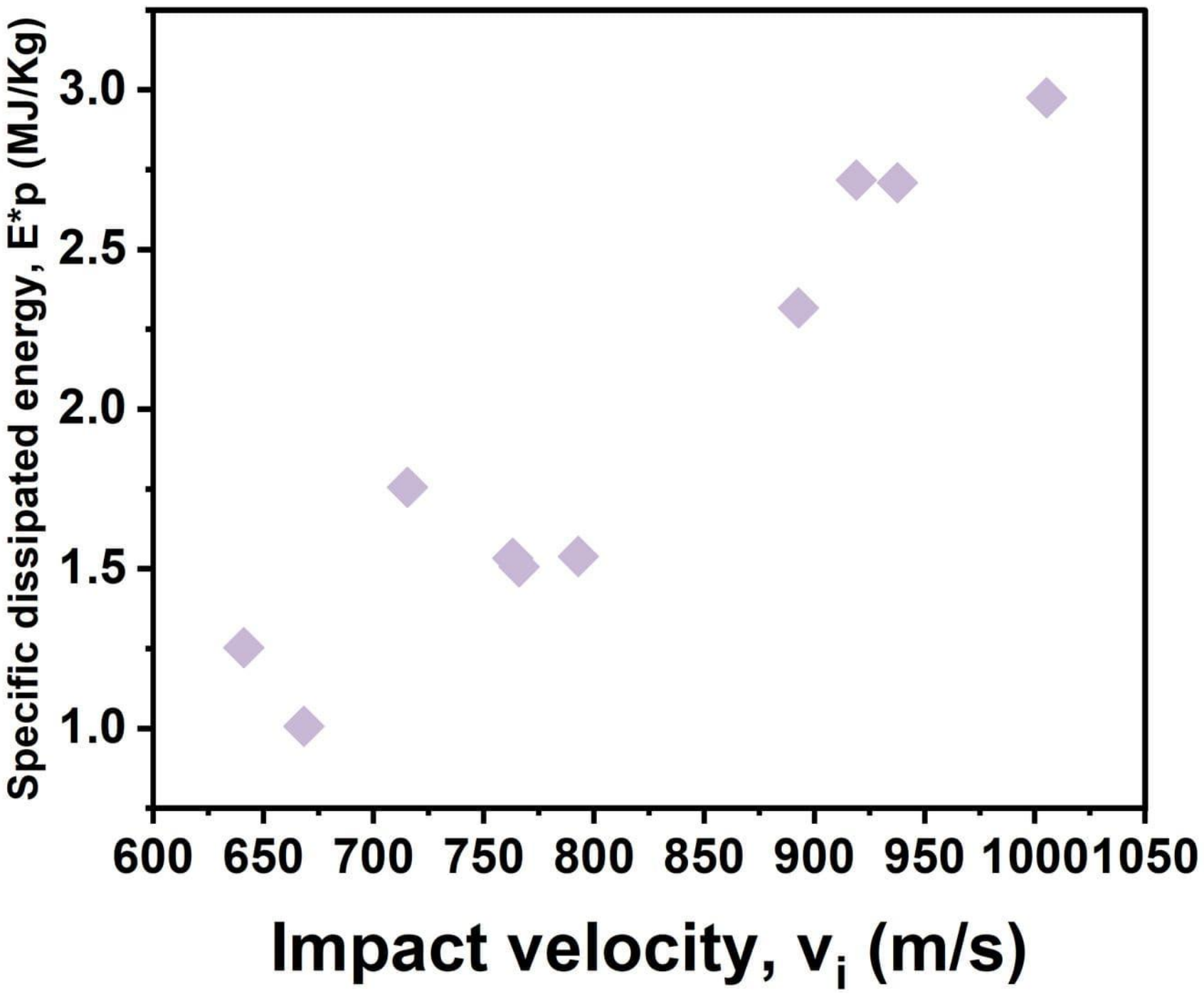

**Supplementary Fig. 74 |** Specific dissipated energy of 2D HOMN-L0 films.

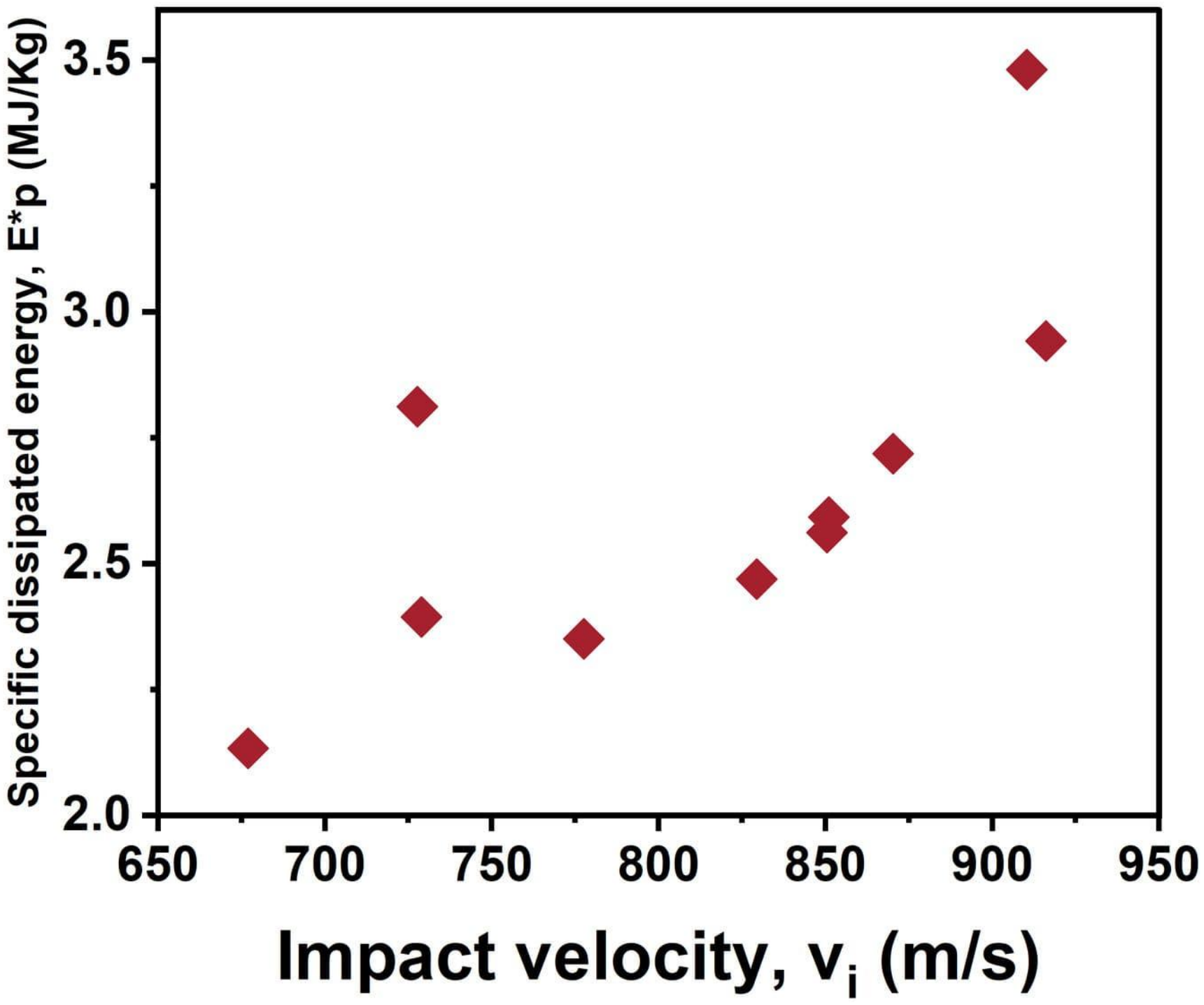

**Supplementary Fig. 75 |** Specific dissipated energy of 2D HOMN-L3 films.

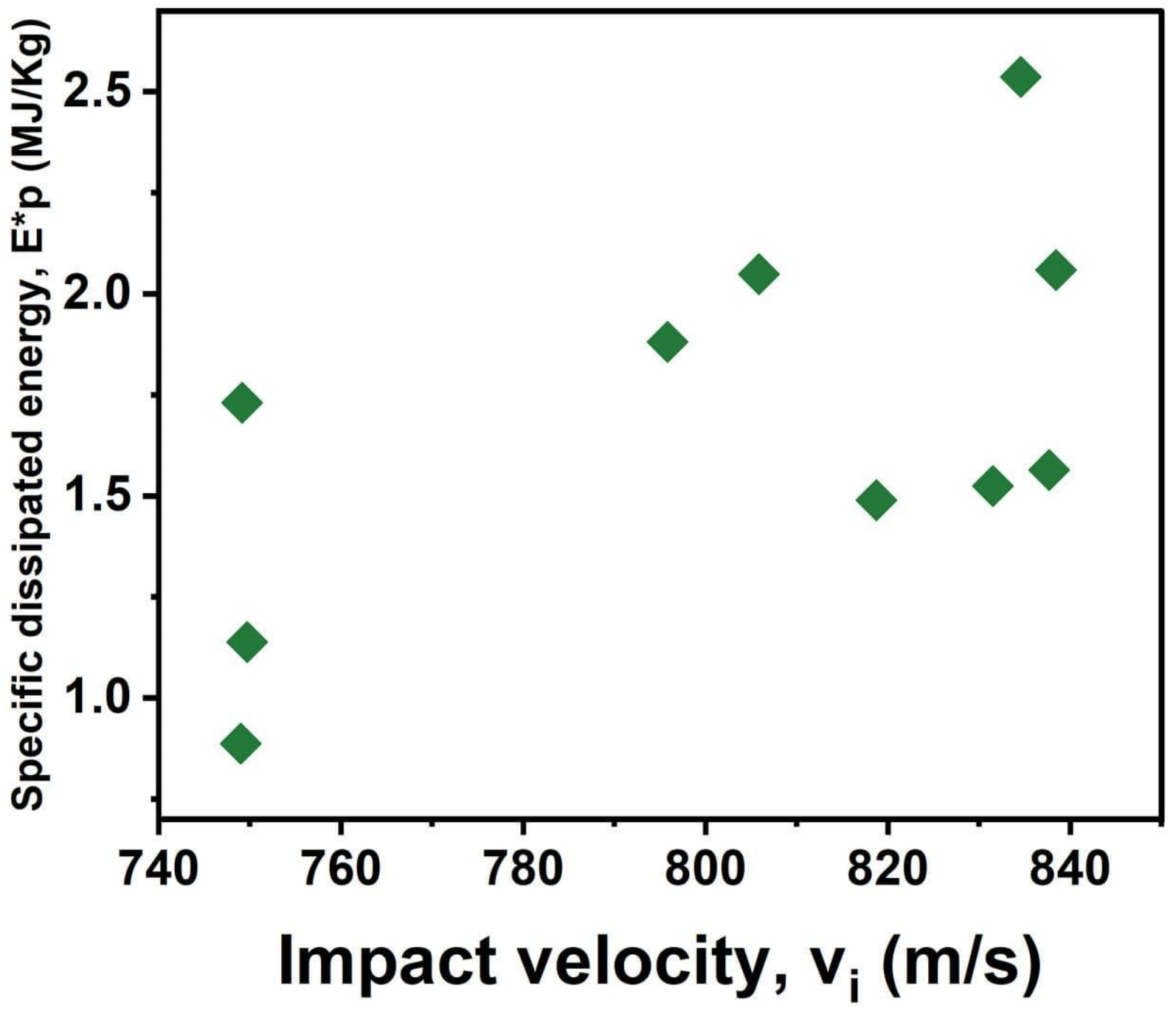

**Supplementary Fig. 76 |** Specific dissipated energy of 2D HOMN-L5 films.

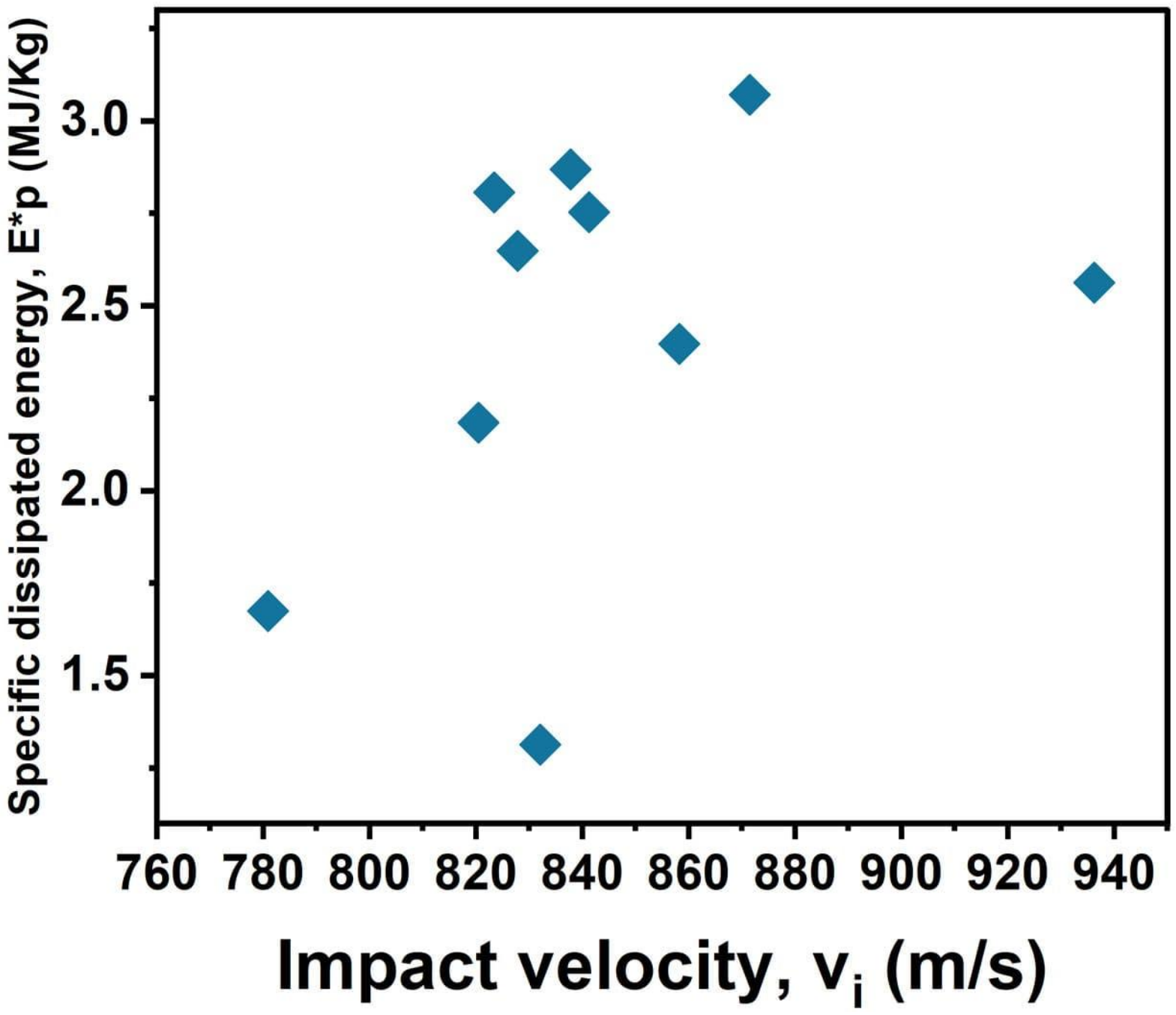

**Supplementary Fig. 77 |** Specific dissipated energy of 2D HOMN-L6 films.

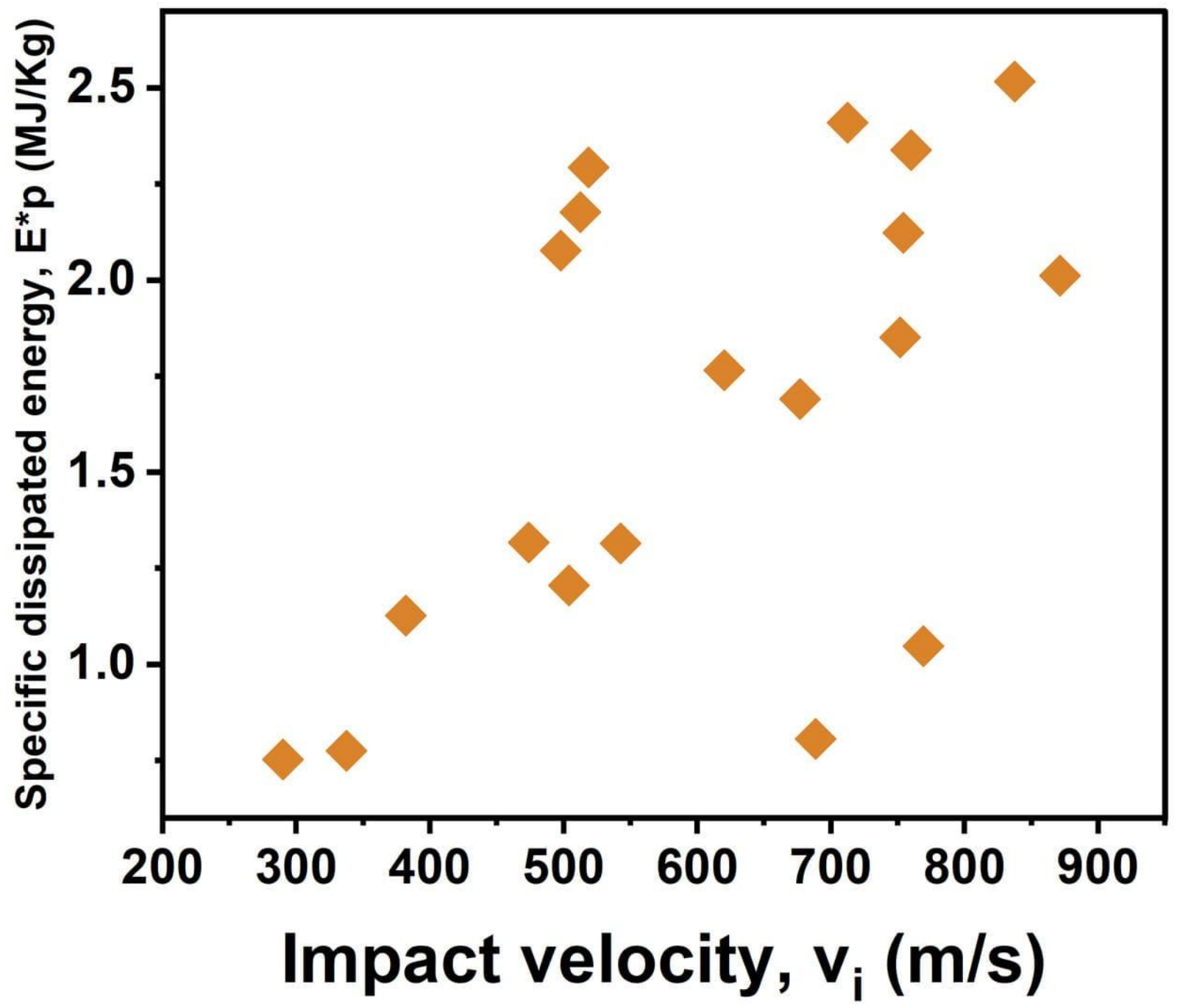

**Supplementary Fig. 78 |** Specific dissipated energy of 2D HOMN-L8 films.

**Supplementary Table 4 |** Comparison of experimental parameters of LIPIT and macroscopic ballistic tests on various materials.

| Target materials | Projectile ($D_p$) | $D_p/h$ | Impact speed range ($v_i$, m/s) | Specific Dissipation Energy (MJ/kg) | References |
|---|---|---|---|---|---|
| Gold ($h$ = 14-52 nm) | $SiO_2$ (3.7µm) | >10 | 880 – 900 | 0.529 – 0.578 | 16 |
| Stainless steel ($h$ = 2 µm) | Glass (32µm) | >10 | 450 – 1000 | 0.192 – 0.534 | 58 |
| Stainless steel ($h$ = 0.4 mm) | Steel (8mm) | >10 | 176 – 592 | 0.116 – 0.259 | 31 |
| Aluminium (2024) ($h$ = 1.27 mm) | Steel (13mm) | >5 | 152 – 869 | 0.11 – 0.56 | 59 |
| Aluminium (1100) ($h$ = 1 mm) | Steel (13mm) | >10 | 92 – 115 | 0.138 – 0.150 | 60 |
| Kevlar/PVP composite ($h$ = 4.0 mm) | Tungsten carbide (12.7mm) | >3 | 298 – 422 | 0.445 – 0.536 | 30 |
| Ni60Ta40 Metallic glass ($h$ = 60 nm) | Al (25µm) | >10 | 185 – 540 | 0.33 – 0.56 | 61 |
| Multi-layer graphene ($h$ = 10-100 nm) | $SiO_2$ (3.7µm) | >10 | 600 – 900 | 1.10 – 1.26 | 16 |
| Nanoarchitected SU-8 ($h$ = 40 µm) | $SiO_2$ (14µm) | / | 448 – 750 | 0.44 – 1.45 | 62 |
| Polystyrene ($h$ = 75–150 nm) | $SiO_2$ (3.7µm) | >10 | 350 – 800 | 0.40 – 1.50 | 32 |
| PMMA ($h$ = 4–6 mm) | Steel (7.8mm) | >2 | 600 – 641 | 0.178 – 0.238 | 63 |
| PMMA ($h$ = 42–352 nm) | $SiO_2$ (3.7µm) | >10 | | | 16 |

## Multifunctional characterization

### Hydrogen permeability

### Hydrogen Gas Barrier Measurement

Gas permeation testing was performed using a sweep-gas apparatus configured according to GB/T 1038.2-2022. Membrane specimens were hermetically sealed between adhesive aluminium foil gaskets within the permeation cell. All measurements were conducted at 298.15 K. High-purity $H_2$ (50 mL min$^{-1}$) and Ar sweep gas (20 mL min$^{-1}$) were regulated by calibrated mass flow controllers. The upstream pressure was maintained at 0.2 MPa (absolute). The system was equilibrated for ≥1 h prior to automated data acquisition to ensure steady-state transport. Permeate composition was quantified by gas chromatography (Shimadzu GC-2014C). The steady-state $H_2$ permeability coefficient (P) was calculated as:

$$P_{H_2} = \frac{N_{H_2}}{\Delta P_{H_2} A} t = \frac{F_{H_2}}{\Delta P_{H_2}} t \quad (18)$$

where $F_{H_2}$ (mol·s$^{-1}$·m$^{-2}$) defined as the molar flux of $H_2$ on the permeation side, $N_{H_2}$ (mol·s$^{-1}$) is the $H_2$ permeation rate, t (m) is the membrane thickness, A (m$^2$) is the effective permeation area, and $\Delta pH_2$ (Pa) is the transmembrane partial pressure difference, defined as:

$$\Delta P_{H_2} = P_{feed} \times 1 - P_{perm} \times X_{perm} \quad (19)$$

Here, $P_{feed}$ and $P_{perm}$ denote absolute pressures on the feed and permeate sides, respectively, and $X_{perm}$ is the $H_2$ mole fraction in the permeate stream. Permeability is reported in Barrer (1 Barrer = 3.35 × 10$^{-16}$ mol mm$^{-2}$ s$^{-1}$ Pa$^{-1}$).

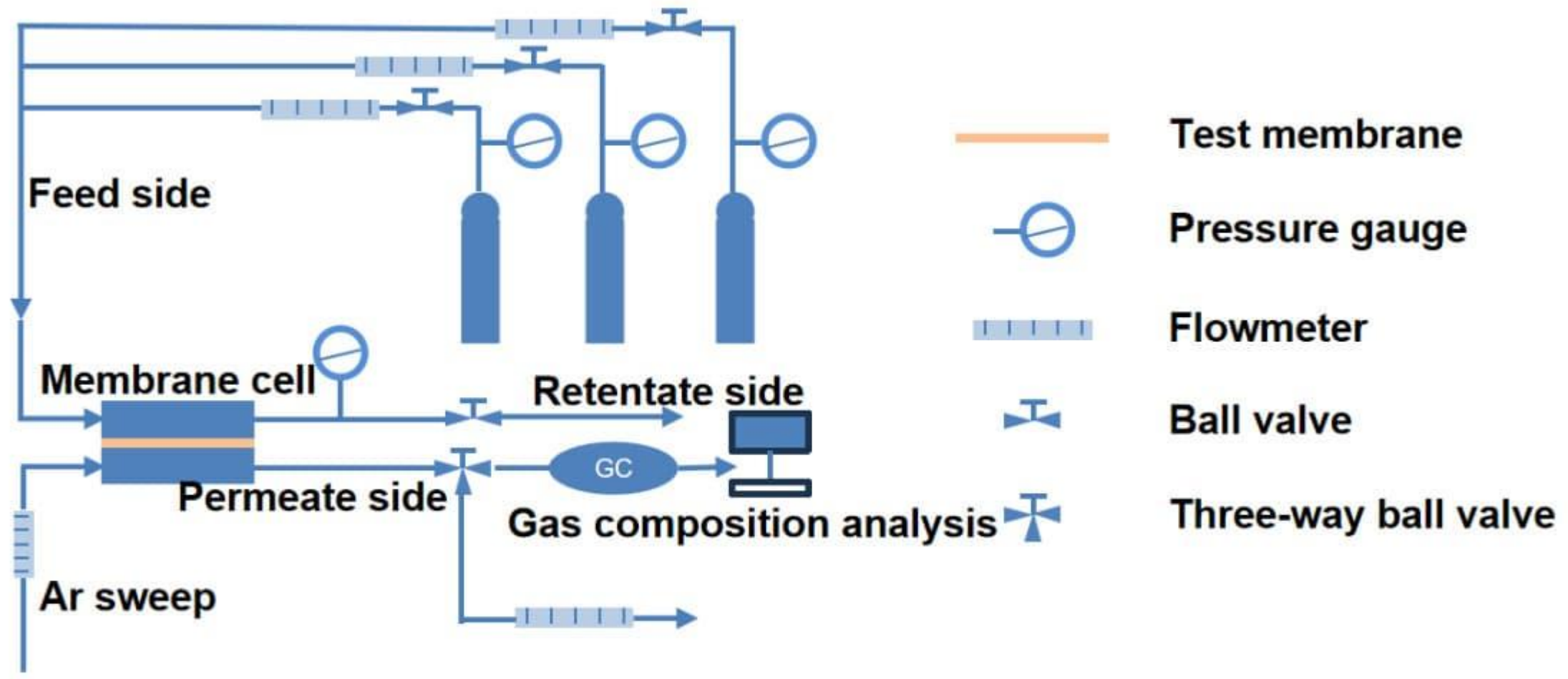

**Supplementary Fig. 79 |** Schematic diagram of the gas permeation testing apparatus.

**Supplementary Table 5 |** Literature gas-permeability comparator values.

| Polymer membrane | $P(O_2)$ cm$^3$ µm/(m$^2$ day atm) | $P(O_2)$(Barrer) | $P(H_2)$ cm$^3$ µm/(m$^2$ day atm) | $P(H_2)$(Barrer) |
|---|---|---|---|---|
| EVOH 32 | 6 | 9.13743E-05 | 600 | 0.009137427 |
| EVOH 44 | 38 | 0.000578704 | 4,000 | 0.060916179 |
| PA 6 | 400 | 0.006091618 | 65,664 | 1 |
| PET | 1,000 | 0.015229045 | 32,832 | 0.5 |
| PP | 50,000 | 0.761452242 | 131,3280 | 20 |
| HDPE | 40,000 | 0.609161793 | 65,6640 | 10 |

Sources for the comparator values are listed in refs.[37,38].

**Water vapour and oxygen barrier testing of 2DPA-IPA films**

Preparation and testing

2DPA-IPA was dissolved at 15 mg $ml^{-1}$ in equal proportions of acetone and trifluoroacetic acid. Films were spin coated at 2,000 r.p.m. for 60 s, directly onto polymer substrates or onto silicon before transfer. Bare substrates and supported films were submitted for external barrier testing. Testing was performed by the Packaging Safety Testing Center of Jinan Labthink Electromechanical Technology Co., Ltd. Water vapour transmission rate (WVTR) was measured using C390M (GB/T 26253-2010) at 38 °C and 90% relative humidity. Oxygen transmission rate (OTR) was measured using C230H (GB/T 19789-2021) at 23 °C and 0% relative humidity.

**Supplementary Table 6 |** Water vapour and oxygen transmission. WVTR, g $m^{-2}$ $d^{-1}$; OTR, $cm^3$ $m^{-2}$ $d^{-1}$.

| Report | Specimen | WVTR | OTR |
|---|---|---|---|
| CT-202405060 | Bare substrate (PI) | 31.52 ± 1.25 | 174.97 ± 1.28 |
| CT-202405060 | Composite (PI-2DP) | 31.79 ± 1.40 | 173.67 ± 2.81 |

Labthink®

技术咨询：400-608-8368

报告编号：CT-202405060

# 检测报告

| | | | |
|---|---|---|---|
| 检验项目 | 水蒸气透过量 | 检测标准 | GB/T 26253-2010 |
| 检测仪器 | C390M 水蒸气透过率测试系统 | | |
| 测试条件 | 1. 根据客户要求进行检测；<br>2. 试验条件：38℃，90%RH。 | | |

**检验结果汇总**

| 试样名称 | 水蒸气透过量 g/(m²•24h) |
|---|---|
| IP 空白样 | 32.813 |
| | 30.315 |
| | 31.421 |
| PI-2DP | 30.249 |
| | 32.149 |
| | 32.981 |

济南兰光机电技术有限公司 包装安全检测中心

备注：结果仅对来样有效。

------------------以下空白----------------

**Supplementary Fig. 80 |** Original stamped water-vapour report CT-202405060.

English translation: water-vapour transmission rate; C390M; GB/T 26253-2010; 38 °C, 90% RH. IP 空白样, bare substrate; PI-2DP, composite. Numerical results are translated in Supplementary Table 6. Report note: results apply only to the submitted samples.

Labthink®
技术咨询：400-608-8368

报告编号：CT-202405060

# 检测报告

| 检验项目 | 氧气透过率 | 依据标准 | GB/T 19789-2021 |
|---|---|---|---|
| 检测仪器 | C230H 氧气透过率测试系统 | | |
| 测试条件 | 1. 根据客户要求进行检测；<br>2. 试验条件：23.0℃，0%RH。 | | |

**检验结果汇总**

| 试样名称 | 氧气透过率 ($cm^3/m^2 \cdot 24h$) |
|---|---|
| IP 空白样 | 176.0111 |
| | 173.5342 |
| | 175.3593 |
| PI-2DP | 174.1560 |
| | 176.2022 |
| | 170.6448 |

备注：结果仅对来样有效。

济南兰光机电技术有限公司
包装安全检测中心

**Supplementary Fig. 81 |** Original stamped oxygen report CT-202405060.

English translation: oxygen transmission rate; C230H; GB/T 19789-2021; 23 °C, 0% RH. Sample labels match Supplementary Fig. 80; numerical results are translated in Supplementary Table 6. Report note: results apply only to the submitted samples.

## Ion permeation

### Ion permeation measurements.

Single-salt diffusion experiments were conducted in a two-compartment H-cell (Supplementary Fig. 82). The 2D HOMN-L4 membrane was clamped between the donor and receiver compartments. For each experiment, the donor compartment contained 100 mL of a 200 mg $L^{-1}$ aqueous solution of LiCl, NaCl, KCl, $MgCl_2$, or $CaCl_2$, and the receiver compartment contained 100 mL of deionized water. Both compartments were magnetically stirred at 300 rpm to minimize concentration polarization.

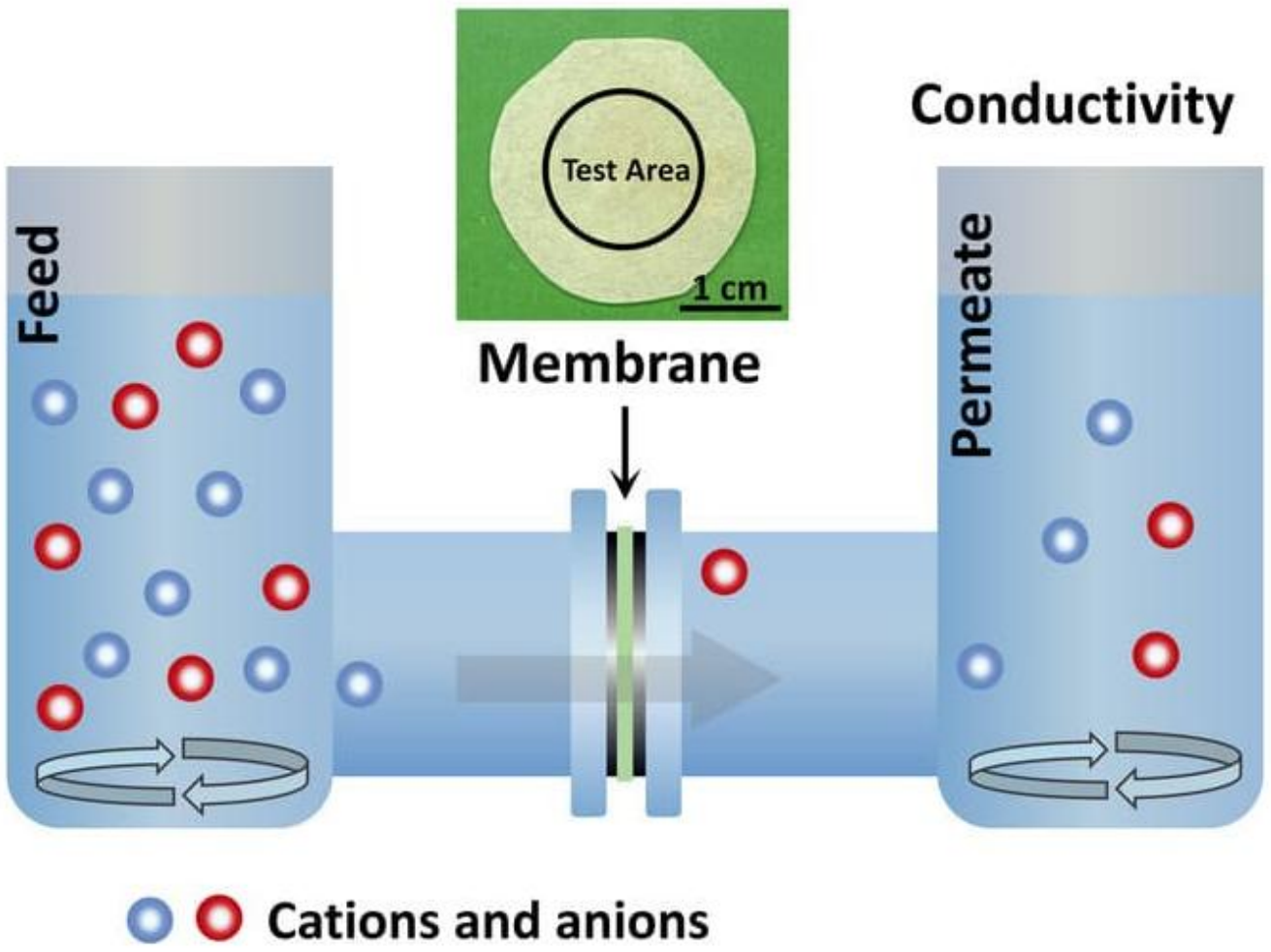

**Supplementary Fig. 82 |** The schematic diagram illustrating the configuration of H-cells for ionic permeation measurement.

Receiver-side conductivity was measured at the actual elapsed times, and each experiment was continued for approximately one week. At every time point, a 5 mL aliquot was removed from the receiver for conductivity analysis, and a 5 mL aliquot was removed simultaneously from the donor. The receiver was then replenished with 5 mL of deionized water, while the donor was replenished with 5 mL of the corresponding 200 mg $L^{-1}$ salt solution. This procedure maintained equal compartment volumes and comparable hydrostatic heads, while limiting changes in the donor concentration and the associated osmotic driving force.
Independent conductivity-concentration calibration curves were constructed for each salt using 0.5, 1.0, 3.0, 5.0, and 10.0 mg $L^{-1}$ standards measured under the same conditions. Receiver conductivity was converted to the molar concentration of the corresponding chloride salt.
Because each receiver aliquot was removed and replaced with deionized water, The average permeation rate, $P_i$, were calculated as follows:

$$P_i = \frac{[C_i - C_{(i-1)'}] \cdot V}{A \cdot (t_i - t_{i-1})} \tag{20}$$

$$C_{(i-1)'} = 0.95 C_{i-1}$$

Where V (0.1 L) is the receiving volume, A ($1.76625 \times 10^{-4}$ $m^2$) is the effective membrane area, $C_i - C_{(i-1)'}$ denote the concentration changes for the $i$-th measurement, where the initial concentration is zero, and $(t_i - t_{i-1})$ is the $i$-th diffusion time interval. All measurements were conducted at 298.15 K.
The five feed solutions had the same mass concentration and the values below are therefore direct comparisons under the experimental conditions. All conductivity data were measured using the DDSJ-308F instrument from Leici.

For comparison with reported membranes, the apparent permeability coefficient was calculated from the ion flux as

$$P_{app} = \frac{P_i L}{\Delta C_0} \tag{21}$$

where L is the membrane thickness and $\Delta C_0$ is the initial concentration difference between the donor and receiver compartments. Because the reported values were calculated using the initial concentration difference and literature-reported fluxes, they are treated here as apparent permeability coefficients.

**Comparison with representative membranes**

Reported ion fluxes and apparent permeability coefficients for representative membranes are summarized in Supplementary Table 7. The ion sequence in the table is ordered by hydrated diameter: $K^+$ (6.62 Å), $Na^+$ (7.16 Å), $Li^+$ (7.64 Å), $Ca^{2+}$ (8.24 Å) and $Mg^{2+}$ (8.56 Å). These hydrated diameters are consistent with values used in previous membrane studies.

**Supplementary Table 7 |** Ion fluxes and apparent permeability coefficients of representative membranes. Values are listed in the order $K^+$, $Na^+$ , $Li^+$ , $Ca^{2+}$and $Mg^{2+}$. $P_i$ ion flux; $P_{app}$, apparent permeability coefficient.

| Membrane | Thickness | Initial salt concentration | Permeantion rate (mol / m$^2$/ h)$K^+$/ $Na^+$/ $Li^+$/ $Ca^{2+}$/ $Mg^{2+}$ | Permeability (cm$^2$ / s)$K^+$/ $Na^+$/ $Li^+$/ $Ca^{2+}$/ $Mg^{2+}$ | References |
|---|---|---|---|---|---|
| rGO | 270 nm | 0.1 mol L$^{-1}$ | 0.00851 / 0.00802 / 0.00611 / 0.00591 / 0.00413 | 6.38E-11 / 6.02E-11 / 4.58E-11 / 4.43E-11 / 3.10E-11 | 39 |
| MXene | 300 nm | 0.2 mol L$^{-1}$ | 0.378 / 0.367 / 0.334 / 0.298 / 0.273 | 1.57E-9 / 1.53 E-9 / 1.39E-9 / 1.24E-9 / 1.14E-9 | 40 |
| 2D HOMN | 400 nm | 200 mg L$^{-1}$ of each salt | 6.4E-5 / 4.5E-5 / 4.5E-5 / 4.75E-5 / 3.62E-5 | 2.64E-11 / 1.46E-11 / 1.06E-11 / 2.93E-11 / 1.91E-11 | This work |
| COFs | 320 nm | 0.2 mol L$^{-1}$ | 0.74 / 0.62 / 0.76 / 0.068 / 0.018 | 3.29E-9 / 2.76E-9 / 3.38E-9 / 3.02E-10 / 8.0E-11 | 41 |
| Nafion 212 | 50 μm | 1.0 mol L$^{-1}$ | 1.53 / 1.08 / 0.776 / 1.15 / 0.626 | 2.13E-7 / 1.50E-7 / 1.08E-7 / 1.60E-7 / 8.69E-8 | 42 |

**Time-dependent permeation.**

The time-dependent permeation rates are shown in Supplementary Figs. 83–87.

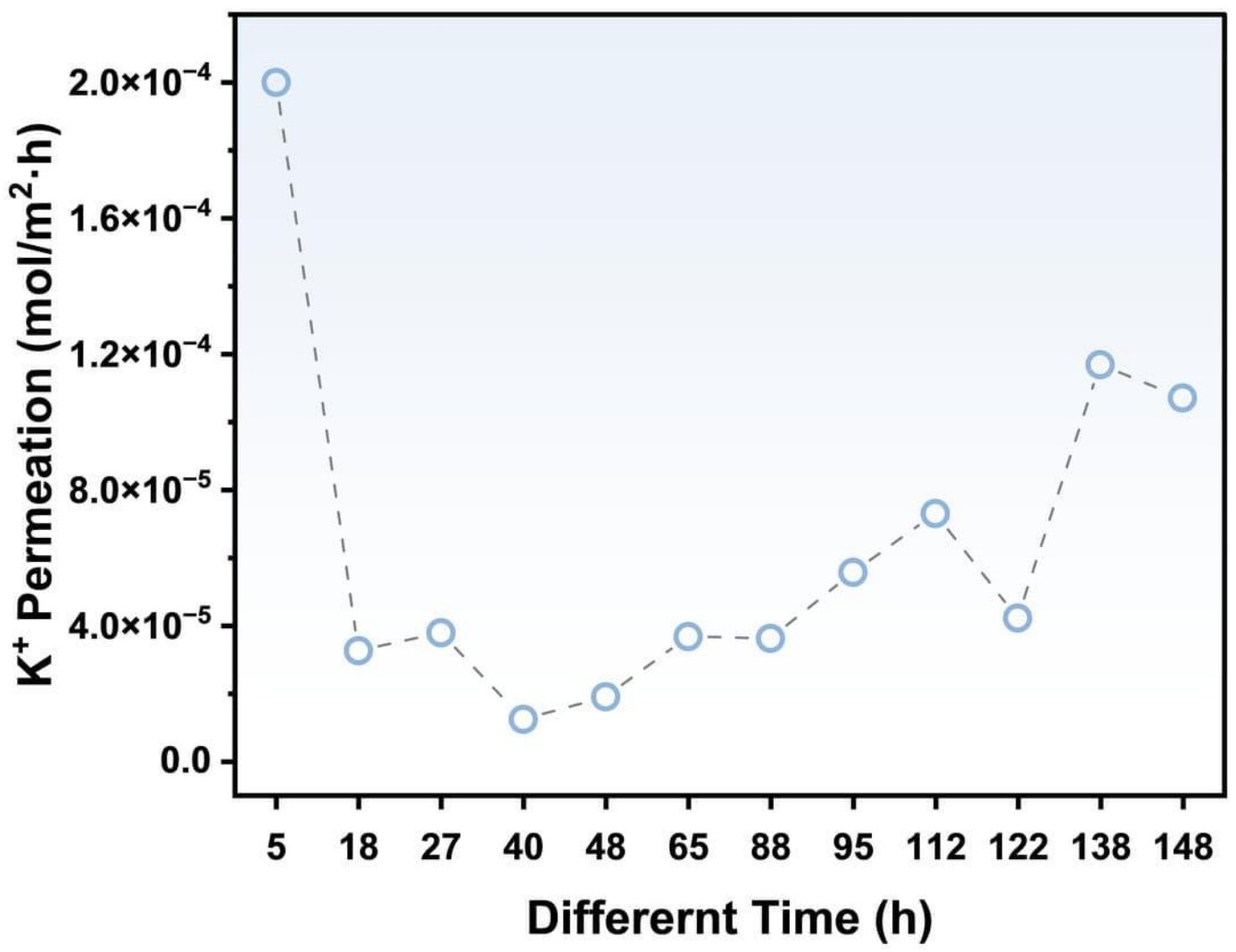

**Supplementary Fig. 83 |** Time-resolved $K^+$ permeation through the membrane.

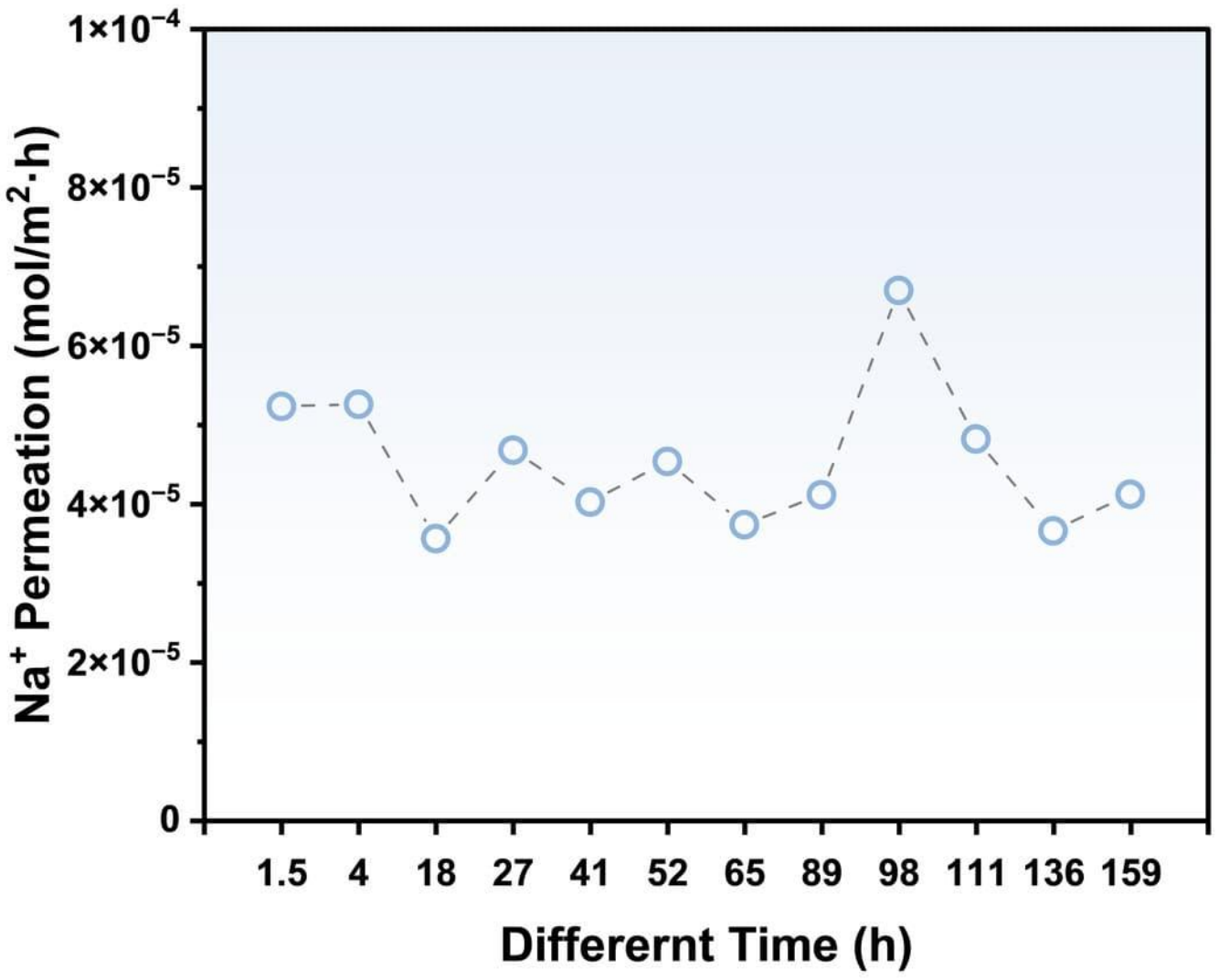

**Supplementary Fig. 84 |** Time-resolved $Na^+$ permeation through the membrane.

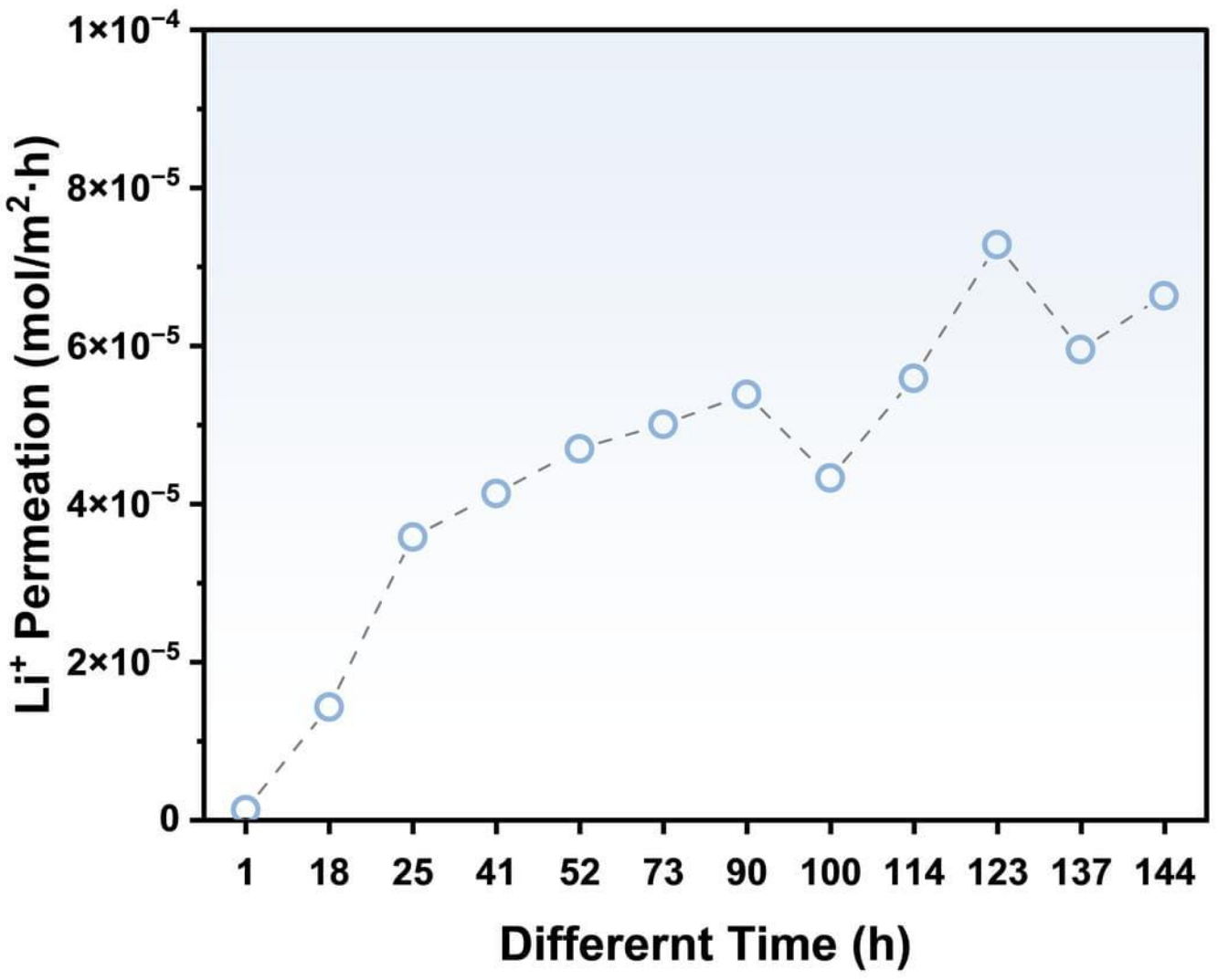

**Supplementary Fig. 85 |** Time-resolved $Li^+$ permeation through the membrane.

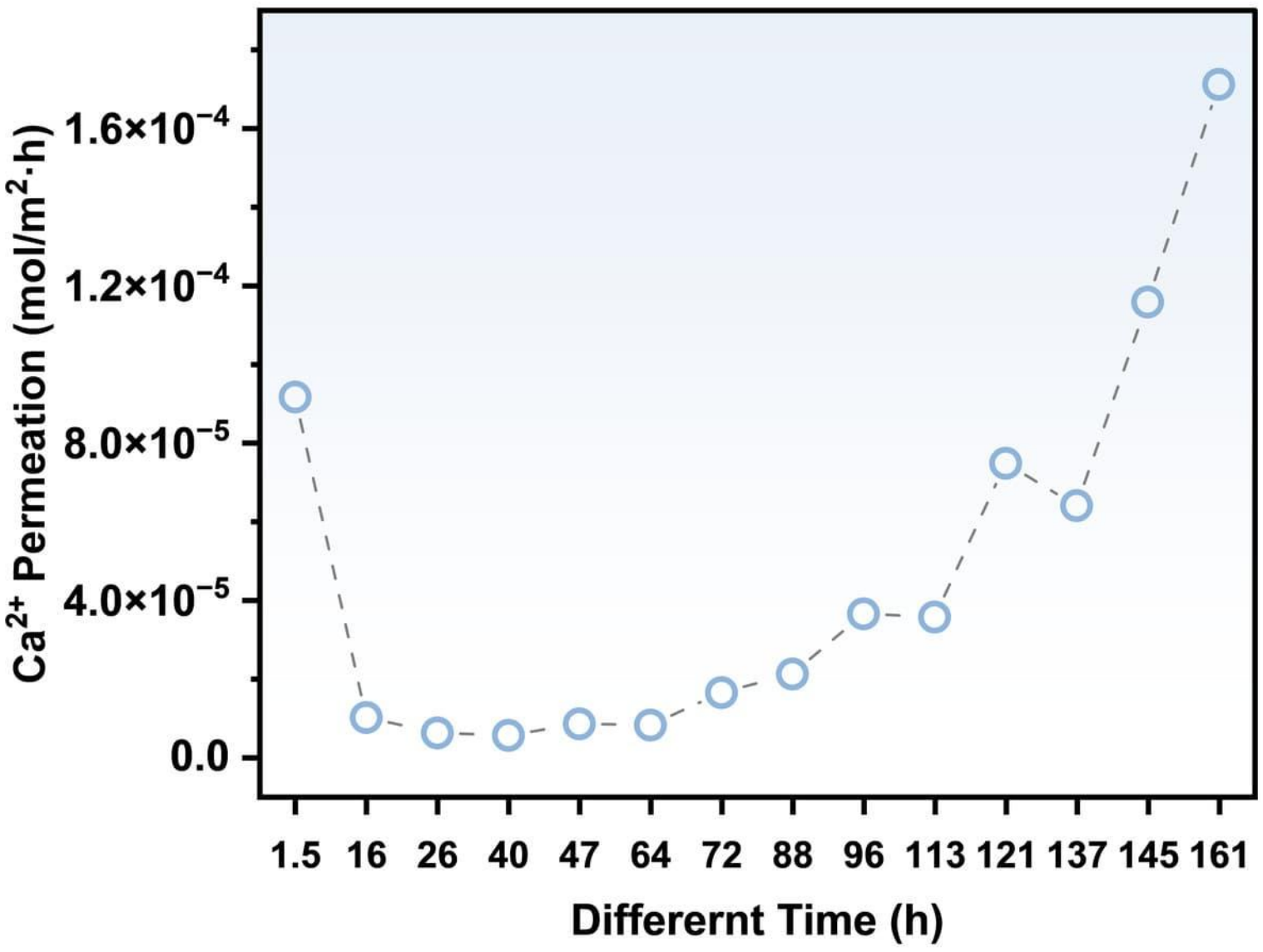

**Supplementary Fig. 86 |** Time-resolved $Ca^{2+}$ permeation through the membrane.

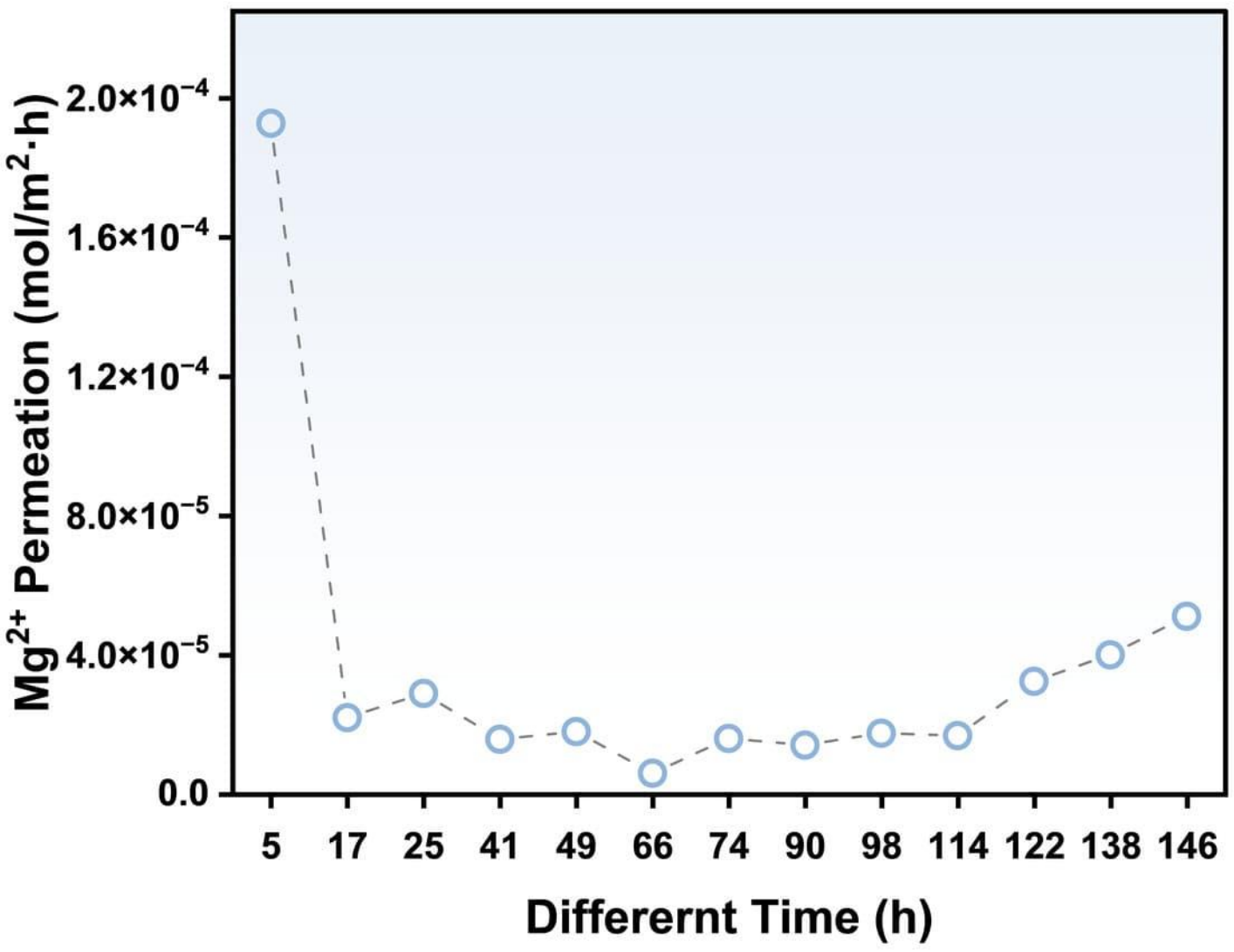

**Supplementary Fig. 87 |** Time-resolved $Mg^{2+}$ permeation through the membrane.

**Proton irradiation stability**

**Proton irradiation facility and beam monitoring**

The irradiation was performed using a tandem accelerator beamline equipped with beam-transport optics, magnetic analysis, a scanning stage and a target chamber. The experimental arrangement is reproduced in Supplementary Fig. 88. The 1 MeV $H^+$ beam was transported to the target chamber through the N15° beamline. Beam stability was checked by preconditioning the beam for 1 h before sample exposure.

Beam current was monitored upstream using a Faraday cup and at the target using a current meter connected to the sample chamber. The target current was nominally 100 nA, with a maximum fluctuation of approximately ±5%.

The target chamber pressure was $6.3 \times 10^{-5}$ Pa. The illuminated area was calibrated using a fluorescent target and set to $10 \times 10$ mm², corresponding to 1.00 cm².

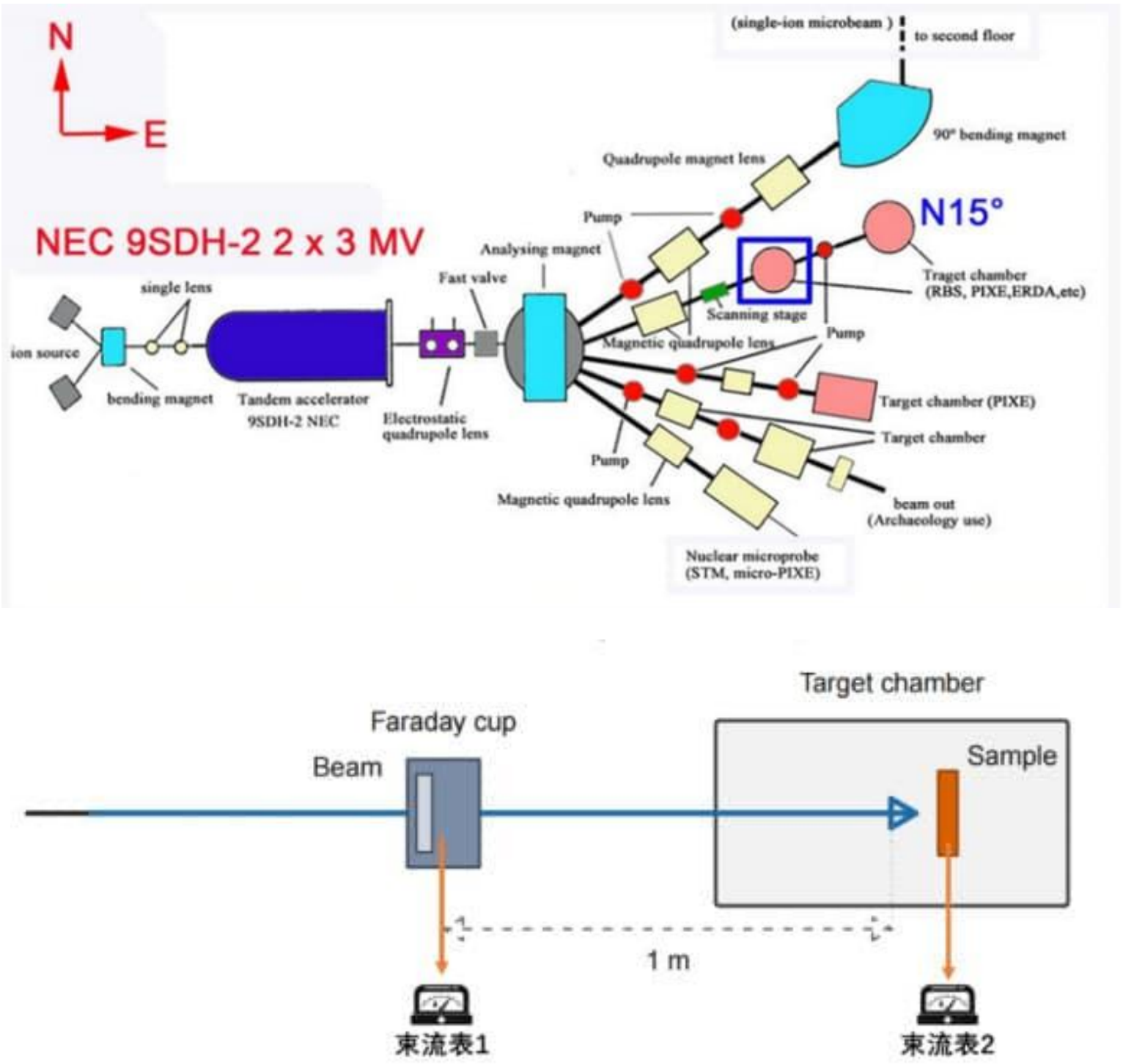

**Supplementary Fig. 88 |** Schematic of tandem-accelerator and beam-transport arrangement used for ion irradiation, and beam-current monitoring. A Faraday cup was used to monitor the transported beam, and the target current was measured at the sample chamber.

**Sample irradiation conditions**

The beam-log parameters are summarized in Supplementary Table 8. Samples were exposed to 1 MeV $H^+$ at nominal fluences of $1 \times 10^{14}$, $2 \times 10^{14}$, $6 \times 10^{14}$, $1 \times 10^{15}$ and $1.4 \times 10^{15}$ ions $cm^{-2}$, alongside an unirradiated control. The provisional dose calculation used a film density of 1.53 g $cm^{-3}$. Measured thicknesses for samples 1–5 were 33.1, 32.4, 32.1, 31.3 and 32.0 nm, respectively.

**Supplementary Table 8 |** Proton irradiation conditions from the beam log.

| Sample | Scan area ($mm^2$) | Beam current | Nominal fluence (ions $cm^{-2}$) | Beam-log time |
|---|---|---|---|---|
| 1 | 10 × 10 | 100 nA, ±5% | $1 \times 10^{14}$ | 3 min |
| 2 | 10 × 10 | 100 nA, ±5% | $2 \times 10^{14}$ | 6 min |
| 3 | 10 × 10 | 100 nA, ±5% | $6 \times 10^{14}$ | 18 min |
| 4 | 10 × 10 | 100 nA, ±5% | $1 \times 10^{15}$ | 30 min |
| 5 | 10 × 10 | 100 nA, ±5% | $1.4 \times 10^{15}$ | 42 min |

## Morphological analysis

Atomic force microscopy was used to examine the surface morphology of 2D HOMN-L4 films before and after irradiation. Topographic images were acquired over a scan area of 30 × 30 µm, and three-dimensional (3D) images were reconstructed from the height data. Both the topographic and 3D images reveal that the film retains a smooth surface and shows no obvious morphological change after irradiation.

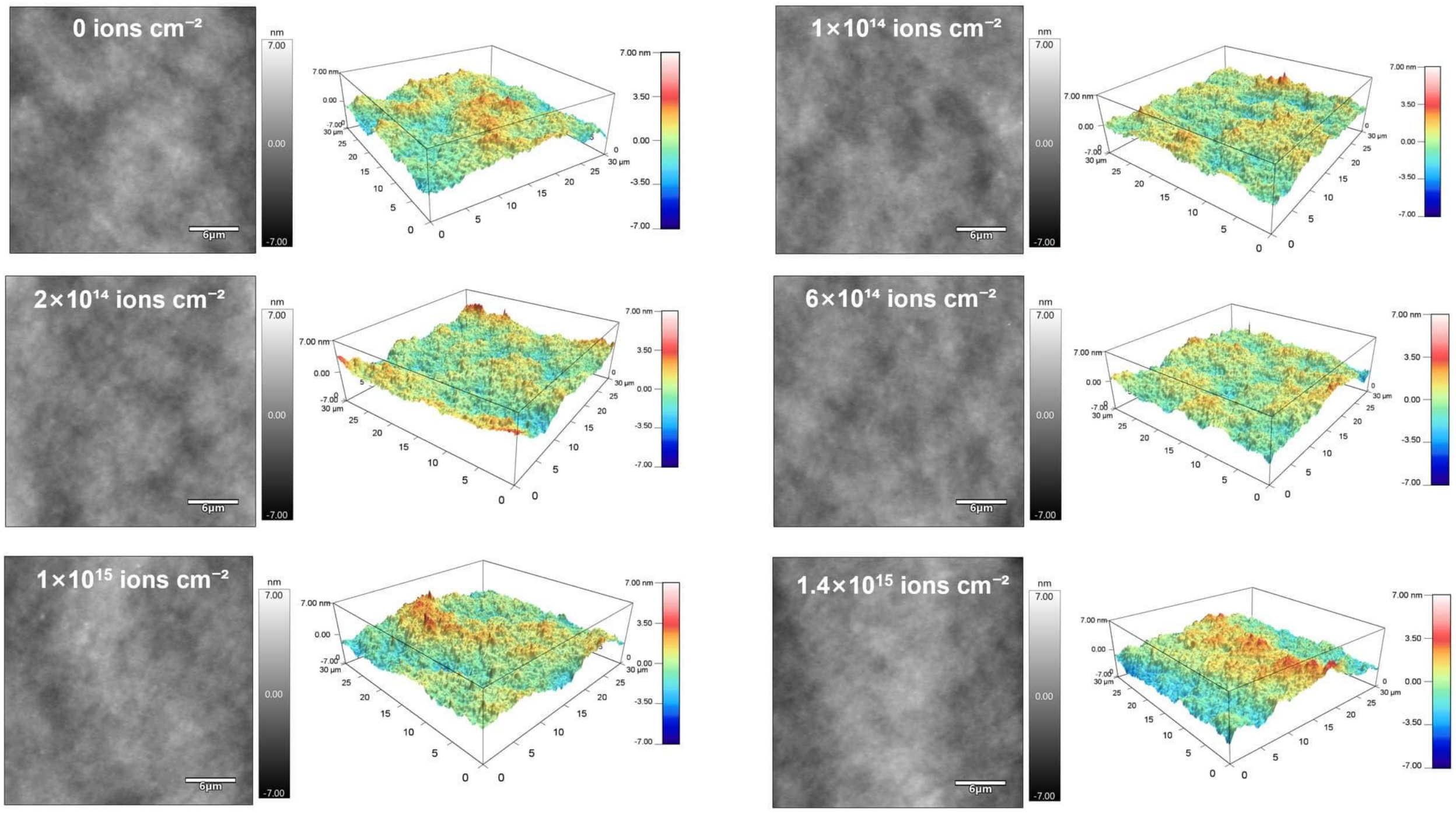

**Supplementary Fig. 89 |** Surface morphology of 2D HOMN-L4 films after proton irradiation.

**Mechanical-property analysis**

Mechanical testing revealed a pronounced reduction in both stiffness and strength after proton exposure. The average elastic modulus decreased from approximately 69.8 to 53.5 GPa. The median breaking strength decreased from approximately 3.9 to 2.4 GPa.

The non-monotonic changes across exposure conditions suggest competition between radiation-induced processes rather than a single degradation pathway. Radical recombination and intermolecular covalent coupling may partially preserve mechanical integrity at intermediate exposure. Radiation-induced densification could further increase the apparent modulus. By contrast, chain scission and post-irradiation oxidation would reduce the effective load-bearing connectivity of the network and decrease the breaking strength.

Previous proton-irradiation studies report both chain scission and crosslinking in polycarbonate and PS/LDPE, whereas 1 MeV irradiation of PVDC produces $H_2$ and chlorine-containing volatile species[64–66]. These studies make competing scission and crosslinking plausible in 2D HOMN-L4. The present AFM and mechanical measurements cannot identify the radiation-induced chemical products or distinguish the contributions of these processes. The retained mechanical response may reflect load transfer through surviving network connections.

Because the film thickness is only approximately 32 nm, it is much smaller than the penetration range of 1 MeV protons in organic materials. The primary energy deposition should therefore be approximately uniform across the active layer rather than concentrated at a Bragg peak. Nevertheless, secondary electrons and interface-mediated energy transport may remain important. Substrate-dependent damage has been observed in polymer films thinner than approximately 40 nm, indicating that the measured response may combine with substrate constraint and interfacial radical chemistry.

## Electron irradiation stability

## Electron Irradiation Stability Characterization

Continuous electron irradiation tests on 2D polymer membranes were performed on a JEOL CryoARM 300 cryo-electron microscope operating at 300 kV, equipped with a Gatan K3 direct electron detector and controlled by SerialEM (version 4.1.17). Typical imaging parameters were: magnification of 100,000×, pixel spacing of 0.4755 Å/pixel, exposure time of ~1 s (0.9987 s), dose rate of ~25.09 $e^{-}$/pixel/s (corresponding to ~111 $e^{-}/Å^2/s$), and a total accumulated dose of ~111 $e^{-}/Å^2$ per exposure. Additional settings included spot size 1, defocus of ~−0.22 μm, and a tilt angle of 0°(normal incidence). Sequential 1-second exposures were acquired to track the surface morphological evolution of the 2D HOMN-L4 film under continuous electron irradiation for radiation stability evaluation.

No discernible morphological changes were observed during irradiation. The films remained continuous and intact, without visible cracks, holes, wrinkles, or aggregation. The FFT patterns also remained essentially unchanged, with no new diffraction features or appreciable changes in the characteristic scattering distribution. These results demonstrate the excellent electron-irradiation stability of the 2D HOMN-L4 films within the investigated dose range.

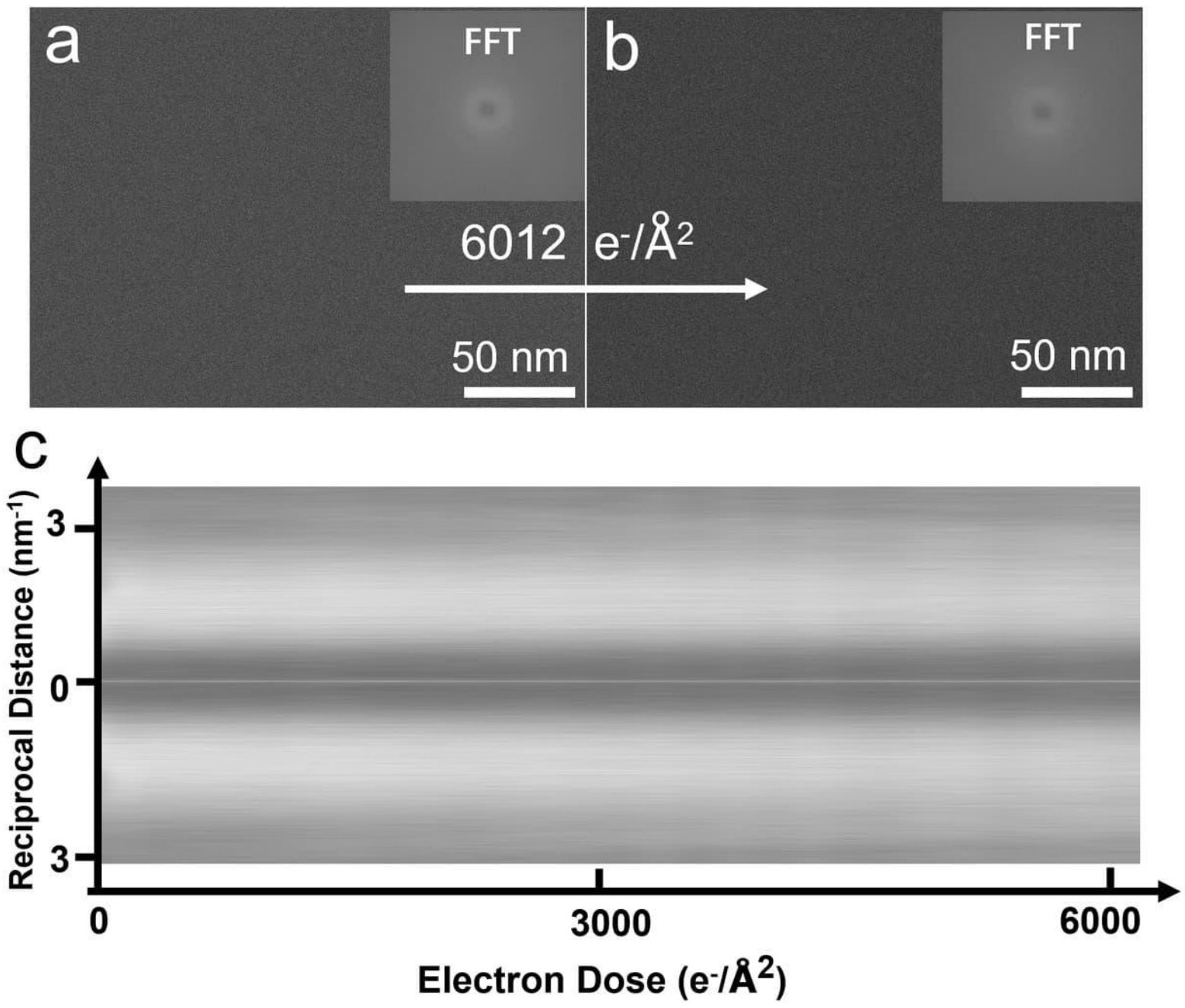

**Supplementary Fig. 90 |** TEM images of a 68-nm-thick 2D HOMN-L4 film before **(a)** and after **(b)** irradiation with 300-keV electrons at a fluence of 6,012 $e^{-}$ $Å^{-2}$. Insets, corresponding fast Fourier transform (FFT) patterns. **c,** Dose-resolved FFT evolution from 0 to 6,012 $e^{-}$ $Å^{-2}$.

## Thermal conductivity measured by FDTR

FDTR Thermal Conductivity Measurement Method

The cross-plane thermal conductivity of the samples was measured using frequency-domain thermoreflectance (FDTR) at room temperature. FDTR is a non-contact optical pump–probe technique that determines thermal properties by measuring the frequency-dependent phase lag between the modulated pump signal and the corresponding thermoreflectance response. The measured phase lag is governed by the thermal properties of the multilayer structure and is analyzed using a thermal diffusion model[67–69].

The optical configuration and measurement principle of the FDTR system are illustrated in Supplementary Fig. 91. A 488 nm continuous-wave pump laser was modulated over frequencies ranging from 10 kHz to 10 MHz to generate periodic heating at the sample surface, while a coaxially aligned 532 nm continuous-wave probe laser monitored the resulting surface temperature oscillation via the thermoreflectance response of the transducer. The reflected probe signal was collected by a photodetector, and the phase lag between the pump modulation and probe response was measured using a lock-in amplifier.

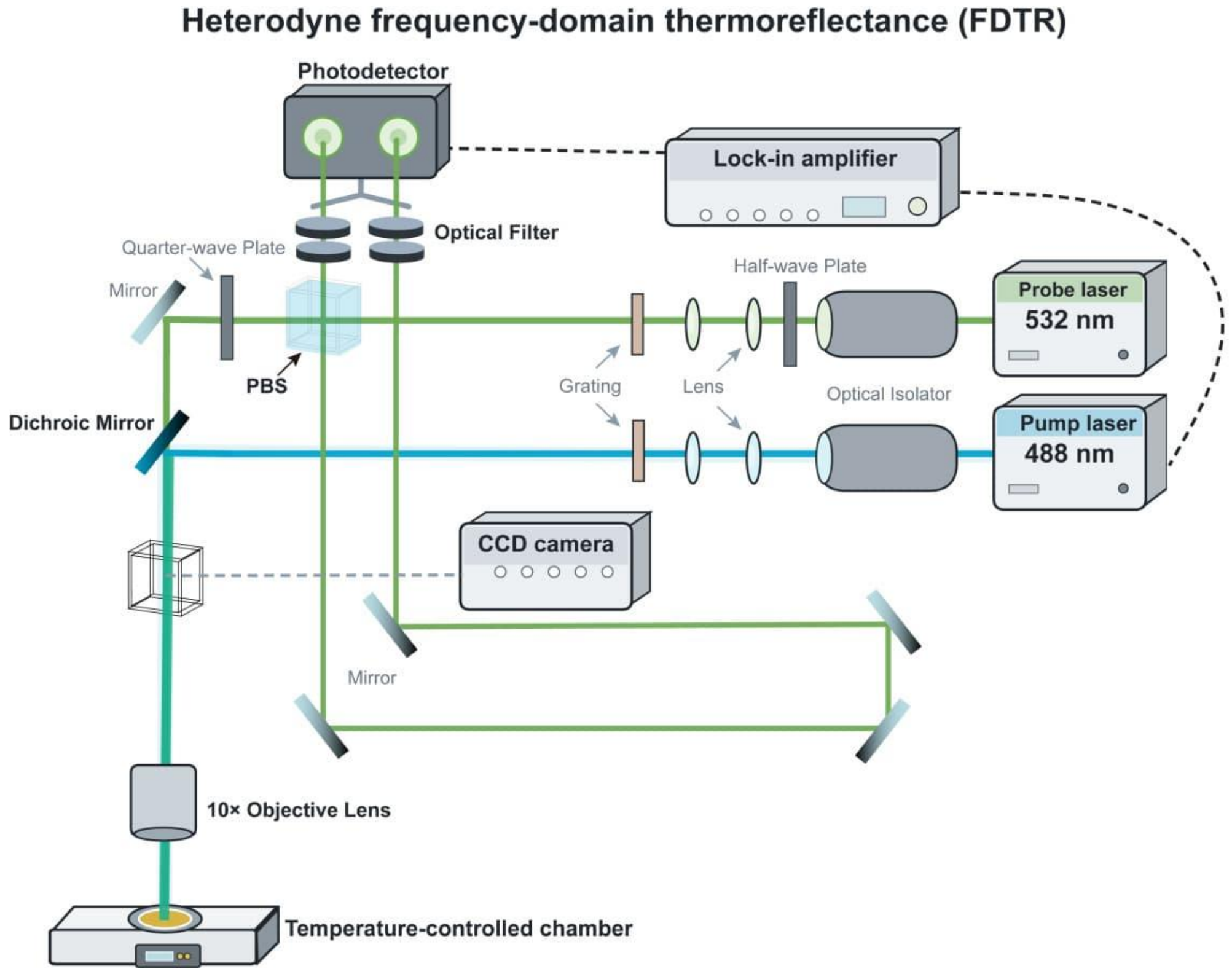

**Supplementary Fig. 91 |** Optical configuration of the frequency-domain thermoreflectance measurement.

Before FDTR measurements, an 100 nm thick Au transducer layer was deposited onto the sample surface by magnetron sputtering. The Au layer served as both the optical absorber for pump excitation and the thermoreflectance transducer for detecting surface temperature variations. The resulting measurement structure was modeled as an Au/polymer film/Si substrate multilayer stack (Supplementary Fig. 92).

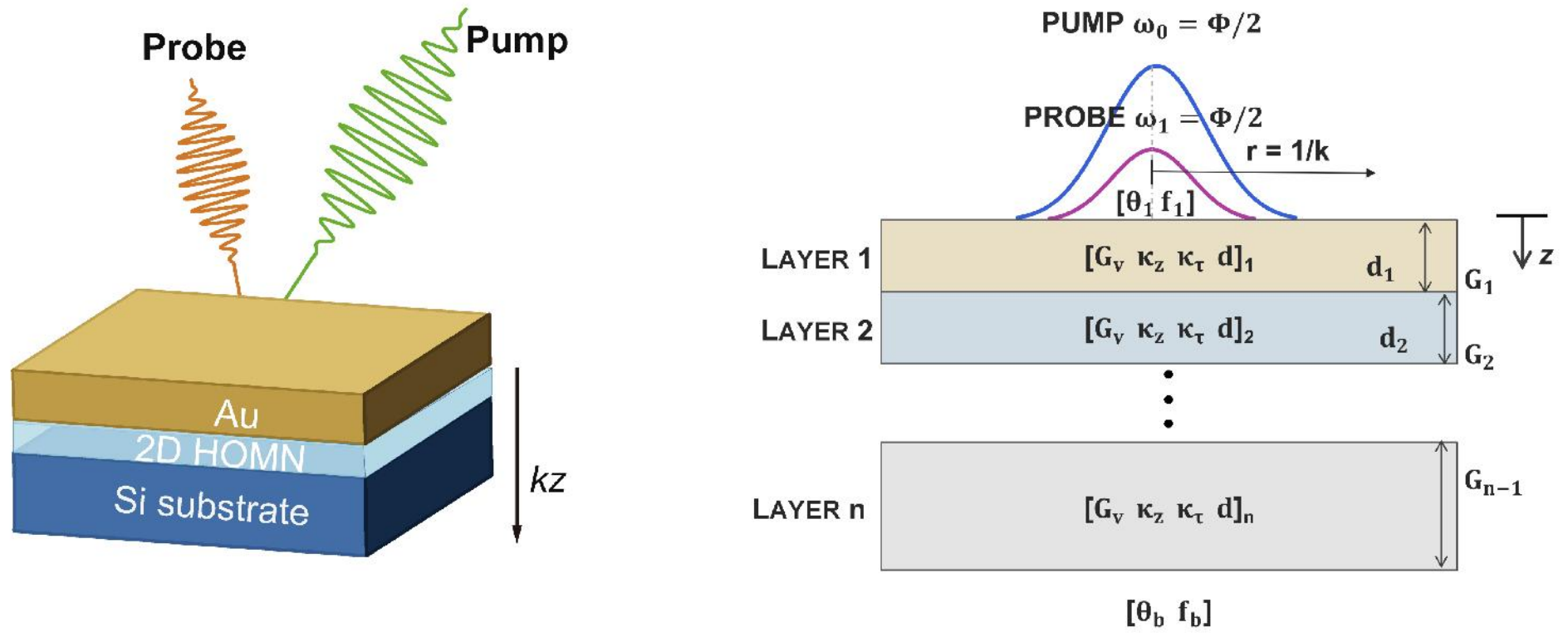

**Supplementary Fig. 92 |** A modulated pump beam produces a periodic heat flux, in the Au transducer deposited on the 2D HOMN-L4 film. A coaxially focused probe beam detects the corresponding surface-temperature oscillation, through the thermoreflectance response of Au. The measured frequency-dependent phase response is fitted using a three-dimensional multilayer heat-diffusion model to extract the effective out-of-plane thermal conductivity of the 2D HOMN-L4 film.

The measured frequency-dependent phase responses were fitted using a multilayer thermal diffusion model that incorporates the thermal conductivity ($\kappa$), volumetric heat capacity ($C_v$), thickness ($d$), and interfacial thermal conductance ($G$) of each layer. The corresponding input parameters used for the FDTR thermal model are summarized in Supplementary Table 9. The thickness of the Au transducer layer was controlled during deposition and independently characterized by stylus profilometer, while the sample thicknesses were determined from AFM. The initial guess $C_v$ of the polymer films were obtained from powder differential scanning calorimetry (DSC) measurements.The effective Gaussian laser spot radius, defined at the $1/e^2$ intensity level, was independently determined through beam-offset measurements and fixed at 9.14 μm in the multilayer thermal diffusion model.

**Supplementary Table 9 |** Input parameters used for the FDTR thermal model. Parameters shown in bold italics represent initial values used for fitting and were optimized during the fitting process. Other parameters were fixed based on independent measurements or literature values. The interfacial thermal conductance ($G$) between Au/polymer and polymer/Si interfaces was fixed as 50 MW $m^{-2}$ $K^{-1}$, and the effective beam radius was measured to be 9.14 μm.

| | $C_v$ (MJ $m^{-3}$ $K^{-1}$) | $\kappa$ (W $m^{-1}$ $K^{-1}$) | $d$ (nm) |
|---|---|---|---|
| Au | 2.48 | 180 | 100 |
| 2DPA-L4/2D HOMN-L4 | ***1.47**** | ***0.2*** | 390/220 |
| Silicon | 1.64 | 145 | Bulk |

*The volumetric heat capacity of the polymer films was measured by powder DSC.

The sensitivity analysis was performed to evaluate the reliability of parameter extraction and determine the fitting strategy for the FDTR thermal model[67,70]. As shown in Supplementary Fig. 93, the phase sensitivities of the fitting parameters exhibit distinct frequency-dependent characteristics. The sensitivity of the *G* is negligible over the measured frequency range, indicating that variations in the assumed *G* value within a reasonable range have minimal influence on the extracted thermal conductivity. Therefore, *G* was fixed during the fitting process based on widely accepted values 50 MW $m^{-2}$ $K^{-1}$ reported in previous studies[71–73]. In contrast, the sensitivities of the $C_v$ and $\kappa$ exhibit different frequency-dependent behaviors with opposite trends. Such distinct sensitivity characteristics reduce parameter correlation and enable the independent extraction of $C_v$ and $\kappa$ through the fitting procedure[68,74].

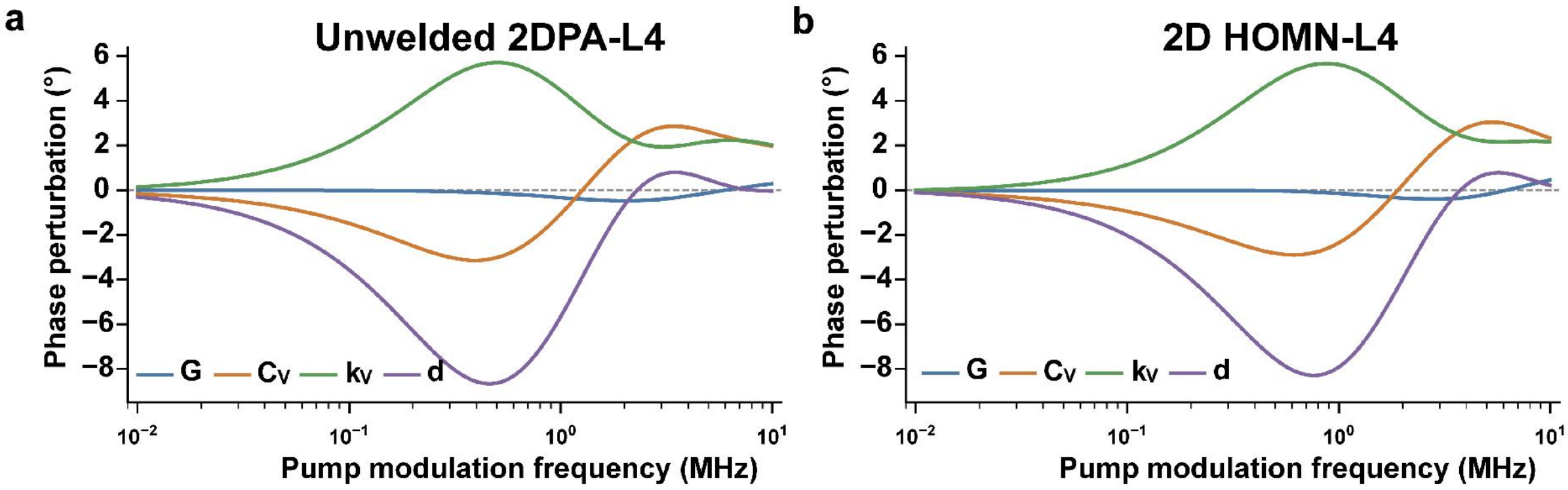

**Supplementary Fig. 93 |** Parameter sensitivity of the FDTR measurements. a,b, Frequency-dependent phase responses produced by perturbing the interfacial thermal conductance *G* of Au/sample, volumetric heat capacity $C_v$, film thermal conductivity $\kappa$ and film thickness *d* for the unwelded 2DPA-L4 films and 2D HOMN-L4 films, respectively.

During fitting, the parameters with independently determined values were kept constant, whereas the unknown thermal properties of interest, $C_v$ and $\kappa$, were optimized by minimizing the difference between the experimental phase response and the calculated phase response from the multilayer thermal diffusion model. As shown in Supplementary Fig. 94a, the measured phase lag, averaged from ten independent measurements to reduce error, exhibits a continuous frequency dependence, and the experimental data agree well with the calculated phase responses from the multilayer thermal model. The single point frequency-sweep measurement obtained the $\kappa$ is 0.24±0.03 and 0.39±0.05 W $m^{-1}$ $K^{-1}$ for unwelded 2DPA-L4 and 2D HOMN-L4 respectively and the $C_v$ is 1.59±0.17 and 1.62±0.14 MJ $m^{-3}$ $K^{-1}$ respectively.

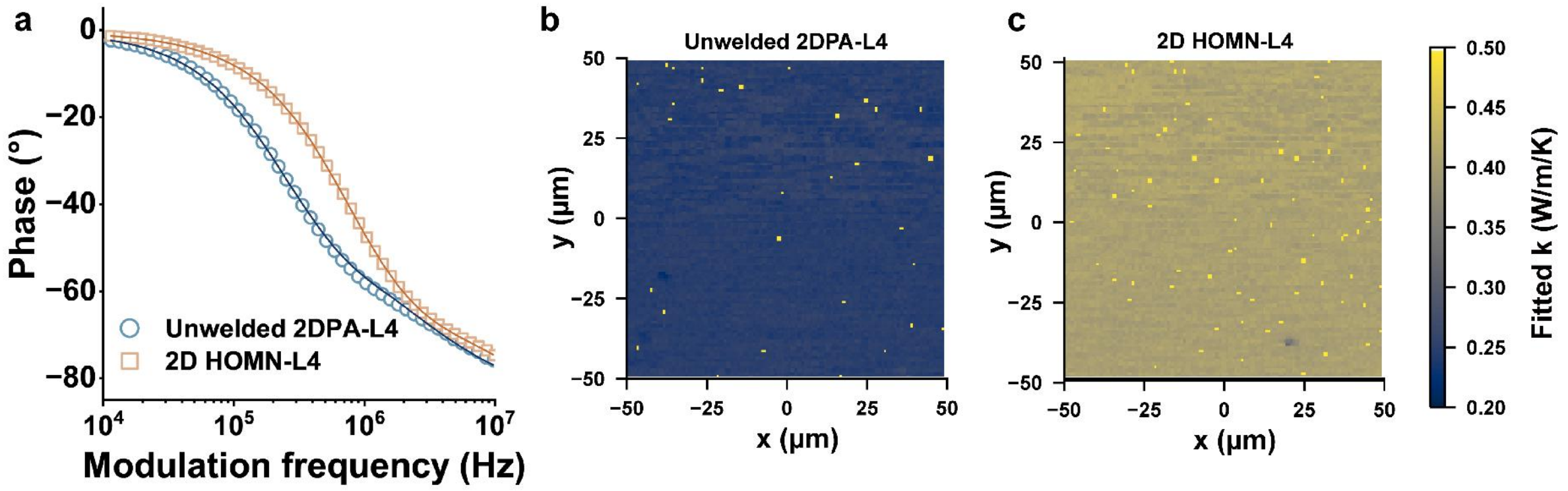

**Supplementary Fig. 94 | FDTR characterization of the unwelded 2DPA-L4 films and 2D HOMN-L4 films.** a, Representative phase lag as a function of modulation frequency. Symbols denote experimental data and solid lines denote the corresponding multilayer heat-diffusion fits. b,c, Spatial maps of the fitted effective out-of-plane thermal conductivity for the unwelded 2DPA-L4 films (b) and 2DHOMN-L4 films (c), displayed using the same color scale.

To further evaluate the spatial uniformity of the films and obtain statistically representative out-of-plane thermal properties, two-dimensional FDTR mapping measurements were performed over an area of 100×100 $\mu m^2$ with a spatial step size of 1 μm. As shown in Supplementary Fig. 94b,c, the extracted thermal conductivity distributions exhibit relatively uniform spatial profiles across the measured regions. The averaged out-of-plane thermal conductivities obtained from the FDTR mapping are 0.25±0.005 and 0.41±0.008 W $m^{-1}$ $K^{-1}$ for unwelded 2DPA-L4 and 2D HOMN-L4, respectively, with corresponding fitted $C_v$ of 1.61±0.06 and 1.64±0.06 MJ $m^{-3}$ $K^{-1}$. The thermal conductivity distributions obtained from the mapping measurements were further analyzed using Gaussian fitting. The narrow distributions and small spatial variations indicate good thermal property uniformity of the films, demonstrating the high reproducibility and reliability of the sample preparation process.

**Thermal conductivity measured by TDTR**

TDTR Thermal Conductivity Measurement Method

Time-domain thermoreflectance (TDTR) measures thermal transport by monitoring the optical response of a sample to pulsed laser heating. The technique has been used to characterize materials with a wide range of thermal conductivity[75–82]. In this work, TDTR was employed to measure the out-of-plane thermal conductivity and interfacial thermal conductance of the unwelded 2DPA-L4 films and 2D HOMN-L4 films at 296.15 K. Each film was coated with an Al film that absorbed the pump energy and acted as a thermoreflectance transducer. For small temperature changes, the change in Al reflectance is approximately proportional to its temperature change, allowing the probe beam to monitor the thermal response.

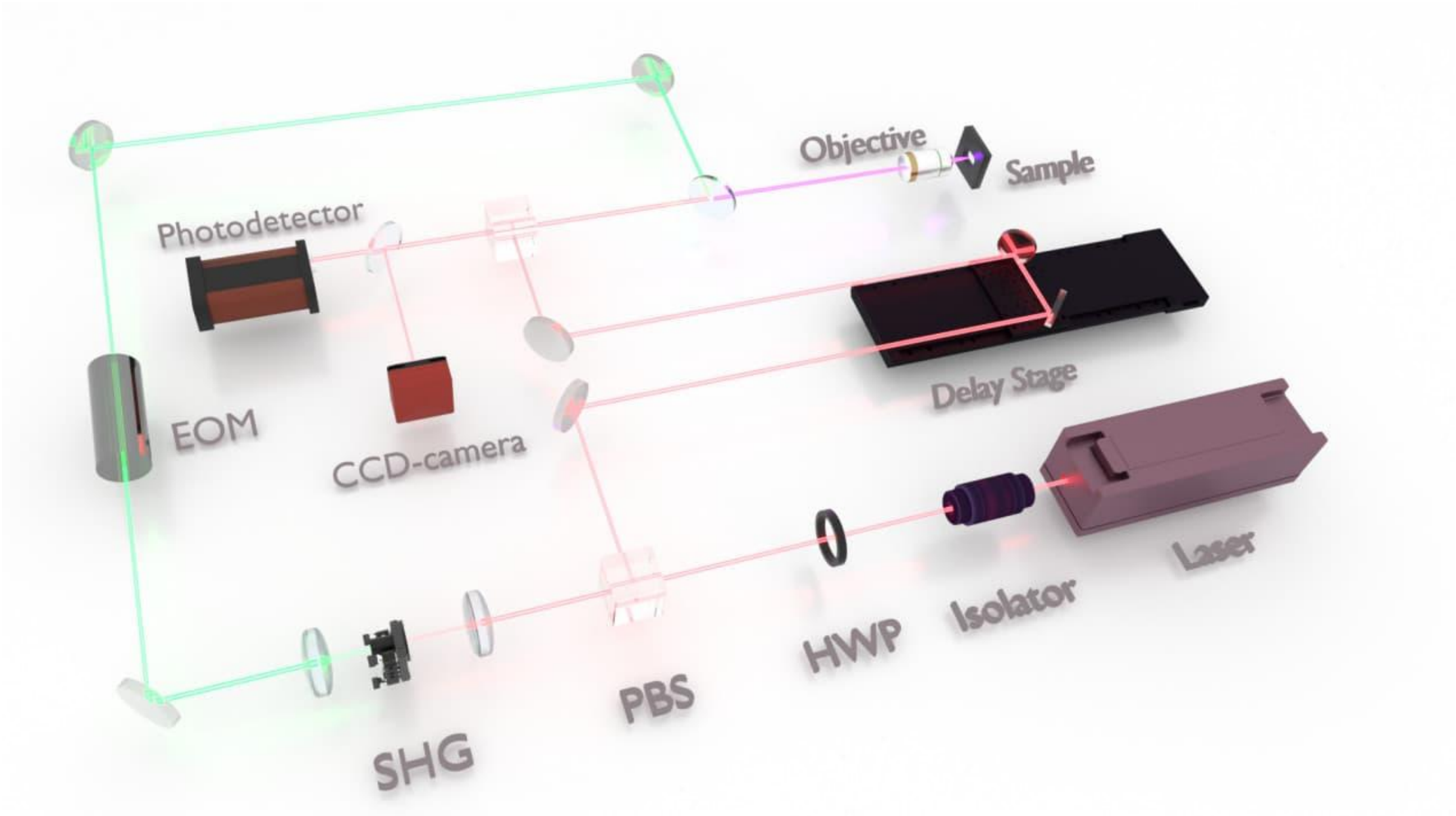

**Supplementary Fig. 95 |** Overview of the ultrafast pump-probe setup based on time-domain thermoreflectance. PBS, polarizing beam splitter; HWP, half-wave plate; CCD camera, charge-coupled device camera; SHG, second-harmonic generator; EOM, electro-optic modulator.

The experimental setup is shown in Supplementary Fig. 95. In our TDTR setup, a fiber-based femtosecond laser (ALCOR 1064-2, Spark Lasers) operating at a central wavelength of 1064 nm, with a nominal repetition rate of approximately 80 MHz and a specified pulse duration below 100 fs. The laser output was divided into pump and probe beams. The pump beam was first frequency-doubled to 532 nm through second-harmonic generation in a β-barium borate (BBO) crystal and subsequently modulated at 4 MHz using an electro-optic modulator (EOM). After being focused onto the Al-coated sample surface, the pump pulses were absorbed by the transducer and produced a rapid temperature rise. The subsequent thermal response was monitored through the temperature-dependent reflectance of Al using the probe beam. A mechanical delay stage controlled the relative arrival time of the pump and probe pulses, allowing the response to be recorded as a function of delay time. The reflected probe was detected by a photodetector, and a lock-in amplifier referenced to the pump modulation measured the in-phase and out-of-phase components. Following instrumental phase calibration, the phase-delay curves were fitted to a multilayer heat-diffusion model to determine the out-of-plane thermal conductivity and interfacial thermal conductance.

The thermal response of the sample was calculated using a multilayer heat-diffusion model[83,84] that accounts for periodic laser heating and heat transport across the layers and interfaces. The modulated component of the pump power can be expressed in the frequency domain as $Q(\omega)$:

$$Q(\omega) = \frac{2\pi Q_m}{T} \sum_{n=-\infty}^{\infty} \delta(\omega - \omega_m - k\omega_s) e^{-in\omega_m T_0} \tag{22}$$

where $Q_m$ is the power of each pump pulse, $T$ is the interval between each pulse, $\omega_m$ is the angular modulation frequency, $\omega_s$ is the laser angular repetition frequency, $T_0$ is the arbitrary delay time between the pump and probe pulses. The symbol $\delta$ denotes the Dirac delta function. For each layer, the temperature response and heat flux at the upper and lower surfaces were related using a transfer matrix. The relations for the material layers and interfaces are given in Supplementary Equations 23 and 24:

$$\begin{bmatrix} \theta_b \\ f_b \end{bmatrix} = \begin{bmatrix} \cosh(qd) & \frac{-1}{\sigma_z q}\sinh(qd) \\ -\sigma_z q \sinh(qd) & \cosh(qd) \end{bmatrix} \begin{bmatrix} \theta_t \\ f_t \end{bmatrix} \tag{23}$$

$$\begin{bmatrix} \theta_b \\ f_b \end{bmatrix} = \begin{bmatrix} 1 & G^{-1} \\ 0 & 1 \end{bmatrix} \begin{bmatrix} \theta_t \\ f_t \end{bmatrix} \tag{24}$$

where $\theta_b$ and $\theta_t$ denote the temperatures at the top and bottom surfaces of a layer, $f_b$ and $f_t$ represent the corresponding heat fluxes. The out-of-plane thermal conductivity and layer thickness are denoted by $\sigma_z$ and $d$. For 3D heat conduction, $q^2 = (\xi_x \sigma^2 + \xi_y \sigma^2 + i\omega C_v)/\sigma$, where $\xi_x$ and $\xi_y$ are the spatial angular frequencies associated with the Fourier transforms in the two in-plane directions. The volumetric heat capacity is denoted by $C_v$. The temperature and heat flux are related through the following matrix equation in the multilayer system:

$$\begin{bmatrix} \theta_b \\ f_b \end{bmatrix} = M_n M_{n-1} \ldots M_1 = \begin{bmatrix} A & B \\ C & D \end{bmatrix} \begin{bmatrix} \theta_t \\ f_t \end{bmatrix} \tag{25}$$

The frequency-domain response function of the whole system can be written as:

$$H(\omega)=\frac{A_m A_s}{4\pi^2}\iint_{-\infty}^{\infty}\left(-\frac{D}{C}\right)\exp\left(-\frac{\left(w_{m,x}{}^2+w_{s,x}{}^2\right)\xi_x{}^2+\left(w_{m,y}{}^2+w_{s,y}{}^2\right)\xi_y{}^2}{8}\right)d\xi_x d\xi_y \tag{26}$$

where $w_{0,x}$ and $w_{s,x}$ are the $1/e^2$ intensity radii of the pump and probe beams along the x direction. $A_m$ is the power absorbed, and $A_s$ is the reflected probe power. The temperature response to pump excitation is given by $\Theta(\omega)=H(\omega)Q(\omega)$. The delayed probe pulses sample the temperature-dependent change in surface reflectance. Including the laser repetition harmonics, the transfer function demodulated at $\omega_m$ is:

$$Z(\tau)=K\sum_{n=-\infty}^{\infty}H(\omega_{\mathrm{m}}+n\omega_s)e^{in\omega_s\tau} \tag{27}$$

where $\tau$ is the pump–probe delay, positive when the probe arrives after the pump. The delay-independent factor $K$ contains the pulse energies, thermoreflectance coefficient, and detection gain. The lock-in amplifier measures the in-phase and out-of-phase components of the complex signal, defined as $X=\mathrm{Re}\{Z(\omega_0)\}$ and $Y=\mathrm{Im}\{Z(\omega_0)\}$, respectively.

Before the TDTR measurements, an Al film was deposited onto each sample surface by electron-beam evaporation using a Lesker LAB18 system. The Al layer served as both the absorber for the pump pulses and the thermoreflectance transducer for probe detection. Its thickness was independently determined from the acoustic echo measured by picosecond ultrasonics[75,85]. The electrical conductivity of the Al film was measured using a Seebeck coefficient measurement system (SBA 458 Nemesis, NETZSCH). The thermal conductivity was then estimated using the Wiedemann–Franz law, $\kappa_{\mathrm{Al}}=L\sigma T$, giving a value of 150 W $m^{-1}$ $K^{-1}$, where $L$ is the Lorenz number, $\sigma$ is the electrical conductivity, and $T$ is the temperature. The spatial overlap of the pump and probe beams was characterized using the beam-offset method[86]. Gaussian fits to the beam-offset profiles yielded a convoluted $1/e^2$ diameter of 4.64 μm. Supplementary Fig. 96a shows the profiles measured along the *x* and *y* directions together with their Gaussian fits. The similar fitted diameters along the two directions indicate a nearly circular pump–probe overlap profile.

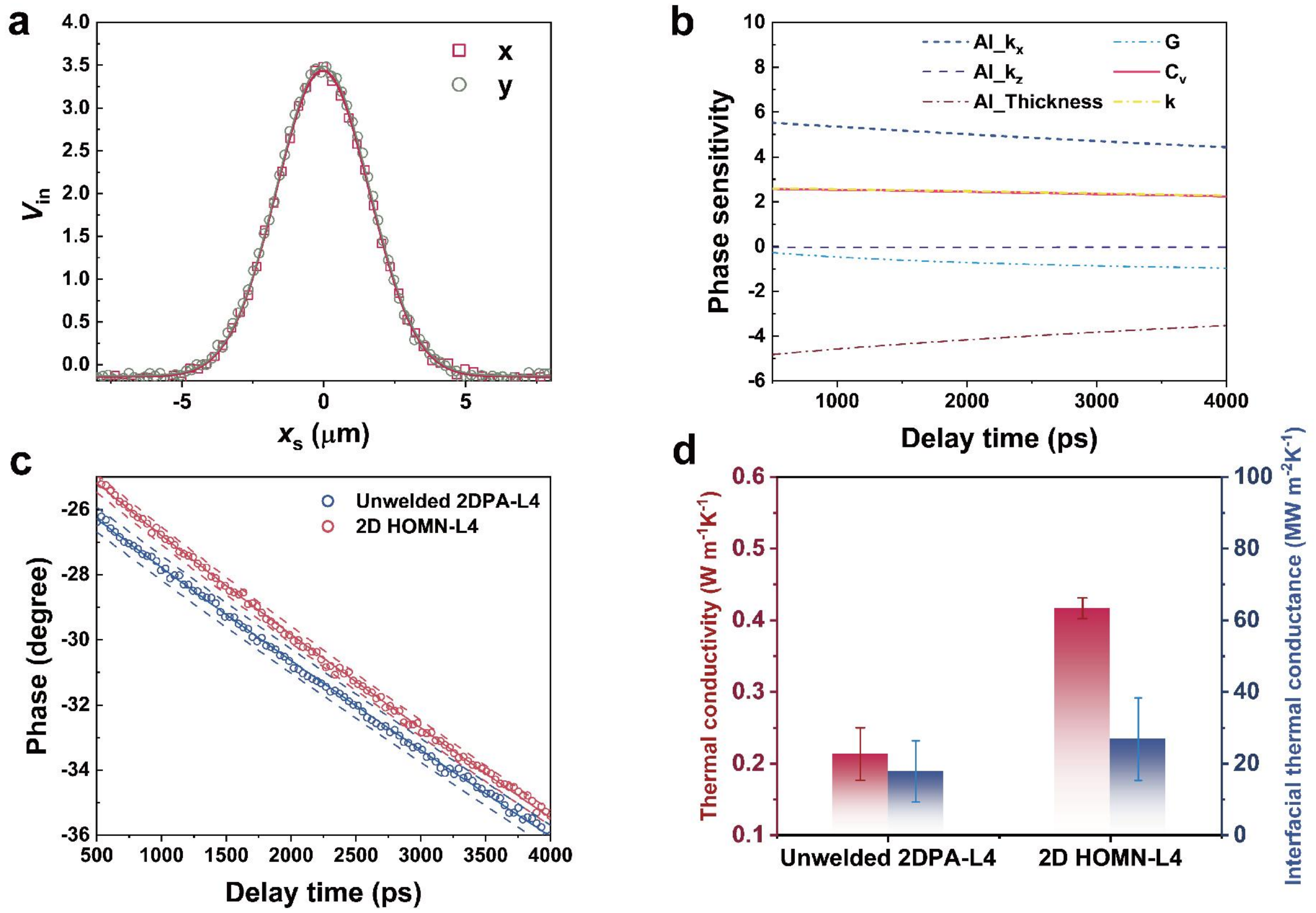

**Supplementary Fig. 96 |** (a) Beam-offset signal $V_{\text{in}}$ as a function of displacement (μm) along the *x* (squares) and *y* (circles) directions. (b) Time-dependent phase responses produced by perturbing the thermal boundary conductance, film volumetric heat capacity, and film thermal conductivity, respectively. (c) Measured thermoreflectance phase signal as a function of delay time (circles) and the best fit to the thermal transport model (solid line). Dashed lines show model predictions for thermal conductivities ±10% from the best-fit value. (d) Comparison of out-of-plane thermal conductivity and interfacial thermal conductance between the unwelded 2DPA-L4 and 2D HOMN-L4 films.

Under the present measurement conditions, the sensitivity curves for thermal conductivity and volumetric heat capacity show similar trends, making it difficult to fit both parameters simultaneously, as shown in Supplementary Fig. 96b. Therefore, the volumetric heat capacities of the unwelded 2DPA-L4 and 2D HOMN-L4 films were determined from FDTR measurements to be 1.61 and 1.64 MJ $m^{-3}$ $K^{-1}$, respectively. Supplementary Fig. 96cshows the measured phase as a function of pump–probe delay and the corresponding model fits at a modulation frequency of 4 MHz. The out-of-plane thermal conductivity and interfacial thermal conductance were extracted by fitting the phase data to the three-dimensional multilayer heat-diffusion model described above. The unwelded 2DPA and 2D HOMN-L4 films had thicknesses of 1.17 and 1.36 μm, respectively. Both thicknesses substantially far exceeded the corresponding out-of-plane thermal penetration depths at 4 MHz, estimated to be approximately 0.103 and 0.142 μm. Measurements were performed at five randomly selected locations on each sample, and the results are shown in Supplementary Fig. 96d. The mean out-of-plane thermal conductivity of 2D HOMN-L4 was 0.4168±0.0143 $Wm^{-1}K^{-1}$, approximately 95.31% higher than that of unwelded 2DPA-L4, 0.2134±0.0367 $Wm^{-1}K^{-1}$. The mean interfacial thermal conductance *G* was 26.81±11.51 MW $m^{-2}$ $K^{-1}$ for 2D HOMN-L4 and 17.85±8.54 MW $m^{-2}$ $K^{-1}$ for unwelded 2DPA-L4.

## Supplementary Video

**Supplementary Video 1 |** Macroscopic load-bearing behaviour of 78 nm thick free-standing 2D HOMN-L8 nanofilm

A 6 cm diameter suspended nanofilm is loaded sequentially with 17 individual 0.3 g metal spheres. No rupture or catastrophic crack expansion is observed throughout full loading, demonstrating outstanding centimeter-scale mechanical integrity of the covalent welded network.

**Supplementary Video 2 |** Macroscopic load-bearing behaviour of 96 nm thick free-standing 2D HOMN-L4 nanofilm

A 6 cm diameter suspended nanofilm is loaded sequentially with 8 individual 0.15 g metal spheres. No rupture or catastrophic crack expansion is observed throughout full loading, demonstrating outstanding centimeter-scale mechanical integrity of the covalent welded network.

**Supplementary Video 3 |** Tear resistance demonstration of damaged 2D HOMN-L4

Despite local penetration by the spheres, the film continued to bear load while the damage remained localized. Video recordings showed no appreciable crack propagation under sustained loading, consistent with covalent welding preserving residual load-bearing capacity after localized structural damage.

**Supplementary References**